%% file: master_thesis_michael.tex
\documentclass[11pt,a4paper,twoside,english]{memoir}
\usepackage[bookmarks, pagebackref=false]{hyperref}
\usepackage{amsmath}
\usepackage{amssymb}
\usepackage{amsbsy}
\usepackage[utf8]{inputenc} 
\usepackage[english,danish]{babel}
\usepackage{listings}
\pdfoutput=1
\usepackage[pdftex]{epsfig,graphicx}
\usepackage{epstopdf}
\graphicspath{{./figures/clusters/}{./figures/data/}{./figures/method/}{./figures/results/ampl/}{./figures/results/cosmo/}{./figures/results/degen/best/}{./figures/results/degen/cosmo/}{./figures/results/degen/grillo/}{./figures/results/maps/}{./figures/results/mass/}{./figures/results/multimg/}{./figures/results/syst/}{./figures/theory/}}
\usepackage{amsthm, gensymb, empheq}
\usepackage{url, adjustbox, float}
\usepackage[usenames,dvipsnames]{xcolor}
\definecolor{orange}{cmyk}{0,0.5,1,0}
\definecolor{rossoCP3}{cmyk}{0,.88,.77,.40}
\definecolor{moerkeroed}{cmyk}{0,.88,.77,.40}
\definecolor{graa}{rgb}{0.8,0.8,0.8}
\definecolor{moerkegraa}{cmyk}{0.67,0.58,0.54,0.09}
\definecolor{blaa}{rgb}{0.2,0.2,0.6}
\definecolor{scienceGroen}{RGB}{70,116,60}
\hypersetup{
      	colorlinks, 
	bookmarksopen, 
	bookmarksnumbered,
	citecolor=blaa, 		
	linkcolor=moerkeroed,	                
	urlcolor=blaa,			
	}
\usepackage{bm}
\usepackage{bbm}
\usepackage{pxfonts}

\usepackage{placeins}
\usepackage{xspace}
\usepackage{cancel} 

\usepackage[pass]{geometry}
\usepackage{youngtab}

\usepackage{slashed}
\usepackage[caption=false, labelformat=simple, listofformat=subsimple, labelfont=default, margin=5pt, justification=raggedright]{subfig}

	\makeatletter
		\renewcommand{\p@subfigure}{}
	\makeatother

\usepackage[style=alphabetic,natbib=true,backend=bibtex,maxnames=10,maxcitenames=4,backref=false]{biblatex} 

\bibliography{master-litt} 

\usepackage{longtable}			

\newcommand{\HRule}{{\color{moerkegraa} \rule{\linewidth}{0.5mm}}} 

\allowdisplaybreaks[1]

\renewcommand{\vec}[1]{\boldsymbol{#1}} 

\newcommand{\beq}{\begin{eqnarray}}
\newcommand{\eeq}{\end{eqnarray}}

\newcommand{\bmp}{\noindent\begin{minipage}{16cm}}
\newcommand{\emp}{\end{minipage}\vskip 7mm} 

\newcommand{\gradient}{\nabla}
\newcommand{\XI}{\vec{\xi}}
\newcommand{\scalpha}{\vec{\alpha}}
\newcommand{\BETA}{\vec{\beta}}
\newcommand{\THETA}{\vec{\theta}}
\newcommand{\ALPHA}{\hat{\vec{\alpha}}}
\newcommand{\ETA}{\vec{\eta}}
\newcommand{\mtrx}{\mathcal}
\newcommand{\REAL}{\mathbb{R}}
\newcommand{\map}{\mapsto}

\DeclareMathOperator{\tr}{Tr}

\usepackage{hyphenat}

\def\lsim{\mathrel{\rlap{\lower4pt\hbox{\hskip1pt$\sim$}}
    \raise1pt\hbox{$<$}}}                
\def\gsim{\mathrel{\rlap{\lower4pt\hbox{\hskip1pt$\sim$}}
    \raise1pt\hbox{$>$}}}                

\usepackage{fancyhdr}
\usepackage{pdfpages}

\include{chapterCP3} 

\newcommand{\nnchapter}[1]{\chapter*{#1}\fancyhead[LO]{\slshape \MakeUppercase{#1}} \fancyhead[RE]{\slshape \MakeUppercase{#1}}\addcontentsline{toc}{chapter}{#1}}
\newcommand{\nnchapters}[1]{\chapter*{#1}\fancyhead[LO]{\slshape \MakeUppercase{#1}} \fancyhead[RE]{\slshape \MakeUppercase{#1}}}

\newcommand{\thistitle}{Gravitational Lensing in Clusters}
\newcommand{\thisauthor}{Michael F. Hansen}

\title{\texorpdfstring{\Large\color{rossoCP3}\thistitle}{\thistitle}} 
\author{\thisauthor}

\newcommand{\abs}[1]{\lvert#1\rvert} 		
\newcommand{\starteqn}{\begin{equation}} 		
\newcommand{\sluteqn}{\end{equation}} 		
\newcommand{\lrarrow}{\Leftrightarrow} 		
\newcommand{\rarrow}{\Rightarrow} 		
\newcommand{\smrarrow}{\rightarrow}

\newcommand{\bsm}{\left[ \begin{smallmatrix}} 	
\newcommand{\esm}{\end{smallmatrix} \right]}	
\newcommand{\bpm}{\begin{pmatrix}}
\newcommand{\epm}{\end{pmatrix}}
\newcommand{\PD}{\partial} 			
\newcommand{\leftparan}{\left(} 			
\newcommand{\rightparan}{\right)} 			
\newcommand{\Sun}{\ensuremath{\odot}} 		
\newcommand{\MSun}{M_{\Sun}} 
\newcommand{\maths}{\,\mathrm}

\newcommand{\onehalf}{\frac{1}{2}}

\newcommand{\angstrom}{\text{\normalfont\AA}}
\newcommand{\mosdrizzle}{\emph{MosaicDrizzle}\xspace}

\newcommand{\iraf}{\emph{IRAF}\xspace}

\newcommand{\macs}{MACS J0416.1-2403\xspace}
\newcommand{\pct}{$\%$\xspace}
\newcommand{\chandra}{\emph{Chandra X-Ray Telescope}\xspace}
\newcommand{\convergence}{\emph{convergence}\xspace}

\newcommand{\lenstool}{\emph{Lenstool}\xspace}
\newcommand{\glee}{\emph{GLEE}\xspace}
\newcommand{\bayesys}{\emph{BayeSys$^{TM}$}\xspace}

\renewcommand{\cite}{\citep}
\renewcommand{\[}{\begin{equation}}
\renewcommand{\]}{\end{equation}}

\def\arcsec{\hbox{$^{\hbox{\rlap{\hbox{\lower2pt\hbox{$\,\prime\prime$}}
          }\hbox{$\;\;\;$}}}$}}

\def\arcmin{\hbox{$^{\hbox{\rlap{\hbox{\lower2pt\hbox{$\;\prime$}}
          }\hbox{$\;\;\;$}}}$}}
          
\def\arcdotsec{\hbox{$^{\hbox{\rlap{\hbox{\lower2pt\hbox{$\,\prime\prime$}}
          }\hbox{$\;\;\;$}}}$}}
         
\def\mean#1{\left< #1 \right>}

\def\median#1{\tilde{#1} }

\begin{document}

\pagenumbering{alph}
\selectlanguage{english}

\newgeometry{margin=3.775cm}
\include{frontmatter}
\restoregeometry

\setboolean{@twoside}{false} 
\thispagestyle{empty}
\clearpage
\ \thispagestyle{empty}\clearpage
\pagenumbering{roman}

\nnchapters{Acknowledgements}
First and foremost, I would like to thank my family for their great patience during the writing of this thesis. Especially my wife, who at the first half of my thesis work, had to take care of our son alone, while pregnant, into late evening and at the second half, also take care of our newborn girl. There is no doubt in my mind that she has made it possible for me to concentrate my time and effort on this thesis. I will also thank my son for constantly forcing me to explain complicated matters in a simple language, by insisting on asking me simple, yet complicated questions.

I especially want to thank my parents. Despite their inability to help me do my homework in the final years of primary school and throughout high school, they taught me never to give up and supported me in every choice I made. I also want to thank them for creating a environment at home inspiring curiosity and knowledge. I owe everything to them.

I would like to thank my advisor for his great patience during this thesis. He has devoted a lot of time to advise and guide me. Even though it was not always pleasant to be put on the spot, reciting theory I had just read, I have no doubt that it has helped me to remember the theory better. I would also like to thank him for insisting on a strict meeting plan. It has certainly helped me to get a grasp in this vast subject and to be on top of both the modelling and writing part of this thesis.
I would also like to thank Mario Bonamigo. Although he has not formally been a part of my thesis work, he has contributed a great deal to my understanding of the results I got.

Finally I would like to thank M.E. Tallica, S. Lipknot, K. Orn and T.I. Esto for their continuous inspiration during my reading and writing sessions, as well as during the 
sometimes tedious job processing of data. It has also helped from time to time, to take long trips through the Corridors of Power, where I constantly met a guy called Johan.

%
\thispagestyle{empty}\clearpage
\ \thispagestyle{empty}\clearpage 

\include{abstract}

\thispagestyle{empty}\clearpage
\ \thispagestyle{empty}\clearpage 

\clearpage
\begin{KeepFromToc} 
  \tableofcontents
\end{KeepFromToc}
\fancyhead[LO]{\slshape \MakeUppercase{Contents}} \fancyhead[RE]{\slshape \MakeUppercase{Contents}} 

\thispagestyle{empty}\clearpage
\ \thispagestyle{empty}\clearpage 

\begin{KeepFromToc}
\listoffigures
\end{KeepFromToc}

\thispagestyle{empty}\clearpage
\ \thispagestyle{empty}\clearpage 

\begin{KeepFromToc}
\listoftables
\end{KeepFromToc}
\thispagestyle{empty}

\selectlanguage{english}

\setboolean{@twoside}{true} 
\cleardoublepage
\pagenumbering{arabic}
\include{introduction}\fancyhead[LO]{\slshape \rightmark}\fancyhead[RE]{\slshape \leftmark} 

\input{strong-lensing}

\input{clusters}

\input{data}

\input{method}

\input{results}

\input{discussion}

\appendix

\include{Appendix}

\fancyhead[LO]{\slshape \MakeUppercase{Bibliography}} \fancyhead[RE]{\slshape \MakeUppercase{Bibliography}}  
\printbibliography[title={Bibliography}]

\end{document}

%% file: frontmatter.tex
%
%
\begin{center}
\textsc{\LARGE Master Thesis}\\[0.5cm] 
\HRule \\[2mm] 
{\Huge \bfseries \textsc{Strong Lensing in Galaxy Clusters \\[1mm]
}}
%
%
\HRule \\[1.5cm] 

\makebox[\textwidth][c]{\includegraphics[width=1.3\textwidth,keepaspectratio=true]{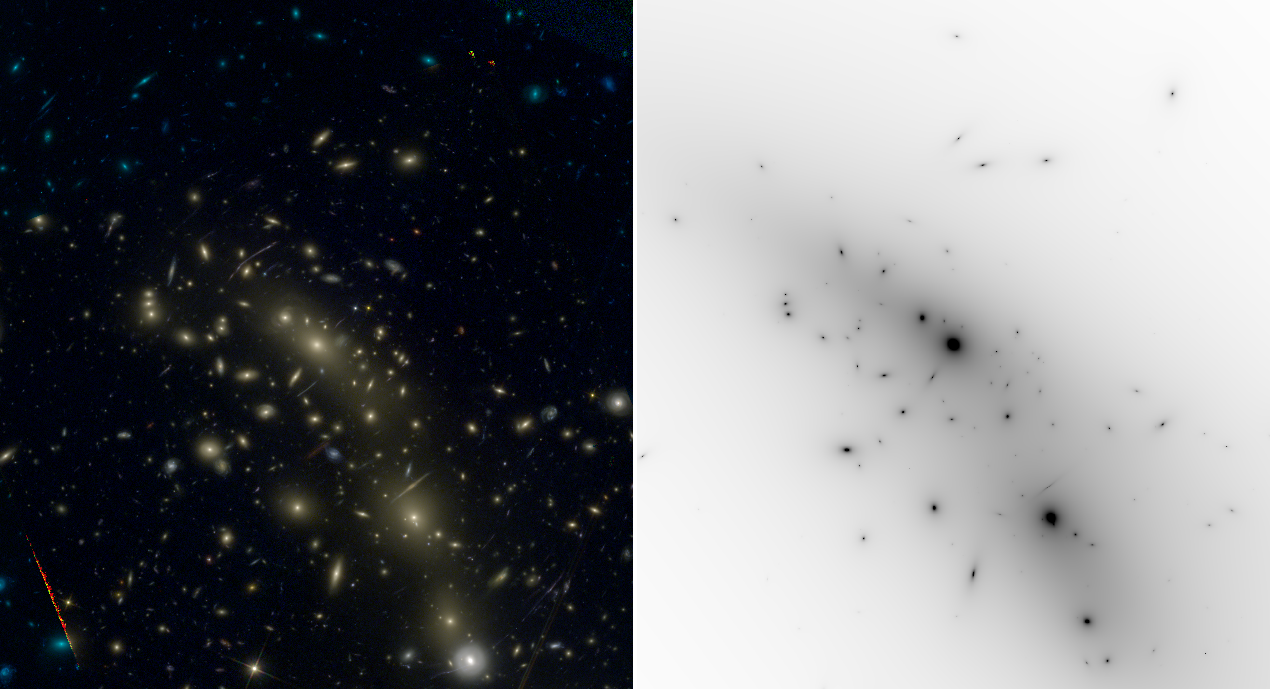}}\\[2cm]

\vfill 

\textsc{\huge Michael F. Hansen}
 
\end{center}
\thispagestyle{empty}

\clearpage
{\phantom{} \vspace{-7mm}}
\vspace{7cm}
\begin{center}
\textsc{
Somewhere, something incredible is waiting to be known.
}
\ \\
\ \\

\textsc{
- Carl Sagan, 1934-1996
}
\end{center}
\thispagestyle{empty}

\clearpage


\hfill \includegraphics[width=0.15\textwidth]{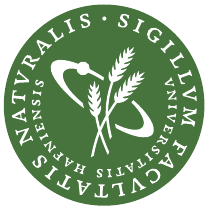} 

\begin{center}
\HRule \\[0.4cm] 
{\LARGE \bfseries 
Strong Lensing 
in Galaxy Clusters  \\[2mm]
\Large First Steps Toward a Method for Estimating Cosmological Parameters using Strong Lensing, X-ray and Dynamics Total Mass Estimates \\[3mm]} 
\HRule \\[1cm] 

\large
\begin{tabular}{>{\hspace{1.5pc}} r >{\hspace{1.3pc}} l} 
Author &  Michael Flemming Hansen \\
Advisor &  Claudio Grillo \\
\end{tabular}

\vfill

\includegraphics[width=1.0\textwidth,keepaspectratio=true]{frontimage_MACS}

\vfill

\large \textit{Thesis for the degree of Master in Physics}\\[1cm] 
{\color{moerkeroed} \Large Dark Cosmology Centre}\\[0.2cm]
Niels Bohr Institute\\[1cm] 

Submitted
 
{\large 28th February 2017} 
 
\end{center}
\thispagestyle{empty}

%% file: abstract.tex
\nnchapters{Abstract}
One of the fundamental tasks in astronomy and astrophysics remains the determination of the mass of various astrophysical objects. Of these objects, galaxy clusters are the most massive gravitational bound objects and an accurate understanding of the total mass and mass distribution in these objects, provides vital understanding of both the galaxy clusters themselves and the universe on large scales. Strong lensing has proven a valuable tool in the estimation of the total mass in galaxy clusters and a vast number of galaxy clusters has already been analysed \cite{Kneib1996,Sand2002,Sand2004, Ardis2007, Sand2008,Limousin2008,Richard2009,Zitrin2013,Richard2014,Johnson2014,Jauzac2014,Grillo2015,Jauzac2015,Caminha2016}. Of these clusters, \macs has been analysed thoroughly since its discovery in 2001 \cite{Ebeling2001}.

In our current understanding, the evolution of the universe and formation of structure can be described as a parametrization of the Big Bang model. In order to find the values of these parameters, a host of different observational method has been used: the Cosmic Microwave Background (CMB) radiation, type Ia supernova, galaxy cluster scaling relations and gravitational lensing, to name a few. This has given us strong indication that our universe can be described by a $\Lambda$ Cold Dark Matter ($\Lambda$CDM) formalism. Recently, the Planck satellite provided very detailed observations of the CMB which made a highly accurate estimate of the cosmological parameters possible \cite{Planck2015}, reinforcing this view. Still, it is an ongoing investigation, employing more detailed observations and new analysis methods. 

In this thesis we want to introduce the first steps towards realising a new method to investigate the cosmological parameters and conduct a detailed analysis of the galaxy cluster \macs. Toward this end, we use the current model from \citet{Grillo2015} as a template and the publicly available lensing code \lenstool. This code has previously been used by \citet{Jauzac2014, Richard2014, Jauzac2015, Caminha2016} to model \macs (\citet{Grillo2015} used \glee). 

We created $10$ different models to cover a reasonable set of different approaches. In addition to the replication of the \citet{Grillo2015} models, with two cluster scale halos and 175 circular cluster member mass-density profiles, we created models using elliptical mass-density profiles for the cluster members and models where we optimize the cluster member scaling relation slopes. 
In order to investigate the viability of using the projected total mass estimate from different cosmological models to estimate the cosmological parameter values, we created 49 models each representing a different set of cosmological parameters.

Like \citet{Grillo2015}, we find that a model using two PIEMD cluster scale halos and 175(+1)dPIE cluster member halos (where +1 accounts for the foreground galaxy \cite{Johnson2014}) provide the best constraints. Although we find that a model using elliptical cluster member mass-density profiles constrain the data better in terms of smaller $\chi^2$ and $\log{(E)}$, we conclude that the difference between it and the model using circular mass profiles for the cluster members, is too small to justify the extra time needed for optimization.

Our best model (2PIEMD + 175(+1)dPIE$_c$ $(M_TL^{-1} = v)$) gives $\chi^2 = 486$ which is better than the results from \citet{Grillo2015} $(\chi^2 = 915)$ and our 2PIEMD + 175(+1)dPIE$_c$ $(M_TL^{-1} \sim 0.2)$ ($\chi^2 = 715$). Our two best models produce better constraints than the best model from \citet{Grillo2015}. We can confirm that \macs most likely contains two cluster scale halos with a flat inner core. 
We find that the cluster members in \macs are best described by a scale radius slope $\zeta_{r_{\mathrm{s,gal}}} =  5.11^{+2.33}_{-1.92}$ and a velocity dispersion slope $\zeta_{\sigma_{\mathrm{gal}}} = 1.69^{+0.32}_{-0.62}$.

By comparing the mass estimates from the different cosmological models, we find that we have the necessary prerequisites in order to combine strong lensing, X-ray and dynamics mass estimates. We find a total mass difference from lowest to highest mass, when comparing our cosmological models with the reference model, of $\sim 49\%$. When comparing with the mean of the mass estimates, we find $\sim 49\%$ difference and comparing with the median of the mass estimates, we find $\sim 50\%$ difference. The next step toward realising this method is to combine the projected total mass estimates with estimates from X-ray and dynamics.

%% file: introduction.tex

\nnchapter{Introduction}

Gravitational lensing has proven to be one of the foremost astrophysical observational methods for investigating the properties of both the lenses and the lensed sources themselves.

Even though Einstein proposed his correction to the deflection angle in 1917 \cite{Einstein1917} and \citet{Dyson1920} confirmed this correction by observing the deflection of stellar light around the limb of the Sun, it would take almost 65 years before the first gravitational lens was observed at cosmological scales.  

In 1979, \citet{Walsh1979} observed a pair of quasars in the 
radio band. They had an angular separation of $6\maths{arcsec}$ and was located at the same redshift $(z_s = 1.41)$. The following year, a massive elliptical galaxy with $z_l = 0.36$ 
was discovered between the two quasars. Subsequent imaging in a wide range of wavelength, determined that it was indeed two images of the same source, lensed by the elliptical galaxy.

This led to a wide range of different discoveries using lensing effects: Micro-lensing, which is the lensing of small object (like a star) by another small object (like another star or planet); weak lensing, which is the lensing of a source into a single image; strong lensing, which is the lensing of a source into multiply lensed images and giant luminous arcs. Furthermore, it led to the development and usage of sophisticated mass density profiles. We have the NFW profile, which is considered a general mass density profile for Dark Matter \cite{ Navarro1997} and the PIEMD/dPIE profile \cite{Kassiola1993,Ardis2007} which is considered to be an excellent profile for mass distribution in galaxies, but has also proven useful in describing the mass distribution in clusters as well. Especially cluster with highly elliptical DM halos.

In particular, the strong lensing regime has led to the examination of many clusters of different sizes and configurations \cite{Kneib1996, Limousin2007, Ardis2006, Limousin2008, 
Richard2009} which have provided us with a detailed information about the distribution and shape of the mass in galaxy clusters, the estimated total mass and the surface mass density. It has also shown a connection between the number of cluster mass-density profiles used in the models and the relaxation of and number of DM halo(s) in a cluster. Here a single mass-density profile is consistent with a relaxed unimodal cluster \cite{Limousin2008,Richard2009} and two mass-density profiles seems to be consistent with the non-relaxed bimodal clusters \cite{Ardis2007, Grillo2015}. Recent studies even indicate tri- \cite{Caminha2016}, quadri- or even quintamodal \cite{Richard2014} structure.

Strong lensing has also proved to be useful in the detection and investigation of massive field galaxies \cite{Grillo2011,Grillo2013}, which have proven to generate a 
lot stronger lensing effects and image separations $(\gtrsim 2\maths{arcsec})$, than originally expected. The generalisation \cite{Golse2002} of the NFW profile \cite{Navarro1997} led to investigation \cite{Sand2002, Sand2004, Sand2008} into whether the NFW profile could truly be 
considered a universal density profile for DM and also a direct comparisons between different density profiles \cite{Eliasdottir2007}.


From the time that the Swiss astronomer Fritz Zwicky \cite{Zwicky1933,Zwicky1937} first suggested that a large fraction of the mass in the universe was constituted by an unseen 
substance, which he called Dark Matter\footnote{In the original German terminology, the term was \emph{dunkle Materie}.}, the quest for determining the nature of this matter has 
set in motion, a wide variety of scientific investigations. 

As mentioned before, \citet{Navarro1997} used N-Body simulations to determine a universal density profile for DM. They constructed several simulations from which they derived the NFW density profile. Recent simulations using supercomputers, giving room for a substantial increase in number of particles, has revealed insight into the large scale structure formation of DM \cite{Springel2005,Gao2005}, but also detailed insight into the DM distribution at the galaxy scale level. Currently the Large Suite of Dark Matter Simulations (LasDamas) is running, creating mock SDSS catalogues to 
compare with the actual galaxy distribution observed, but also to study the detailed properties of DM halos. The Illustris project has enhanced our understanding of the 
co-evolution of dark and baryonic matter \cite{Vogelsberger2014}, combing with tracer particles. At the microscopic level, the \citet{Atlas2013} project at the LHC facility (CERN) investigates the particle nature of DM from measurements of W and boson decay.

To summarize, multiple branches within physics are engaged in the search for the nature of dark matter and gravitational lensing is no exception. Since DM only 
interacts via gravity, gravitational lensing is an important tool in determining the distribution and content of DM in both galaxies and galaxy clusters. Here, a precise estimation of the values of the cosmological parameters is a key component.

The primary goal of this thesis is to investigate the viability of using projected total mass estimates from strong lensing modelling, of the cluster \macs \cite{Ebeling2001, Zitrin2013, Johnson2014, Grillo2015}, to estimate the cosmological parameter values. The secondary goal is to conduct a comprehensive analysis of \macs, particularly analysing the projected cumulative total mass and average surface mass-density. Toward this end, we will investigate whether optimization of the cluster member scaling relations and whether adapting elliptical cluster member profiles, instead of spherical profiles, will yield better results. Both are methods that have not previously been performed on \macs.

The thesis is organized in the following way. In Chapter \ref{chap:theory} we introduce the theory necessary for our strong lensing analysis. In Chapter \ref{chap:clusters} we give a basic introduction to clusters of galaxies in general and the cluster \macs. In Chapter \ref{chap:obs-data} we present the data we are using. In Chapter \ref{chap:method} we introduce the lensing program \lenstool and describe our modelling method. In Chapter \ref{chap:results} we present our results and analysis. In Chapter \ref{chap:discussion} we will discuss the results, summarize and present prospects for future studies of \macs.

Throughout this thesis we adopt the following notation. Vectors will be in bold font $(\vec{a})$ and matrices will be in calligraphic fonts $(\mtrx{A})$. We generally assume a locally Minkowski flat space, that the Robertson-Walker-Lemaítre-Friedmann metric applies and that cosmological parameters are given by the values $H_0 = 70\maths{km s^{-1} Mpc^{-1}}$, $\Omega_M = 0.3$ and $\Omega_{\Lambda} = 0.7$, unless otherwise stated.

%% file: strong-lensing.tex

\chapter{Strong Lensing}\label{chap:theory}
Gravitationally lensed systems are usually divided into three separate types: Strong lensing, where we have more than one lensed image from the same 
source and often see large deformations i.e. arcs; weak lensing where we only see one image from the same source and generally weak to no deformation; micro lensing, 
which deals with situations where an object is lensed by a relatively small object (like a star deflected by another star). 

In Section \ref{sec:basictheory} we give an introduction to strong gravitational lensing and in Section \ref{sec:lensmodels} we introduce some simple lensing models. The theory behind strong lensing will mainly be derived from 
\cite{Schneider1992}, \cite{Schneider2006}, \cite{Narayan1997} and \cite{Grillo2007}.

\section{What is Gravitational Lensing?}\label{sec:basictheory}
The main idea is that a gravitational potential will bend a ray of light around it. Even though Newton's Law 
of Gravity describes the force between two objects with mass
\[ F = \frac{GMm}{r^2}\]
Newton speculated that an object with mass could bend light around it. The law of gravity can be combined with the connection between force and acceleration
\[ F = ma \]
to give an acceleration that is independent of the mass of the smaller object, here the photon
\[ a = \frac{GM}{r^2}\]
If we now consider a mass M where a photon passes closely by at distance r, the deflection can be described by
\[ a = \frac{GM}{r^2+z^2}\] 
where z is along the direction of motion for the photon. We can decompose the attraction felt by the photon into an element along $z$ and an element perpendicular to $z$. The attraction felt along $z$ will be positive when the photon moves towards M and negative when the photon moves away from M. This means that the only constant attraction felt by the photon will be perpendicular to $z$. This gives that the 
acceleration towards M is given by
\[ a_{perp} = \frac{GMr}{(r^2+z^2)^{3/2}} \]
The change in velocity felt by the photon will then be the total integration of the acceleration
\[v_{perp} = \int \frac{GMr}{(r^2+z^2)^{3/2}} dt = \frac{2GM}{cr}\]
where $dt = dz/c$. The deflection-angle $\hat{\alpha}$ will then be given by $\hat{\alpha} \propto \frac{v_{perp}}{c}$ which gives
\[
 \label{eqn:newton_deflect}
 \hat{\alpha}_{Newton} = \frac{2GM}{c^2r}
\]
This seems to imply that in order to find the deflection around an object, we will need to integrate over the entire path. As we shall see later, this is fortunately not the case.

Newtonian gravity is a special case of General Relativity (GR) and so it could seem that also we need to apply solutions from GR in order to use gravitational 
lensing. That is, we would have to solve the Einstein Field Equations (EFE)
\[
\label{eqn:EFE}
 R_{\mu \nu} - \onehalf Rg_{\mu \nu} = \frac{8\pi G}{c^4}T_{\mu \nu}
\]
where $R_{\mu \nu}$ is the Ricci curvature tensor, $R$ is the curvature scalar, $g_{\mu \nu}$ is the metric and $T_{\mu \nu}$ is the Energy-Momentum tensor. These equations 
tells us how space-time will behave in the presence of energy-momentum or in the words of John Wheeler: "Mass tells space-time how to curve and space-time tells mass how to move". Although gravitational lensing can be derived from GR \cite{Schneider1992}, we shall see later that we do not need to solve the EFE in our particular case.

If we consider the Poisson equation for the gravitational potential of an object with density 
$\rho$
\[ \gradient^2 \phi = 4\pi G \rho \rarrow \phi = -\frac{GM}{r}\]
we find that if $\phi / c^2 \ll 1$, we are in the weak regime and the EFE \eqref{eqn:EFE} can then be linearized. A linearization of the EFE would imply that we can treat gravitational lensing as a geometrical problem.

If we consider the typical velocity distribution in galaxy clusters 
$(v \approx 1500\maths{km\;s^{-1}})$ and assume that the clusters are in virial equilibrium, it follows that $v^2 \sim \phi$. So we see that $\phi / c^2 \sim v^2/c^2 \ll 
1$ which means we are in the weak regime and hence, that we can treat gravitational lensing as a purely geometrical problem. All we really need from GR is the corrected 
deflection angle
\[
 \label{eqn:deflectangle_gr}
 \hat{\alpha}_{GR} = \frac{4GM(\xi)}{c^2 \xi}
\]
where $M(\xi)$ is the enclosed spherically symmetrical mass and $\xi$ is the impact parameter of the photon. This angle is only 
valid in a situation where $\xi$ is much larger than the Schwarzschild radius $\xi \gg R_S \equiv 2GMc^{-2}$.

Since gravitational lensing belongs to the weak regime, we can say that the condition $\xi \gg R_S$ implies that the deflection angle is small $\hat{\alpha} \ll 1$ and we can use 
the small angle approximation $\sin{\alpha} = \alpha$. Had we been in the strong regime $(\phi/c^2 \sim 1)$ we would observe large bending angles. In 
this case we would have to apply a full solution of \eqref{eqn:EFE}. This is the case around Black Holes and Neutron Stars.

\subsection{Thin Lens Approximation}
\label{subsec:thin_lens_approx}
\begin{figure}[h!]
 \centering
 \includegraphics[width=0.9\textwidth,keepaspectratio=true]{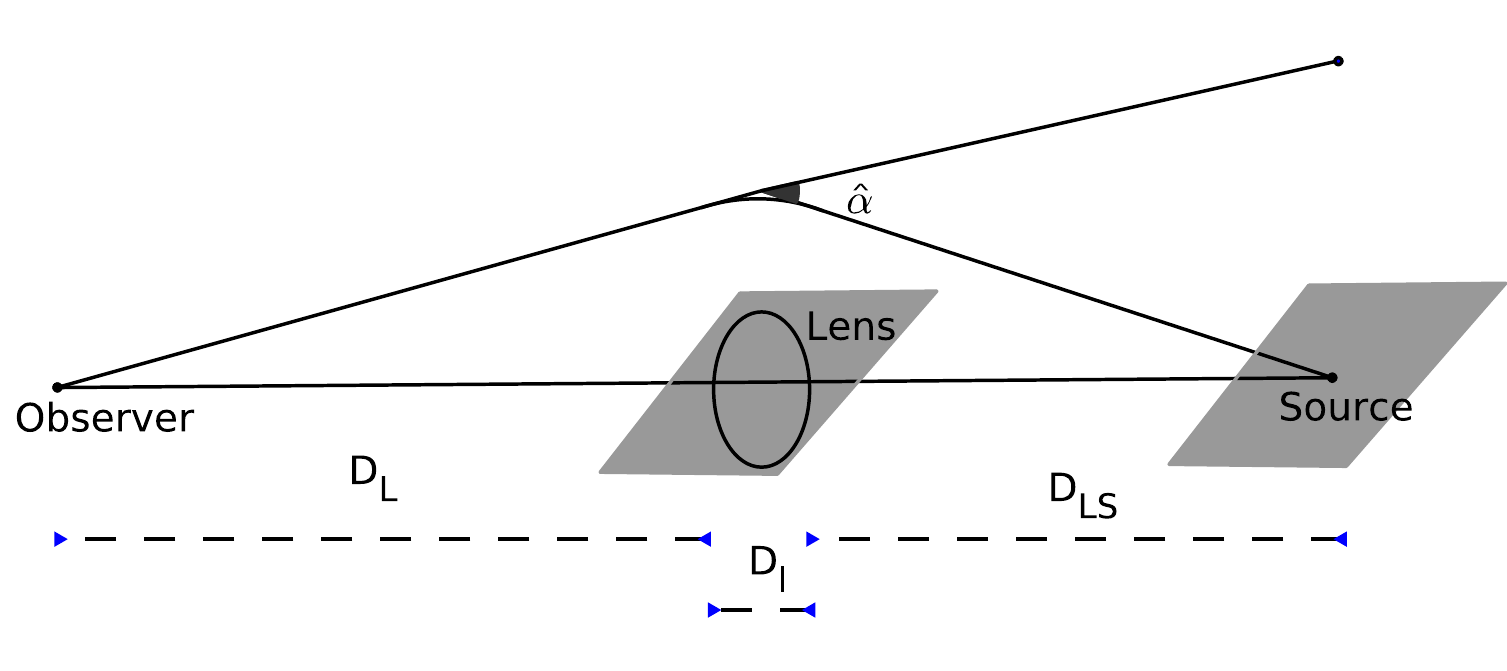}
 \caption{An illustration of the thin lens approximation. While the real deflection from the gravitational potential is in three dimensions and therefore the
 deflection is an arc over the entire depth of the lens $(D_l)$, since $(D_{LS} \gg D_l \ll D_{SL})$ the deflecting mass can be collapsed into a plane and hence the deflecting 
angle is sharp.}
 \label{fig:thin_lens_approx}
\end{figure}

Now that we have shown that we are in the weak regime and can use the small angle approximation, we have to consider the geometry of the lens itself. If 
we compare the distances $D_{LS}$, $D_L$ and $D_l$ on figure \ref{fig:thin_lens_approx} we see that if the distances $D_L$ and $D_{LS}$ are much larger than $D_l$, the latter is almost insignificant. Here we have that $D_L$ is the distance from the observer to the lens, $D_{LS}$ is the distance from the lens to the source and $D_l$ is the physical extent along 
the line of sight of the lens. Since $D_L$ and $D_{LS}$ are in the Gpc regime and $D_l$ is in the Mpc regime, we have that $D_L \gg D_l$ and $D_{LS} \gg D_l$. 

When this is the case we can allow ourselves to collapse the physical size along the line of sight into a plane or surface
\footnote{We shall later see (sect.\ref{subsec:surf_mass_dens}) that we actually integrate the mass density along the line of sight, into a surface mass density. So this is not 
just a qualitative property of the lens, but indeed also a quantitative property.} i.e. the lens is \emph{thin}. 
This means that the deflection angle $\hat{\alpha}$ can be considered a sharp angle instead of an arc. 

These conditions are satisfied in almost all relevant astrophysical situations, except in cases where the size of the deflecting mass constitute a significant amount of the distance from the source to the observer.

\newpage
\subsection{The Lensing Equation}
\label{subsec:lens_eq}
\begin{figure}[h!]
 \centering
 \includegraphics[height=5cm,keepaspectratio=true]{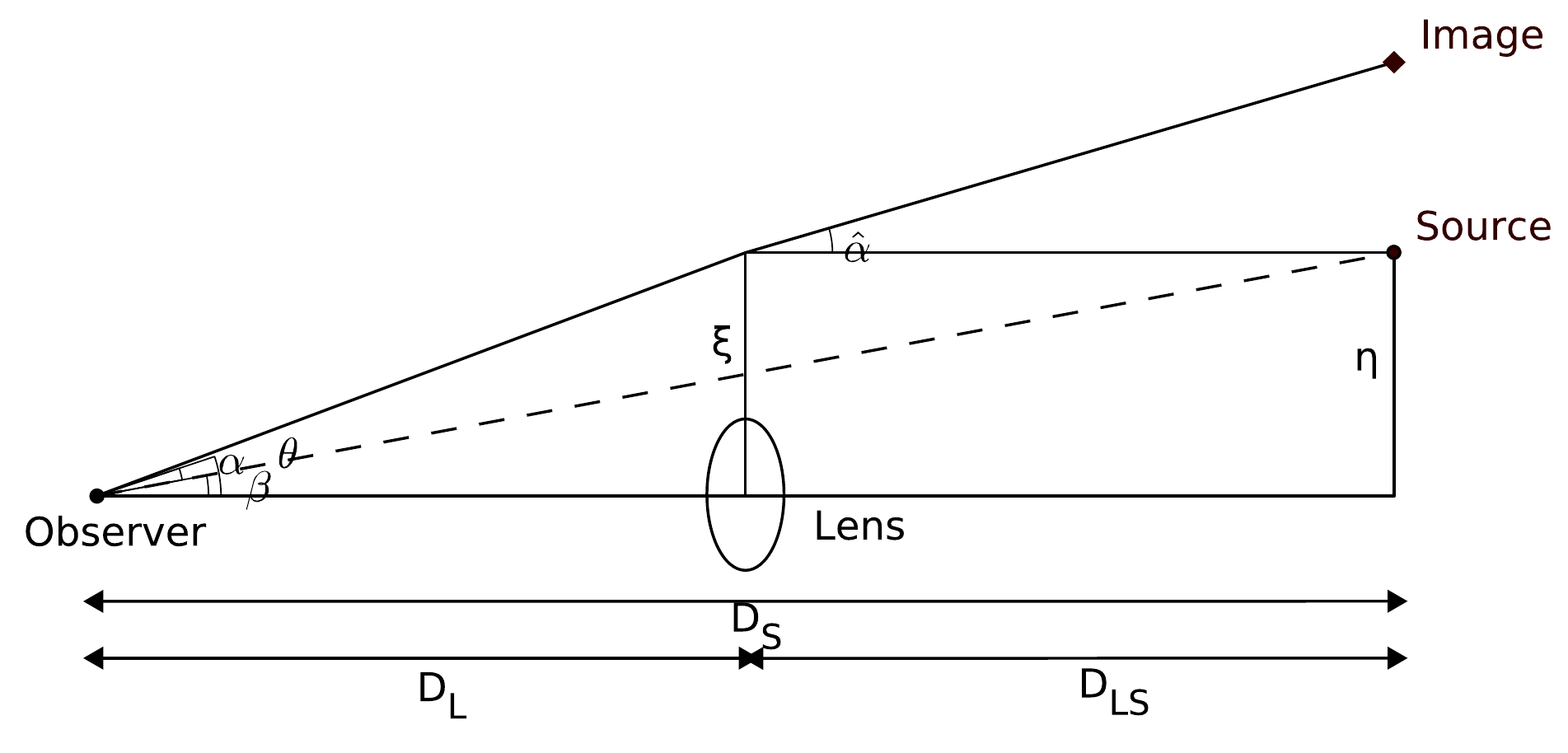}
 \caption{An illustration of the geometrical configuration for a typical lensing system.}
 \label{fig:lens_setup}
\end{figure}

We are now prepared to consider a geometrical configuration for a typical lensing configuration. 
In Figure \ref{fig:lens_setup} we show such a configuration, where the angular position of a source is $\vec{\beta}$ (on the source plane), the deflection angle 
is $\ALPHA$, the scaled deflection angle is $\scalpha$, $\vec{\eta}$ is the two-dimensional position of a source, $\vec{\xi}$ is the two-dimensional position of an image and $\vec{\theta}$ is the angular position of an image (on the lens plane). We further have the angular diameter distances from the observer to the source $(D_{S})$, 
from the observer to the lens $(D_{L})$ and from the lens to the source $(D_{LS})$. In general we have that $D_{S} \neq D_{L} + D_{LS}$ which means that the distances do not scale 
additively. This comes from the fact that distances depend on cosmology.

By looking at Figure \ref{fig:lens_setup} we find the following relations to be true
\begin{align}
 \vec{\beta}(\vec{\theta}) &= \vec{\theta} - \vec{\alpha}(\vec{\theta}) \label{eqn:lensequation} \\ 
 \vec{\alpha}(\vec{\theta}) D_{S} &= \ALPHA (\vec{\theta}) D_{LS} \\
 \vec{\xi} &= \vec{\theta} D_{L} \\
 \vec{\eta} &= \vec{\beta} D_{S} 
\end{align}
where equation \eqref{eqn:lensequation} is the lensing equation in its most general form. It should be mentioned, that although Figure \ref{fig:lens_setup} represents the source and image position in one dimension, the position of images and sources are in two dimensions. So the lensing equation is a vector equation.

We can also see that $\THETA = \BETA(\THETA) + \vec{\alpha}(\THETA)$ cannot be true, since it is an invalid equation. This tells us that \eqref{eqn:lensequation} is non-linear. Therefore, the mapping from the lens plane to the source plane $\THETA \mapsto \BETA(\THETA)$ is straightforward for any given mass distribution. It is simply a matter of saying that if we have a lensed image we know that a 
source must exist. The problem is the inverse mapping $\BETA(\THETA) \mapsto \THETA$, which is a mapping of the possible images from a given source. Because of the non-linearity of \eqref{eqn:lensequation} we cannot know the number of images from the knowledge of the position of a source.

We now want to combine the lensing equation \eqref{eqn:lensequation} with the angular diameter distances. By first realising that $\vec{\xi} = \vec{\theta}D_{L}$ we can rewrite \eqref{eqn:deflectangle_gr} into the form
\[
 \label{eqn:reducedalpha2}
 \hat{\alpha}(\vec{\theta}) = \frac{4GM(\THETA)}{c^2 D_{L}}\frac{1}{\vec{\theta}}
\]
and by looking at Figure \ref{fig:lens_setup} we see that
\[
 \label{eqn:anglerelation}
 \vec{\alpha}(\vec{\theta}) D_{S} = \ALPHA (\vec{\theta}) D_{LS} \lrarrow \vec{\alpha}(\vec{\theta}) = \ALPHA (\vec{\theta}) \frac{D_{LS}}{D_S}
\]
We can combine \eqref{eqn:lensequation} with \eqref{eqn:anglerelation} to get
\[
 \label{eqn:lensequation2}
 \vec{\beta}(\vec{\theta}) = \vec{\theta} - \ALPHA (\vec{\theta}) \frac{D_{LS}}{D_S}
\]	

It would now seem natural to combine \eqref{eqn:reducedalpha2} with \eqref{eqn:lensequation2}, but in order to do that we need to make assumptions about the shape and size of the mass $M(\THETA)$. This will be done in Section \ref{sec:lensmodels} where we introduce simple lensing models.

\subsection{Surface Mass Density}
\label{subsec:surf_mass_dens}
From the deflection angle \eqref{eqn:deflectangle_gr} with a 2-dimensional impact parameter $\XI$ and the thin lens approximation (sect. \ref{subsec:thin_lens_approx}), we can introduce the 
deflection angle for a lens made of several point-mass 
components. This can be described as the sum of the deflections of the individual components, given by
\[
 \label{eqn:deflection-sum}
 \ALPHA(\vec{\xi}) = \sum_{i} \ALPHA_i (\vec{\xi})
\]
where $\ALPHA_i$ is the deflection of component i.

By introducing the mass components as a three-dimensional system, where the three-dimensional density is given by $\rho(\vec{r}) = \rho(\xi_1, \xi_2, r_3)$, we can collapse the 
density distribution into a surface mass density by integrating
\[
 \label{eqn:surfmassdens}
 \Sigma(\vec{\xi}) = \int \rho(\xi_1, \xi_2, r_3) dr_3
\]
From this we can find the deflection angle for an arbitrary density distribition
\[
 \label{eqn:scaled-integral}
 \ALPHA(\XI) = \frac{4G}{c^2} \int \Sigma(\vec{\xi}') \frac{\XI - \XI'}{\abs{\abs{\XI - \XI'}}^2} d^2 \xi'
\]
and define the scaled deflection angle
\[
 \label{eqn:3dscaled}
 \scalpha(\THETA) = \frac{1}{\pi} \int \kappa(\THETA') \frac{\THETA - \THETA'}{\abs{\abs{\THETA - \THETA'}}^2} d^2 \theta'
\]
where $\kappa(\THETA')$ denotes a quantity called convergence, which is defined as
\[
 \label{eqn:convergence}
 \kappa(\THETA) \equiv \frac{\Sigma(D_L \THETA)}{\Sigma_{cr}}
\]
and $\Sigma_{cr}$ is the the critical surface mass density
\[
 \label{eqn:critsurfmass}
 \Sigma_{cr} \equiv \frac{c^2}{4\pi G} \frac{D_S}{D_L D_{LS}}
\]


\subsection{Magnification}
\label{subsec:magnitication}
A gravitational lens do not only create one or more images of a source, the lens also affects the physical properties (the shape and orientation) of 
the individual images. Such an effect is called magnification.

Although a gravitational lens do not affect the colors of the images \footnote{The lens do not have chromatic aberration since the light will be displaced equally for all wavelength. Whether the photon represents red, green or blue, it will feel the same force of the potential. The force is independent on the frequency of the photon.}, it will affect the flux of an image. Here we have the monochromatic flux $S_{\nu}^*$ from a source with surface brightness $I_{\nu}$, where $\nu$ denote frequency and $d\omega^{*}$ is the solid angle of the light without gravitational lensing
\[
 \label{eqn:fluxfromsource}
 S_{\nu}^{*} = I_{\nu}\, d\omega^{*} 
\]
If the light is deflected by a lens, the solid angle $d\omega$ will differ from $d\omega^{*}$ and hence the observed deflected light will be described by
\[
 \label{eqn:flux-observed}
 S_{\nu} = I_{\nu}\, d\omega
\]
This means that the observed change in the flux can be found as the ratio between the flux with and without lensing
\[
 \label{eqn:flux-change}
 \frac{S_{\nu}}{S^{*}_{\nu}} = \frac{d\omega}{d\omega^{*}} = \abs{\mu}
\]
which is completely independent of the frequency of the light from the source, since $I_{\nu}$ cancels out. We define this difference as the magnification $\abs{\mu}$.

\begin{figure}[h!]
 \centering
 \includegraphics[height=6cm,keepaspectratio=true]{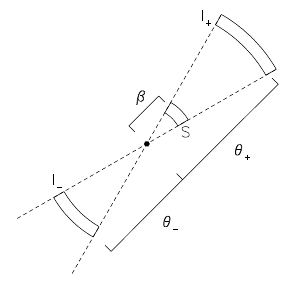}
 \caption{An illustration of magnification, including the effects of how parity will influence the magnification of the image. We see that a positive parity will shape the 
image similar to the source, whereas a negative parity will shape the image opposite that of the source. It is important to say that a negative parity image can be mirrored or 
reversed or both. From \cite{Narayan1997}.}
 \label{fig:magnification}
\end{figure}

Since the ratio of the solid angles can also be related to the second-order derivative of the angular positions $\THETA$ and $\BETA$, we find that the change in flux can be described as

\[
 \label{eqn:Jacobian}
 \frac{S_{\nu}}{S^{*}_{\nu}} = \frac{d\omega}{d\omega^{*}} = \frac{d^{2}\theta}{d^{2}\beta}
\]
and it can now be shown that the magnification factor $\abs{\mu}$ can be obtained from the determinant of the Jacobian matrix. By defining the Jacobian 
matrix as

\[
 \label{eqn:JacDet}
 \mtrx{A}(\THETA) = \frac{\PD\BETA}{\PD\THETA}
\]
we find that the magnification factor is given by
\[
 \label{eqn:magni-factor}
 \mu (\THETA) = (\det{\mtrx{A}(\THETA)})^{-1}
\]
which means that an infinitesimal small source is either brightened or dimmed by a factor $\abs{\mu(\THETA)}$ by the gravitational lens. This is illustrated in 
figure \ref{fig:magnification} where we see two images that are larger than their source i.e. magnified. We shall later see that there are three possible configurations for images relative to their source. 

At certain values of $\THETA$, $\det{\mtrx{A}}$ can vanish and $\mu(\THETA)$ will diverge. When this happens we say 
that $\mu(\THETA)$ has hit a \emph{critical point}.


\subsection{Convergence and Shear}
\label{subsec:conv_shear}
A more detailed view on magnification can be found by expanding the theory of \emph{convergence} and \emph{shear}, which was mentioned briefly previously. In order to do that, we first need to define the lensing scalar potential
\[
 \label{eqn:deflect_potential}
 \psi(\THETA) = \frac{1}{\pi} \int_{\REAL^2} \theta' \kappa(\THETA') \ln{\abs{\THETA - \THETA'}}\; 
 d^2 \theta'
\]
By further introducing the identity $\gradient \ln{\abs{\THETA}} = \frac{\THETA}{\abs{\THETA}^2}$, which is 
valid for any two-dimensional vector $\THETA$, we can rewrite the scaled deflection angle
\[
  \label{eqn:deflect_angle_gradient}
  \scalpha (\THETA) = \gradient \psi (\THETA)
\]
using the gradient of the deflection potential. This means that the mapping $\THETA \map \BETA$ is a gradient mapping.
From here we can use the identity
$\gradient^2 \ln{\abs{\THETA}} = 2\pi \delta_{D} (\THETA)$ to find
\[
 \label{eqn:def_convergence}
 \gradient^2 \psi (\THETA) = 2\kappa (\THETA)
\]
which is the \emph{Poisson equation} in two dimensions, $\delta_{D}$ is the two-dimensional Dirac delta function and $\kappa$ is the \convergence. By writing the \convergence as 
\[
 \kappa (\THETA) = \onehalf (\psi_{11} + \psi_{22})
\]
and defining the \emph{shear} in the form of complex numbers
\[
 \label{eqn:def_shear_complex}
 \gamma \equiv \gamma_1 + i\gamma_2 = \abs{\gamma} e^{2i\phi}
\]
with the real part
\[
 \label{eqn:def_shear_real}
 \abs{\gamma} = \sqrt{(\gamma_{1}^{2} + \gamma_{2}^{2})}
\]
where $\gamma_{1}(\THETA) = \onehalf (\psi_{11}-\psi_{22}) \equiv \gamma(\THETA) \cos{\left[ 2\phi \right]}$ 
and $\gamma_{2} = \psi_{12} = \psi_{21} \equiv \gamma(\THETA) \sin{\left[ 2\phi \right]}$, we find that the Jacobian can be rewritten as

\begin{align}
 \label{eqn:def_Jacobian}
 \mtrx{A}(\THETA) &= \bpm 1-\kappa(\THETA) - \gamma_1(\THETA) & -\gamma_2(\THETA) \\ -\gamma_2(\THETA) & 1-\kappa(\THETA) + \gamma_1(\THETA) \epm \nonumber \\
          &= (1-\kappa(\THETA)) \bpm 1 & 0 \\ 0 & 1 \epm -\gamma(\THETA) \bpm \cos{2\phi} & \sin{2\phi} \\ \sin{2\phi} & -\cos{2\phi} \epm \\
\end{align}
From here we can see that the first term of the Jacobian describe an isotropic focusing (convergence) and the second term describe introduced astigmatism or distortion into 
the image (shear), where $\gamma$ defines the magnitude and $\phi$ define the orientation. Be aware that this $\phi$ is not the same as in \eqref{eqn:def_shear_complex}.

\begin{figure}[h!]
 \centering
  \includegraphics[height=5cm,keepaspectratio=true]{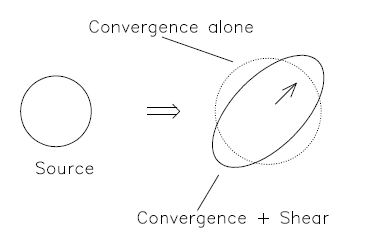}
  \caption{Illustration of how an extended spherical source will experience convergence and shear. Notice that the convergence alone only produces magnification of 
the image, while adding shear introduces elliptical distortion. This is directly related to the parity of the magnified image(s). Since real sources rarely are perfectly circular, but have more complex shapes, the resulting image from shear will never have a nice elliptical shape. From \cite{Narayan1997}.}
  \label{fig:conv-shear}
\end{figure}

We can now explain in more detail, what convergence and shear do (see Fig. \ref{fig:magnification} and Fig. \ref{fig:conv-shear}). Convergence alone will only result in an 
isotropic magnification or focusing of the lensed image, where the shear will introduce astigmatism in the image, usually in an elliptical shape. The shape of the distortion is driven 
by a gravitational tidal field and the magnification itself is driven in parts by the isotropic focusing $(\kappa)$ from the local matter density and the anisotropic 
focusing $(\gamma)$ from the shear. 

This tells us that strong lensing will not only provide multiple images, but will also magnify and distort the images. Here are arcs are the most prominent 
examples of shear and the Einstein ring is an example of perfect uniform shear. Both magnification and distortion are dependent on the mass density of the lens and the position of 
the source, relative to the lens. We will later show a tight connection between convergence and shear and the critical lines and caustics. It is important to note here, that magnification means that an image can also be demagnified. 

When dealing with models that are not singular in nature, a center-located, highly demagnified image will occur in certain configurations. Since these images are also very close to the lens, we usually do not see them.

Finally we can derive the determinant in the form 
\[
 \label{eqn:JacobianDet}
 \det{\mtrx{A}} = (1-\kappa(\THETA))^2 - \gamma(\THETA)^2
\]
the trace
\[
 \label{eqn:JacobianTrace}
 \tr{\mtrx{A}} = 2(1-\kappa(\THETA))
\]
and the eigenvalues
\[
 \label{eqn:JacobianEigval}
 a_{i} = 1-\kappa(\THETA) \mp \gamma(\THETA)
\]
of $\mtrx{A}(\THETA)$. The Jacobian matrix has two eigenvalues $(i = \{1,2\})$ and the sign of these two eigenvalues define the \emph{partial parity} of an image. The product of the eigenvalues defines the \emph{total parity}. We shall later see that the parity is very useful in describing qualitative properties of the images.

\subsection{The Fermat Potential}
The Fermat potential is one of the fundamental equations of gravitational lensing, and in some ways, a more proper approach to the lensing formalism \cite{Schneider2006}. The Fermat potential is usually defined as the 
scalar function
\[
 \label{eqn:fermat}
  \tau(\THETA;\BETA) = \onehalf (\THETA - \BETA)^2 - \psi(\THETA)
\]
which is a function of the lens plane coordinate $\THETA$ with the source position $\BETA$ as a parameter. The term $\psi(\THETA)$ is 
the deflection potential \eqref{eqn:deflect_potential}. The \emph{Fermat potential} is up to some affine transformation, the light travel time along a ray starting at $\BETA$, 
traversing the lens plane at position $\THETA$ and arriving at the observer. We shall see in the next section that the \emph{Fermat potential} is a very powerful tool for 
classifying multiple images. It should also be noted that 
\[
 \gradient \tau (\THETA;\BETA) = 0
\] 
is the equivalent of the lens equation \eqref{eqn:lensequation}.

\begin{figure}[h!]
 \centering
  \includegraphics[height=0.6\textwidth,keepaspectratio=true]{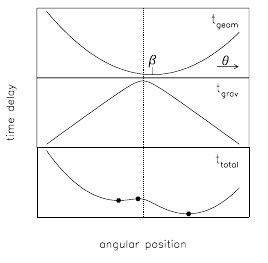}
 \caption{When defining the time-delay function using the two terms $t_{geom}$ and $t_{grav}$, the results can be seen as a sum of the 
two different contributions. The upper panel show the geometrical contribution from the path itself $(t_{geom})$. The middle panel show gravitational contribution $(t_{grav})$ as time is delayed in a gravitational field. The bottom panel show the combined effect. The source is lensed into three images, where we can define a local minimum, a global maximum and a global minimum, respectively from left to right. From \cite{Schneider2006}.}
 \label{fig:timedelay-geometry}
\end{figure}

The time it takes for a ray of light to travel from the source to the observer, through the lens, is defined by the time-delay function
\begin{align}
 t(\THETA;\BETA) &= \frac{1+z}{c}\frac{D_L D_S}{D_{LS}} \left[\onehalf (\THETA - \BETA)^2 - \psi(\THETA) \right] + C(\BETA) \\
 \label{eqn:time-delay}
                 &= t_{geom}(\THETA;\BETA) + t_{grav}(\THETA)
\end{align}
where $C(\BETA)$ is an additive constant. The terms in the bracket can be understood as the geometrical contribution $t_{geom}$ due to the extra path length travelled and $t_{grav}$ is the contribution in the form of the deflection from a gravitational potential, which is also called the \emph{Shapiro delay}. This relation is illustrated in Figure \ref{fig:timedelay-geometry} showing the time-delay for a circular symmetric lens with a slightly offset 
source $\BETA$. Here the upper panel show the geometric time-delay, the middle panel show the gravitational time-delay and the bottom panel show the combined time-delay and the 
relative position of the images $\THETA$.

One of the great advantages of using the Fermat principle for gravitational lensing is that it can tell us a lot about a given lensing configuration without modelling it\cite{Blandford1986}. We can find out the number of images, the position, the magnification and the parities for the images. 

First we extend the discussion on the parities a bit more. We know already that the total parity of an image is the sign of the product of the two eigenvalues, the partial parities. From this we can say that at a \emph{minimum} the partial parities are positive so the total parity is also positive. At a \emph{maximum} the partial parities are negative so the total parity is also positive. For a \emph{saddle point} the partial parities are mixed, so the total parity is negative.

For a given configuration of images, there are certain parity orderings that are allowed. The allowed configurations are displayed in Table \ref{table:time-delay-allowed-conf} for three and five image configurations according to total parity, where only the first and the fourth represents a unique topology for the five-image configurations. The first image to arrive (least time-delay) will always be a global minimum and hence have total positive parity\cite{Blandford1986}. If present, we will then have a local minima. 

\begin{table}[h!]
 \centering
 \caption{Table over allowed configurations. The total parities are arranged in sequence of arrival time. L is minima, S is saddle point and H is maxima.}
 \begin{tabular}{c c}
 \toprule
 Allowed configurations & Possible orderings \\
 \midrule
 $++-$ & LLS \\
 $+-+$ & LSH\\
 $+++--$ & LLLSS\\ 
 $++-+-$ & LLSLS/LLSHS\\
 $++--+$ & LLSSH/LHSSH \\
 $+-++-$ & LSLHS \\
 $+-+-+$ & LSHSH/LSLSH\\ 
 $+--++$ & LSSHH\\
 \bottomrule 
 \end{tabular} 
 \label{table:time-delay-allowed-conf}
\end{table}

\begin{figure}[h!]
 \centering
 \includegraphics[width=0.8\textwidth,keepaspectratio=true]{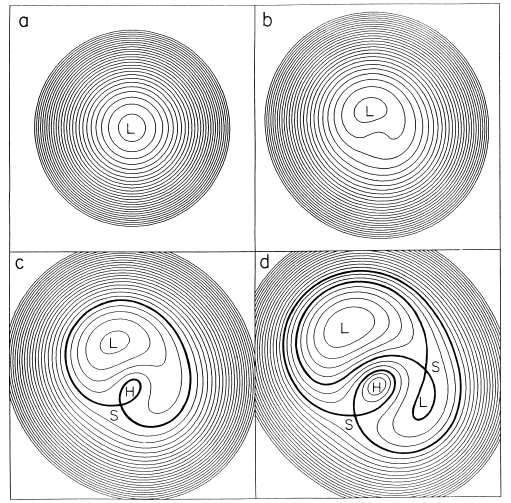}
 \caption{Time-delay surface-plots. Panel a: A source with no lens present. We will just see the image of the source in the middle at a global minimum and there are no time-delay. Panel b: Adding some mass will shift the image a bit and hence add a little time-delay, but no additional images are created. Panel c: Adding more mass will create a two additional images, one at a global maximum and one at a saddle point. The time-delay is arranged so that the minimum arrives first, then the maximum and lastly the saddle point. Panel d: Adding more mass, two additional images are created at a local minimum and another saddle point.}
 \label{fig:time-delay-surface}
\end{figure}

If we consider the three-image configurations in Table \ref{table:time-delay-allowed-conf} we see that the first configuration must be a global minimum (L), local minimum (L) and saddle point (S) and the second must be a global minimum (L), saddle point (S), global maximum (H). So for a three-image configuration it seems fairly simple to derive the time-delay orderings. Looking at Figure \ref{fig:time-delay-surface} panel c, we can see the time-delay surface for the $(+-+)$ configuration where the ordering is the global minimum, the saddle point and lastly the global maximum.  For five-image configurations it is more complicated. Here we have six allowed configurations and it is not as easy to derive the time-delay orderings. In Table \ref{table:time-delay-allowed-conf} we see some possible time-delay orderings in relation to minima, maxima and saddle points. In order to find the time-delay ordering we would have to plot the time-delay surface or calculate the time-delay, for a given configuration. 
An example of time-delay surface plot of a five image configuration is shown in panel d of Figure \ref{fig:time-delay-surface}. In this particular case we have first the global minimum (top L), local minimum (bottom L), saddle point (S closest to local minimum), saddle point (S closest to H) and finally the global maximum (H).
\begin{figure}[h!p]
 \centering
 \includegraphics[width=0.3\textwidth,keepaspectratio=true]{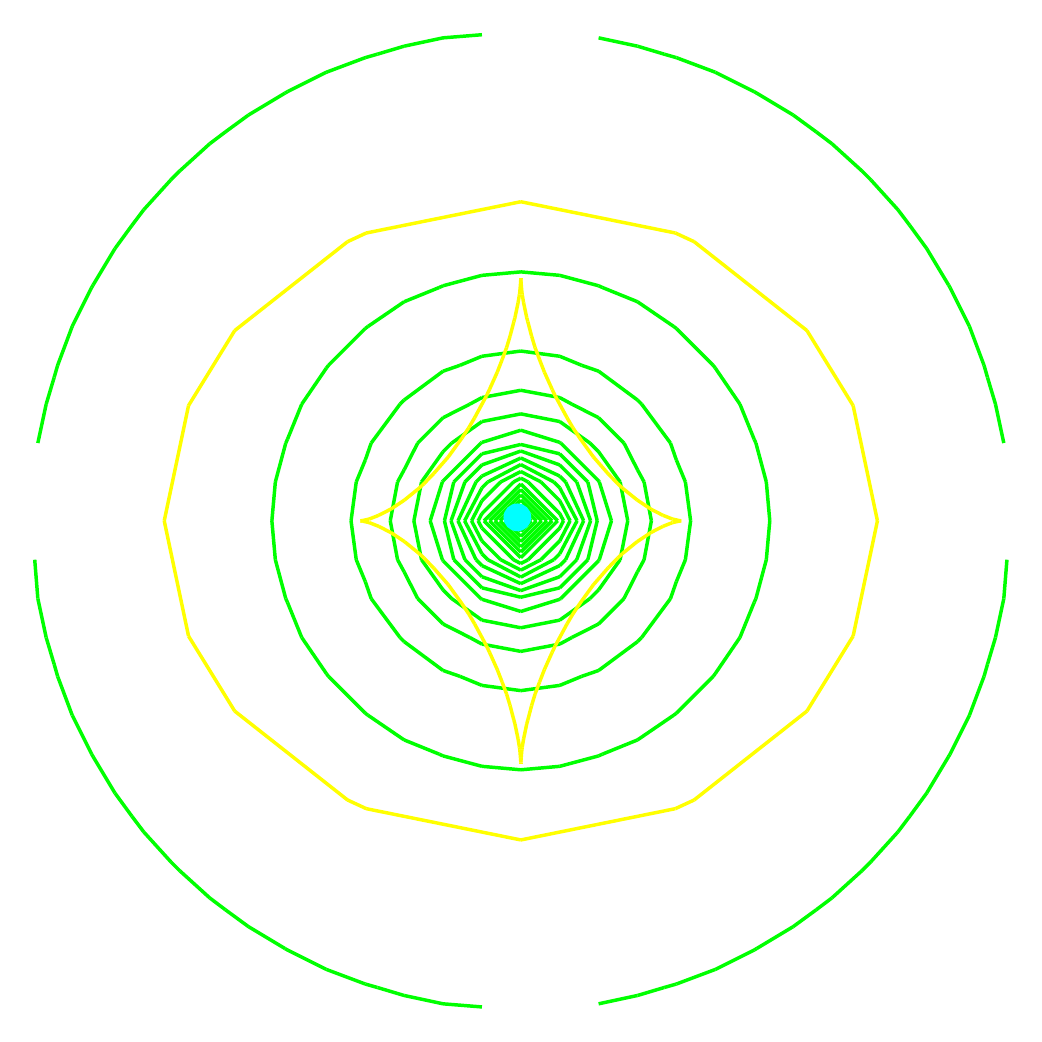}
 \includegraphics[width=0.3\textwidth,keepaspectratio=true]{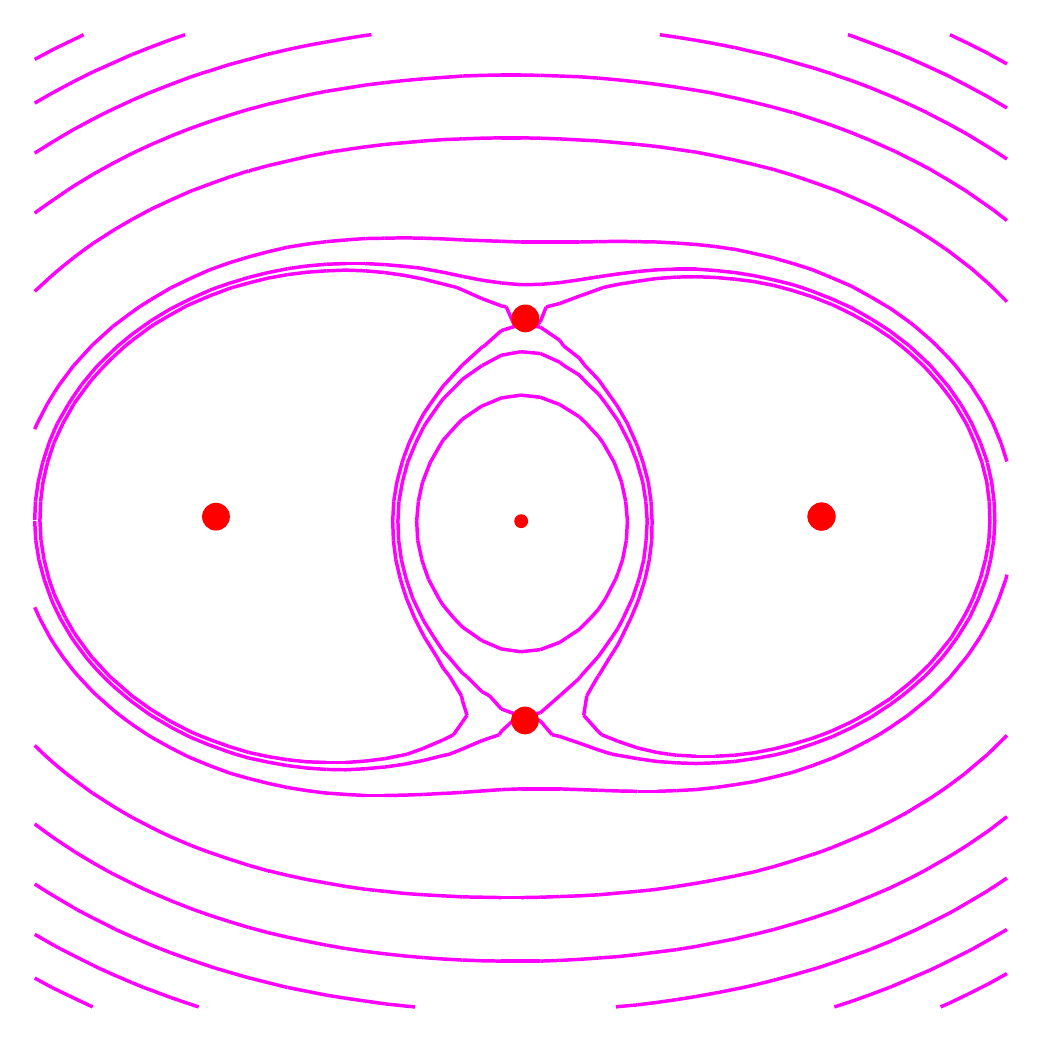} \\
 \includegraphics[width=0.3\textwidth,keepaspectratio=true]{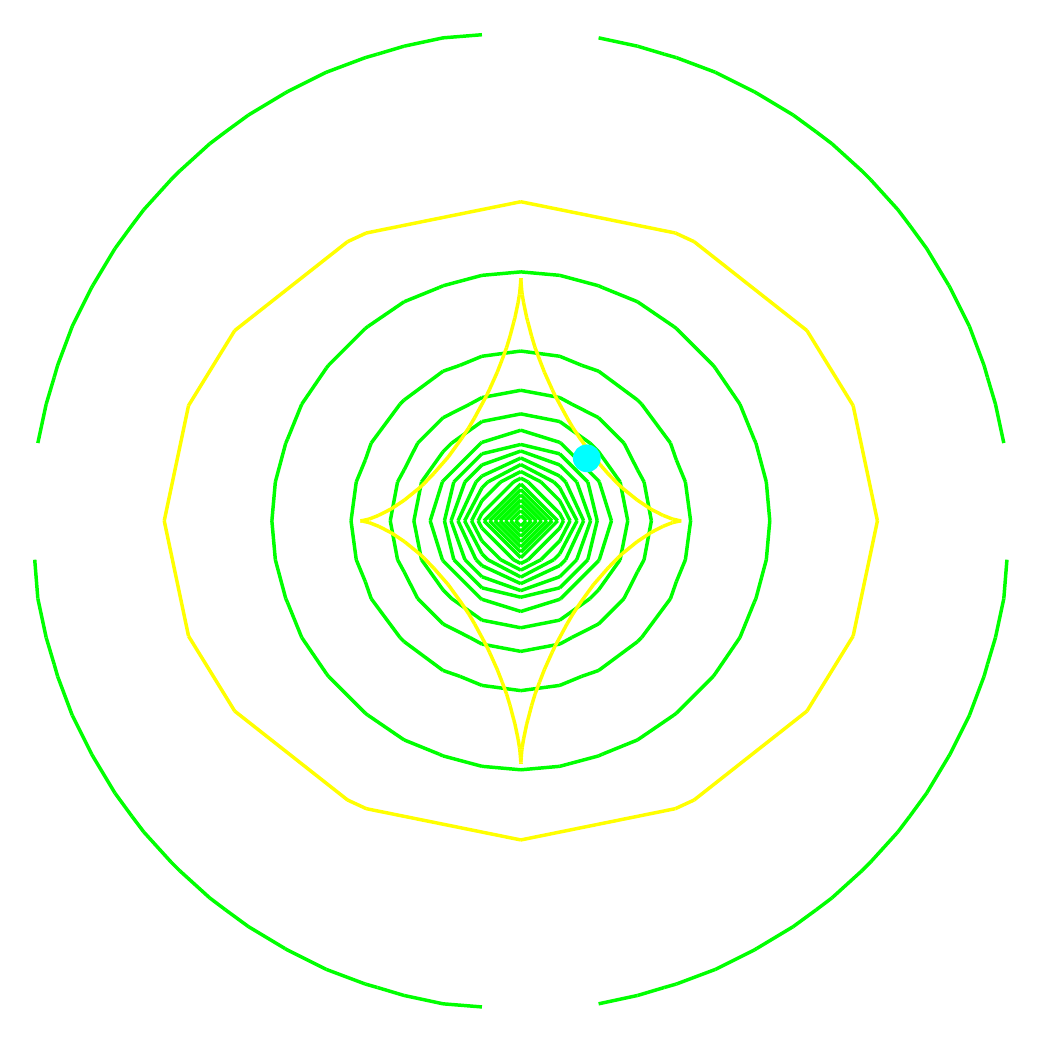} 
 \includegraphics[width=0.3\textwidth,keepaspectratio=true]{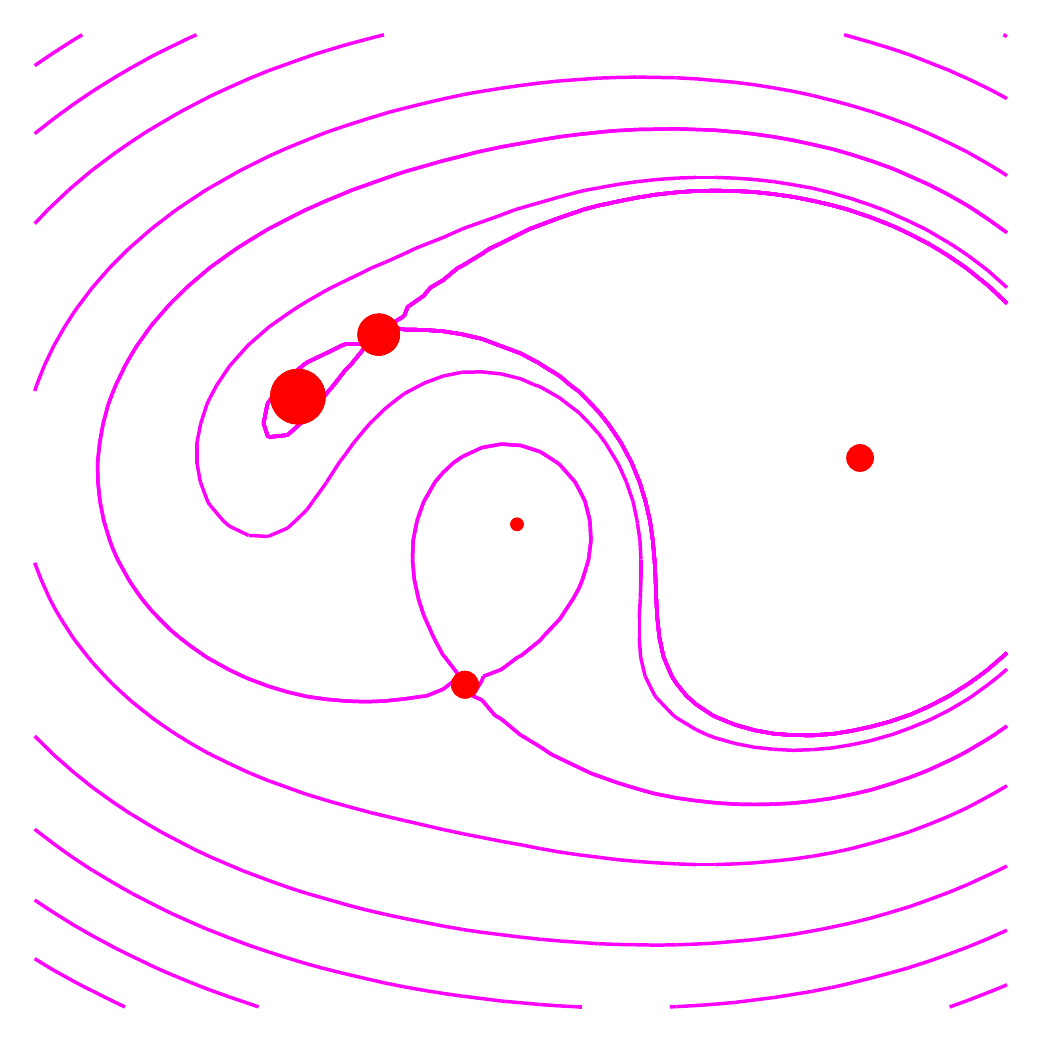} \\
 \includegraphics[width=0.3\textwidth,keepaspectratio=true]{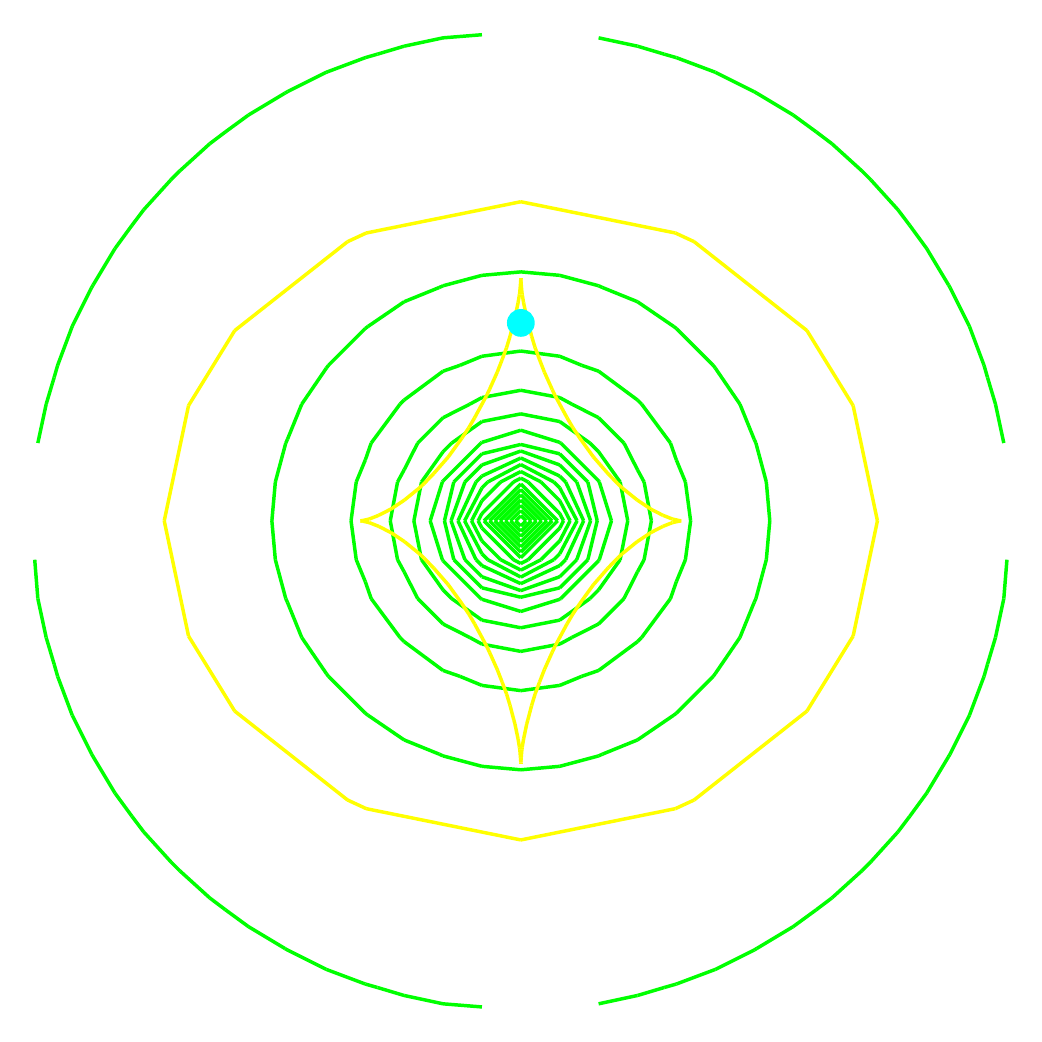} 
 \includegraphics[width=0.3\textwidth,keepaspectratio=true]{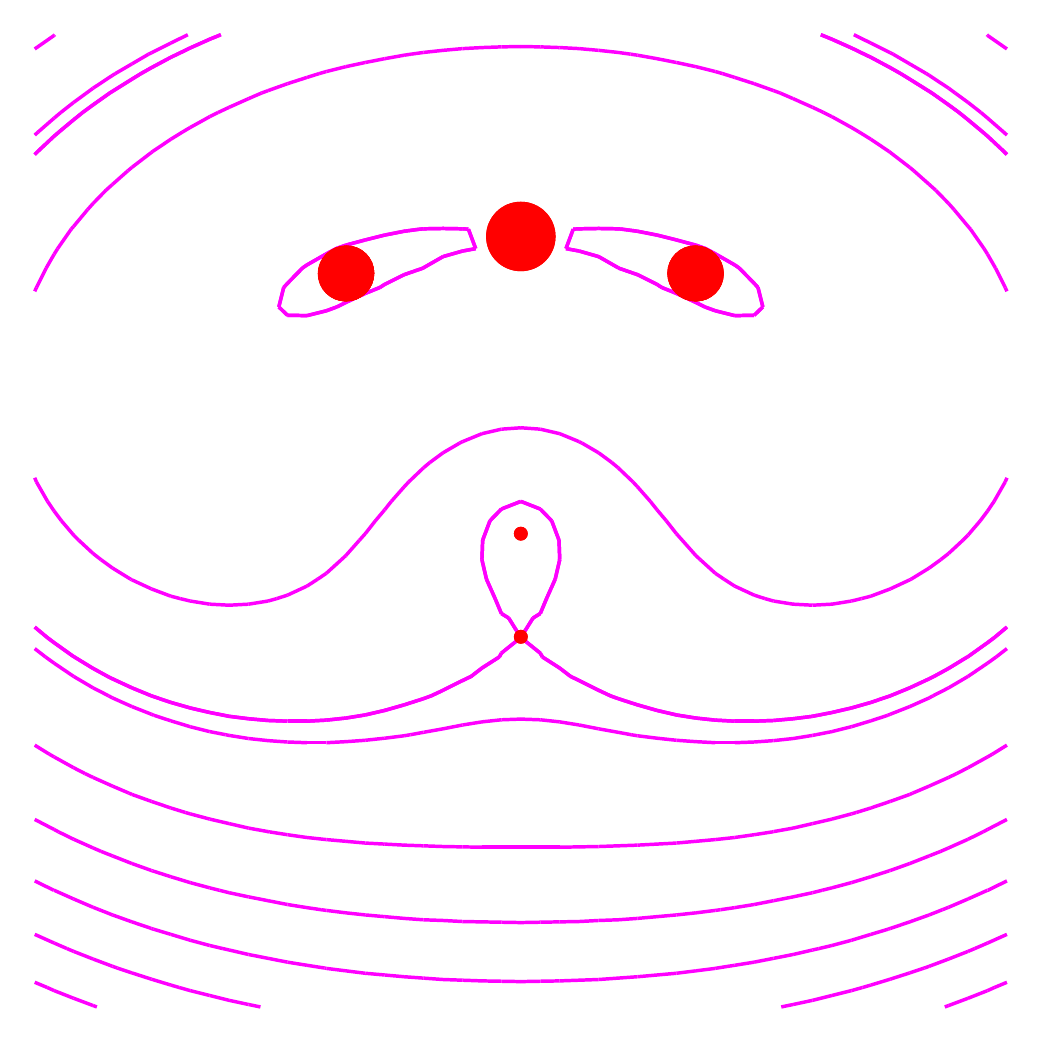} \\
 \includegraphics[width=0.3\textwidth,keepaspectratio=true]{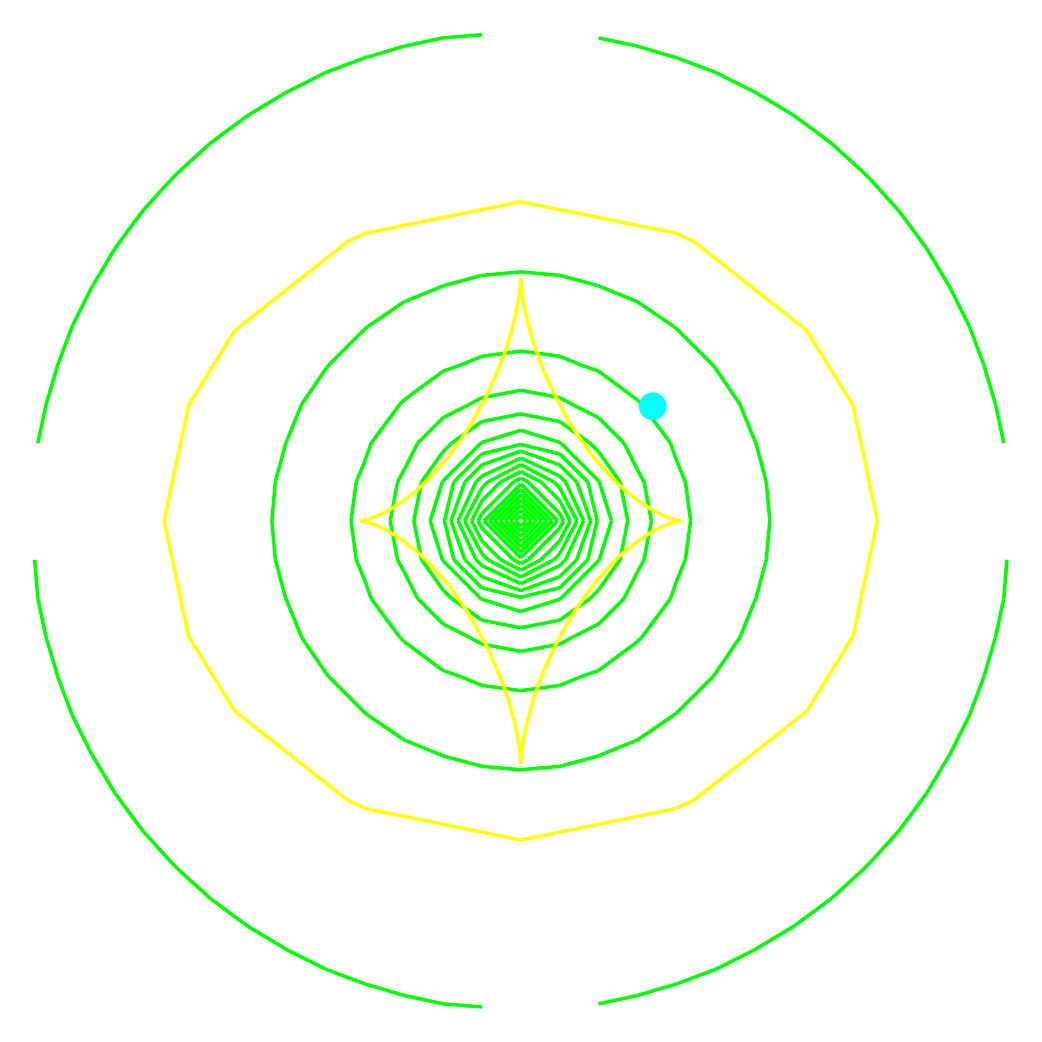} 
 \includegraphics[width=0.3\textwidth,keepaspectratio=true]{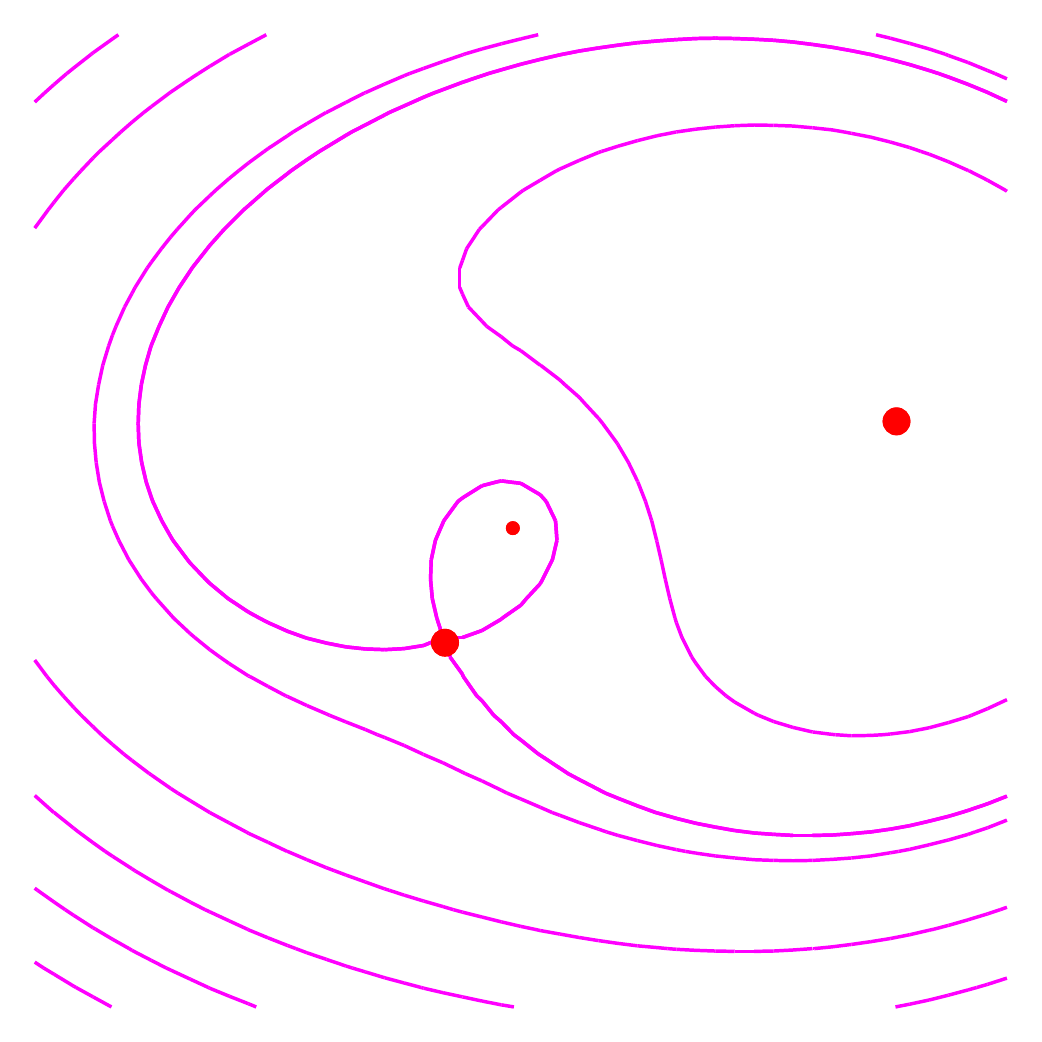}
 \caption{Source position and time-delay surfaces from \emph{SimpLens} software. From the top: Row 1: The source is centered, which gives two minima to the left and right, two saddle-points at top and bottom and a global maximum in the center. Row 2: The source is positioned slightly to the right, close to the inner fold caustic, which gives a global minimum to the right, a local minimum to the left followed by a saddle-point, then another saddle point at the bottom and again a global maximum at the center. Row 3: The source is positioned just close to the inner cusp caustic, which gives two minima at the top, right and left side, followed by a saddle-point in the middle, a saddle-point at the bottom followed by the global maximum just above. Row 4: The source is positioned to the right, just outside the inner fold caustic, which gives a global minimum to the right, followed by a saddle-point to the left and finally the global maximum just above, slightly to the right. The size of the images represents their magnification.}
 \label{fig:model_arrival_time}
\end{figure}
By using the software package \emph{SimpLens} \cite{Saha2003} we can generate vanilla models and move the source around, in order to generate different examples of time-delay surfaceplots. This can be seen in Figure \ref{fig:model_arrival_time} where we show examples for four different source positions. \emph{SimpLens} also shows the magnification of the images and the caustics, which we will explain in the next section. 

\newpage
\subsection{Critical Curves and Caustics}
\label{subsec:crit_curv}
\begin{figure}[h!]
 \centering
 \includegraphics[height=8cm,keepaspectratio=true]{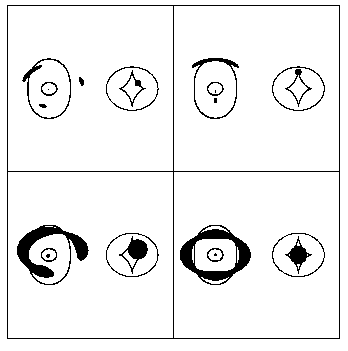}
 \caption{An illustration of multiple images created from an elliptical lens. The upper left panel show the images created when a point source is located close to a \emph{fold caustic}. The upper right panel show the images created when a source is located close to a \emph{cusp caustic}. The lower left panel show the same as the upper left panel, only with an extended source. The lower right panel show the images created from an extended source place at the center of the lens. From \cite{Narayan1997}.}
 \label{fig:critcaust}
\end{figure}

\begin{figure}[h!]
 \centering
 \includegraphics[height=7cm,keepaspectratio=true]{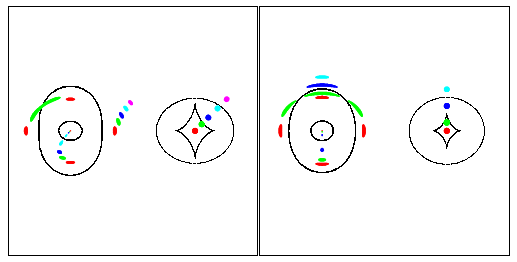}
 \caption{An illustration of a point-like source with different positions relative to the critical lines / caustics and the number of images created or destroyed. In each panel we have the image plane with critical lines to the left and the source plane with the caustics to the right. From \cite{Schneider2006}.}
 \label{fig:multimages}
\end{figure}

For all lenses, \emph{critical curves} can be defined as closed smooth curves and appear where $\det{\mtrx{A}(\THETA)} = 0$. A further set of lines can be defined by mapping these 
critical curves onto the source plane, which are called \emph{caustics}. Caustics are not necessarily smooth but can develop cusps. The critical lines and caustics are important 
in order to describe the positions of multiple images and mapping in general. An illustration of critical curves and caustics can be seen in Figure \ref{fig:critcaust} and 
\ref{fig:multimages}.

The critical curves and caustics can provide a qualitative understanding of the lens mapping:
\begin{enumerate}
 \item The magnification diverges for an image on a critical curve. 
 In general, a source placed close to a caustic will produce a highly magnified image close to the corresponding critical line.
 \item The number of images a lens will produce depends on the position of the source relative to the caustics. 
\end{enumerate}
In general we can say that a change in the position of a source leads to a change in the number of images when the source crosses a caustic. As a source traverse a 
caustic, a pair of images are either created or destroyed, depending of the direction of the crossing. If the source is moving towards the center of the lens from a distance i.e. the 
source traverse from the position producing a single image (large $\BETA$) towards $\BETA \smrarrow 0$, two new images are created whenever the source traverse a caustic. It was shown by \citet{Burke1981} that for any lens with a smooth surface mass density decreasing faster than $\abs{\THETA}^{-1}$ as $\THETA \rightarrow \infty$, the number of images equals the number of extrema of $\tau$. This means that the number of images with positive parity equals the number of saddle points $+1$. Furthermore, at least one of the images will correspond to a minimum of $\tau$. This is only valid when the source is not located on a caustic. We also say the the critical curves occur where the lensing equation \eqref{eqn:lensequation} is not locally invertible.

When looking at Figure \ref{fig:multimages} it also becomes apparent that, in contrary to the critical lines, the caustics need not bo smooth. 
In order to show this more formally, we can define a parametrization of the critical curve as $\THETA(\lambda)$. In order to find the shape of the critical curve at a given point 
we find the tangent to the critical curve by taking the derivative $d\THETA(\lambda)/d\lambda = \dot{\THETA}$. Since the caustic is defined as the mapping of the 
critical curve from the image plane to the source plane, the parametrization of the caustic will then become $\BETA(\THETA(\lambda))$ and the derivative can be defined as
\[
 \frac{d(\BETA(\THETA(\lambda)))}{d\lambda} = \frac{\PD \BETA}{\PD \THETA} \frac{\PD \THETA}{\PD \lambda} = \mtrx{A}(\THETA(\lambda))\dot{\THETA}(\lambda)
\]
where we see that the tangent to the caustic depends on the matrix $\mtrx{A}(\THETA(\lambda))$ and the derivative $\dot{\THETA}$. If we find the eigenvectors of $\mtrx{A}$ then a 
point can arise where the eigenvector whose value is $0$ (for a critical curve there is always one eigenvector with a value of $0$), is parallel to the tangent of the critical 
curve. At that point, the shape of the caustic will no longer be smooth, but abrupt. In mathematical terms we say that the function is no longer continuous.

These points are called \emph{cusps} and can be seen in the upper right panel in Figure \ref{fig:critcaust} and right panel in Figure \ref{fig:multimages} , where the source is close to a cusp. The smooth part is then called 
a \emph{fold} and can be seen in the upper left panel in Figure \ref{fig:critcaust} and in the left panel in Figure \ref{fig:multimages} where the source is close to a fold. More generally, we could name the two types of caustics by their shapes, so that a smooth caustic (fold only) is a \emph{circular caustic} and a caustic with four 
cusps is an \emph{astroid caustic}.

The side of a caustic where the number of images will increase are called the \emph{inner side} and the side of a caustic where the number of images will decrease is called the \emph{outer side}. An illustration of a source crossing from the inner side to the outer side is shown in Figure \ref{fig:multimages}. 


In the left panel in Figure \ref{fig:multimages} we first see 4 images, pairwise symmetric, from the central source (red). When the source moves toward the outer side and gets close to the fold of the astroid caustic (green), we see two of the images merge on the outer critical line to the west, the southern image moves westward, toward the lens and the eastern image eastward, away from the lens. The two merging images will be highly magnified and because the lens is elliptical, they experience shear that distorts them into arcs. There is also a central image, which is highly demagnified.

When the source has crossed the fold of the astroid caustic (blue), the two images to the west have merged and vanished. We are now left with the image to the south of the lens, the one east to the lens and the central (demagnified) image, which has also moved slightly. As the source moves close to the circular caustic (turquoise) the image closest to the inner critical line begins to merge with the central image, which then disappears when the source has moved beyond the outer caustic (purple) and we are left with one image, the purple image to the east. Notice also that both the southern and eastern images moves "along" with the source, as it moves from the inner to the outer side. This, we will later see, is generally consistent with simpler models like the singular isothermal sphere.

The same applies to the right panel, except that when a source crosses the cusp of the astroid caustic (green) we see three images merging and becoming magnified. After the source has crossed the cusp caustic, there is one image left.

We can derive the following qualitative understanding of the lens mapping. The critical lines divide the lens plane into positive parity regions $(\mu > 0)$ and negative parity regions $(\mu < 0)$. The caustics divide the source plane into regions of different image multiplicity. Whenever a source crosses a 
caustic, the number of images change by $\pm 2$. This tells us that the number of critical lines determines how many images we can expect to see. In the case with Figure \ref{fig:multimages} we expect to see maximally 5 images (which is also the case). It is important to point out here that the position of the caustics/critical lines changes with the redshift of the source, since the critical lines appears when $\det{\mtrx{A}(\THETA)} = 0$ and $\mtrx{A}(\THETA) = \frac{\PD \BETA}{\PD \THETA} = \frac{\PD (\ETA / D_S)}{\PD (\XI / D_L)}$.

Further, since $\det{\mtrx{A}} = 0$ implies that at least one 
of the two eigenvalues of $\mtrx{A}$ vanish, the image of a circular source can become highly elongated and distorted. This also denotes the connection between convergence and 
shear and the critical lines and caustics. 

This elongation and distortion of images is the origin of the giant arcs found in clusters.

\subsection{Multiple Images}
\label{subsec:mult_images}
From the Fermat potential \eqref{eqn:fermat} we can derive a way to classify images, by determining whether an image $\THETA$ is located at a minimum, maximum or saddle 
point of $\tau$. Since the Jacobian matrix is the Hessian of $\tau$, we have
\[
 A_{ij} = \frac{\PD^2 \tau}{\PD \theta_i \PD \theta_j}
\]
and from this we can determine the image type in relation to the signs of the two eigenvalues $a_i$ of $\mtrx{A}(\THETA)$, presented in \eqref{eqn:JacobianEigval}. From the trace \eqref{eqn:JacobianTrace} and determinant \eqref{eqn:JacobianDet} of the Jacobian matrix, we can derive the following:
\begin{itemize}
 \item At a minimum of $\tau$, both eigenvalues $a_i$ are positive, which means that $\det{\mtrx{A}} > 0$ and $\tr{\mtrx{A}} > 0$. This comes 
from $\gamma(\THETA)<1-\kappa(\THETA)\leq 1,\, a_i > 0,\, \mu \geq \frac{1}{1-\gamma(\THETA)^2} \geq 1$
 \item At a maximum of $\tau$, both eigenvalues $a_i$ are negative, which means that $\det{\mtrx{A}} > 0$ and $\tr{\mtrx{A}} < 0$. This comes from $(1-\kappa(\THETA)^2) > 
\gamma(\THETA)^2,\, \kappa(\THETA) > 1,\, a_i < 0$.
 \item At a saddle point of $\tau$, the signs of $a_i$ are different, which means that $\det{\mtrx{A}} < 0$. This comes from $(1-\kappa(\THETA)^2) < \gamma(\THETA)^2,\, a_2 > 0 > 
a_1$.
\end{itemize}

\begin{figure}[h!]
 \centering
 \includegraphics[height=0.6\textwidth,keepaspectratio=true]{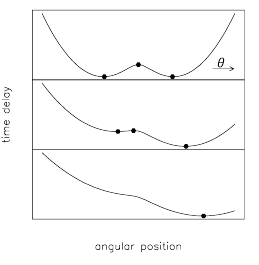}
 \caption{Illustration of the Fermat potential where two of the three different type of images are seen (minimum and maximum) in relation to 
the position of the source. Saddle points can not be determined on a 2	-dimensional plot. The top panel shows the images when $\BETA$ is in the center (an Einstein ring). The 
middle panel shows the images for a slightly offset $\BETA$ to the right, but still within the inner caustic. The bottom panel shows the images when $\BETA$ is offset beyond the 
inner caustic, where we only have one image. From \cite{Schneider2006}.}
 \label{fig:timedelay-center-source}
\end{figure}

The plots in Figure \ref{fig:timedelay-center-source} show the position of images in relation to minima and maxima. There are no saddle points, since in order to show a saddle point we would need 
a 3-dimensional plot. In the top panel we see the images from a centered source $\BETA$. One global maximum and two local minima, which could also be one global minimum 
going all the way around, giving rise to an Einstein ring. The middle panel show the images for a slightly offset source to the left. We have two images moved to 
the left and closer to each other in a local minimum and global maximum and one image moved to the right in a global minimum. The bottom panel show the images for a source 
that has moved further to the left beyond the inner caustic. Here we only see one image in a global minimum.

\newpage
\subsubsection{Conditions for Multiple Images}
\label{subsec:condictions}
In general, a lens will not produce any multiple images if the lens equation is globally invertible. This means that there must be a point $\THETA$ where $\det{\mtrx{A}(\THETA)} < 
0$ for any multiple images to occur, whereas if $\det{\mtrx{A}(\THETA)} > 0$ for all $\THETA$, the lens equation is globally invertible.

This can also be stated as follows. If there exist a point $\THETA_0$ for which $\det{\mtrx{A}(\THETA_0)} < 0$, then a source at $\BETA_0 \equiv \BETA(\THETA_0)$ will have an 
image which will correspond to a saddle point and from the odd-number theorem\cite{Burke1981}, at least two more images must exist corresponding to the extrema.

Another sufficient, but not necessary condition for multiple images, is using the dimensionless surface mass density. If there exist a point $\THETA_0$ so that 
$\kappa(\THETA_0) > 1$, then a source at $\BETA_0 \equiv \BETA(\THETA_0)$ has an image which cannot correspond to a minimum, since for these $\kappa < 1$. Therefore the source 
must have at least one additional image, corresponding to this minimum.

This also explains why lenses with $\kappa > 1$ are called strong lenses. Although $\kappa > 1$ does not represent a necessary condition for the occurrence of multiple images, 
$\Sigma_{cr}$ does represent the characteristic scale for the occurrence of strong lensing features.

Since the $\Sigma_{cr}$ is highly dependent of the redshift of the source, the strength of the lens will increase with distance (from lens to source), since $\Sigma_{cr}$ gets smaller. This is another way of explaining why the critical lines has different positions for difference sources.

\section{Lensing models}
\label{sec:lensmodels}
In order to describe the physical properties of different lensing configurations, we fit a mass profile to the image configuration i.e. a model. The basic understanding of a model is that it compare the predicted positions of the multiple images/source, given a certain mass-density profile, with the actual observed images/barycenter position of the source(s). Lensing software and optimization methods will be discussed in more detail in Section \ref{sect:Modelling}.

The mass profiles can generally be divided in two different types: Axisymmetric and non-axisymmetric mass profiles. Axisymmetric mass profiles have one axis of symmetry and include the point-mass lens and the singular isothermal sphere (SIS). The point-mass lens is useful for galactic microlensing and the SIS is generally used for simple models of matter distributions in galaxies and galaxy clusters. 

Since galaxies and galaxy clusters are not generally expected to have axisymmetric gravitational potentials, the SIS is not considered a realistic mass distribution. We then have to apply mass distributions with two axes of symmetry, like the elliptical mass distribution and/or add external shear to the model.

\subsection{Axisymmetric Lenses}
For axisymmetric lenses in general, the matter distribution can be defined as
\[
 \Sigma(\XI) = \Sigma(\abs{\XI})
\]
and the deflection angle can be defined as
\[
 \label{eqn:SIS-deflectangle}
 \ALPHA(\XI) = \frac{\XI}{\abs{\XI}^2} \frac{4G}{c^2} 2\pi \int_{0}^{\xi} d\xi'\; \xi'\; \Sigma(\xi') \equiv \frac{4GM(\abs{\XI})}{c^2 \abs{\XI}^2}\XI
\]
where $M(\XI)$ is the projected mass within a circle of radius $\abs{\XI}$. This means that when considering a geometrically thin axisymmetric mass distribution at a point 
$\XI$, the deflection angle is simply \eqref{eqn:deflectangle_gr} for $M(\abs{\XI})$ enclosed in a circle with radius $\abs{\XI}$. With a working assumption about the mass distribution, we can insert this deflection angle \eqref{eqn:SIS-deflectangle} into \eqref{eqn:lensequation2}, using $\XI = \THETA D_L$.

Since both $\scalpha$ and $\BETA$ are co-linear with $\THETA$ when using \eqref{eqn:lensequation}, we have that if the position of a source can be described by $\BETA = \beta 
\vec{e}$, then $\THETA = \theta \vec{e}$ follows and the lens equation now is one-dimensional
\[
 \label{eqn:1dlensequation}
 \beta(\theta) = \theta - \alpha(\theta)
\]

By using the one dimensional lens equation \eqref{eqn:1dlensequation} and setting $\beta = 0$ which means that the source is exactly on the optical axis and therefore perfectly aligned with the observer and the lens, along line-of-sight, we then get
\begin{align}
 \nonumber
 0 = \theta - \frac{D_{LS}}{D_{L} D_{S}} \frac{4GM}{c^2 \theta} \\
 \nonumber
 \theta = \frac{D_{LS}}{D_{L} D_{S}} \frac{4GM}{c^2 \theta} \\
 \nonumber
 \theta^2 = \frac{D_{LS}}{D_{L} D_{S}} \frac{4GM}{c^2} \\
 \label{eqn:einstein_radius}
 \theta_{E} = \sqrt{\frac{D_{LS}}{D_{L} D_{S}} \frac{4GM}{c^2}}
\end{align}
where in the last term we have introduced the expression $\theta_{E}$ for the Einstein radius. This means that we will see an "infinite" number of multiple images with the exact same distance from the lens i.e. an \emph{Einstein ring}.

Although in perfect agreement with lensing theory, these rings were not expected to be observed at the time they were proposed. Primarily because lenses were not expected to be perfectly axisymmertric or sources perfectly aligned with the lens and observer along line-of-sight. Fortunately, the theory was confirmed by the discovery of numerous rings. $\theta_E$ corresponds to the critical line for the lens which when mapped into the source plane, the caustic becomes a point. 

\begin{figure}[h!]
 \centering
 \includegraphics[width=8cm,keepaspectratio=true]{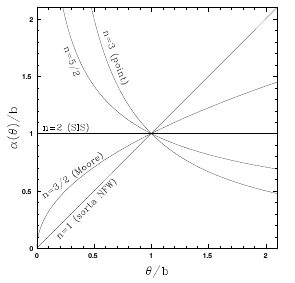}
 \caption{Deflection angle for the power law lens models. From the plot it can be seen that the more centrally concentrated profiles $(n<2)$ have divergent deflection angles and the more extended profiles $(n>2)$ have deflection angles that become $0$ at the center. The SIS $(n=2)$ has a constant deflection angle. From \cite{Schneider2006}.}
 \label{fig:power-law-lens}
\end{figure}

Both the point-mass lens and the SIS belongs to a wider family of lensing models, called \emph{power-law lenses}. These are defined by the scaled deflection angle
\[
 \label{eqn:power-law-lens}
 \alpha(\theta) = b\leftparan \frac{\theta}{b} \rightparan^{2-n}
\]
where we have assumed a lens with a density distribution of $\rho \propto r^{-n}$ and $b$ is a constant. These kind of models are particularly relevant since most clusters and galaxies are believed to have centrally density cusps rather than core radii.
For these models we have the convergence profile
\[
 \kappa(\theta) = \frac{3-n}{2}\leftparan \frac{\theta}{b}\rightparan^{1-n}
\]
and the shear profile
\[
 \gamma(\theta) = \frac{n-1}{2} \leftparan \frac{\theta}{b} \rightparan^{1-n}
\]
An illustration of the deflection angle for the power-law lenses are shown in Figure \ref{fig:power-law-lens}. 
For the point-mass lens we have a scaled deflection angle of
\[
 \alpha(\theta) = \frac{b^2}{\theta}
\]
where we later see that $b^2 = \frac{4GM}{c^2}\frac{D_{LS}}{D_L D_S}$. Here we also have that the convergence is $\kappa(\theta) = 0$ and shear $\gamma(\theta) = \frac{b^2}{\theta^2}$. Similarly for the SIS we have a scaled deflection angle of
\[
 \alpha(\theta) = b
\]
where we will also later see that $b = 4\pi\frac{\sigma_{v}^{2}}{c^2}\frac{D_{LS}}{D_S}$ and the convergence and shear $\kappa(\theta) = \gamma(\theta) = \frac{b}{2\theta}$

That last model that we will mention briefly here is the $n=1$ model. Since this model has a density profile of $\rho = r^{-1}$ it resembles the popular NFW profile that is thought to be a universal density profile for dark matter halos \cite{Navarro1997}. The NFW and the $n=1$ differs from the fact that the convergence is $\kappa(\theta) = \ln{(\theta)}$ for the $\rho = r^{-1}$ profile rather than a constant, as with the $n=1$ profile. The deflection angle for the $n=1$ profile is $\alpha(\theta) = b\leftparan \frac{\theta}{b} \rightparan^1 = \frac{b\theta}{b} = \theta$. The NFW profile will be introduced in more details, in Chapter \ref{chap:method}.
 
\subsubsection{Point-Mass Lens}
If we consider a point mass M or the outside of a spherically symmetric mass M, we can define the surface mass density as
\[
 \Sigma(\XI) = M\delta_{D}(\XI)
\]
and this leads to the deflection angle
\[
 \label{eqn:point_deflectangle}
 \ALPHA(\XI) = \frac{4GM}{c^2}\frac{\XI}{\abs{\XI}^2}
\]
where we have used \eqref{eqn:scaled-integral}. This shows that the deflection angle for this mass distribution agrees with \eqref{eqn:deflectangle_gr}.

By specializing the lens equation \eqref{eqn:lensequation} to this lens model, we get
\[
 \BETA(\THETA) = \THETA - \frac{4GM}{c^2}\frac{D_{LS}}{D_{S}D_{L}} \frac{\THETA}{\abs{\THETA}^2} = \THETA - \theta_{E}^{2} \frac{\THETA}{\abs{\THETA}^2}
\]
where we have used the definition of the Einstein radius \eqref{eqn:einstein_radius} in the last step. 

We can here select the position of the source $\BETA$ to be on the positive $\beta_1$-axis, without any loss of generality, and it then follows that the position $\THETA$ will be on the $\theta_1$-axis 
as well \cite{Schneider2006}. This means that the lens equation becomes one-dimensional and can be reduced to
\[
 \label{eqn:1d_pointmass_lenq}
 \beta = \theta - \theta_{E}^{2} \frac{1}{\theta}
\]
which is consistent with 
\[
 \beta = \theta - b^2 \frac{1}{\theta}
\]
where $b^2 = \frac{4GM}{c^2}\frac{D_{ls}}{D_l D_s} = \theta_{E}^{2}$.

This lensing equation has the following two solutions
\[
 \theta_{\pm} = \onehalf \left( \beta \pm \sqrt{\beta^2 + 4\theta_{E}^{2}} \right)
\]
These solutions shows that the point-mass lens will always have two images, one on each side of the lens. Furthermore, since it is clear that $\abs{\theta_{+}} \geq \abs{\theta_{-}}$, the 
image that is closest to the source will be further away from the lens, unless the source is at $\beta = 0$. Here we have an Einstein ring.

\begin{figure}[h!]
 \centering
 \includegraphics[width=8cm,keepaspectratio=true]{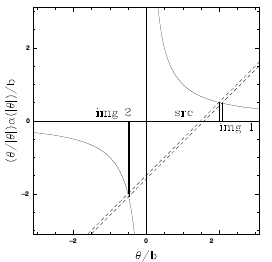}
 \caption{Graphical solutions for the point-mass lens. The two solid curves represent the solutions. The slanted solid line represents a particular position of the source with respect to the lens and the two vertical lines the corresponding images. The distance between the solid and dashed lines represents the size of the source and images. From \cite{Schneider2006}}
 \label{fig:graph-solution-pointmass}
\end{figure}

This can be illustrated nicely by plotting solutions to \eqref{eqn:power-law-lens}. In Figure \ref{fig:graph-solution-pointmass} we see the possible solutions to the point-mass lens represented as the two solid curves. We can see from these curves that the point-mass is singular, since the solutions tends to $\infty$ and $-\infty$ when $\theta/b \rightarrow 0$. When introducing a source at a particular position (slanted solid line) we can derive the position of two images from that particular solution (vertical solid lines). A magnified image (img1) and a less magnified image (img2). The magnification is represented by the size of the image (distance between the vertical solid and dashed lines), which is given from the size of the source (distance between the slanted solid and dashed lines). We can see that as the source moves further away from the center, one of the images will disappear. From the plot we also see one of the characteristics of a singular model. There are no solutions near $\theta/b \sim 0$ i.e. we have no third image. Without going into further details about non-singular models, we can instantly see from this plot alone that a non-singular model demands a continuous function.

For a circular symmetric lens, the magnification is given by
\[
 \mu = \frac{\theta}{\beta}\frac{d\theta}{d\beta}
\]
and by inserting the lens equation \eqref{eqn:1d_pointmass_lenq} we get
\[
 \mu_{\pm} = \left[ 1 - \left( \frac{\theta_E}{\theta_{\pm}} \right) \right]^{-1}
\]

For the point-mass lens, the image separation will usually be $\Delta \theta = 2\theta_E$ and in this case we have
\[
 \Delta \theta = 2\theta_E \sqrt{1+u^2/4} \gtrsim 2\theta_E
\]
where $u = \beta \theta_{E}^{-1}$ which means that the image separation will be just a slightly larger than $2\theta_E$.

\subsubsection{Singular Isothermal Sphere}
A simple model for the mass distribution in galaxies, assume that the stars behave like an ideal gas. This means that they are confined by their combined
spherical-symmetric gravitational potential. This gives us the expression
\[
 p = \frac{\rho k T}{m}
\]
where $\rho$ and $m$ is the mass density and the mass of the stars, respectively. By putting this into thermal equilibrium we get
\[
 m\sigma_{v}^{2} = kT
\]
where $\sigma_{v}^{2}$ is the velocity distribution of the stars in the galaxy. Here it is generally 
assumed that the gas is isothermal, which means that $\sigma_{v}$ is constant throughout the galaxy.
The equation for hydrostatic equilibrium then gives

\[
 \frac{\rho'}{\rho} = -\frac{GM(r)}{r^2} 
\]
and
\[
 \label{eqn:mass_sis}
 M'(r) = 4\pi r^2 \rho
\]
where the prime denotes a derivative with respect to $r$.

A simple solution to these equations is
\[
 \label{eqn:iso-solution}
 \rho(r) = \frac{\sigma_{v}^{2}}{2\pi G} \frac{1}{r^2}
\]
which is the density distribution for a \emph{singular isothermal sphere}.

By projecting along the line of sight we can obtain the surface mass density

\[
 \label{eqn:iso_surfmassdens}
 \Sigma(\xi) = \frac{\sigma_{v}^{2}}{2G} \frac{1}{\xi}
\]
and from that we find the scaled deflection angle to be the Einstein radius

\[
 \label{eqn:iso_einst_rad}
 \alpha = 4\pi \frac{\sigma_{v}^{2}}{c^2} \frac{D_{LS}}{D_{S}} = \theta_E
\]

Because this system is circular symmetric, the lens equation becomes one-dimensional, which means that 
we will only find multiple images if we have a source within the Einstein radius.
\[
 \beta = \theta - \theta_E \frac{\theta}{\abs{\theta}}
\]
When this condition is satisfied, we will find the following two solutions to the lens equation

\[
 \theta{\pm} = \beta \pm \theta_{E}
\]
We can see that this solution is consistent with the solution from \eqref{eqn:power-law-lens}, where if we use that $b = \alpha = \theta_E$ and that $0 < \beta < b$ we get the following solutions
\[
 \theta_1 = \beta + \theta_E \; \text{with} \; \theta_1 > \theta_E
\]
which is a minimum and the magnification is $\mu_1 = 1+\theta_E / \beta$ and the other solution
\[
 \theta_2 = \beta - \theta_E \, \text{with} \; -\theta_E < \theta_2 < 0
\]
which is a saddle point, where the magnification is $\mu_2 = 1-\theta_E / \beta$. We also see that the image separation is given by $\abs{\theta_1 - \theta_2} = 2\theta_E$, which is constant for all source positions within $\theta_E$.

\begin{figure}[h!]
 \centering
 \includegraphics[width=8cm,keepaspectratio=true]{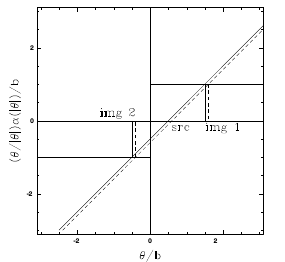}
 \caption{Graphical solution-plot for the SIS with two images. The horizontal solid lines represent the solutions, the slanted solid and dashed line represent the position and size of the source which gives the position and magnification of the images, represented by the vertical solid and dashed lines. From \cite{Schneider2006}.}
 \label{fig:SIS-two-image}
\end{figure}

This is illustrated in a graphical solution-plot in Figure \ref{fig:SIS-two-image}. Here we see two horizontal solid lines representing the possible solutions for that particular model. The slanted solid line presents the position of the source, where the size of the source is represented by the distance between the slanted solid and dashed line. From that particular position of the source, we get two solutions ie. two images, presented by the vertical solid lines. Like with the point-mass profile, the distance between the vertical solid and dashed lines represents the magnification of the images. We can see here that the distance between the images is always the same $(2\theta_E)$, unlike the point-mass lens.

\begin{figure}[h!]
 \centering
 \includegraphics[width=8cm,keepaspectratio=true]{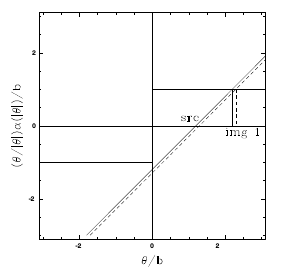}
 \caption{Graphical solution-plot for the SIS with a single image. The horizontal solid lines represent the solutions, the slanted solid and dashed line represent the position and size of the source which gives the position and magnification of the image, represented by the vertical solid and dashed lines. From \cite{Schneider2006}.}
 \label{fig:SIS-one-image}
\end{figure}

If we have a situation where $\beta > \theta_E$, then we will only have one solution, which is the minima solution that tells us that the image will be placed on the same side the lens as the source. This can also be illustrated with a graphical solution, which can be seen in Figure \ref{fig:SIS-one-image}. We have the same parameters as with the two-image solution, only we see just one image. The magnification is the same as the with the minima solution.

\subsection{Revisiting parity}
We can now expand the understanding of the magnification a bit, by taking a closer look at the parities. As mentioned previously (Sect. \ref{subsec:conv_shear}) we have two eigenvalues $a_i$. The sign of the eigenvalues do not only define whether we are having a minimum with two positive eigenvalues $(++)$, a saddle point with one positive and one negative eigenvalue $(-+)$ or a maximum with two negative eigenvalues $(--)$. The sign of the eigenvalues also defines the orientation of the images. 

\begin{figure}[h!]
 \centering
 \includegraphics[width=8cm,keepaspectratio=true]{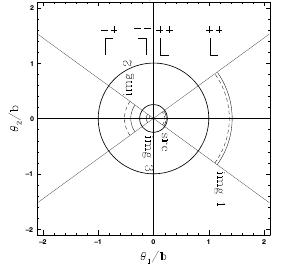}
 \caption{The parity and magnification for a three-image solution to the Moore profile. The size of the images describes the magnification and the signs of the eigenvalues describe the tangential and radial orientation of the images. From \cite{Schneider2006}.}
 \label{fig:eigvalue-parity}
\end{figure}

If we introduce the tangential orientation as the tangent to the critical line and the radial orientation as the normal to the critical line, we can define the tangential eigenvalue as the first $(a_1)$ and the radial eigenvalue as the second $(a_2)$ where the eigenvalues are $(a_1,a_2)$. If we consider a three image solution to the non-singular \emph{Moore} profile $(n=3/2)$, we can get a better understanding on how the parities affect the images. First we define the orientation and shape of the source as having the eigenvalues $(++)$. In Figure \ref{fig:eigvalue-parity} we see that for a minimum $(++)$ the image will have the same tangential and radial orientation as the source and be highly magnified. For a maximum $(--)$ the image is mirrored both radially and tangentially and is highly demagnified. For the saddle point we see that the image can either be mirrored radially or tangentially. In this particular case $(-+)$ the image is only mirrored 
tangentially. The saddle point is magnified, although less than the minimum.

\subsection{Non-Symmetric Lenses}
Even though the axisymmetric models can give a simple and reasonable description of the characteristics of a given lensing configurations, in order to get a more realistic idea about these characteristics, we have to use more complex mass profiles. Especially since we rarely can allow ourselves to exclude the angular structure of the gravitational potential \cite{Schneider2006}. This will however also mean that the models may no longer have any analytical solution. 

One way to include the angular perturbations is to generalize an axisymmetric profile to an elliptical profile. For a lens with an axis ratio of $q$ we can have an ellipticity of $\varepsilon = 1-q$ or an eccentricity of $e = (1-q^2)^{1/2}$. What counts here is the elliptcity of the gravitational potential rather than the surface density \cite{Schneider2006}. 

For instance, the generalization of the SIS, the singular isothermal ellipsoid (SIE), is described by replacing the radial coordinate $\xi$ with $\zeta = \sqrt{q^2\xi_{1}^{2}+\xi_{2}^{2}}$ \cite{Grillo2007}. This gives us the ability to include the angular structure.

Another way is to introduce some form of external tidal perturbations from nearby objects. Since galaxies are typically not isolated, but members of groups or clusters, the member galaxies and/or the dark matter halos from the group or cluster can break the symmetry of the main lens. Since we are interested in the astrophysical cases, we can assume that the perturbing gravitational field does not change much over the length scale of the main lens. At the lowest order, the perturber will add a uniform sheet of matter and an external shear. So, in lensing studied we will usually accompany a simple model, like the SIS, with an external shear (SIS+ES) \cite{Grillo2007}, but the mass ellipticity and the shear are usually difficult to disentangle. Likewise we can introduce an external shear to the SIE model (SIE+ES), but this model is often to degenerate to be properly constrained \cite{Grillo2007}.

One of the consequences of breaking the symmetry of the gravitational potential (of the lens) is that the centrally caustic point will be extended into an extended curve with cusps, most likely an astroid caustic. So, contrary to the axisymmetric (singular) models, a point-like source inside this astroid caustic will produce five images. Further, since axisymmetric models with monotonically decreasing densities can produce three images at most and systems may have four (visible) images, we can see that non-symmetric models are necessary to explain observations \cite{Grillo2007}.

%% file: clusters.tex
\chapter{Clusters of Galaxies}\label{chap:clusters}
In this Chapter we will introduce a vital part of this thesis, clusters of galaxies. In Section \ref{sect:what_are_clusters} we will give a short introduction into the nature of clusters of galaxies. In Section \ref{sect:why_are_clusters_interesting} we will show why clusters of galaxies are interesting, with respect to cosmology. In Section \ref{sect:our_cluster} we will introduce the cluster we model and analyse in this thesis, \macs.

\section{What are Clusters of Galaxies}
\label{sect:what_are_clusters}
\begin{figure}[!htb]
 \centering
 \includegraphics[width=0.49\textwidth,keepaspectratio=true]{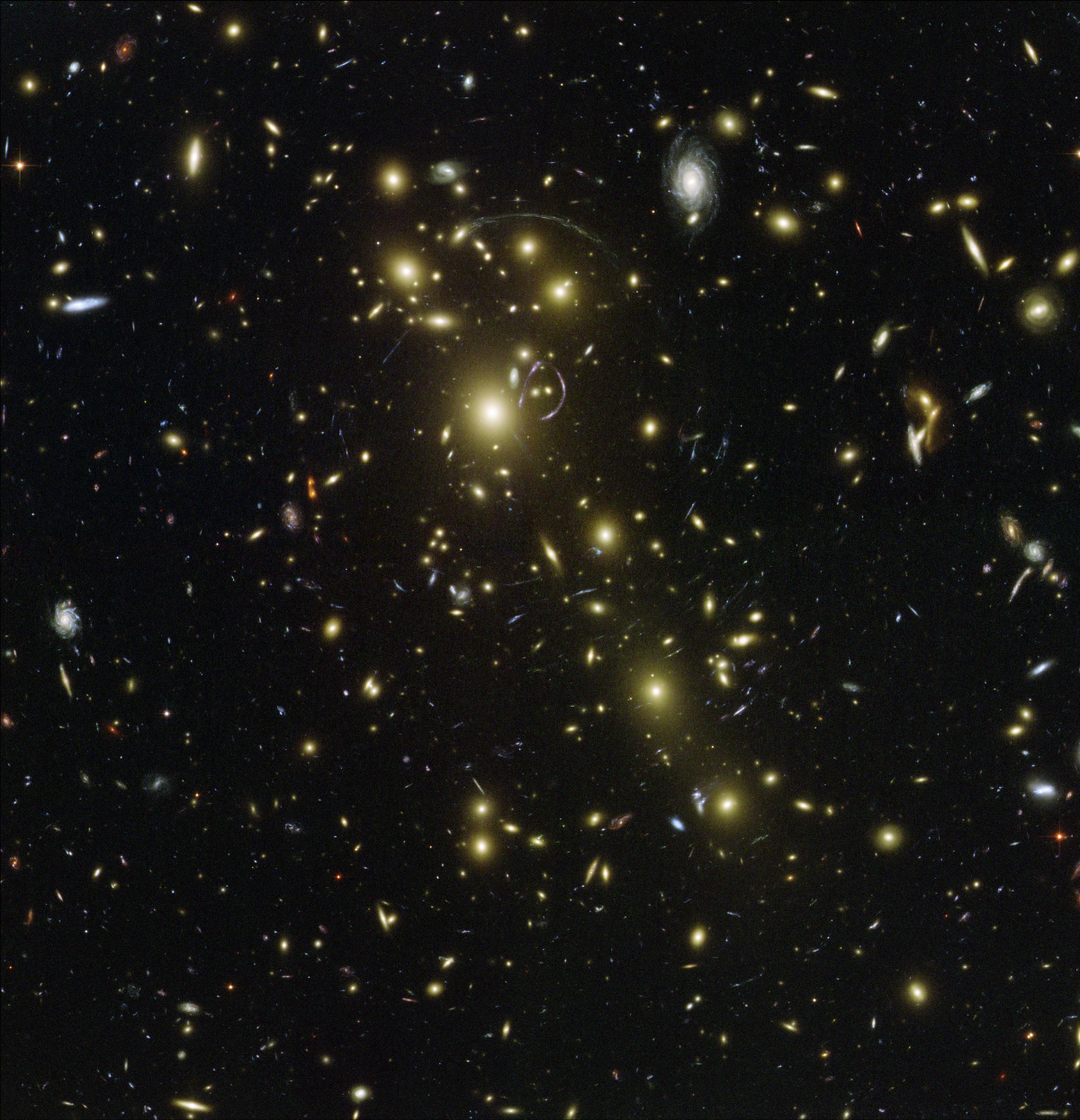}
 \includegraphics[width=0.49\textwidth,keepaspectratio=true]{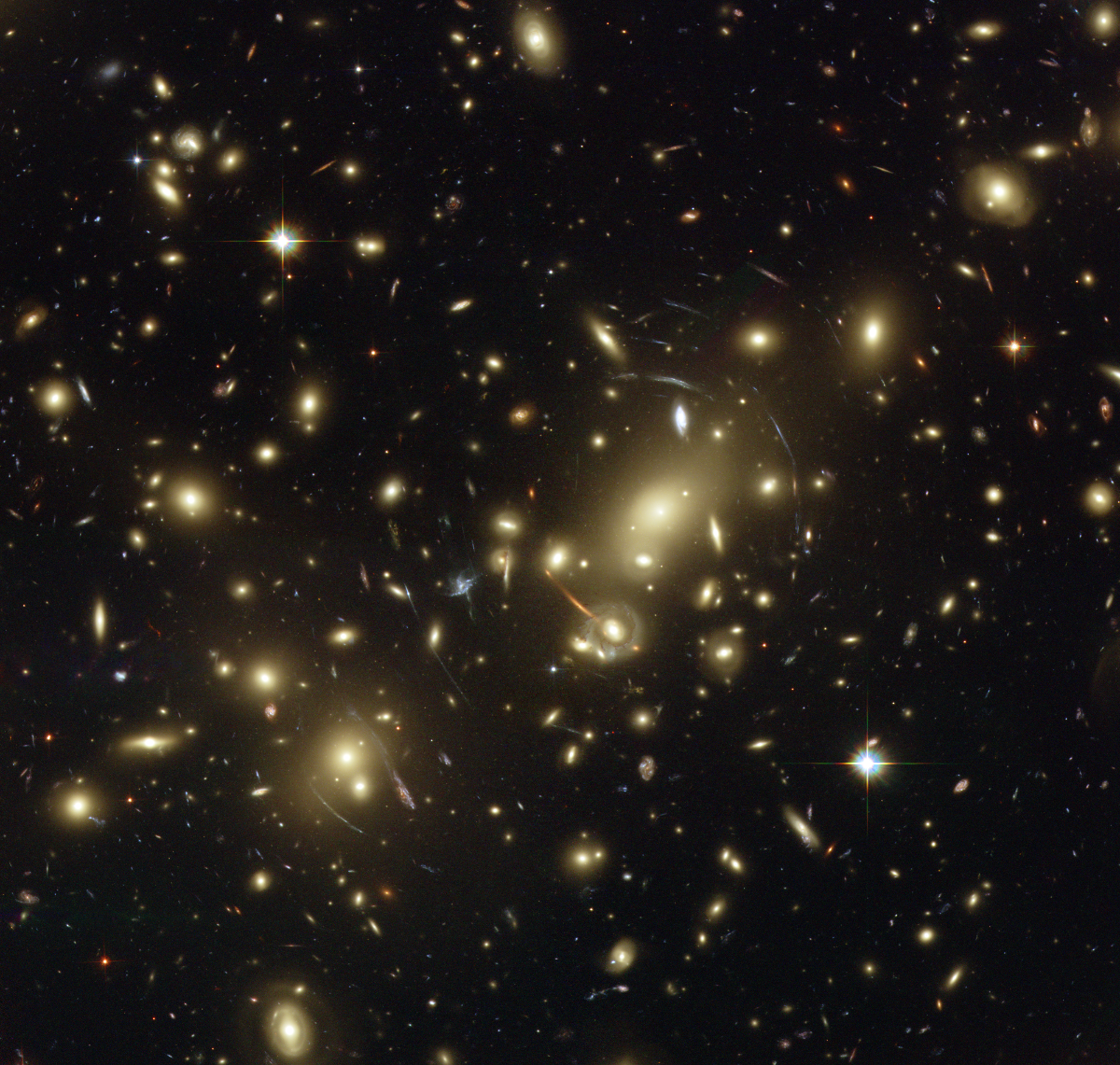}
 \caption{Color images of Abell 1703 cluster (right) and Abell 2218 (left). Abell 1703 is an example of a relaxed unimodal cluster while Abell 2218 is an example of a bimodal non-relaxed clusters. From the Hubble Legacy Archive.}
 \label{fig:clusters_in_color}
\end{figure}

Cluster of galaxies, henceforth clusters, can in general be defined as a gravitationally bound system \cite{Mo2010} consisting of galaxies, intracluster gas (Intracluster Medium - ICM) and dark matter. The galaxies themselves primarily consists of dark matter, then followed by interstellar gas (Interstellar Medium - ISM) and finally stars and planets \cite{Schneider2015}. Two examples of clusters are shown in Figure \ref{fig:clusters_in_color} where we see Abell 1703 and Abell 2218.

\subsection{The Definition of Clusters}
Classically, clusters have been defined by looking at the surface number density $\sigma$ over the uniform background number density $\sigma_{bg}$ \cite{Bahcall1977}
\[
 \langle \sigma/\sigma_{bg} \rangle \leq N
\]
where $N$ is specified manually. Obviously, setting $N$ too low will result in virtually all galaxies belonging to extremely large groups and setting $N$ too high will result in virtually no galaxies belonging to groups, except for a very few, tightly related galaxies. In compiling catalogues, one can either use an extent growing  with a fixed linear scale and a minimum density within it (Abell) or a variable growing extent determined directly by the observed density at its border (Zwicky Catalog of Galaxies and Clusters of Galaxies - CGCG). Differences in methods are also evident, which is why the Zwicky CGCG contains fewer rich systems than those of Abell \cite{Bahcall1977}. So, although quantitative methods do exist to distinguish clusters from field galaxies, the classical approach is in some degree, a matter of taste.

To simplify and standardize the selection criteria, we can simply say that we have a group of galaxies if it consists of $N \lesssim 50$ members within a sphere of diameter $D \lesssim 1.5h^{-1}\maths{Mpc}$ and a cluster if it consists of $N \gtrsim 50$ members within a sphere of diameter $D \gtrsim 1.5h^{-1}\maths{Mpc}$ \cite{Schneider2015}, where $h = 0.7$. Everything else can be considered field galaxies.

Clusters are also defined by the richness, which is a measure of the number of member galaxies within a given radius measured from the center. This means that it is also a measure of the mean number density of galaxies. The richness vary hugely from the rich Coma cluster containing thousands of member galaxies to low-density groups like our Local Group. The estimated number of galaxies within a cluster depends obviously on the assumed extent of the cluster, which is a non-trivial decision. \citet{Zwicky1968} defined the population as the number of galaxies visible on a red plate, corrected for the mean field count, that are located within a line defined by twice the field density. \citet{Abell1958}, on the other hand, introduced a method that is largely distance independent. He selected all galaxies within a fixed magnitude range and a circle of radius $R_A = 1.7/z\maths{arcmin} = 3h_{50}^{-1}\maths{Mpc}$. The selected population was then corrected by a background count in a nearby field. The classical indication is that only a few percent of all galaxies $(\sim 10\%)$ are members of rich clusters \cite{Bahcall1977}. It should be mentioned here that this selection criteria is only valid for clusters at low redshifts $(z \leq 0.1)$. At higher redshifts this is no longer valid.

Clusters can be classified in types and the usual approach resembles the method of classifying galaxies in sequences. Here we have early- to late type or equivalently, regular to irregular clusters. The early or regular clusters are believed to be systems more dynamically evolved compared to the late or irregular counterparts. Many of the cluster properties, like shape, presence of bright galaxies and X-ray emission, are correlated with the position in the cluster sequence \cite{Bahcall1977}. One of the early systematic classifications divided clusters into compact, medium-compact and open systems. Here compact clusters has a single outstanding concentration of bright members with at least ten galaxies in the neighbourhood. A medium-compact cluster has either single concentration, where at least ten galaxies are separated by more than their own diameter or several apparent concentrations. An open cluster has no apparent concentration. Comparing this classification with the regular-irregular scheme, we find that regular clusters have a population of at least $10^3$ in the brightest six magnitude range, high central concentration and circular symmetry. Irregular clusters, on the other hand, show neither circular symmetry nor central concentration. The distinction is also apparent with respect to the galaxy content. Regular clusters mainly contain E and S0 galaxies, irregular clusters contain galaxies of all types, with a substantial amount of spirals and irregular galaxies \cite{Bahcall1977}. 

Another classification scheme, developed by Rood \& Sastry, divide the classification into cD-, B- (binary), L- (line), C- (core), F- (flat) and I-Type (irregular):
\begin{itemize}
 \item cD-Type: The cluster is dominated by a cD galaxy (BCG).
 \item B-Type: The cluster is dominated by a bright "binary"\footnote{Not gravitational bound binary, like binary stars} or dual system.
 \item L-type: Three or more of the ten brightest members are arranged in a line.
 \item C-Type: At least four of the ten brightest members are located with comparable separation in the core.
 \item F-Type: Several of the ten brightest galaxies are distributed in a flattened configuration.
 \item I-Type: An irregular distribution with no well-defined center.
\end{itemize}
They also find that the I types are the most frequent, followed by cD, F, C, B and L, respectively, where B and L occur equally frequently \cite{Bahcall1977}. The final classification scheme we wish to introduce, is the Bautz \& Morgan scheme, which simply divide clusters into three major classifications:
\begin{itemize}
 \item Type I: The cluster is dominated by a single, centrally located cD galaxy.
 \item Type II: The brightest members are intermediate in appearance between cD galaxies and normal giant ellipticals.
 \item Type III: The cluster contain no dominant galaxies.
\end{itemize}
This scheme also has two intermediates, Type I-II and Type II-III. As is apparent here, there is no single valid classification system, but there are similarities. All systems agree on the division between centrally dominated (cD or BCG) clusters, clusters in between and clusters with no apparent dominating galaxies. 

Another important aspect of clusters are their sizes. As mentioned before, one way to define the size of a cluster is to use the selection limit, but since this is a parameter not directly related to the cluster itself, another method is needed. The size of a cluster is a matter of definition \cite{Bahcall1977} since clusters do not exhibit a sharp edge. One extreme view is that each cluster merges into the low-density envelopes of other clusters. Classically, cluster sizes have usually been defined as a gravitational radius, which is the radius at which the gravitational energy approximately equals the kinetic energy of a galaxy moving in the cluster and a core radius, which is the innermost characteristic size of the cluster, for which several definitions has been proposed. One method has been to fit the observed density profile to an \emph{bounded} isothermal model, but a systematic study of 15 rich clusters \cite{Oemler1974} has suggested a core radius defined by $R_c = (0.25\pm 0.04) \times h_{50}^{-1}\maths{Mpc}$. As mentioned before, the typical distinction now, are diameters $\gtrsim 1.5h^{-1}\maths{Mpc}$.

Lastly we will also mention the masses of clusters. Clusters are some of the most massive gravitationally bound structures in the universe, with typical values of $M \gtrsim 3 \times 10^{14}\maths{\MSun}$ for massive clusters and $M \sim 3 \times 10^{13}\maths{\MSun}$ for groups. The total mass ranges $10^{12}\maths{\MSun} \lesssim M \lesssim 10^{15}\maths{\MSun}$ \cite{Schneider2015}. Stars (and planets, moons etc.) are believed to constitute $\lesssim 5\%$ of the total mass in clusters \cite{Schneider2015}.

\subsection{Dark Matter in Clusters}
One of the key questions with respect to clusters was the missing mass problem \cite{Schneider2015}. If one finds the characteristic mass of a cluster $m \sim M / N$ where $M$ is the total mass of the cluster and $N$ is the total number of galaxies in the cluster or the mass-to-light M/L ratio
\[
 \left( \frac{M}{L_{tot}} \right) \sim 300h \left( \frac{\MSun}{L_{\Sun}} \right)
\]
where $L_{\Sun}$ is the total optical luminosity of the cluster galaxies, one finds that the characteristic mass is very high $(m \sim 10^{13}\maths{\MSun})$ or that the M/L ratio of clusters greatly exceeds the M/L ratio of early-type galaxies. This led \citet{Zwicky1933} to conclude that clusters had to contain vast amounts of \textit{Dunkle Materie} or Dark Matter, which is defined as invisible non-collisional and non-luminous matter halos enveloping both galaxies and cluster, only interacting through the force of gravity. This is confirmed by both dynamical studies \cite{Balestra2016}, X-ray studies \cite{Ogrean2015, Balestra2016} and both weak- \cite{Umetsu2014} and strong lensing \cite{Zitrin2013, Johnson2014, Grillo2015} studies. 

\subsection{Hot Gas in Clusters}
\begin{figure}[!htb]
 \centering
 \includegraphics[width=0.9\textwidth,keepaspectratio=true]{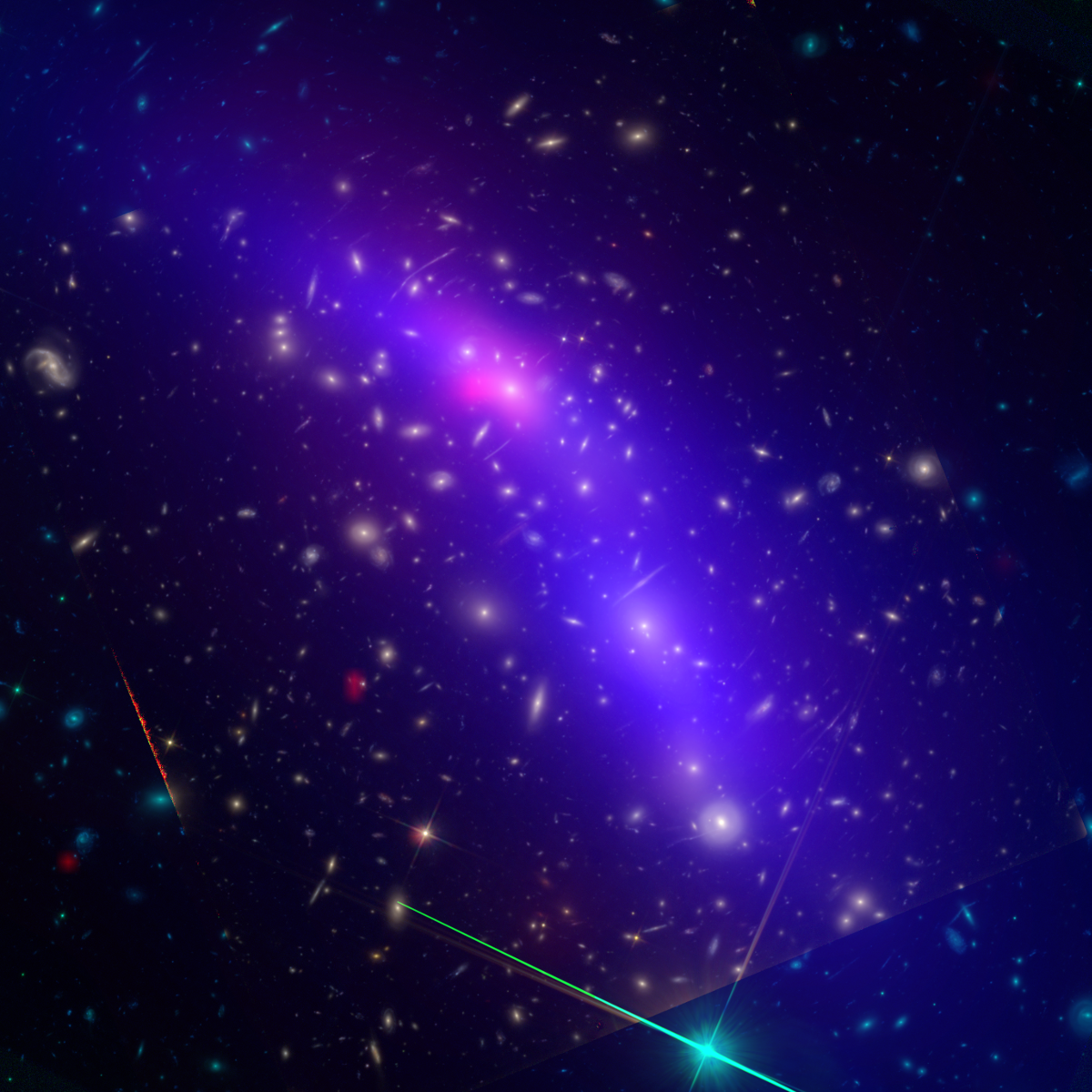}
 \caption{Color image of \macs with X-ray gas emission in red and strong lensing mass estimates in blue, superimposed.}
 \label{fig:macs_gas_mass}
\end{figure}

Another aspect of clusters are their content of hot gas $(T \sim 3 \times 10^7\maths{K})$. Even though Dark Matter is the main constituent of the total mass in clusters $(\sim 80\%)$, the hot gas is believed to comprise $\sim 15\%$ of the total mass, making it a significant portion of the total mass \cite{Schneider2015}. The hot gas can be observed in the clusters due to the high-energy X-ray emission at energies $k_BT \sim 5\maths{keV}$ which is also the typical temperature measurement scale. In Figure \ref{fig:macs_gas_mass} we show a color image of \macs with the hot gas emission superimposed as red and the projected mass estimate from strong lensing, superimposed as blue.

\begin{figure}[!htb] 
 \centering
 \includegraphics[width=0.9\textwidth,keepaspectratio=true]{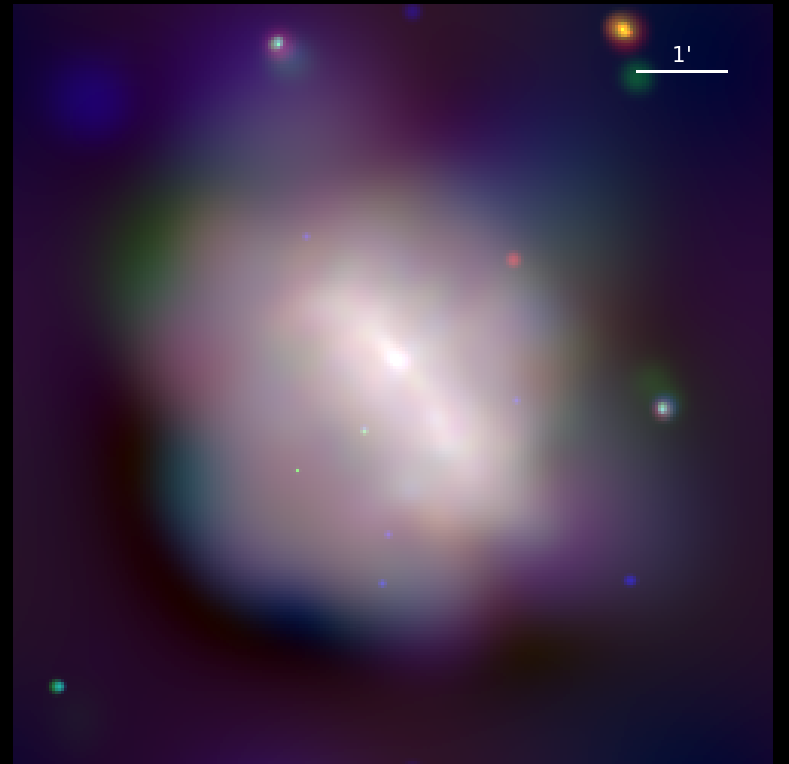}
 \caption{Color image of the X-ray emission from \macs where the energy-levels are represented as colors. Red represents $0.5-1.2\maths{keV}$, Green represents $1.2-2\maths{keV}$ and Blue represents $2-7\maths{keV}$.}
 \label{fig:color_xray}
\end{figure}

We know that the X-ray emission comes from hot gas in the ICM and not individual galaxies, since it is spatially extended. This can be seen in both Figure \ref{fig:macs_gas_mass} and Figure \ref{fig:color_xray}, where in the latter we show a color image of the hot gas emission from \macs in different energy-levels. Here we clearly see that the hot gas is spatially extended with very little emission from the individual galaxies, except the two brightest ones. 
The X-ray emission process is optically thin bremsstrahlung emission due to the acceleration of electrons in the Coloumbfield of protons and atomic nuclei \cite{Schneider2015}.

Regular clusters show a smooth brightness distribution centered on the optical center (typically a BCG), which then decrease outwards. Regular clusters also have high X-ray brightness and high temperatures. Irregular clusters may have several brightness maxima, typically centered on cluster galaxies or subgroups of cluster galaxies.

In this regard, a more recent method to detect and classify clusters was conducted by selecting via X-Ray emission from intracluster gas. By using data from the ROSAT All-Sky Survey (RASS), \citet{Ebeling2001} selected $850$ galaxies at all redshifts and specifically, $101$ clusters at $z > 0.3$, which was the primary goal. This project was named MAssive Cluster Survey (MACS) from which the galaxy used in this project originates. At present time, the MACS project has led to the discovery of $124$ confirmed clusters at $0.3 < z < 0.7$.

Further studies has used the X-ray emission to estimate the mass of clusters \cite{Balestra2016} and to determine the dynamical status of clusters (pre- or post-merging) \cite{Ogrean2015}. 

\subsection{Other Ways to Detect Clusters}
One particular interesting way to detect clusters is the Sunyaev-Z'eldovich (SZ) effect. This effect is comes from the inverse Compton scattering of the photons from the Cosmic Microwave Background radiation (CMB) which is scattered by the electrons is the hot ICM gas. The probability of scattering, or cross section, is very low, but nevertheless, the effect is observable \cite{Schneider2015} and catalogs of clusters has been produced by the observation of this effect \cite{Allen2011}.

Although the scattering of the photons from the CMB means that a given photon will move away from us, statistically another photon will be scattered towards us. This means that the total number of photons are preserved, but the energy of the photons will change, by the transfer of energy from the hot ICM to the CMB photons. 

The SZ can be spatially resolved using interferometry, which means that the spatially extent and temperature of the hot ICM can be measured. The SZ effect can also be used to determine the distance to clusters independent of the redshift \cite{Schneider2015}.

\section{Why are Clusters of Galaxies actually interesting?}
\label{sect:why_are_clusters_interesting}
One of the greatest mysteries of contemporary astrophysics is to answer the questions regarding the nature and behaviour of dark matter and dark energy. Part of these answers are related to the values of the cosmological parameters, since these quantify the expansion of our universe and rule the formation of structures \cite{Mo2010, Allen2011}.

One method to explore the values of the cosmological parameters, is to investigate the nature of clusters. Clusters can be viewed as buoys for the density peaks in the large-scale matter density \cite{Allen2011} and therefore knowledge of the current structure of clusters can provide insight into the formation of structure in the universe on large scales. The cosmological evolution of clusters are directly related to the growth of cosmic structure \cite{Schneider2015}. One particular example is to look at merging clusters which provide the means to test physical models of the nature of dark matter \cite{Allen2011}. Clusters provide multiple observable signals across the electromagnetic spectrum. X-Rays from bremsstrahlung in the ICM, stellar and intracluster light in optical and near infrared, inverse Compton scattering at millimetre wavelengths and finally we have gravitational lensing, which gives a unique way to determine the total mass distribution in clusters \cite{Allen2011}.

In the current understanding of the universe, the consensus concordance model $\Lambda$CDM, it is believed that dark matter is comprised of non-relativistic, weakly interacting particles, only observed through their interaction via gravity \cite{Mo2010} and that dark energy is associated with a small, non-zero vacuum energy equivalent to the cosmological constant \eqref{eqn:EFE}. Dark energy could, however, also be a light scalar field that evolves over cosmic time or an apparition that actually signifies that the Einstein Field Equations \eqref{eqn:EFE} are not valid over cosmic length- and timescales \cite{Allen2011}. 

A way to test our current understanding of the universe using massive clusters could be to look at the three present-epoch densities for Baryons $\Omega_B$, Dark Matter $\Omega_{DM}$ and Dark Energy $\Omega_{\Lambda}$, where $\Omega_B + \Omega_{DM} = \Omega_M$ \cite{Allen2011} and the density is defined as $\Omega_X = \frac{\rho_X}{\rho_{crit}}$. Another way is to investigate the dark energy equation of state parameters $w_0$ and $w_a$, where $w(a) = w_0 + w_a(1-a)$ is the linearly evolving dark energy equation of state. These parameters define the fate of the universe, where $w > -1$ means that the Dark Energy density will decrease and $w < -1$ means an increasing energy density. Usually we assume a flat universe with means $\Omega_M + \Omega_{\Lambda} = 1$ and $w(a) = -1$ \cite{Allen2011}. These quantities can in principle be measured in strong lensing analyses of clusters \cite{Jullo2010}.


An important step in estimating the cosmological parameter values, also in our thesis, is to estimate the total mass of the cluster. One way to estimate the total mass, is using the X-ray emission. In \citet{Evrard1996} we find a method used to calculate the total mass from the X-ray data. The mass estimation, assuming hydrostatic equilibrium and isothermality, is defined as
\[
 \label{eqn:X-ray-mass}
 M(r) = -\frac{kT(r)}{G\mu m_p}r \left[ \frac{d\log{\rho(r)}}{d\log{r}} + \frac{d\log{T(r)}}{d\log{r}} \right]
\]
where $k$ is Boltzmann's constant, $T(r)$ is the measured gas temperature and $\mu m_p$ is the mean molecular weight of the gas.

Another way is to use dynamics. One such way, MAMPOSSt, described in \citet{Mamon2013}, use the Jeans equation
\[
 \label{eqn:MAMPOSSt}
 \frac{d(\nu \sigma_{r}^{2})}{dr} + 2\beta \frac{\nu \sigma_{r}^{2}}{r} = -\nu(r) \frac{GM(r)}{r^2}
\]
where $\nu(r)$ is the tracer density profile, $\sigma_{r}^{2}$ is the radial velocity anisotropy and $\beta$ is defined as
\[
 \beta(r) = 1- \frac{\sigma_{\theta}^{2}(r) + \sigma_{\phi}^{2}(r)}{2\sigma_{r}^{2}(r)}
\]
where we must have $\sigma_{\theta} = \sigma_{\phi}$ in the case of spherical symmetry \cite{Mamon2013}, which is assumed. Observations of clusters do however indicate various degrees of ellipticity, especially in merging clusters. 

Gravitational lensing is completely free of assumptions regarding the dynamical state of the clusters, but triaxiality can introduce scattering in the deprojected mass estimates \cite{Allen2011}. 

\section{\macs}
\label{sect:our_cluster}
As mentioned previously, we can view clusters as large Dark Matter (DM) halos containing numerous smaller cluster member halos. Classically, the approach has been to distinguish between unimodal relaxed clusters \cite{Limousin2008, Richard2009} and bimodal non-relaxed clusters \cite{Kneib1996, Ardis2007}. This is usually also confirmed by X-Ray imaging of the hot gas, since peaks in gas emission are believed to coincide with the DM peaks. Due to advances in the ability to select and confirm both multiple images and cluster members, which have increased the number of both significantly, we now have evidence of cluster with tri, quadra, even pentamodal structure, which is not directly evident from the gas emission \cite{Jauzac2014, Richard2014, Jauzac2015, Balestra2016, Caminha2016}.

One of the clusters derived from the work by \citet{Ebeling2001} is the cluster \macs. It is situated at redshift $z = 0.396$ and is comprised of almost $1000$ cluster members \cite{Umetsu2014}. The core of the cluster is characterized by two main parts evident by the two bright cluster galaxies \cite{Zitrin2013} with $193$ cluster members \cite{Caminha2016} and two bright clumps or subclusters in X-ray emission (see Fig. \ref{fig:macs_gas_mass}). One subcluster is situated at NE and the other at SW and hence are named thereafter. They are both situated close to the respective BCGs. The cluster is also characterized by a bright foreground galaxy at $z = 0.114$ which is not part of the cluster, but still have significant effects on strong lensing multiply images \cite{Grillo2015}. 

\macs seems to be best described as a bimodal non-relaxed cluster \cite{Zitrin2013, Ogrean2015, Grillo2015} undergoing merging in either a pre-merging or post-merging phase, where \citet{Jauzac2015} suggested two possible scenarios. A pre-merging scenario where the SW subcluster comes behind the NE subcluster and are seen at first passages. A post-merging scenario where the SW subcluster approached the NE subcluster from above and are now is now seen near its second core passage. \citet{Ogrean2015} find evidence via X-ray and radio emission for a pre-merging scenario. 

Newer analyses with better constraints due to a vast increase in multiple images, and a better selection of cluster members, have indicated that \macs has a third halo situated North-East relative to the NE subcluster \cite{Caminha2016} which is consistent with a concentration of galaxies, indicating that the NE subcluster is merging with this third substructure \cite{Ogrean2015}. This further favours a pre-merging scenario. This substructure is however not directly observable in the X-ray data (see Fig. \ref{fig:color_xray}) which indicates that it does not contain a hot gas halo, thus has far less mass than the two main halos. This is also indicated by the \citet{Caminha2016} model where while the third halo have a comparable core radius to the NE halo, the velocity dispersion is much lower. This could indicate a trimodal structure.

\begin{figure}[h!t]
 \centering
 \includegraphics[width=0.49\textwidth,keepaspectratio=true]{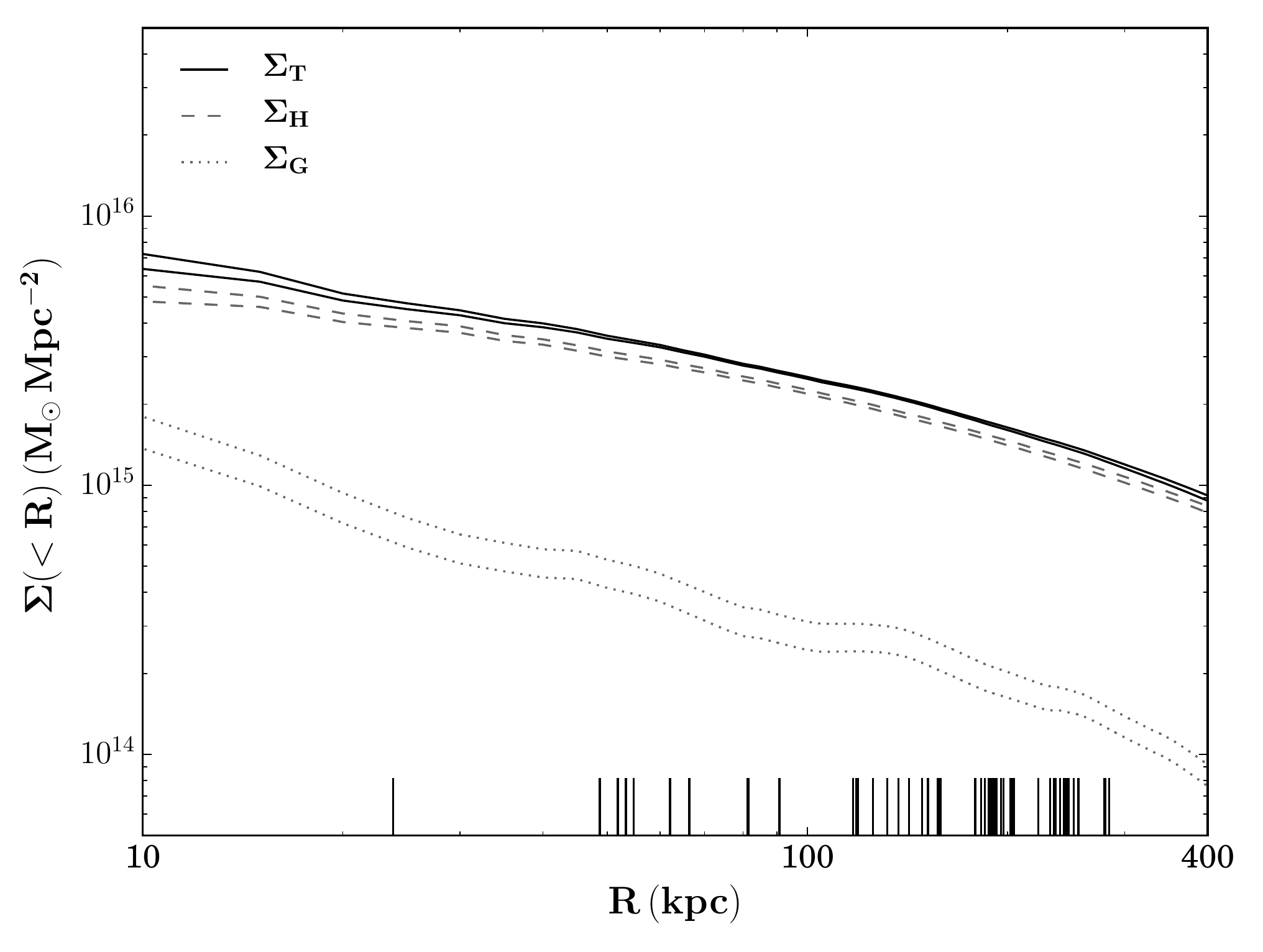}
 \includegraphics[width=0.49\textwidth,keepaspectratio=true]{2dpie_Grillo_averageMassDensity_1sigma.pdf}
 \caption{Average surface mass-density maps of Abell 1703 (derived from \citet{Richard2009}) and \macs. The flat inner core of \macs is clearly evident and opposed to the more cuspy inner core found in Abell 1703.}
 \label{fig:inner_cores}
\end{figure}

Analyses of the mass distribution in \macs shows that the cluster contains a flat inner core \cite{Jauzac2014, Grillo2015} which is typical for merging (bimodal) clusters. Unimodal relaxed clusters typically display cuspy inner cores \cite{Limousin2008, Richard2009}. This is evident in Figure \ref{fig:inner_cores} where we clearly see the flat inner core of \macs opposed to the more cuspy inner core of Abell 1703. The results suggesting a trimodal structure of \macs does not change the indication of a flat inner core.

Even though \macs was only discovered during the last decade \cite{Ebeling2001} it has already been the subject of several  strong lensing studies \cite{Zitrin2013, Johnson2014, Jauzac2014, Richard2014, Grillo2015, Ogrean2015, Jauzac2015, Balestra2016, Caminha2016} and yet, there are still new discoveries to be conducted in this cluster. Not the least, future methods in modelling gravitational lensing using more than one lensing plane. 

Since the bright foreground galaxy in \macs influence the shape and position of the lensed images, any attempt to estimate the cosmological parameters directly, via optimizing them in a single-lensing plane software, will give unreliable results. Therefore we have chosen to investigate a method to estimate the cosmological parameters using mass estimates from X-ray, dynamics and strong lensing, in order to circumvent this dependence on the foreground galaxy. In this thesis we will present the first step.

%% file: data.tex
\chapter{Observations and Data}
\label{chap:obs-data}
In this chapter we will present the data we have used for this thesis. The primary data source is the Hubble Space Telescope legacy survey \emph{Cluster Lensing and Supernova Survey with Hubble} or CLASH \cite{Postman2012} and the ESO VLT extension of that survey, called CLASH-VLT \cite{Rosati2014}. We have also used data from the more recent follow-up survey, the Hubble Frontier Fields or HFF, and X-Ray data from the Chandra Space Telescope. 

This chapter is in overall a summary of \citet{Postman2012} and \citet{Rosati2014}.

\section{CLASH}
\label{sect:clash}
The Hubble Space Telescope (HST) has provided us with high resolution, deep data on various massive cluster and lensed sources and has played a key role in providing confirmation and constraints on the properties, of dark matter (DM). Especially observations using the Advanced Camera for Surveys (ACS) have provided the best and highest resolution maps of DM distribution in massive clusters. This is particularly important since clusters, by virtue of their position at the high end of the cosmic mass power-spectrum, provide a powerful way to constrain the frequency of high amplitude perturbations in the primordial density field \cite{Postman2012}.

\subsection{Scientific Justification}
N-Body large-scale structure formation simulation have shown that cold dark matter (CDM) dominated halos of all masses, evolve to have a so-called universal density profile that steepens with radius, for example the NFW profile \cite{Navarro1997}. The core densities or "concentrations" of both simulated and observed halos are shown to be highly dependent on the background density of the universe at the formation time. Halos that form later, including the most massive galaxy clusters, thus have the least relatively dense cores. The core density is measured as the concentration $c_{vir} = r_{vir} / r_{-2}$ ratio between the virial and inner radius
\footnote{The virial radius is defined as the radius at which the density is 200 times the critical density and for the NFW profile, the inner radius is equivalent to the scale radius $R_S$.}. 
The density slope of the fitted profile is here isothermal $\rho \propto r^{-2}$. Analyses of a wide range of radii will allow us to map the matter
\footnote{Primarily dark matter.}
profiles of the observed halos and measure their concentrations. 

The CLASH program has been particularly successful in mapping DM profiles in the cores of clusters, using multi-bandpass imaging for strong lensing analysis \cite{Postman2012}. Interestingly, the best studied galaxy clusters, using both strong- and weak lensing, have shown an overly high concentration for the cores (dense cores) compared with the results from N-Body simulations of halos with similar masses \cite{Broadhurst2008b, Sereno2010}. Similarly, clusters have also been shown to have relatively larger Einstein radii than expected \cite{Broadhurst2008}. This might be connected with the fact that the best studied clusters are some of the strongest gravitational lenses known. Such clusters are prone to have halos with higher concentrations, both intrinsically and as projected along the line of sight. To investigate further on these topics, the CLASH program has targeted a large and unbiased sample of clusters \cite{Postman2012}. 

In order to properly select a catalogue of background galaxies and avoid dilution of the weak-lensing signal with un-lensed foreground galaxies, multi-bandpass images are required. Although stacked weak-lensing analysis have been performed on larger samples of clusters, combining strong and weak lensing can provide the most accurate total mass estimates to be compared with those from other techniques. For instance, X-ray analyses can also provide an estimate of the cluster mass concentrations, but they are subject to uncertainties due to the assumption of hydrostatic equilibrium \cite{Postman2012}. At the time the paper by \citet{Postman2012} was written, only five massive clusters had been well-studied using both strong and weak lensing. With the CLASH data we now have 20 well-studied clusters \cite{Umetsu2014}.

\subsection{The goals of CLASH}
After the 2009 fourth service mission to HST, with the installation of the Wide Field Camera 3 (WFC3), the Hubble Multi-Cycle Treasury Program (MST) was initiated. Also, ambitious multi-cycle programs ($>500$ orbits) with broad scientific potential that could not be accomplished within the constraints of a single HST observation cycle, took full advantage of the upgraded and refurbished HST. CLASH was such a programme. The four primary goals for CLASH were
\begin{itemize}
 \item Measure the mass profiles and substructures of DM in galaxy clusters with unprecedented precision and resolution.
 \item Detect Type Ia supernovae (SNe Ia) out to redshift $z \sim 2.5$ in order to measure the time dependence of the dark energy equation of state and potential evolutionary effects in the SNe themselves.
 \item Detect and characterize some of the most distant galaxies yet discovered $(z >7)$.
 \item Study the internal structure and evolution of galaxies in and behind these clusters.
\end{itemize}

\subsection{Observational Method}
\begin{figure}[h!t]
 \centering
 \includegraphics[width=0.99\textwidth,keepaspectratio=true]{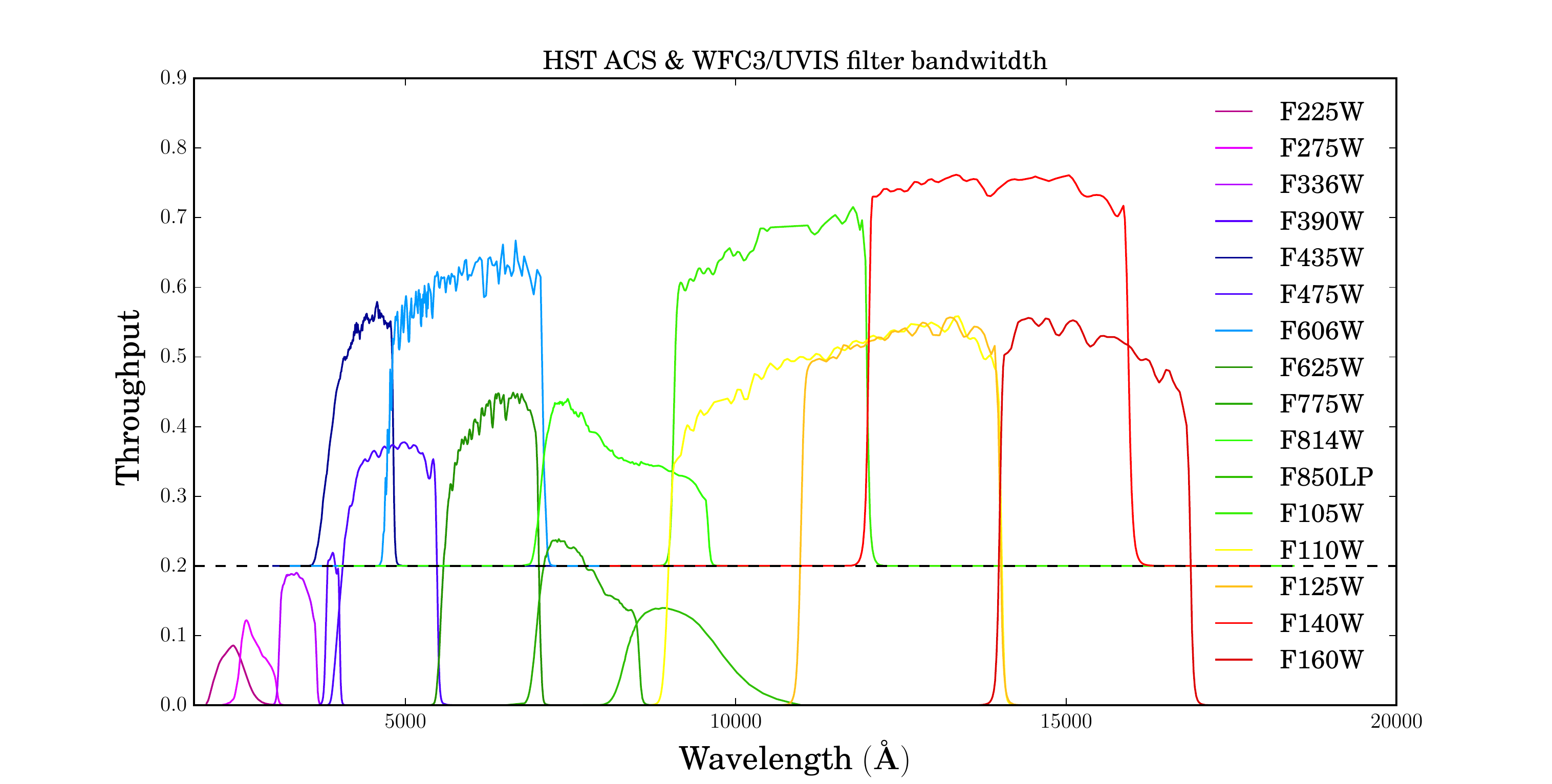}
 \caption{Troughput and bandpass for the individual filters. Note that this plot differs from that in \cite{Postman2012} because these throughputs are from the Hyperz package \cite{Bolzonella2000}. Some of the filters are offset by $0.2$ for clarity.}
 \label{fig:hst-bandpass}
\end{figure}

To reach this goal, CLASH targeted 25 massive galaxy clusters and imaged them in 16 broad-range bandpass filters using WFC3/UVIS, WFC3/IR and ACS/WFC. The range of those filters span $2000-17000\maths{\angstrom}$ (near-ultraviolet to near-infrared) as illustrated in Figure \ref{fig:hst-bandpass}. ACS was used in the $~4000-9000\maths{\angstrom}$ range because of its higher throughput efficiency and larger field-of-view. CLASH was allocated 524 orbits of observations time where 474 orbits were used for cluster and SN observations and 50 orbits were reserved for SN follow-up observations. The individual observations are listed in Table \ref{table:clash-obs}. 

\begin{table}[h!t]
 \centering
 \caption{Listings of the individual HST CLASH observations in camera and channel, filters, orbits, average exposure time and $5\sigma$ limits.}
 \begin{tabular}{c c c c c}
   \toprule
   Camera/Channel & Filter & Orbits & Avg. exp. time (s) & $5\sigma$ limit (AB) \\
   \midrule
   WFC3/UVIS & F225W  & 1.5 & 3558 & 26.4 \\
   WFC3/UVIS & F275W  & 1.5 & 3653 & 26.5 \\
   WFC3/UVIS & F336W  & 1.0 & 2348 & 26.6 \\
   WFC3/UVIS & F390W  & 1.0 & 2350 & 27.2 \\
   ACS/WFC   & F435W  & 1.0 & 1984 & 27.2 \\
   ACS/WFC   & F475W  & 1.0 & 1994 & 27.6 \\
   ACS/WFC   & F606W  & 1.0 & 1975 & 27.6 \\
   ACS/WFC   & F625W  & 1.0 & 2008 & 27.2 \\
   ACS/WFC   & F775W  & 1.0 & 2022 & 27.0 \\
   ACS/WFC   & F814W  & 2.0 & 4103 & 27.7 \\
   ACS/WFC   & F850LP & 2.0 & 4045 & 26.7 \\
   WFC3/IR   & F105W  & 1.0 & 2645 & 27.3 \\
   WFC3/IR   & F110W  & 1.0 & 2415 & 27.8 \\
   WFC3/IR   & F125W  & 1.0 & 2425 & 27.2 \\
   WFC3/IR   & F140W  & 1.0 & 2342 & 27.4 \\
   WFC3/IR   & F160W  & 2.0 & 4910 & 27.5 \\
   \bottomrule
  \end{tabular}
  \label{table:clash-obs}
\end{table}

In more details, the CLASH sample included 20 massive clusters from X-ray-based catalogues of dynamically relaxed systems. Sixteen of these were taken from the \citet{Allen2008} catalogue of massive relaxed cluster. The clusters in the CLASH sample have temperatures $T_X \geq 5\maths{KeV}$ and have a high degree of dynamic relaxation shown by \emph{Chandra X-Ray Observatory} observations. These observations also indicate that these clusters deviate very little from hydrostatic equilibrium. Most of the clusters are also smooth and mildly elliptical $(\mean{\varepsilon} = 0.19)$ according to the X-ray emission and have a BCG within a projected distance of $23\maths{kpc}$ of the X-ray centroid. These clusters are an important sample for DM distribution studies. Additional five clusters was selected solely on their properties as strong lenses $(\theta_E > 35\maths{arcsec})$ for $z_s = 2$, which allowed a detailed quantifying analysis of the lensing selection bias towards higher concentrations. \macs is one of those five. Strong lensing analyses on these clusters enable us to measure of some of the highest resolution DM maps \cite{Postman2012}. The primary motivation for the selection of these clusters was the increased likelihood of finding highly magnified high-redshift galaxies. The cluster sample is primarily drawn from the Abell and MACS catalogues.

As mentioned previously, a large sample of multiple images with accurate redshifts are crucial in order to break lensing degeneracies and for tightening the constraints on cluster mass profiles. Typical source magnitudes are $23 < I < 28$, so only the largest and brightest arcs yields spectroscopic redshifts when observed with even the largest ground-based facilities. By using the broad bandpass range from CLASH, the photometric redshift should be very accurate $(\Delta z \sim 0.02(1+z))$ for $80\%$ of the objects with F775W mag $<26$ (AB). More importantly, CLASH enabled acquirement of $~6$ times as many photometric redshifts, as spectroscopic, for objects at $z > 1$. It should be noted that these estimates are based on BPZ simulations.

The coverage of the 16 filters allowed the Lyman-limit features to be photometrically traced to $z \sim 1.5$ and Ly$\alpha$ to be detected out to $z \sim 10$. By including NUV photometry, the most common photometric redshift degeneracies between the Balmer break in $z \sim 0.2$ galaxies and the Lyman-break in $z \sim 3$ galaxies could be resolved.

\begin{figure}[h!t]
 \centering
 \includegraphics[width=0.5\textwidth,keepaspectratio=true]{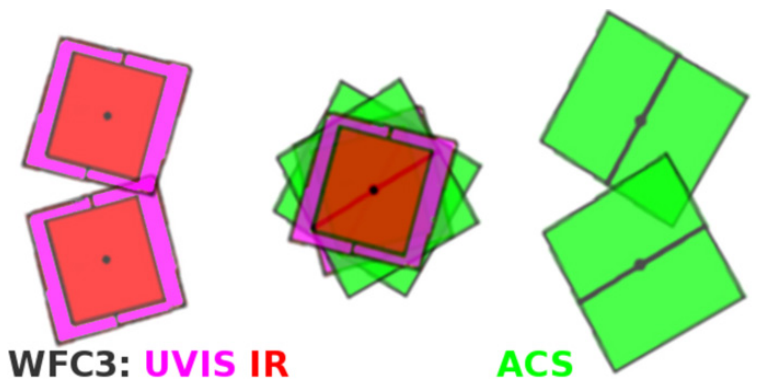}
 \caption{The different rotation angles for the CLASH HST observations. Each observation is rotated $\approx 30\degree$ and shifted so both instruments are in use at the same time, but one images the center field and the other the parallel field. From \cite{Postman2012}.}
 \label{fig:hst-rotation-angle}
\end{figure}

CLASH observed in two orientations $\approx 30\degree$ apart and with both instruments active at the same time, where one observed the cluster center and the other a parallel field. This is illustrated in Figure \ref{fig:hst-rotation-angle}. 

In each orbit, a compact four-point dither pattern was applied in order to achieve half-pixel sampling along both detector axes which served to improve the spatial sampling of the point spread function (PSF) and to help remove hot pixels and other detector imperfections that may not be accounted for in the calibration files. The improvement of the PSF was particularly important for the WFC3 with its relatively large pixel scale $0.128\maths{\arcsec \, pixel^{-1}}$. One other important advantage of observing using the roll-angle, was that the effect of the gap between the two ACS sensors (Fig. \ref{fig:hst-rotation-angle}) would be minimized to a small diamond at the center of rotation.

\subsection{Data reduction}
In order to get the mosaics and co-added images used for the bulk of the analyses, the \mosdrizzle pipeline was used to combine the FLT images, which have been traditionally reduced with bias and flats. Before combining the images with \mosdrizzle, the ACS/WFC data are corrected for bias-stripping and charge transfer efficiency (CTE) degradation effects. The CTE corrections reduces the effects of charges being trapped and trailed across the image during readout. A similar process has been developed for the WFC3/UVIS images. Trails from cosmic rays can leak into photometric apertures of non-detections and boost their observed flux. This can be mitigated by applying a more aggressive rejection of cosmic rays and their trails.

\mosdrizzle then carries out steps in order to align the exposures from the different camera/filter combinations from each visit and also across visits, using a combination of catalogue matching and cross-correlations \cite{Postman2012}. The catalogue shifts can solve for shifts and rotations, giving an accuracy of $0.1-0.2\maths{pixel}$ and shifts are then further refined using cross-correlations yielding an accuracy of $0.02-0.05\maths{pixels}$. 

The cosmic ray detection and removal routines are then used for the creation of a cosmis ray and bad pixel mask. These are then weighted according to the sky level in each input exposure, along with readnoise and accumulated dark current, to form an inverse variance image for each exposure. The drizzle combination use these inverse variance maps as weights, using a square kernel as well as a "pixfrac" parameter set typically to $0.8$. 

For each cluster, two set of mosaics was created. One with a resolution of $30\maths{mas \, pixel^{-1}}$ and one with a resolution of $65\maths{mas \, pixel^{-1}}$, for both ACS and WFC3 images. All images are oriented with North up.

\section{CLASH-VLT VIMOS Programme}
\label{sect:clash-vlt}
The CLASH-VLT VIMOS project was a large ESO programme initiated as an extension of the CLASH program. The goal was to carry out a comprehensive campaign to obtain spectroscopic data on 13 massive galaxy clusters in the southern sky, at a median redshift of $z = 0.4$ \cite{Rosati2014}. At the time the \citet{Rosati2014} paper was published, 95\pct of the observations was completed, yielding spectra for 500-1000 members pr. cluster and over 200 background lenses at $z < 7$. At completion, the programme should yield $~30.000$ spectra of which $~7000$ belongs to cluster members, giving us a long lasting legacy for studies of galaxy evolution in different environments.

\subsection{Scientific motivation}
Galaxy clusters have for a very long time served as a connection between astrophysics and cosmology. Their number distributions at varying masses is an important tool to constrain cosmological models, since their abundance with respect to redshift is extremely sensitive to both the underlying geometry of the Universe and the large scale structure growth \cite{Rosati2014}. The mass budget of typical massive galaxy clusters consists of $85-90$\pct dark matter and $10-15$\pct baryons, of which the major part is hot X-ray plasma. Only $1$\pct accounts for stars. Multi-wavelength investigations of meaningful samples of clusters will lead to a better understanding of the physical mechanisms driving the formation of galaxies and clusters, over a large range of environmental densities.

As mentioned before, Cold Dark Matter (CDM) drives the formation and dynamical evolution of structure, as shown in cosmological simulations \cite{Springel2005,Vogelsberger2014}. This leads to specific predictions in the mass density profile of DM halos from galaxy to cluster scale. Significant deviations from the theoretical predictions could suggest a different nature for dark matter which can be revealed in the inner, high-density core of clusters, where a non-collisionless behaviour of the dark matter could modify the inner slope of the mass distribution.

Such an investigation requires very accurate determination of the mass density profile of a representative sample of clusters with radii ranging from kpc to Mpc. This can only be achieved by using all all the available tools, namely gravitational lensing, galaxy dynamics and X-rays. These methods are sensitive to different radial ranges and hence, are prone to different systematic errors. For lensing we have structures along the line-of-sight, for dynamics we have substructures and the velocity anisotropy of orbits and for X-ray we have deviations from hydrostatic equilibrium. We also have that lensing results, for a handful of clusters, suggest that the mass concentration is significantly higher than expected \cite{Rosati2014}. For strong lensing to produce accurate total mass estimates in the inner regions of clusters down to galaxy scale, a sufficient number of multiple images with reliable distance information (i.e. spectroscopic) must be available \cite{Rosati2014}.

\subsection{Method}
Building on the CLASH survey, the VIMOS Large Programme (CLASH-VLT) was approved in period 86 to carry out an unprecedented spectroscopic campaign in the 14 CLASH clusters accessible from the Very Large Telescope. The main idea with the project was to provide a third dimension to the CLASH survey by finding a large sample of spectroscopically confirmed cluster members and background lensed galaxies, especially giant arcs and multiple images.

CLASH-VLT was granted 225 hours of observational time, distributed over 200 hours of multi-object spectroscopy and 25 hours of pre-imaging. The primary objective were:
\begin{itemize}
 \item Obtain spectroscopically confirmation of $\sim 500$ cluster members in each cluster out to at least twice the virial radius.
 \item Measure redshifts for over 200 lensed galaxies in the cluster cores, including several highly magnified galaxies out to $z\approx 7$ to provide confirmation of multiply imaged systems.
\end{itemize}

To obtain the spectra, the VIMOS low-resoution LR-blue grism was primarily used, but when lensed sources with high photometric redshifts were present, the medium-resolution (MR) grism was used instead. The coverage of approximately $10\maths{Mpc}$ at $z \sim 0.4$ by the VIMOS field-of-view and the multiplexing capabilities, proved to be excellent for this project. The Subaru Suprime-Cam was also used for the spectroscopic target selection and slit-mask design. 

Eight to twelve VIMOS pointings were used for each cluster, spanning an area of $15-20\maths{arcmin^2}$ with one quadrant locked on the cluster core to increase the total exposure time on faint lensed sources \cite{Rosati2014}. 

\subsection{Data reduction}
The data was reduced using the VIMOS Interactive Pipeline and Graphical Interface (VIPGI) software package \cite{Rosati2014}. A significant amount of time had to be invested in finding the correct position on the sky for each spectrum in an automated fashion. The risk of the VIPGI algorithm misidentifying an object, in a crowded field,  was not negligible. All of the data has been fully reduced and the positions and redshifts identified.

\subsection{Results}
\begin{figure}[h!t]
 \centering
 \includegraphics[width=0.8\textwidth,keepaspectratio=true]{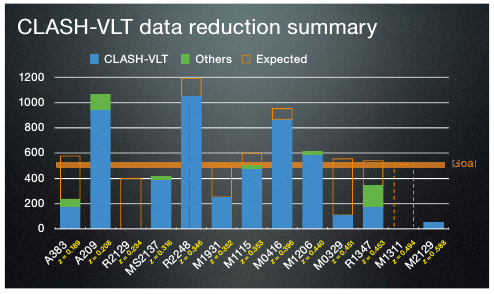}
 \caption{Data reduction summary for the CLASH-VLT showing the number of confirmed cluster members for each cluster. Notice that \macs has over 800 spectroscopically confirmed cluster member. From \cite{Rosati2014}.}
 \label{fig:clash-vlt-data-reduction-summary}
\end{figure}

In Figure \ref{fig:clash-vlt-data-reduction-summary} the success rate in identifying cluster members is shown. We see that the goal of 500 cluster members already at the time the article \citet{Rosati2014} was written had been achieved. For instance, \macs had over 800 indentified cluster members with more to come. 

More than thousand of these spectroscopic redshifts, with the bulk in the $z = 0.1-1.2$ range but also extending to $z \sim 5$, have been used to calibrate and check the accuracy of the CLASH-HST photometric redshift \cite{Jourvel2014}. These can be used as an important ingredient for strong lensing models.

\begin{figure}[h!t]
 \centering
 \includegraphics[width=0.8\textwidth,keepaspectratio=true]{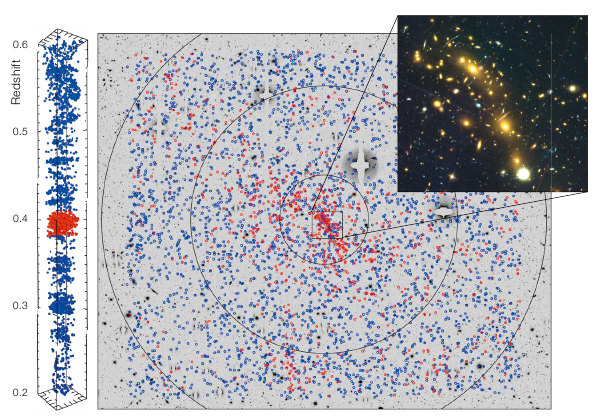}
 \caption{Spatial distribution of galaxies in \macs with reliable redshifts. The galaxy positions are overlaid on a Subaru R-band image. The red symbols indicate the 880 confirmed cluster members (galaxies with a restframe velocity within $\pm3000\maths{km s^{-1}}$ from $\median{z} = 0.396$) and blue are the other $3307$ galaxies over the z-range $0.02-4.15$. From \cite{Rosati2014}.}
 \label{fig:clash-vlt-spatial-dist}
\end{figure}

An example of the results for \macs is shown in Figure \ref{fig:clash-vlt-spatial-dist}. The final product for \macs which is included in the Hubble Frontier Fields survey (sect. \ref{sect:hff}), should yield almost $1000$ confirmed cluster members (Fig. \ref{fig:clash-vlt-data-reduction-summary}). 

\section{Hubble Frontier Fields}
\label{sect:hff}
The Hubble Frontier Fields (HFF) was a part of the Directors Discretionary Time and, like CLASH, a multi-cycle programme. The main goal of the HFF survey was to provide detailed deep imaging in 7 filters of 6 selected cluster with focus on the discovery of high-redshift multiply lensed images.

The observations were performed in Cycles 21 and 22 and the final observation in Cycle 23 was contingent on the results from the two previous cycles. The reduced number of clusters (25 in CLASH, 6 in HFF) and number of filters (16 in CLASH, 7 in HFF) allowed longer exposure time in the individual filters. A summary of the filters, orbits and exposure time is shown in Table \ref{table:hff-obs}. A total of 140 orbits for each cluster was used with 560 orbits in Cycle 21/22.

\begin{table}
 \centering
 \caption{The observational details for the Hubble Frontier Fields survey.}
 \begin{tabular}{ccccc}
 \toprule
 Camera/Channel. & Filter & Orbits & Tot. Exp. Time (s) & $5\sigma$ lim. (AB) \\
 \midrule
 ACS/WFC & F435W & 18 & 54512 & 28.8 \\
 ACS/WFC & F606W & 10 & 33494 & 28.8 \\
 ACS/WFC & F814W & 42 & 129941 & 29.1 \\
 WFC3/IR & F105W & 24 & 67341 & 28.9 \\
 WFC3/IR & F125W & 12 & 33071 & 28.6 \\
 WFC3/IR & F140W & 10 & 27559 & 28.6 \\
 WFC3/IR & F160W & 24 & 66141 & 28.7 \\
 \bottomrule
 \end{tabular}
 \label{table:hff-obs}
\end{table}

The data has been reduced in much the same way as CLASH, using the HST pipelines for reduction and mosaic merging (see Sect. \ref{sect:clash}). We have used the data from HFF in the creation of color images due to their increased depth.

\section{Chandra X-Ray Data}
\label{sect:chandra}
As a supplement to the HFF survey, \macs was also observed with the \chandra for $324\maths{ks}$ between 2009 June and 2014 December. \citet{Ogrean2015} reduced the data using CIAO v. 4.7 and CALDB v.4.6.5 calibration files. A summary of the observations can be found in Table \ref{table:chandra-obs}. The observations with ObsID 10466 contained a relatively large flare and instead of trying to model this flare, \citet{Ogrean2015} excluded it. We have chosen to do the same. 

\begin{table}[h!t]
 \centering
 \caption{Summary of the \chandra observations used by \citet{Ogrean2015}. The observation IDs with an asterisk denote the observatios that we have used.}
 \begin{tabular}{ccccc}
 \toprule
 ObsID & Obs. Mode & Start Date & Total Time (ks) & Clean Time (ks) \\
 \midrule
 10466       & VFAINT & 2009 Jun 07 & 15.8 & 15.8 \\
 16236$^{*}$ & VFAINT & 2014 Aug 21 & 39.9 & 38.6 \\
 16237$^{*}$ & FAINT  & 2013 Nov 20 & 36.6 & 36.3 \\
 16304$^{*}$ & VFAINT & 2014 Jun 10 & 97.8 & 95.2 \\
 17313$^{*}$ & VFAINT & 2014 Nov 28 & 62.8 & 59.9 \\
 16523$^{*}$ & VFAINT & 2014 Dec 19 & 71.1 & 69.1 \\
 \bottomrule
 \end{tabular}
 \label{table:chandra-obs}
\end{table}

The remaining five observations were reprojected to a common reference frame and merged images were created in the energy bands $0.5-2.0,0.5-3.0,0.5-4.0,0.5-7,$ and $2.0-7.0\maths{keV}$. These images are both exposure- and vignette corrected. In order to detect point sources, exposure map-weighted point-spread function images with ECF $= 90$\pct was created for each of the energy bands and the task \emph{wavdetect} was used on the merged maps with wavelet scales of $1,2,4,8,16,$ and $32\maths{pixels}$, a sigma threshold of $5\times 10^{-7}$ and ellipses with $5\sigma$ axes. 

All point sources detected with \emph{wavdetect} were removed from the data conservatively, using the elliptical regions that covered the largest area among the five elliptical regions. It seems that \citet{Ogrean2015} decided on using a simple copy procedure to remove the point sources all together, evident by the black holes in Figure 2 of their paper.

We have used this approach as a template. First we recalibrated the images with CIAO v.4.8 and CALDB v.4.7.2 with the reprocessing script \emph{chandra\_repro} in order to get new event files. We then performed the same reduction as \citet{Ogrean2015} with the reprojection, but for the $0.5-1.2$, $0.5-2$, $0.5-4$, $0.5-7$, $0.5-10$, $1.2-2$ and $2-7\maths{keV}$ energy bands and with a binning value of 4. We then created exposure-map weighted point-spread function maps for merged data first using the \emph{dmimgcalc} task, then using these maps to detect point-source using \emph{wavdetect}. We then sorted the catalog for the point sources that we did not want to include (the two point-sources detected at the peak emissions of the cluster) and then created source and background profiles using the \emph{roi} task. The advantage of this method is that the detected sources can be loaded into DS9, where one can remove any unwanted sources and then save the new catalog as a region file. It is however important to use correct units (image coordinates). From the DS9 catalogs a new set of catalogs, that \emph{roi} can read, can be created using the \emph{dmmakereg} task.

These new regions are then removed from the two energy band images using the task \emph{dmfilth}. The advantage of this method is that the task tries to preserve the image by replacing the point-source with information about the background. This gives a much smoother image and since in this work we use the image for tracing the X-ray mass distribution only, any additional information introduced into the image (\emph{dmfilth} tries to guess the correct background values) is not crucial. We especially noticed a new point-source introduced at the top of the image and suspect that this was a result of the \emph{dmfilth} routine. 

\begin{figure}[h!t]
 \centering
 \includegraphics[width=0.8\textwidth,keepaspectratio=true]{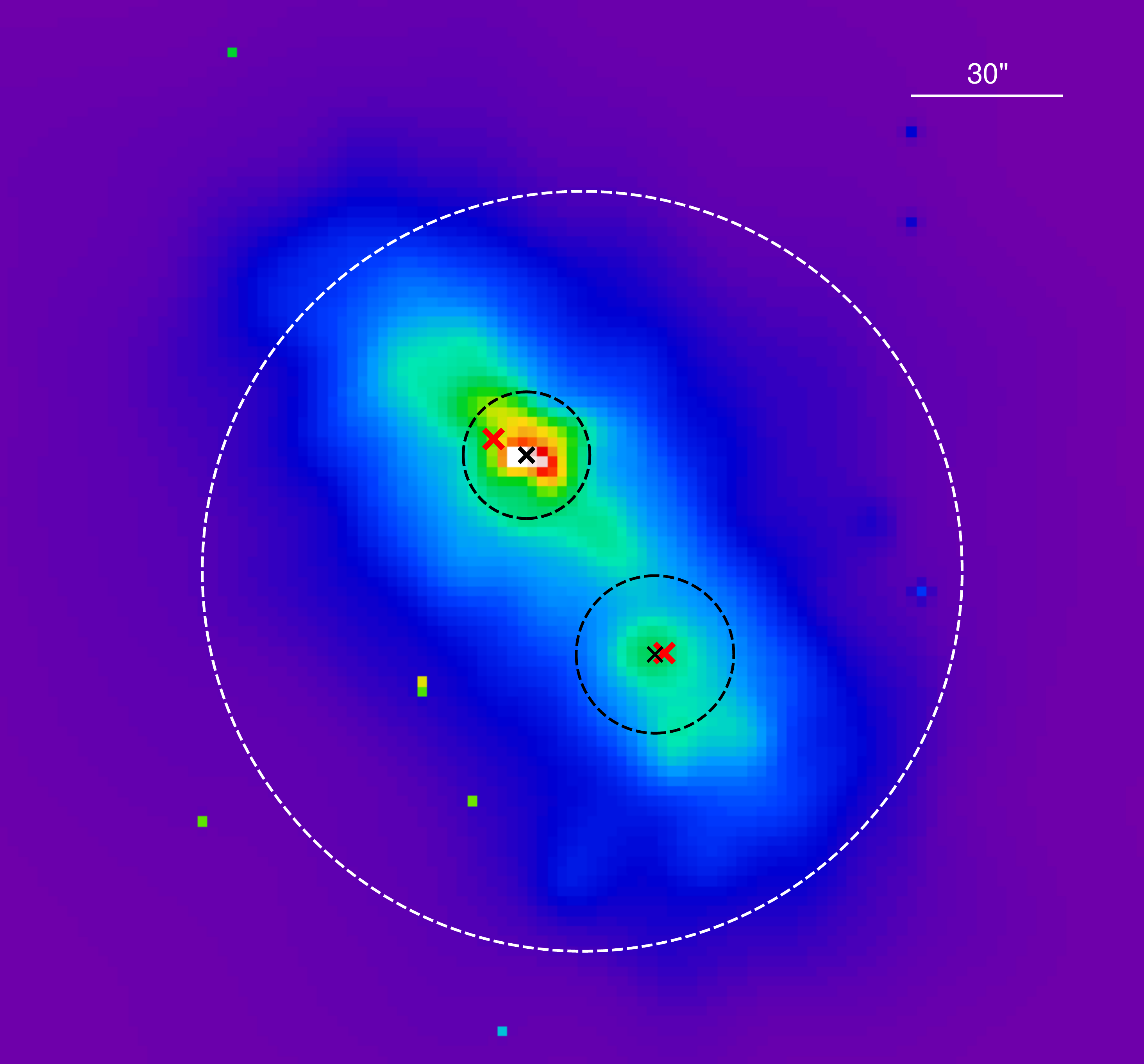}
 \caption{False color image of Chandra data in the $0.5-4\maths{keV}$ energy band. The white dashed circle represents the limit for our mass calculations $(400\maths{kpc}$, the two red X's represents the two sources found by the CIAO software and the two black X's represent the centroids found using the \emph{imexamine} package in \iraf. The black dashed circle represents the FWHM from the \emph{imexamine} analysis. We adopt the standard of North is up and East is to the left.}
 \label{fig:chandra-false-color-image}
\end{figure}

The images are then finally smoothed with \emph{csmooth} using a Gaussian convolution kernel, a Fast-Fourier Transform (FFT) convolution method and a minimum S/N of $3\sigma$. Any parts of the image with pixels greater than $5\sigma$ remains unsmoothed. We also tried using the \emph{slide} convolution method, but considering the end result and the extra computational time needed, we preferred the FFT method. Unlike \citet{Ogrean2015} we did not remove any point-sources by hand. The reduced data can be seen in Figure \ref{fig:chandra-false-color-image}. 

The data clearly shows a elongated bimodal structure. This is confirmed by the fact that \emph{wavdetect} found two point-sources in the core of the cluster, one to the NE and one to the SE. These are marked with a red X in Figure \ref{fig:chandra-false-color-image}. We crossed checked these positions using the \iraf package \emph{imexamine} to find the peaks. These are marked with a black X. We notice that there is a discrepancy between the peaks found for the NE part of the cluster and while bearing in mind that the NE core is believed to have an AGN \cite{Ogrean2015}, we find that the centroids found by \emph{imexamine} best marks the true centroid for the cluster. The elongated structure indicates that the cluster halos are highly elliptical in shape.

%% file: method.tex
\chapter{Method}
\label{chap:method}

In this chapter we present the strong lensing software we use to model \macs and the models we create. In Section \ref{sect:Software} we introduce gravitational modelling software in general. In Section \ref{sect:lenstool} we introduce \lenstool, the modelling software we use. In Section \ref{sect:Modelling} we explain the details of our models in this thesis. 

\section{Gravitational Lensing Software}
\label{sect:Software}
Software for modelling strong gravitational lensing has traditionally been divided into two subgroups: parametric and non-parametric models. The major distinction between the two is whether the calculation is model-based (parametric) or model-free (non-parametric) at start. The distinction between parametric and non-parametric is however somewhat of a misnomer, since all models use parameters. A later and more appropriate distinction is to divide into models where light traces mass (LTM) and Non-LTM \cite{Lefor2013}. 

In general we can divide the modelling problem into either a "forward" or "reverse" problem. The forward problem is the prediction of images, using the position of a source and some lensing mass. These models always assume a physical model which fits to the data with relatively few defined parameters. This problem is usually solved by LTM models where the data is fitted to a physical object (e.g. Point Mass, SIS, SIE, De Vacouleurs, NFW, etc.) \cite{Lefor2013}. A disadvantage of the LTM models is that some educated guess about the cluster mass distribution has to be made, like for instance, that dark matter traces luminous matter. This approach differs from many other aspects of astrophysics and cosmology, where one test for rather than assume, the underlying physics \cite{Diego2005}. The standard approach of using LTM models for strong lensing in clusters, comes from the fact that clusters usually do not have more than a few arcs. The general approach is to assume that a smooth dark matter halo coincides with the centroid of the luminous matter in the cluster, which remains the Achilles heal of LTM modelling \cite{Diego2005}.

The more complex reverse problem is often solved using non-LTM models where the lens inversion are done by predicting the nature of the lensing mass from the lensed image(s). This problem is more complicated due to huge degeneracies in the parameter space, meaning that a vast number of models may fit the same set of data. The non-LTM models usually apply a larger number of parameters in the form of basis functions, making the term non-parametric paradoxical, and reconstructs the lens potential/mass distribution on a map defined as a grid of pixels \cite{Lefor2013}. Especially when the number of parameters in both LTM and non-LTM models are comparable, non-LTM models are appropriate to use since they do not rely on these assumptions. The huge disadvantage of non-LTM models is that they may give unphysical results, whereas the LTM models inherently gives physical meaningful quantities. \citet{Diego2005} argues that non-LTM models can be usuful in many situations and provide physical meaningful insights into strong lensing in clusters where LTM models are usually applied.

For all LTM or parametrized models we have two important values. The number of constraints and the number of free parameters. The number of free parameters $F$ is simply the sum of the number of optimized parameters in the model. The number of constrains $C$ can be found as 
\[
 \label{eqn:N-constraints}
 \sum_{i=0}^{N} 2(n-1)
\]
where $n$ is the number of images in a given system $N$. In order to have a reliable model we must have that $F \leq C$. If $F > C$ the model is considered under-constrained and unreliable. This calculation only works when the we have reliable redshifts for the images.

In our work we will use the LTM approach since we are interested in the physical quantities provided with the LTM method. We have chosen to use the \lenstool software package which will be introduced in the following section.

\section{Lenstool}\label{sect:lenstool}
\lenstool is a publicly available\footnote{\url{https://projets.lam.fr/projects/lenstool/wiki}} software package that works in Linux and OS environments \cite{Kneib1996, Jullo2007, Jullo2009}. Although \lenstool generally belongs the to LTM type, it has the ability to include methods from the non-LTM type, in a hybrid approach \cite{Lefor2013, Jullo2009}. 

\lenstool works by taking a number of character-based input files where each line describe commands and appropriate data, such as number of total number of profiles, how many profiles to optimize and their properties etc. The keywords are divided into primary and secondary identifiers. At least two input files are mandatory. 

\subsection{Input Files}
The main parameter file (\textit{.par}) gives the optimization method, initial priors for the profiles to be optimized and in general, the basic parameters for the model. We show underneath an exempt of the main parameter file, illustrating the initial parameters and priors for a cluster halo profile:
\begin{lstlisting}
potentiel Cluster1
   profil             81
   x_centre           0.0
   y_centre           0.0
   ellipticite        0.856
   angle_pos          148.713
   core_radius_kpc    95.005
   cut_radius_kpc     1000.000
   v_disp             770.710
   z_lens             0.3960
   end

limit Cluster1
   x_centre          1 -15.0 15.0 0.010
   y_centre          1 -15.0 15.0 0.010
   ellipticite       1 0.01 0.99 0.001
   angle_pos         1 0.0 180.0 0.100
   core_radius_kpc   1 10.0 500.0 0.10
   v_disp            1 100.0 2000.0 0.10
   end
\end{lstlisting}
The input in the \emph{Limit} part defines the parameter-space that \lenstool explores. The first value defines the name of the parameter. The second value defines the optimization method, where $0$ means that \lenstool should not optimize the parameter, $1$ means that \lenstool should use flat priors and $3$ means that \lenstool should use Gaussian priors. For flat priors the format is:
\begin{lstlisting}
 parameter 1 lower-limit upper-limit precision
\end{lstlisting} 
For Gaussian priors the format is:
\begin{lstlisting}
parameter 3 mean deviation precision
\end{lstlisting}
For the flat priors, any value in the parameter-space can be equally likely, where for the Gaussian priors the mean is the most likely parameter and the code penalize any parameter deviating from that value. If one choose not to optimize the code the values from \emph{Potentiel} will be used.

A list of multiple images / arcs is also mandatory. These can be given in either absolute coordinates (Ra and Dec) or in relative coordinates, from a given reference-point. Usually one use the same reference-point as defined in the main parameter-file. The list can have any name and any extension. We show an exempt of such as list:
\begin{lstlisting}
#REFERENCE 3 64.038142 -24.067472
1.1 -8.626 21.147 1. 1. 0. 1.892 0.
1.2 -17.549 14.213 1. 1. 0. 1.892 0.
1.3 -30.285 -4.312 1. 1. 0. 1.892 0.
\end{lstlisting}
where we have here used relative coordinates. The format of the list is: 
\begin{lstlisting}
id x y a b angle-pos z mag
\end{lstlisting}
where id is the name of the image ($1.1$ in the exempt), x and y is the position of the image, a and b defines the major and minor axes, where $a = b = 1$ defines a circular image. Angle-pos or $\theta$, is the angle position of the image where $\theta = 0$ is along the East-West axis (which is not relevant with a circular image), z is the redshift of the image and L is the magnitude of the image. 

The \#REFERENCE input defines the reference-point of choice. \#REFERENCE 3 $\alpha$ $\delta$ gives the list in relative coordinates with respect to $\alpha$ and $\delta$. \#REFERENCE 0 means that the list is in absolute coordinates.

When modelling clusters one can input the cluster members as individual optimized parameters in the main parameter file or as an optional file containing the relevant information about the cluster members. The advantage of using a list for the cluster members is that one can optimize using galaxy scaling relations. The format of the list is similar to the multiple image list, with some exceptions:
\begin{lstlisting}
#REFERENCE 3 64.0381417 -24.0674722
9148 -37.7242890 -1.2027600 1.0 1.0 0.0 23.6149 0
4290 48.6257355 -32.3708400 1.0 1.0 0.0 22.9752 0
2317 10.8689114 -84.2180400 1.0 1.0 0.0 22.9668 0
\end{lstlisting}
where the format is
\begin{lstlisting}
id x y a b angle-pos mag lum
\end{lstlisting}
where id is the name of the cluster member, x and y are the position of the cluster member, a and b defines the ellipticity $(a = b = 1$ is a circle), angle-pos or $\theta$ define the angle-position, mag defines the magnitude and lum defines the luminosity. Usually the magnitude is used and the luminosity ignored, as is the case in the exempt. 

If we where to include all the subhalos in a galaxy cluster as individually optimized potentials, we would quickly get an under-constrained model. In order to include the cluster members in a model and have the number of free parameters comparable with the number of constraints, we have to apply certain assumptions about the mass. A connection between the mass profile of a galaxy and the light has been shown, which is why it is assumed that, to a first approximation, that the subhalo position, ellipticity and orientation match the luminous counterpart \cite{Jullo2007}. Further, since strong lensing will not constrain the details about the individual subhalos, we can use the well known scaling relations to model the cluster members. This will be explained further in Section \ref{sect:Modelling}.

\citet{Kneib1996} showed that it is necessary to include the cluster members, in a strong lensing model, in order to reproduce the observed multiple images. The standard approach for selecting cluster members is to measure the photometry of the galaxies in the cluster and then plot color-magnitude diagrams i.e. the red sequence \cite{Ardis2007, Limousin2007, Limousin2008}. Although this method is simple and fast, it has shown to exclude bluer galaxies which is part of the cluster and include redder galaxies which is not part of the cluster. This is evident when plotting a color-color diagram of the photometrically selected cluster members against spectroscopically confirmed cluster members \cite{Richard2014, Jauzac2015}. With the advent of better selection methods \cite{Grillo2015} and new technology, like the MUSE instrument, allow us to include even the smallest cluster members as long as they are bright enough to be measured \cite{Caminha2016}. One should however also have computational time in mind, which depends heavily on the number of cluster members and their shape. 

Another reason to include the cluster members is that a large number of subhalos in a cluster may increase the projected surface mass density and hence, the strength of lens. Although the contribution to the total mass from the perturbers are limited, the effect is systematic and thus must be accounted for \cite{Jullo2007}. 

\subsection{Output Files}
As output, \lenstool provides a number of files. The primary outputfiles are the \emph{burnin.dat} and \emph{bayes.dat} files. These two represent the core of the model optimization. From the \emph{bayes.dat} \lenstool extracts the \emph{best.par}, giving the best model parameters, which can be used to produce a variety of maps, further output files and predictions of new images. This file will not optimize the model further. The \emph{bestopt.par} files is also extracted, which gives the best possible optimization parameters in Gaussian priors. Running this file will optimize the model, albeit very fast due to the restricted parameter space.

Secondary output-files are the \emph{chires.dat} defines the optimized image error position and $\chi^2$ (which is also given in the \emph{best.par} file), the \emph{dist.dat} giving information about image parity, amplification and time-delay, the \emph{ci.dat} and \emph{ce.dat} defining the critical lines and caustics, the \emph{pot.dat} giving information about the optimized potential position.

Tertiary output files are the \emph{para.out} simply outputting the used parameters and \emph{restart.dat}, which is a relatively new addition to \lenstool, providing the opportunity to restart the optimization from the last point calculated. Both secondary and tertiary output files depend to some extent on selections done in the main parameter file. Further information can be found in the \lenstool online manual.

\subsection{Optimization}\label{subsect:optimization}
When we model a cluster with parametric lensing codes we are seeking to constrain the lensing potential, defined by a set of parameters and priors, using a list of observable images from given source at some redshift. In this Section we will introduce the optimization approach used in \lenstool. This approach is not exclusive for \lenstool, but used in a large variety of strong lensing codes.

\subsubsection{$\chi^2$ Minimization}
In order to find the best fitting model, we need a way to compare the predicted images from the model with the observed images. In \lenstool, the basic method is $\chi^2$ minimization. The overall contribution to the $\chi^2$ from the multiple images in system $i$ is
\[
 \chi_{i}^{2} = \sum_{j=1}^{n_i} \frac{[\vec{x}_{obs}^{j} - \vec{x}^j(X)]^2}{\sigma_{ij}^{2}}
\] 
where $\vec{x}^j(X)$ is the position of the predicted image $j$ from the model with the parameter $X$, $\vec{x}_{obs}^{j}$ is the position of the observed image $j$ and $\sigma_{ij}$ is the error on position of image $j$. The position error parameter $\sigma_{ij}$ is defined globally for all images in \lenstool. An accurate determination of the position-error can only be done by a pixelated approach (for an extended image) because of the dependence on the S/N ratio. This approach takes every pixel within the image into account and hence is very time consuming. So as a first approximation one can assume that the error in image position is Gaussian and hence can be determined by fitting a 2-dimensional Gaussian profile to the image surface brightness. In order to do this we have to assume that the source galaxy is compact, the surface brightness profile is smooth and that the peak in brightness in the image corresponds to the peak in the source \cite{Jullo2007}.  

One of the greatest problems with the $\chi^2$ computation is how to match the predicted and observed images one by one. It is especially problematic in the beginning of the optimization where the position of the predicted and observed images can differ substantially \cite{Jullo2007}. To this date, there has been many attempts to find the roots of the lens equation, but the matching of predicted images with the observed one by one becomes problematic when their respective positions do not match closely. This always happens during the first steps of the optimization process and \citet{Jullo2007} have found no algorithm that performs this matching automatically.

In \lenstool, the observed image is coupled to the predicted image all along the iterative refinement of the predicted position, which makes the $\chi^2$ easy to compute. But in a model producing different configurations (e.g. a tangential system where a radial is expected), the method will fail and hence the model is completely rejected. This primarily happens when the model is not well constrained. This will slow the optimization process.  

One way to circumvent this problem entirely, is to compute the $\chi^2$ in the source plane. Here we compute the difference in the source position for a given parameter sample $X$. The source plane $\chi^2$ is computed as
\[
 \chi^{2}_{S_{i}} = \sum_{j=1}^{n_j} \frac{\left[\vec{x}_{S}^{j}(X) - \langle \vec{x}_{S}^{j}(X) \rangle \right]^2}{\mu_{j}^{-2} \sigma_{ij}^{2}}
\]
where $\vec{x}_{S}^{j}$ is the source position of the observed image $j$, $\langle \vec{x}_{S}^{j}(X) \rangle$ is the barycenter position of the $n_i$ source positions and $\mu_j$ is the magnification for image $j$ which is important since a magnified image extends over a larger area than a un-magnified or negatively magnified image. So for a magnified image, the position-error has to be reduced by the that factor squared. 

Here we have no need for solving the lens equation and hence, the optimization is very fast. Because the image plane optimization has to test (and reject) a sizeable amount of models before focusing on the best-fit region, the standard approach is to size-up the best-fit region with source plane optimization and then refine this region using image plane optimization \cite{Jullo2007}.

The overall problem with $\chi^2$ minimization, is the risk of hitting a local minimum and hence not find the true best model. One way to circumvent this problem is to apply Bayesian Monte Carlo Markov Chain as part of the optimization routine \cite{Jullo2007}. This will be explained in the following Section.

\subsubsection{Bayesian MCMC Optimization}
\label{subsect:Bayesian}
\lenstool employs the Bayesian Markov Chain Monte Carlo package \bayesys \cite{Skilling2004} which supports both source-plane and image-plane optimization. Theoretically, the Bayesian approach is better suited than standard regression methods in cases where the data does not constrain the model by itself i.e. in cases where we risk hitting a local minimum. Specifically, the Bayesian method is well suited for strong lens modelling since such a model usually have few constraints available \cite{Jullo2007}. 

The Bayesian method offers two levels to deduce the best parameters. First we have parameter space exploration which is done using the unnormalized posterior probability distribution function (PDF), essentially equal to the product of the likelihood and the prior. This is followed by model comparison or sampling which is done by calculating the normalisation of the posterior or evidence. These quantities are governed by the Bayes theorem
\[
 P(X | D,M) = \frac{P(D|X,M)^{\lambda} P(X|M)}{Pr(D|M)}
\]
Here we have that $P(X | D,M)$ is the posterior PDF, $P(D|X,M)$ is the likelihood of getting the observed data $D$ given the parameters $X$ of model $M$, $P(X|M)$ is the prior PDF and $Pr(D|M)$ is the evidence. We also have the parameter $\lambda$ which is specific for the implementation in \lenstool. 

\begin{figure}[htb]
 \centering
 \includegraphics[width=0.8\textwidth,keepaspectratio=true]{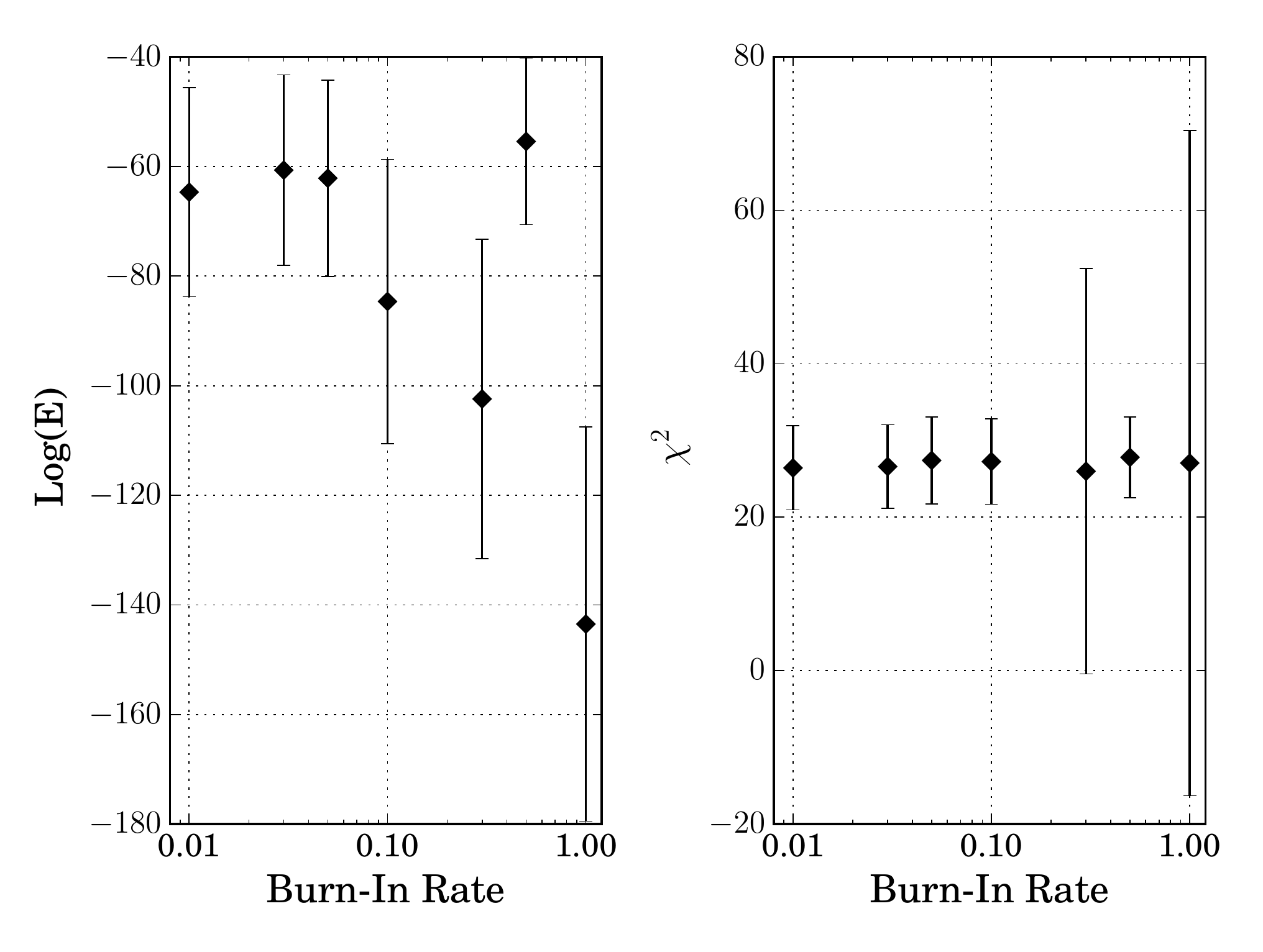}
 \caption{The results from the burn-in test, using different $R$ values.}
 \label{fig:burnintest}
\end{figure}

\lenstool runs ten interlinked Markov Chains at the same time, so that no chain will fall into a local minimum. The convergence to the posterior PDF is performed using \emph{selective annealing}. At each step, ten new samples (one per chain) are drawn randomly from the current posterior PDF. This roughly means that the samples with the worst likelihood are deleted and the ones with the best are duplicated, so that ten chains are always running at any given moment. \bayesys then provides eight methods for exploring the parameter space and keeping the chains uncorrelated. The speed of the annealing comes into play with the $\lambda$ term, which is the cooling factor for the annealing. During the 'burn-in' phase, the likelihood influence is increased incrementally from $\lambda = 0$ to $\lambda = 1$ in steps of $\delta \lambda \propto R$, where $R$ is a constant provided in the primary parameter file \cite{Jullo2007}. The constant $R$ also determines the convergence speed and the resolution of the parameter space exploration. The higher the value of $R$ the faster the 'burn-in' phase will converge, but will also decrease the resolution of the parameter space. This is because a higher value of $R$ will result in fewer samples in the chains and hence, the parameter space will be more coarsely explored. Essentially we might miss the best-fit region. According to \citet{Jullo2007} a $R$ parameter between $R = 0.5-0.1$ is sufficient. Newer models \cite{Limousin2008} however suggest that $R = 0.05$ gives a reasonable resolution. When we test this assumption using our base model with a reduced cluster member catalog ($1/2$ size), we conclude that $R = 0.05$ is a good compromise between parameter space resolution and convergence speed as it also offers the best $\log{(E)}$ value even though the $\chi^2$ is higher than both $R=0.10$ and $R=0.03$ (see Fig. \ref{fig:burnintest}). 

The MCMC sampler does not look for the best sample of parameters, but instead samples the posterior PDF. This means that it will draw more samples where the posterior PDF is higher. The more samples collected after the 'burn-in' phase determines the resolution of the posterior PDF. This is important since we use the posterior PDF to draw 2D histograms to represent the marginalised posterior PDFs $P(\THETA_i|M)$ and $P(\THETA_i,\THETA_j|M)$ whose resolution are limited by the number of samples. More importantly, we also use the 2D histograms to illustrate the degeneracies that will occur in parametrized models. In general we can say that with the MCMC sampler we do not only have the means to find the best-fit model, in terms of the $\chi^2$, but also the most likely model in terms of the Evidence \cite{Ardis2007}. The MCMC routine will also penalize unnecessary complicated models meaning that a model with a good fit from more free parameters alone, will give a lower Evidence and suggest that a more simple model is better \cite{Ardis2007}.

\section{Modelling}
\label{sect:Modelling}
The models for this thesis are based on the best model from \citet{Grillo2015}. Here the lensing program \glee was used, which offers different mass-density profiles and parameter input/output. \citet{Grillo2015} created eight models containing two mass density profiles for the cluster halo and 175 mass density profiles for the cluster members, where two models where with the cluster halos only. 

In any parametric gravitational lensing model of galaxy clusters, two different mass component types are ordinarily used. One or more mass profiles representing the cluster halo and a catalog of cluster members optimized using some kind of scaling to model the galaxy mass profiles. Individual galaxies are usually pulled out of the cluster member catalog and optimized as individual components if multiple images are detected around them. This is the case in \cite{Limousin2008}, \cite{Richard2009} and \cite{Caminha2016}.

In this section we present a detailed description the different models we use to model \macs and the scientific justification for them. In all, we have 10 different models. A list of all the models with the components they use can be found in Table \ref{table:model_list}, where we adopt the same naming convention as \citet{Grillo2015}. For all models we use $\alpha: 64.0381417, \delta:-24.0674722$ as reference point. In general, it is assumed that the mass of the cluster members follows the light so that the shape of the cluster member halos should be elliptical \cite{Jullo2009}, but up until now all studies of \macs has been utilizing circular mass profiles \cite{Johnson2014, Grillo2015, Caminha2016}. Therefore we have chosen to compare models using elliptical and circular cluster member mass profiles. Likewise, all studies of \macs has up until now used fixed slopes parameters for the cluster member scaling relations \cite{Richard2014, Jauzac2014, Caminha2016}, where we have chosen to use both the slope parameters from \citet{Grillo2015} as fixed parameters and also optimize the slope parameters. From this we should be able to see whether elliptical cluster member mass profiles and optimizing the slope parameters provide better results. 

\begin{table}[htb]
 \centering
 \caption{Overview of the different models we use with their respective components. $M_TL^{-1} = v$ indicates that the slope parameters are variables in the optimization. The $e$ or $c$ subscript indicates whether circular or elliptical cluster member profiles, respectively, are used.}
 \label{table:model_list}
 \begin{tabular}{cc}
 \toprule
  Name & Cluster member profile \\
 \midrule
 2PIEMD & \\
 2NFW  &  \\
 1PIEMD + 175(+1)dPIE$_c$ $(M_TL^{-1}\sim L^{0.2})$ & Circular (no slope opt). \\
 2PIEMD + 175(+1)dPIE$_c$ $(M_TL^{-1}\sim L^{0.2})$ & Circular (no slope opt). \\
 2PIEMD + 175(+1)dPIE$_e$ $(M_TL^{-1}\sim L^{0.2})$ & Elliptical (no slope opt). \\
 2PIEMD + 175(+1)dPIE$_c$ $(M_TL^{-1} = v)$ & Circular (slope opt.) \\
 2PIEMD + 175(+1)dPIE$_e$ $(M_TL^{-1} = v)$ & Elliptical (slope opt.) \\
 2NFW + 175(+1)dPIE$_c$ $(M_TL^{-1}\sim L^{0.2})$ & Circular (no slope opt.) \\
 2NFW + 175(+1)dPIE$_c$ $(M_TL^{-1} = v)$ & Circular (slope opt.) \\
 2NFW + 175(+1)dPIE$_e$ $(M_TL^{-1} = v)$ & Elliptical (slope opt.) \\ 
 \bottomrule
 \end{tabular}
\end{table}

\citet{Zitrin2013} used elliptical NFW and elliptical Gaussian smoothed PIEMD profiles and found mass densities in agreement with weak lensing (WL) results. They also found a connection between the number of multiple images, the ellipticity of the mass profiles and the separation of the two mass profiles. Higher ellipticity and/or lower separation increased the number of images, for two merging clumps. They conclude that the observed critical area size in \macs is due to a merger scenario. \citet{Johnson2014} used PIEMD profiles for modelling both cluster scale halos and cluster members. They fixed the cluster-size halo scale radius to $1500\maths{kpc}$ since the strong lensing regime is to small to constrain the scale radius properly. They also included the foreground galaxy at $\alpha: 64.028417, \delta: -24.085681$ for which they optimize the core radius and velocity dispersion. The scale radius is fixed to $1500\maths{kpc}$. Finally they fix the parameters for the scaling relations to $r_{c,G} = 0.15\maths{kpc}$, $r_{s,G} = 30\maths{kpc}$ and $\sigma_{G} = 120\maths{km\;s^{-1}}$. They found that their model could not be well constrained without those settings. \citet{Johnson2014} used $10$ spectroscopically confirmed image systems and 5 optimized image systems in their model. 

\citet{Caminha2016} found evidence that a small sub-group of galaxies at $\alpha:64.034056,\delta:-24.066851$ should be included due to multiple images around them. At the same time, the number of multiple images used has increased drastically from $10$ spectroscopically confirmed images systems \cite{Grillo2015} to $40$ spectroscopically confirmed images systems due to better imaging from the new MUSE instrument on the VLT \cite{Caminha2016}. Although \citet{Jauzac2014} have more image systems $(57)$, it is only a handful that are spectroscopically confirmed $(8)$. The number of cluster members has also increased from 175 \cite{Grillo2015} to 193, where 144 $(75\%)$ are spectroscopically confirmed \cite{Caminha2016}.

As \citet{Johnson2014}, we include the foreground galaxy at $\alpha: 64.028417, \delta: -24.085681$ for which we fix the core radius to $r_{c,G} = 0.15\maths{kpc}$ and optimize the scale radius and velocity dispersion. Likewise for the cluster members, we fix the core radius to $r_{c,G} = 0.15\maths{kpc}$ and allow the scale radius and velocity dispersion to be optimized. For the \citet{Grillo2015} models we fix the slopes. In all other models the slopes are optimized as well. This will be explained in more details in the following sections.

\subsection{Multiple Images}
The list of multiple images have been selected from \citet{Grillo2015} and are listed in Table \ref{table:multiple_images}. Although more images has recently become available \cite{Jauzac2014, Caminha2016}, because of time constraints, since we would have to redo all our models, we have decided to use this catalogue of multiple images. Future models, however, would undoubtedly benefit from the additional multiple images and confirmed cluster members. 

The multiple images have been selected using the spectroscopic data from \emph{CLASH-VLT}. This means that all the images used in the models are spectroscopically confirmed and we have no need for optimization of image redshifts. This reduce the number of free parameters in the models.

In \lenstool we use one single position error $(0.065\arcsec)$ for the images as opposed to \glee, where the position errors are given individually for the images. We have chosen to retain the practice from \citet{Grillo2015} and show the position errors for the images individually (Table \ref{table:multiple_images}).

\begin{table}[!htb]
 \centering
 \caption{List of multiple images}
 \label{table:multiple_images}
 \begin{tabular}{ccccc}
 \toprule
  ID & R.A. & Decl. & $z_{\mathrm{spec}}$ & $\delta_{x,y}$  \\
     & (J2000) & (J2000)  &  & $(\arcsec)$ \\
  \midrule
  1.1 & 04:16:09.784 & $-$24:03:41.76 & 1.892 & 0.065  \\
  1.2 & 04:16:10.435 & $-$24:03:48.69 & 1.892 & 0.065  \\
  1.3 & 04:16:11.365 & $-$24:04:07.21 & 1.892 & 0.065  \\ 
  \midrule
  2.1 & 04:16:09.871 & $-$24:03:42.59 & 1.892 & 0.065 \\
  2.2 & 04:16:10.329 & $-$24:03:46.96 & 1.892 & 0.065 \\
  2.3 & 04:16:11.395 & $-$24:04:07.86 & 1.892 & 0.065 \\
  \midrule
  3.1 & 04:16:09.549 & $-$24:03:47.08 & 2.087 & 0.065 \\
  3.2 & 04:16:09.758 & $-$24:03:48.90 & 2.087 & 0.065 \\
  3.3 & 04:16:11.304 & $-$24:04:15.94 & 2.087 & 0.065 \\
  \midrule
  4.1 & 04:16:07.385 & $-$24:04:01.62 & 1.990 & 0.065 \\
  4.2 & 04:16:08.461 & $-$24:04:15.53 & 1.990 & 0.065 \\
  4.3 & 04:16:10.031 & $-$24:04:32.62 & 1.990 & 0.065 \\ 
  \midrule
  5.1 & 04:16:07.390 & $-$24:04:02.01 & 1.990 & 0.065 \\
  5.2 & 04:16:08.440 & $-$24:04:15.57 & 1.990 & 0.065 \\
  5.3 & 04:16:10.045 & $-$24:04:33.03 & 1.990 & 0.065 \\ 
  \midrule
  6.1 & 04:16:06.618 & $-$24:04:21.99 & 3.223 & 0.065 \\
  6.2 & 04:16:07.709 & $-$24:04:30.56 & 3.223 & 0.065 \\
  6.3 & 04:16:09.681 & $-$24:04:53.53 & 3.223 & 0.065 \\
  \midrule
  7.1 & 04:16:06.297 & $-$24:04:27.60 & 1.637 & 0.065 \\
  7.2 & 04:16:07.450 & $-$24:04:44.23 & 1.637 & 0.065 \\
  7.3 & 04:16:08.600 & $-$24:04:52.76 & 1.637 & 0.065 \\ 
  \midrule
  8.1 & 04:16:06.246 & $-$24:04:37.76 & 2.302 & 0.065 \\
  8.2 & 04:16:06.832 & $-$24:04:47.10 & 2.302 & 0.065 \\
  8.3 & 04:16:08.810 & $-$24:05:01.93 & 2.302 & 0.065 \\ 
  \midrule
  9.1 & 04:16:05.779 & $-$24:04:51.22 & 1.964 & 0.065 \\
  9.2 & 04:16:06.799 & $-$24:05:04.35 & 1.964 & 0.065 \\
  9.3 & 04:16:07.586 & $-$24:05:08.72 & 1.964 & 0.065 \\ 
  \midrule
  10.1 & 04:16:05.603 & $-$24:04:53.70 & 2.218 & 0.065 \\
  10.2 & 04:16:06.866 & $-$24:05:09.50 & 2.218 & 0.065 \\
  10.3 & 04:16:07.157 & $-$24:05:10.91 & 2.218 & 0.065 \\ 
  \bottomrule
  \end{tabular}
\end{table}

\begin{figure}[htb]
 \centering
 \includegraphics[width=0.9\textwidth,keepaspectratio=true]{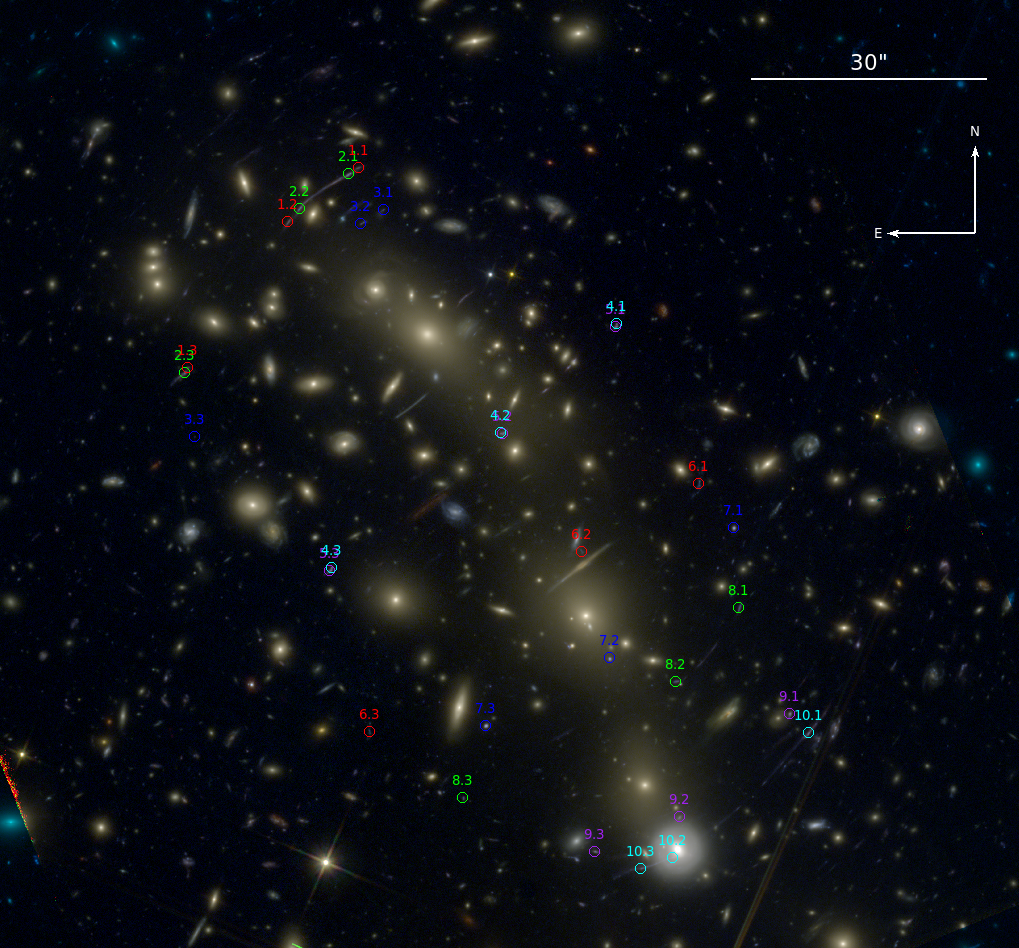}
 \caption{Color-composite image of \macs combining the HFF filters of HST/ACS and WFC3. The thirty multiple images used in our models are presented as coloured circles with labels. Further information can also be found in Table \ref{table:multiple_images}.}
 \label{fig:multiple_images}
\end{figure}

A color-composite image of the multiple images are shown in Figure \ref{fig:multiple_images}. Using the method presented Section \ref{sect:Modelling} we can calculate the number of constraints from the multiple images. We see that we have $10$ systems consisting of $3$ images each. Using \eqref{eqn:N-constraints} we find that we have $40$ constraints in our model. The number of free parameters depends in the profiles in the models and will be introduced in the following sections.

\subsection{The Cluster Scale Halos}
\label{subsect:(Modelling)Cluster-scale-halos}
In \citet{Grillo2015} two different profiles were used to represent the cluster scale halos, the PIEMD and the NFW profile. We have decided to use the same profiles in our models, although the NFW differs.

The NFW profile was originally proposed by \citet{Navarro1997} as a universal mass density profile for dark matter halos and is given by the expression
\[
 \rho(r) = \frac{\rho_c}{(r/r_s)(1+r/r_s)^2},
\]
where $\rho_c$ is the characteristic density and $r_s$ the scale radius.

In the \citet{Grillo2015} \glee model, the NFW profile is defined as a prolate 3-dimensional mass distribution
\[
 \rho(r) = \frac{\rho_{0}}{(r/r_{s})(1+r/r_{s})},
\]
where
\[
 r^2 = c^2 \leftparan \frac{x^2 + y^2}{a^2} + \frac{z^2}{c^2} \rightparan \; a \leqslant c
\]
Here the parameter $a$ and $c$ describe the prolateness of the halo where $a/c=1$ refers to a spherical halo and $a/c=0$ to an extremely elongated halo. The orientation of the halo can be described by the viewing angle responsible for the ellipticity $\varphi$ and the projected major axis position angle $\phi$. \citet{Grillo2015} has further chosen to use the Einstein radius $\vartheta_{E}$ instead of $\rho$ to characterize the strength of the halo. This has been chosen since the Einstein radius can be measured to a high degree of accuracy using strong lensing \cite{Grillo2015}. So for a PNFW profile in \glee, the parameters are the position parameters $x$ and $y$, the prolateness $a/c$, the angles $\varphi$ and $\phi$, the Einstein radius $\vartheta_{E}$ and the scale radius $r_{s}$ which gives 7 parameters total.

The NFW profile in \lenstool behaves somewhat differently. It is in general based on the work by \citet{Golse2002} where a spherical NFW profile is converted into an elliptical profile and they find a 3-dimensional pseudo-elliptical mass distribution. Here they discover that the profile gives a reasonable projected mass distribution for ellipticities $\lesssim 0.4$, although boxyness is introduced around that point.

This profile has further been generalized by \citet{Sand2008}
\[
\rho(r) = \frac{\rho_c \delta_c}{(r/r_s)^{\alpha}(1+(r/r_s))^{3-\alpha}}
\]
where $\delta_c$ is related to the concentration parameter $c_{200}$ through
\[
 \delta_c = \frac{200}{3} \frac{c_{200}^{3}}{\ln{(1+c_{200})}-c_{200} / (1+c_{200})}
\]
and $\alpha$ corresponds to the generalization parameter. The generalised NFW profile is abbreviated gNFW. 

The motivation for the generalisation comes from the fact that dark matter profiles may not follow the original NFW profile \cite{Sand2002, Sand2004, Sand2008}. When $\alpha = 1$, the profile falls back to the original NFW profile. With the line-of-sight or optical axis defined as the $z$ axis, $r$ is defined as $r = \sqrt{R^2+z^2}$ \cite{Golse2002} or $r^2 = R^2 + z^2$ \cite{Sand2008}.

For the gNFW profile in \lenstool the ellipticity is introduced into the potential rather than the surface mass density as opposed to the profile in \citet{Grillo2015}. This approach has been chosen, since it makes the lensing calculations more tractable, as the deflection angle is just the gradient of the scaled lensing potential \cite{Sand2008}. For the gNFW profile, the surface mass density and deflection angle cannot be calculated analytically. This means that when having to explore the parameter space in large parameter hypercubes, the calculation becomes greatly slowed down. In order increase the speed of the calculation, lookup tables for all the necessary integrations are created in advance, from which interpolations can be drawn when solving the lensing equations \cite{Sand2008}. 

As mentioned, the gNFW profile is only valid within relatively small ellipticities. In order to find the degree of boxiness by measuring the distance $\delta r$, \citet{Sand2008} presents several plots of $\delta r/r$ as a function of $\varepsilon$ for several values of the inner slope $\beta$ and a variety of $r/r_{s}$, which they present in Figure 8. In order to have their pseudoelliptical implementation within $10\%$ of true elliptical surface mass distribution for $r/r_s < 10$, the ellipticity must be $\varepsilon \lesssim 0.2$. This means that the NFW profile in \lenstool will break down when constraining highly elliptical DM halos. This is opposed to the prolate NFW profile where the ellipticity is defined as the prolateness, which means that the profile is valid for almost any ellipticity.

In \lenstool the parameters for the gNFW profile are the position $x$ and $y$, the ellipticity of the mass distribution $\varepsilon$, the angular position (in the image plane) $\theta$. the scale radius $r_{s}$, the slope of the profile $\alpha$ and the concentration parameter $c_{200}$ which gives 7 parameters in total. 

\citet{Grillo2015} use the PIEMD profile in \glee, with a dimensionless surface mass density defined as
\[
 \kappa_h(x,y) = \frac{\vartheta_{E,h}}{2\sqrt{R_{\varepsilon}^{2} + r_{c,h}^2{2}}}
\]
where
\[
 R^{2}_{\varepsilon} = \frac{x^2}{(1+\varepsilon)^2} + \frac{y^2}{(1 - \varepsilon)^2}
\]
$\varepsilon$ is the ellipticity defined as $\varepsilon \equiv (1-q_h)/(1+q_h)$ and $q_h$ is the axis ratio. The PIEMD (psudo-Isothermal Elliptical Mass Distribution) profile has a central core $r_{c,h}$. This profile require six parameters in \glee: The position centroid $(x_h,y_h)$, the axis ratio $q_h$, the position angle $\phi_h$, the Einstein radius $\vartheta_{E,h}$ and the core radius $r_{c,h}$. This profile is the equivalent of the original PIEMD profile defined by \citet{Kassiola1993}. The other variant in \glee is called the dPIE profile \cite{Ardis2007} and have both a core radius $(r_{c,h})$ and and truncation or cut radius $(r_{t,h})$. In \glee, these profiles exist as two separate profiles, which is opposed to \lenstool which only have the dPIE profile. In some articles it has become custom not to distinguish between the two names \cite{Limousin2008,Richard2009} but in this work we will denote the PIEMD profile as the profile with just the core radius and the dPIE profile as the profile with both core and cut radius. Furthermore, we will show in the next section that the PIEMD profile can be modelled as a dPIE profile with a pseudo-infinite cut radius.
A similar consideration about whether to put the ellipticity in the potential or the mass distribution can be found in \citet{Kassiola1993} where they find that the PIEMD can represent arbitrary different ellipticities.

The dPIE profile \cite{Ardis2007} is represented in \lenstool, by the 3D density profile
\[
 \rho(r) = \frac{\rho_0}{\leftparan 1+r^2/r_{c}^{2} \rightparan \leftparan 1+r^2/r_{s}^{2} \rightparan}
\]
where $r_{c}$ is the core radius, $r_{s}$ is the scale radius and $\rho_0$ is the central density. In \citet{Ardis2007} it is pointed out that the cut radius is not a true cut or truncation radius but a scale radius for $\rho \sim r^{-4}$ when $r \ll r_{c}$. We therefore choose to adopt a similar terminology. The 2D density profile or surface mass density profile can then be defined as
\[
 \Sigma(R) = \Sigma_0 \frac{r_{c} r_{s}}{r_{s} - r_{c}} \leftparan \frac{1}{\sqrt{r_{c}^{2}+R^2}} - \frac{1}{\sqrt{r_{s}^{2} + R^2}} \rightparan
\]
with 
\[
 \Sigma_0 = \pi \rho_o \frac{r_{c} r_{s}}{r_{s} + r_{c}}
\]
and \cite{Limousin2005}
\[
 \rho_0 = \frac{\sigma_{0}^{2}}{2\pi G} \frac{r_s+r_c}{r_{c}^{2}r_s}
\]

We use the PIEMD profile to represent the cluster scale halos \cite{Grillo2015} which is derived from the dPIE profile with an pseudo-infinite scale radius $r_s \gg R$.
\[
 \Sigma(R) = \frac{\sigma_{v}^{2}}{2G} \leftparan \frac{1}{\sqrt{r_{c}^{2} + R^2}} \rightparan
\]
which in \lenstool has the parameters: the positions $x$ and $y$, the ellipticity $\varepsilon$, the position angle $\theta$, the core radius $r_c$ and the fiducial velocity dispersion $\sigma_v$.

\subsection{The Cluster Member Halos}
\label{subsect:(Modelling)cluster-member-halos}
In order to represent the cluster members we use the dPIE profile with a vanishing core $r_c \rightarrow 0$
\[
 \Sigma(R) = \frac{\sigma_{v}^{2}}{2G} \leftparan \frac{1}{\sqrt{R^2}} - \frac{1}{\sqrt{R^2 + r_{s}^{2}}} \rightparan
\]
which is the profile proposed for galaxy-galaxy lensing \cite{Ardis2007}. The optimization is done using the scaling relations for the luminosity $L$ \cite{Ardis2007}, for the core radius
\[
r_{c,gal} = r_{c,gal}^{*} \leftparan \frac{L}{L^*} \rightparan^{\onehalf}
\]
the scale/truncation radius
\[
r_{s,gal} = r_{s,gal}^{*} \leftparan \frac{L}{L^*} \rightparan^{\frac{2}{\zeta_{r_{t,gal}}}}
\]
and the velocity dispersion
\[
\sigma_{gal} = \sigma_{gal}^{*} \leftparan \frac{L}{L^*} \rightparan^{\frac{1}{\zeta_{\sigma_{gal}}}}
\]
where $\zeta_{r_{s,gal}}$ and $\zeta_{\sigma_{gal}}$ are the slopes and enter as parameters into \lenstool. In \citet{Grillo2015} the scaling relations are defined as
\[
 \vartheta_{E,gal} = \vartheta_{E,gal}^{*} \leftparan \frac{L}{L^*} \rightparan^{0.7}
\]
for the Einstein radius and
\[
 r_{s,gal} = r_{s,gal}^{*} \leftparan \frac{L}{L^*} \rightparan^{0.5}
\]
for the scale radius. This means that we need to convert the \glee parameters into \lenstool parameters. The conversions of the scale radius slope is straight forward
\[
 0.5 = \frac{2}{\zeta_{s_{t,gal}}} \rarrow \zeta_{s_{t,gal}} = 4.0
\]
For the velocity dispersion slope $\zeta_{\sigma_{gal}}$ we need a further conversion from the Einstein radius $\vartheta_E$ to the velocity dispersion $\sigma$. From \citet{Grillo2015} we know that $\sigma \sim \vartheta_{E}^{0.5}$
\[
 \vartheta_{E}^{0.5^{0.7}} = \vartheta^{0.5 \cdot 0.7} = \sigma^{0.35}
\]
so the slope then become
\[
 0.35 = \frac{1}{\zeta_{\sigma_{gal}}} \rarrow \zeta_{\sigma_{gal}} = 2.86
\]

If we would want the cluster members to follow a constant mass/light (M/L) ratio in \lenstool, we would have to use $\zeta_{r_{s,gal}} = 4$ and if we would want them to follow the fundamental plane \cite{Mo2010}, we would have to use $\zeta_{r_{s,gal}} = 2.5$. So in the model parameters from \citet{Grillo2015} we use constant M/L ratio. 

The catalog with the cluster members originate from \citet{Grillo2015}. We will here briefly sketch the selection method used. Because the catalog consist of 175 individual members, we will not present the list in its entirety. As mentioned before, the usual approach in selecting cluster members, is to use a large-scale photometric program (Source Extractor \cite{Bertin1996}) and then plot the red sequence of the photometry in order to select the reddest galaxies \cite{Ardis2007,Limousin2008,Zitrin2013,Johnson2014}. This method has proven to be too imprecise, since in some sitations, cluster members that has later turned out to be spectroscopically confirmed members of the cluster, was not present in the red sequence selection \cite{Jauzac2015}. From \emph{CLASH-VLT} we have 800 spectroscopically confirmed cluster members and of those, 113 are within the HST field-of-view. From the n-dimensional distance of a given galaxy from the color distribution of spectroscopic members, \citet{Grillo2015} assigned membership probability to each galaxy. 
This was done by selecting the 113 spectroscopically confirmed galaxies in the range $z = 0.369 \pm 0.014$ which corresponds to $\pm 3000\maths{km\,s^{-1}}$ rest-frame which have good photometric data. As mentioned in the previous Chapter, F225W, F275W , F336W and F390W data was excluded due to low SNR of the faint member galaxies. From this sample, the average colors and covariance matrix was calculated using a Minimum Covariance Determinant method. 
\citet{Grillo2015} also selected a representative set of field galaxies, that is, galaxies outside the redshift range of the cluster, where the mean and covariance matrix of the colors are calculated. Here it is assumed that the cluster members and field galaxies can be well described by a multivariate normal distribution with the averages and covariances. From this a pure and complete cluster member catalog was produced. The catalog was then further tuned with respect to the probability threshold so that the purity of the cluster members, especially at the bright end of the luminosity function, could be maximized. This is also where the most massive galaxies reside. From this method 109 galaxies was selected as cluster members.

In simple terms, one constructs a N-dimensional matrix consisting of the photometrical data in N filters, selects a group consisting of spectroscopically secure cluster members and another group of spectroscopically secure non-cluster members. From these the distance from the individual galaxies to the secure group of cluster members is determined as a probability of the galaxy belonging to the cluster. 

The catalog was then further refined using the color-magnitude relation (CMR) of the spectroscopic and photometric members. The photometric sample was supplemented with galaxies fainter than the brightest cluster galaxies lying on the cluster sequence of the F606W-F814W versus F606W color-magnitude diagram. The CMR is defined using the biweight estimator on the spectroscopically confirmed members and from here, 66 more galaxies was selected. The F160W magnitude limit was fixed at magnitude 24 AB. The final catalog then constained 175 candidate members. Further analysis by spectroscopically confirming more cluster members \cite{Caminha2016} show that only 4 of the 175 galaxies selected by \citet{Grillo2015} has turned out not to be part of the cluster members. This indicates a very robust selection method. 

Since it is indicated by \citet{Jullo2009} that the cluster member halo shape should follow the distribution of the light, we have divided the cluster member catalog into two parts. One resembles the original from \citet{Grillo2015} where the mass distribution of the halos are assumed to be spherical and a new catalog, using the shape information from Source Extractor, where the shape of the mass distribution follow the shape of the light distribution. We have chosen the brightest galaxy $(\alpha:\,64.000642, \delta:\,-24.067472)$ in the catalog with a magnitude $mag_{F160W} = 17.02\maths{AB}$ as the reference magnitude for the optimization.

\begin{figure}[htb]
 \centering
 \includegraphics[width=0.9\textwidth,keepaspectratio=true]{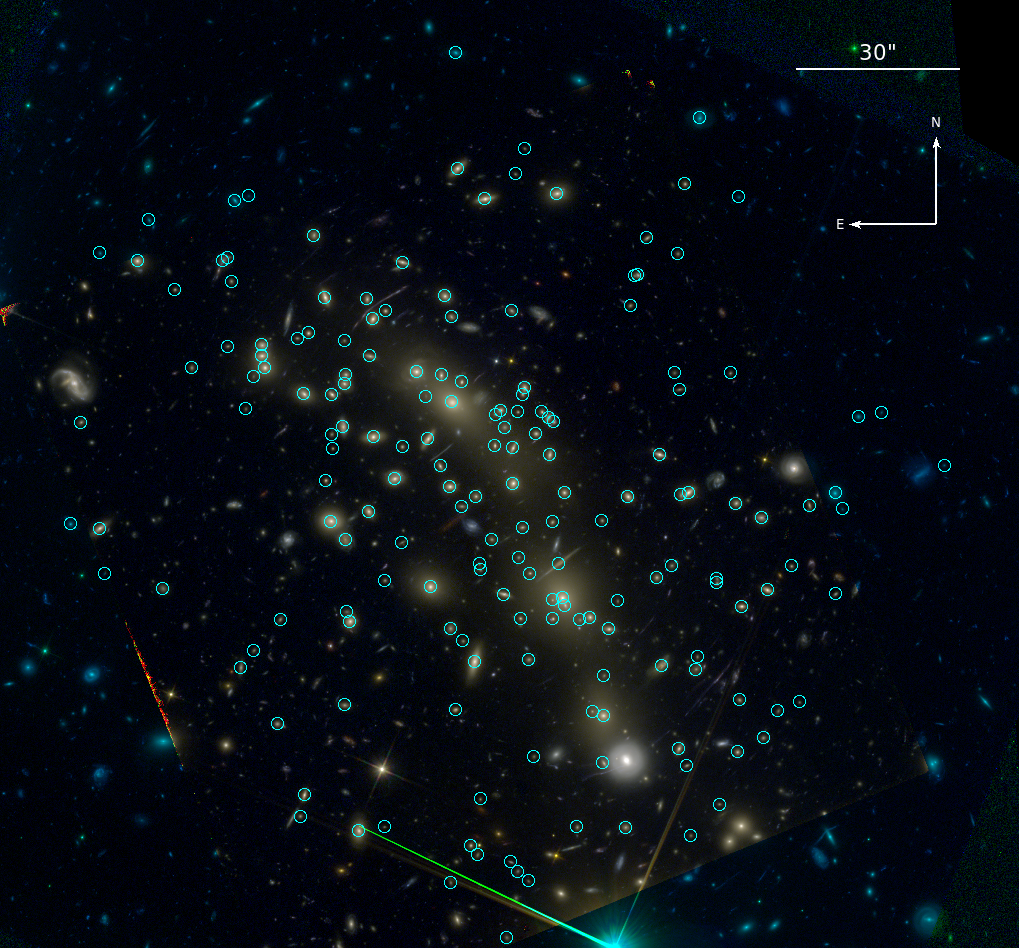}
 \caption{Color-composite image of \macs combining the HFF filter images. The cluster members from \citet{Grillo2015} are shown as cyan circles.}
 \label{fig:cluster_members}
\end{figure}

A color-composite image with the positions the cluster members superimposed, are shown in Figure \ref{fig:cluster_members}.

\newpage
\subsection{Cluster Halo Only}
The most simple model setup we can come about is the one where we include only the cluster scale halos. From \citet{Grillo2015} we have chosen to model the cluster scale halos using both PIEMD and NFW profile, although the NFW profile differs. Since the best-fit parameter-values from \citet{Grillo2015} are not readily available to use in \lenstool, we have chosen to use the parameter priors from \cite{Richard2014} as template values and have added our own constraints. This means that we have widened the parameter space in which \lenstool will find the best model. Using only two cluster scale halos, we have $12$ free parameters in the 2PIEMD model and $14$ free parameters in the 2NFW model.

\subsection{Cluster Halo with Cluster Members}
The ensemble of models resembling the model from \citet{Grillo2015} contains two cluster scale halos and the cluster members. The  models are divided into two models resembling the model from \citet{Grillo2015}, using either PIEMD (dPIE) or NFW profiles, with circular cluster member halos and no optimization of the scaling relation slopes, two models having circular and elliptical cluster member halos, respectively, and scaling relation slope optimization and one model resembling the \citet{Grillo2015}, but with elliptical cluster member halos. We have chosen to create models that optimize the scaling relation slopes in order to see whether a scaling relation slope that lies outside the traditional values (constant M/L or Fundamental Plane) will provide better a better model. Likewise, we have chosen to include elliptical mass profiles for the cluster members, in order to see whether the DM halos trace the light. We expect the 2dPIE \citet{Grillo2015} and 2dPIE models to give a very reasonable result which is in full agreement with \citet{Grillo2015}, but do not expect the 2NFW models to give reasonable results. 

The cluster members are added in \lenstool in a separate file, as mentioned previously. In \glee, the practice is to set the reference magnitude and velocity dispersion from the same galaxy, but since \lenstool optimize these values, they can be selected within a reasonable interval. We have chosen to optimize the scale radius for the cluster members within $50-400\maths{kpc}$ and the velocity dispersion within $100-500\maths{km\,s^{-1}}$. Furthermore, the slopes are optimized within $0-6$ for the scale radius slope and $0-6$ for the velocity dispersion slope. For the 2PIEMD + 175(+1)dPIE$_c$ $(M_TL^{-1}\sim L^{0.2})$, 2PIEMD + 175(+1)dPIE$_e$ $(M_TL^{-1}\sim L^{0.2})$, 2NFW + 175(+1)dPIE$_c$ $(M_TL^{-1}\sim L^{0.2})$ and 2NFW + 175(+1)dPIE$_e$ $(M_TL^{-1}\sim L^{0.2})$ models we have chosen to use the parameters derived from the Planck Mission \citet{Planck2015} $H_0 = 67.04346\maths{km\,s^{-1}\,Mpc^{-1}}$, $\Omega_M = 0.318263$ and $\Omega_{\Lambda} = 0.681737$. For such a cosmology $1\maths{arcsec} = 5.547\maths{kpc}$.

Although we do not expect it to give reasonable results, we have included a model using only a single cluster halo profile, in order to investigate whether \macs is a single or double profile cluster. All results so far, from both strong lensing and X-Ray observations, indicate that \macs is a bi- or trimodal cluster \cite{Grillo2015,Jauzac2015,Ogrean2015,Caminha2016}, indicating that at least two cluster scale halos should be used.

For the 2PIEMD + 175(+1)dPIE models without slope optimization we have $2 \times 6 = 12$ free parameters from the cluster scale halos, $2$ free parameters from the foreground galaxy and $2$ free parameters from the cluster member optimization, giving a total of $16$ free parameters. For the same models with slope optimization we have $2$ additional parameters, giving a total of $18$ free parameters. For the 2NFW + 175(+1)dPIE models we have $2 \times 7 = 14$ free parameters from the cluster halos. The foreground galaxy and cluster member parameters are same as for the 2PIEMD + 175(+1)dPIE models, giving a total of $18$ free parameters for the models without slope optimization and $20$ free parameters for the models with slope optimization. For the 1PIEMD + 175(+1)dPIE model we have $11$ free parameters. 

\section{Cosmological Models}
\label{sec:cosmo_models}
Another interesting aspect of strong lensing is the ability to optimize the cosmological parameters $\Omega_{m}$,  $\Omega_{\Lambda}$ and $w_x$. Given that the position of the multiple images not only depends on the mass distribution of the lens, but also on the angular diameter distances between the lens, source and observer, they provide valuable information about the underlying cosmology \cite{Jullo2010}. In practice this means that we can derive constraints on the underlying cosmology based on the sensitivity of the angular size-redshift with the sources at particular redshifts \cite{Jullo2010}.

In order to determine the cosmological parameters from a lensing model, we need to have a representative model of the mass distribution of the cluster and we need multiple images from sources at different redshifts. If we have a single source at one redshift, we have a degeneracy between the distance $D_{ls}$ and the mass distribution. We also require that there are no structure along the line-of-sight, since any such might perturb the image positions.

Previous studies on the estimation of the values of the cosmological parameters using strong lensing, have been performed on clusters in \citet{Jullo2010} and \citet{Caminha2016} and galaxy-galaxy lensing in \citet{Mandelbaum2013} and \citet{Grillo2008}. For the cluster lensing, \citet{Jullo2010} applied a technique where one finds the angular diameter distance ratio for two images from different sources, from which the constraints on $\Omega_m$ and $w_x$ was extracted on the massive cluster Abell 1689. This cluster has 114 multiple images from 34 unique background sources, which was reduced to 28 images from 12 families. Abell 1689 consists of two groups of galaxies, a dominant located at the center and a secondary group located $\sim 1\arcmin$ to the North. Since this indicates that the cluster might be merging, it holds similarity to \macs. \citet{Jullo2010} modelled Abell 1690 using a set of parametrized pseudo-elliptical potentials and was able to reproduce the images with an average positional accuracy of $2.87\arcsec$. The Hubble parameter was fixed to $74\maths{km\,s^{-1}\,Mpc^{-1}}$ since the test is not sensitive to the value of this parameter. \citet{Jullo2010} find that when combining their results with that from WMAP5 and x-ray clusters their results are in agreement with the combined results from WMAP5, BAO and Supernovae. When combining with all available results, they reduce the error in $w_x$ by $30\%$.

The combination of geometrical probes and statistics depending on the cosmic growth of structure (matter power spectrum) is recognized as a critical parameter in the efforts to measure the global geometry of the Universe \cite{Caminha2016b}. Here gravitational lensing is a powerful tool to investigate this global structure. In addition, the observed position and time delays of multiple images are also sensitive to the geometry of the Universe. This means that these observables are directly dependent on the angular diameter distances and hence can be measured as a function of the redshift of lens and source \cite{Caminha2016b}. It is only recently that the possibility of using the observed position of multiple spectroscopically confirmed images to determine the constraints of the cosmological parameters became available. \citet{Caminha2016b} conducted a  strong lensing model on the massive cluster RXC J2248, which has also been studied using X-ray emission and weak lensing analysis with a general good agreement between these methods, indicating a robust total mass estimate. \citet{Caminha2016b} used a single PIEMD profile to describe the cluster scale halo. They also considered a NFW profile with an elliptical potential, but it provided a significantly worse fit. In order to constrain the cosmological parameters they optimized the parameters $\Omega_m$, $\Omega_{\Lambda}$ and $w_x$. For the flat cosmological model their results were in good agreement with that from CMB probes. They also found a clear modelling degeneracy between $\Omega_m$ and $\Omega_{\Lambda}$ parameters. This suggest that a large sample of spectroscopically confirmed images are essential in constraining the cosmological parameters. Furthermore, they anticipate that repeating this experiment on other CLASH-VLT clusters will help constrain the cosmological parameters even further \cite{Caminha2016b}.

Since we are not dealing with galaxy-galaxy lensing in this work directly, we will only briefly mention the two studies. \citet{Grillo2008} used a novel technique involving the combined measurement of stellar dynamics and gravitional lensing on distant elliptical galaxies, starting from a simple assumption that homology strictly follow these systems. In particular, they explored whether this assumption could lead to constraints on the cosmological parameters. By first investigating the method on simulated galaxies and then moving on to actual data from real galaxies, they found that this method indeed could be used to measure the geometry of the Universe and that with a large number of lenses, that this method could lead to constraints on the cosmological parameters comparable to that of the standard techniques \cite{Grillo2008}. \citet{Mandelbaum2013} used a different technique that involved the autocorrelation of galaxy positions (galaxy clusterings) and the cross-correlation between foreground galaxy positions and background galaxy shears (galaxy-galaxy lensing). Like \citet{Grillo2008} they initially tested the method on simulations (zHORIZON) and later on data from SDSS DR7. They found that the method they applied gave results consistent with previous studies.

For \macs we have both multiple images from different redshifts and a reasonable mass model so it should be possible to optimize the cosmological parameters. This has been attempted previously on \macs, but was abandoned due to problems with structure along the line of sight (C. Grillo, private comm.).

In this thesis we propose a different method for evaluating the cosmological parameters. We run a series of models of \macs using different cosmological parameters. From these models we derive the cumulative projected total mass in order to see whether it could be used to estimate the cosmological parameters. If it is viable, it is the idea that the cumulative projected total mass estimate is to be compared with the results from dynamical and X-ray analyses \cite{Balestra2013, Ogrean2015}.

\begin{figure}[htb]
 \centering
 \includegraphics[width=0.6\textwidth,keepaspectratio=true]{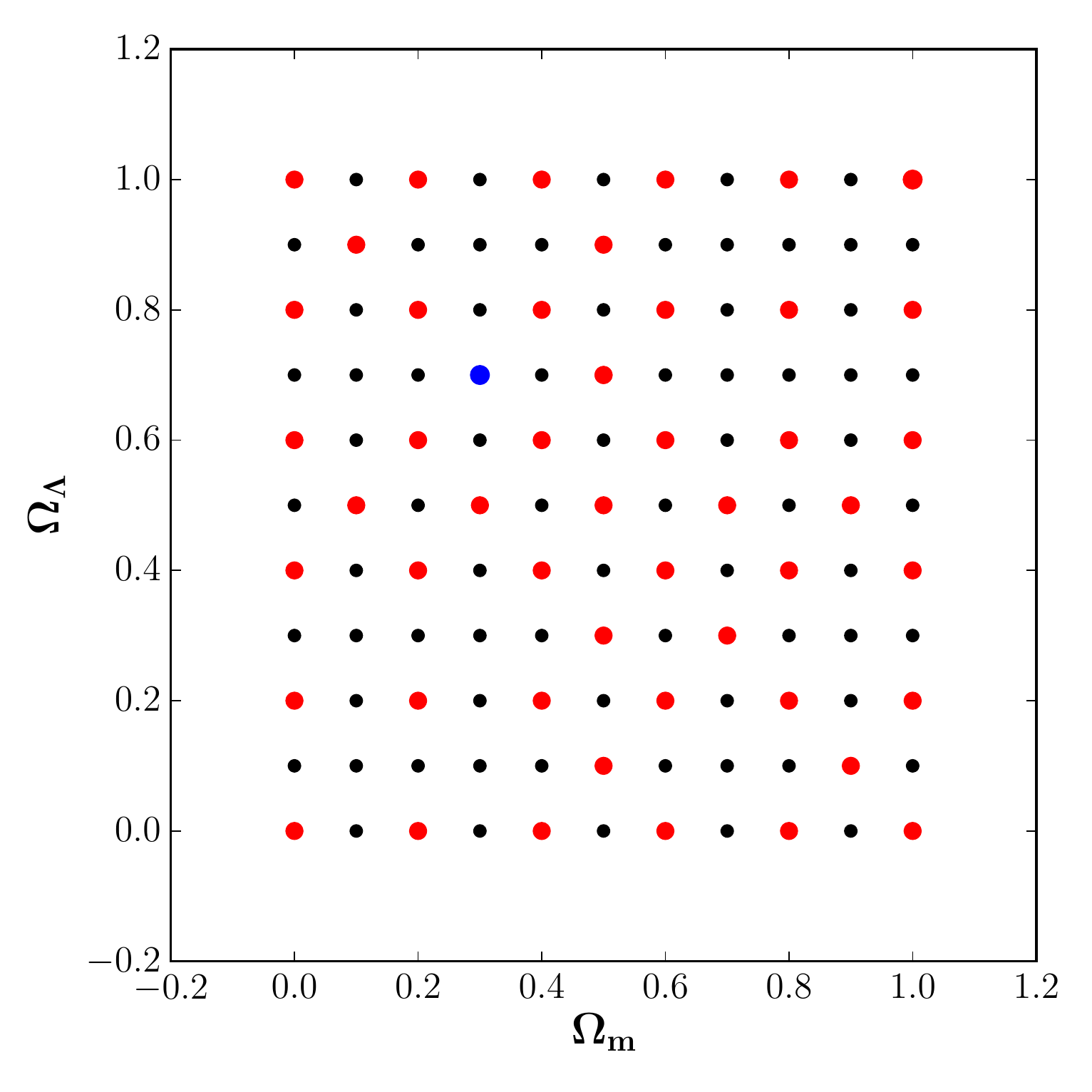}
 \caption{Graphical representation of all the possible permutations of models using a 0.1 precision. The optimal situation with 121 different models are represented with black dots, the actual model situation are represented with red dots and the model from the current cosmology is represented with a blue dot. The filled diagonal running from $\Omega_M = 1.0$ to $\Omega_M = 0.0$ represents the physically meaningful cosmological parameters.}
 \label{fig:cosmo-models}
\end{figure}

Specifically, we systematically change the cosmological parameters for each model we run. Optimally we would create a model for each permutation of the cosmological parameters with a precision of $0.1$ which will give $11 \times 11 = 121$ different models. Because of time constraints we have aimed at running a little more than half of the models, emphasizing the models which make physical sense. These are shown in Figure \ref{fig:cosmo-models} and give us $6 \times 6 = 36$ and with additional $13$ manually selected models, we get in total $49$ different cosmological models. In order to make the models directly comparable to \citet{Grillo2015} we have chosen the GrHa model as the template for this systematic test, regardless of what we find as best model. Since \lenstool will not accept $0$ as an actual value for both cosmological parameters (\lenstool will not converge), we have chosen to accept $0.01$ as a value close enough to $0$. This is represented in the table as values inside parenthesis. The cosmological parameters for the original model are also included, for comparison.

We have chosen not to change the Hubble constant $H_0$. It is a general praxis when modelling gravitational lenses that one use the distance ratio $\frac{D_{ls}}{D_{s}}$ in modelling and since both $D_{ls} \propto H_0$ and $D_{s} \propto H_0$, we have the ratio $\frac{D_{ls}}{D_{s}} \propto \frac{H_0}{H_0}$ which completely eliminates the Hubble constant from the optimization. 

By comparing the statistical distribution of the cumulative mass from the different cosmological models, with that of the results from X-ray and dynamical relations, it is the future hope that we find a significant representation of the true cosmological parameter values. In order to compare the results from X-ray and dynamics with strong lensing, we have to account for two major differences. First, both dynamics and X-ray models find the 3D mass, while strong lensing (lensing in general) find the 2D mass. Second, the X-ray and dynamics models are only calculated for the current cosmology, hence we need to adjust the results from the X-ray and dynamics to represent results for different cosmologies. In order to do that, we need to know how the different results scale, if they scale at all, with cosmology. In \citet{Balestra2013} we find that they use MAMPOSSt method \cite{Mamon2013} to calculate the dynamical mass and the method described in \citet{Evrard1996} to find the mass from the X-ray data. By rearranging \eqref{eqn:MAMPOSSt} we find that the mass can be defined as 
\[
 M(r)_{\mathrm{X-ray}} = -\frac{d(v \sigma_{r}^{2})}{dr} \frac{r}{Gv(r)} - 2\beta v\sigma_{r}^{2} \frac{r}{Gv(r)} \\
\]
and see that the mass scale linearly with $r$. We do the same with the X-ray mass estimate method \eqref{eqn:X-ray-mass}. We redefine  $\frac{d\log{\rho(r)}}{d\log{r}} = \alpha$ and $\frac{d\log{T(r)}}{d\log{r}} = \beta$ and get
\[
 M(r)_{\mathrm{Dyn.}} = -\frac{kT(r)}{G\mu m_p}r \left[ \alpha + \beta \right]
\]
and we find that the X-ray mass scales linearly with $r$. This is important since we want to rescale the mass estimates from these methods to fit the mass estimates from our different cosmological models.

%% file: results.tex
\chapter{Results}\label{chap:results}

\section{Model Ranking}
In order to explore the various possibilities of the best parameter values for \macs, we use a large ensemble of models (see Section \ref{sect:Modelling} and Table \ref{table:model_list}). To reduce the time spent on computation, to get a first impression on the model validity and to narrow down the width of the priors, we first optimize all models in the source plane. If we can exclude models from source plane optimization, we will do that, but the main purpose is to ensure that the priors remain within the limits. Priors excessing the limits could indicate that we need to expand the parameter space.

Our first impression is that the models applying elliptical cluster member profiles seems to provide better results than the models applying circular cluster member profiles (see Table \ref{table:Results-Source-plane-opt}). From the Evidence and $\chi^2$ ranking, we can clearly exclude the 2PIEMD, 2NFW and 1PIEMD + 175(+1)dPIE$_c$ $(M_TL^{-1}\sim L^{0.2})$ models from further analysis. The rest will undergo image plane optimization.

As \citet{Grillo2015}, we find clear evidence that the best results are derived from a model using two cluster halos and 175 cluster member halos, where the 2NFW, 2PIEMD and 1PIEMD + 175(+1)dPIE$_c$ $(M_TL^{-1}\sim L^{0.2})$ gives significantly higher $\chi^2$ and $\log{(E)}$.

\begin{table}[!ht]
 \centering
 \caption{Table over $\Delta \chi^2$ and $\Delta \log{(E)}$ for the different models, optimized in source plane. The models are ranked after best evidence $(\log{(E)})$, but also chosen after lowest $\chi^2$.}
 \label{table:Results-Source-plane-opt}
 \begin{tabular}{c c c}
  \toprule
  Model & $\chi^2$ & $\log{(E)}$ \\
  \midrule
  2PIEMD + 175(+1)dPIE$_e$ $(M_TL^{-1} = v)$						& $350$  & $-100$ \\
  2PIEMD + 175(+1)dPIE$_c$ $(M_TL^{-1} = v)$		         		& $342$  & $-104$ \\
  2PIEMD + 175(+1)dPIE$_c$ $(M_TL^{-1}\sim L^{0.2})$				& $408$  & $-123$ \\
  2PIEMD + 175(+1)dPIE$_e$ $(M_TL^{-1}\sim L^{0.2})$				& $441$  & $-144$ \\
  2NFW + 175(+1)dPIE$_e$ $(M_TL^{-1} = v)$							& $687$  & $-306$ \\
  2NFW + 175(+1)dPIE$_c$ $(M_TL^{-1} = v$                         & $596$  & $-308$ \\
  2NFW + 175(+1)dPIE$_c$ $(M_TL^{-1}\sim L^{0.2})$					& $1555$ & $-734$ \\
  2PIEMD                     		  								& $4144$ & $-1980$ \\
  2NFW                      		  								& $4276$ & $-2079$ \\
  1PIEMD + 175(+1)dPIE$_c$ $(M_TL^{-1}\sim L^{0.2})$				& $11725$ & $-5765$ \\
  \bottomrule
 \end{tabular}
\end{table}

We suspected that the source-plane optimization might provide very different parameter values, than the image-plane optimization, but the differences turn out to be only minor. This is fortunate, since we used the source plane optimization to tune the parameter space.

\begin{table}[!ht]
 \centering
 \caption{Table over $\chi^2$ and $\log{(E)}$ for the selected models, optimized in image plane. The models are ranked according to the evidence $\log{(E)}$.}
 \label{table:Results:Image-plane-opt}
 \begin{tabular}{c c c}
  \toprule
  Model & $\chi^2$ & $\log{(E)}$ \\
  \midrule
  2PIEMD + 175(+1)dPIE$_e$ $(M_TL^{-1} = v)$						& $461$		& $-205$	\\
  2PIEMD + 175(+1)dPIE$_c$ $(M_TL^{-1} = v)$        				& $486$ 	& $-216$ 	\\
  2PIEMD + 175(+1)dPIE$_e$ $(M_TL^{-1}\sim L^{0.2})$				& $715$		& $-349$ \\
  2PIEMD + 175(+1)dPIE$_c$ $(M_TL^{-1}\sim L^{0.2})$				& $742$		& $-349$	\\
  2NFW + 175(+1)dPIE$_c$ $(M_TL^{-1} = v$                         & $740$		& $-354$ \\
  2NFW + 175(+1)dPIE$_e$ $(M_TL^{-1} = v)$							& $708$		& $-422$ \\
  2NFW + 175(+1)dPIE$_c$ $(M_TL^{-1}\sim L^{0.2})$					& $1754$	& $-841$ \\
  \bottomrule
 \end{tabular}
\end{table}

The 2PIEMD + 175(+1)dPIE$_c$ $(M_TL^{-1}\sim L^{0.2})$ (hereafter, GrHa) model gave $\chi^2 = 742$ and an total position error (rms offset) of $\Delta_{rms} = 0.34\arcsec$ which is better than the result from \citet{Grillo2015} ($\chi^2 = 915$ and $\Delta_{rms} = 0.36\arcsec$). \citet{Grillo2015} do not report $\log{(E)}$, presumably the value is not provided by \glee. Compared with the results from \citet{Zitrin2013} ($\Delta_{rms,NFW} = 1.89\arcsec$ and $\Delta_{rms,eGauss} = 1.37\arcsec$) our results are better. We also find our $\Delta_{rms}$ to be better than that from \citet{Johnson2014}, where we have calculate their total $\Delta_{rms} = 0.51\arcsec$ from the reported $\Delta_{rms}$ of the individual images. Of more recent results, \citet{Caminha2016} find a total position error of $\Delta_{rms} = 0.59\arcsec$. Although our results seem to be significantly better than \citet{Caminha2016}, the additional constraints from the increased number of multiple images compared with the free parameters (We have $40/18 = 2.2$ for our best model, \citet{Caminha2016} have $130/26 = 5$), increase the $\chi^2$ and position-error.

Our best model (2PIEMD + 175(+1)dPIE$_e$ $(M_TL^{-1} = v)$) give us an error of $\chi^2 = 461$ and $\log{(E)} = -205$ which is not only better that the original model from \citet{Grillo2015} but also better that our GrHa model. Our best model also has significantly better $\Delta_{rms} = 0.27$ than the \citet{Grillo2015} model. However, if we take a look at our next-best model (2PIEMD + 175(+1)dPIE$_c$ $(M_TL^{-1} = v)$) we find an error of $\chi^2 = 486$ and $\log{(E)} = -217$. The total position offset for this model $(\Delta_{rms} = 0.27)$ is the same as that from our best model. 

Even though the results clearly favour the model using elliptical mass density distributions for the cluster member halos, considering the additional computation-time needed compared with the circular cluster member profiles and the little difference in $\chi^2$, $\log{(E)}$ and $\Delta_{rms}$, we conclude that the additional time needed to optimize using elliptical profiles are not worthwhile. We select the 2PIEMD + 175(+1)dPIE$_c$ $(M_TL^{-1} = v)$ as our best model.

This actually presents an interesting question regarding the cluster member profiles. The standard method has been to use cluster member profiles that trace the light, which inherently are elliptical in shape \cite{Jullo2007, Ardis2007, Limousin2008, Richard2009}, but using circular cluster member profiles seems to give equivalent results and significantly faster computation. This is particular important given the recent increase in available multiple images and confirmed cluster members and is something that should be investigated further, in future studies, both on new and previously modelled systems. 

In order to get realistic information about the uncertainties in the parameter values in our models we change the position error on the images in order to get a final $\chi^2$ that is equal to the number of dof. For the GrHa model we change the position error to $0.361\arcsec$ which gives a final model error of $\chi^2 = 27 \, (\mathrm{dof} = 24)$. For our best model, we change the position error to $0.317\arcsec$ which gives the final model error $\chi^2 = 24 (\mathrm{dof} = 22)$. This gives us final reduced $\chi^2$ values close to one ($\chi^{2}_{\mathrm{GrHa}} = 1.1$ and $\chi^{2}_{\mathrm{best}} = 1.1$). We also find that both our changed position errors are comparable to the position error from \citet{Grillo2015} $(\approx 0.4\arcsec)$. The better resolved model errors will enable us to account for any structure along the line of sight and small dark matter clumps, not included in our models.

\section{Parameter Values}
Since \citet{Jauzac2014}, \citet{Jauzac2015} and \citet{Caminha2016} use \lenstool we can compare the results directly. But since \citet{Grillo2015} use \glee we need to convert the results to \glee value units.

\subsection{\glee parameter values}
The converted parameter values can be found in Table \ref{table:grha-parm-values-glee} for the GrHa model and Table \ref{table:best-parm-values-glee} for our best model. 

Overall we find that for both models our results are reasonably well constrained. For the GrHa model we do find that there are some problems constraining the cut radius and Einstein radius for the cluster members, given the relatively large differences in errors whereas for our best model we find that only the cut radius is unreasonably underconstrained. We do however also find problems constraining the Einstein radius slope in our best model. These values might be correlated, for which a later analysis into the degeneracy of the parameters, might shed light on.

In general we see that our results are reasonably in agreement with the results in  Table 7 from \citet{Grillo2015}, comparing the results at $3\sigma$ CL. The largest discrepancies are found at the Einstein radius and core radius where \citet{Grillo2015} finds $\vartheta_{E,h1} = 21.0^{+8.6}_{-7-1}\maths{arcsec}$ and $r_{c,h1} = 12.9^{+7.5}_{-5.7}\maths{arcsec}$ for halo 1, $\vartheta_{E,h2} = 32.8^{+9.5}_{-7.4}\maths{arcsec}$ and $r_{c,h2} = 14.0^{+4.8}_{-4.0}\maths{arcsec}$ for halo 2 and $\vartheta_{E,gal} = 2.3^{+2.6}_{-1.1}\maths{arcsec}$ and $r_{s,gal} = 21^{+39}_{-17}\maths{arcsec}$ for the cluster members.

\begin{table}[!ht]
 \centering
 \caption{Parameter values for our GrHa model in \glee coordinates.}
 \label{table:grha-parm-values-glee}
 \begin{tabular}{cccccc}
    \toprule
        & Best & Median & $1\sigma$ CL & $2\sigma$ CL & $3\sigma$ CL \\ 
   \midrule
   $x_{\mathrm{h1}}\,(\arcsec)$ & $-8.0$ & $-8.1$ & $^{+1.5}_{-1.3}$ & $^{+3.2}_{-2.7}$ & $^{+5.2}_{-3.8}$ \\ 
   $y_{\mathrm{h1}}\,(\arcsec)$ & $5.2$ & $5.6$ & $^{+1.4}_{-1.5}$ & $^{+2.7}_{-3.0}$ & $^{+3.8}_{-4.4}$ \\ 
   $\varepsilon_{\mathrm{h1}}$ & $0.19$ & $0.21$ & $^{+0.04}_{-0.04}$ & $^{+0.07}_{-0.09}$ & $^{+0.10}_{-0.14}$ \\ 
   $\theta_{\mathrm{h1}}\,(\mathrm{rad})$ & $2.59$ & $2.60$ & $^{+0.03}_{-0.03}$ & $^{+0.06}_{-0.07}$ & $^{+0.08}_{-0.10}$ \\ 
   $r_{\mathrm{c,h1}}\,(\arcsec)$ & $14.7$ & $15.4$ & $^{+1.9}_{-2.1}$ & $^{+3.5}_{-4.5}$ & $^{+5.3}_{-7.4}$ \\ 
   $\vartheta_{\mathrm{E.h1}}\,(\arcsec)$ & $16.87$ & $16.01$ & $^{+1.85}_{-1.97}$ & $^{+3.49}_{-3.97}$ & $^{+4.50}_{-6.03}$ \\ 
   $x_{\mathrm{h2}}\,(\arcsec)$ & $22.9$ & $23.9$ & $^{+0.8}_{-0.8}$ & $^{+1.6}_{-1.6}$ & $^{+2.6}_{-2.3}$ \\ 
   $y_{\mathrm{h2}}\,(\arcsec)$ & $-42.0$ & $-43.0$ & $^{+1.0}_{-1.0}$ & $^{+2.0}_{-2.1}$ & $^{+3.0}_{-3.5}$ \\ 
   $\varepsilon_{\mathrm{h2}}$ & $0.37$ & $0.39$ & $^{+0.03}_{-0.03}$ & $^{+0.05}_{-0.07}$ & $^{+0.08}_{-0.11}$ \\ 
   $\theta_{\mathrm{h2}}\,(\mathrm{rad})$ & $2.23$ & $2.22$ & $^{+0.01}_{-0.01}$ & $^{+0.03}_{-0.03}$ & $^{+0.04}_{-0.04}$ \\ 
    $r_{\mathrm{c,h2}}\,(\arcsec)$ & $16.7$ & $16.7$ & $^{+1.3}_{-1.3}$ & $^{+2.7}_{-2.8}$ & $^{+3.8}_{-4.1}$ \\ 
   $\vartheta_{\mathrm{E,h2}}\,(\arcsec)$ & $26.15$ & $25.06$ & $^{+1.95}_{-2.02}$ & $^{+3.93}_{-4.29}$ & $^{+5.87}_{-5.97}$ \\ 
   $r_{\mathrm{s,gal}}\,(\arcsec)$ & $16.2$ & $26.8$ & $^{+9.7}_{-18.5}$ & $^{+15.3}_{-39.5}$ & $^{+17.1}_{-47.6}$ \\ 
   $\vartheta_{\mathrm{E,gal}}\,(\arcsec)$ & $1.63$ & $1.55$ & $^{+0.19}_{-0.26}$ & $^{+0.35}_{-0.62}$ & $^{+0.48}_{-1.08}$ \\ 
   $\zeta_{r_{\mathrm{s,gal}}}\,$ & $[0.50]$ &  &  &  &  \\ 
   $\zeta_{\vartheta_{\mathrm{E,gal}}}\,$ & $[0.70]$ &  &  &  &  \\ 
   \bottomrule
   \end{tabular}
\end{table}

\begin{table}[!ht]
 \centering
 \caption{Parameter values for our best model (2PIEMD + 175(+1)dPIE$_c$ $(M_TL^{-1} = v)$) in \glee coordinates.}
 \label{table:best-parm-values-glee}
 \begin{tabular}{cccccc}
    \toprule
        & Best & Median & $1\sigma$ CL & $2\sigma$ CL & $3\sigma$ CL \\ 
   \midrule
   $x_{\mathrm{h1}}\,(\arcsec)$ & $-10.8$ & $-11.3$ & $^{+2.2}_{-2.1}$ & $^{+3.5}_{-3.8}$ & $^{+3.7}_{-5.1}$ \\ 
   $y_{\mathrm{h1}}\,(\arcsec)$ & $7.3$ & $7.4$ & $^{+1.7}_{-1.5}$ & $^{+3.1}_{-2.7}$ & $^{+4.1}_{-3.3}$ \\ 
   $\varepsilon_{\mathrm{h1}}$ & $0.12$ & $0.13$ & $^{+0.03}_{-0.03}$ & $^{+0.05}_{-0.07}$ & $^{+0.07}_{-0.10}$ \\ 
   $\theta_{\mathrm{h1}}\,(\mathrm{rad})$ & $2.61$ & $2.59$ & $^{+0.02}_{-0.03}$ & $^{+0.04}_{-0.05}$ & $^{+0.06}_{-0.07}$ \\ 
   $r_{\mathrm{c,h1}}\,(\arcsec)$ & $16.5$ & $17.5$ & $^{+2.0}_{-2.3}$ & $^{+3.6}_{-4.6}$ & $^{+5.3}_{-7.4}$ \\ 
   $\vartheta_{\mathrm{E.h1}}\,(\arcsec)$ & $16.43$ & $17.09$ & $^{+1.46}_{-1.78}$ & $^{+2.72}_{-3.74}$ & $^{+3.64}_{-6.22}$ \\ 
   $x_{\mathrm{h2}}\,(\arcsec)$ & $23.4$ & $24.2$ & $^{+0.9}_{-0.9}$ & $^{+1.7}_{-1.7}$ & $^{+2.5}_{-2.5}$ \\ 
   $y_{\mathrm{h2}}\,(\arcsec)$ & $-42.4$ & $-43.3$ & $^{+1.3}_{-1.2}$ & $^{+2.7}_{-2.4}$ & $^{+3.9}_{-3.4}$ \\ 
   $\varepsilon_{\mathrm{h2}}$ & $0.32$ & $0.34$ & $^{+0.02}_{-0.03}$ & $^{+0.05}_{-0.06}$ & $^{+0.08}_{-0.08}$ \\ 
   $\theta_{\mathrm{h2}}\,(\mathrm{rad})$ & $2.21$ & $2.20$ & $^{+0.01}_{-0.01}$ & $^{+0.03}_{-0.03}$ & $^{+0.05}_{-0.04}$ \\ 
    $r_{\mathrm{c,h2}}\,(\arcsec)$ & $19.4$ & $18.3$ & $^{+1.4}_{-1.5}$ & $^{+2.9}_{-2.9}$ & $^{+4.3}_{-4.0}$ \\ 
   $\vartheta_{\mathrm{E,h2}}\,(\arcsec)$ & $28.42$ & $26.07$ & $^{+1.99}_{-2.33}$ & $^{+4.03}_{-4.69}$ & $^{+5.99}_{-6.67}$ \\ 
   $r_{\mathrm{s,gal}}\,(\arcsec)$ & $19.0$ & $26.7$ & $^{+8.5}_{-13.5}$ & $^{+14.8}_{-33.3}$ & $^{+17.5}_{-44.1}$ \\ 
   $\vartheta_{\mathrm{E,gal}}\,(\arcsec)$ & $2.98$ & $2.97$ & $^{+0.41}_{-0.48}$ & $^{+0.86}_{-1.04}$ & $^{+1.29}_{-1.74}$ \\ 
   $\zeta_{r_{\mathrm{s,gal}}}\,$ & $0.39$ & $0.50$ & $^{+0.12}_{-0.29}$ & $^{+0.16}_{-0.72}$ & $^{+0.17}_{-1.04}$ \\ 
   $\zeta_{\vartheta_{\mathrm{E,gal}}}\,$ & $1.18$ & $1.14$ & $^{+0.12}_{-0.12}$ & $^{+0.30}_{-0.26}$ & $^{+0.39}_{-0.43}$ \\ 
   \bottomrule
  \end{tabular}
\end{table}

\subsection{\lenstool parameter values}
\citet{Richard2014,Jauzac2014,Jauzac2015,Caminha2016} all use \lenstool for their analysis of \macs and we can therefore compare our values directly with theirs. Furthermore, the comparison with these values are interesting because they all use different approaches, regarding number of cluster halos, multiple images and cluster-member selection. In general we can say that new evidence suggests that we need more than two halos in order to correctly model \macs. 

Our results in \lenstool units are presented in Table \ref{table:grha-parm-values-lenstool} for our GrHa model and in Table \ref{table:best-parm-values-lenstool} for our best model. Since the conversion from \lenstool values to \glee values does not entail changes in the model itself but a direct conversion of the final values, we see the same constraining problems here.

Comparing our results with the results from \citet{Richard2014} we find that our results differs considerably within $1\sigma$ CL. When we look at the position parameter values for the cluster halos $x_{h1} = -5.6^{+1.0}_{-0.6}\maths{arcsec}$, $y_{h1} = 2.7^{+0.7}_{-0.7}\maths{arcsec}$, $x_{h2} = 23.7^{+1.4}_{-0.7}\maths{arcsec}$, $y_{h2} = -45.7^{+1.5}_{-1.4}\maths{arcsec}$ from \citet{Richard2009}, we see that ours differ significantly for halo 1 and coincide for halo 2, for both our models. We see that our results agree for the mass-ellipticity of the cluster halos and to some extent the position angle. We find partial agreement for the cluster halo core radii (\citet{Richard2014} finds $r_{c,h1} = 76^{+12}_{-7}\maths{kpc}$ and $r_{c,h2} = 120^{+10}_{-7}\maths{kpc}$) where we find $r_{c,h1} = 78.4^{+10.0}_{-11.3}\maths{kpc}$ and $r_{c,h2} = 89.4^{+7.0}_{-7.1}\maths{kpc}$ which shows that we have agreement with our results for halo 1, but not halo 2. For our best model we have $r_{c,h1} = 91.7^{+11.1}_{-12.8}\maths{kpc}$ and $r_{c,h2} = 107.4^{+7.7}_{-8.1}\maths{kpc}$ we find agreement with both values. \citet{Richard2014} find $\sigma_{h1} = 809^{+46}_{-38}\maths{km\,s^{-1}}$ and $\sigma_{h2} = 1019^{+36}_{-43}\maths{km\,s^{-1}}$ where we find $\sigma_{h1} = 765^{+44}_{-44}\maths{km\,s^{-1}}$ and $\sigma_{h2} = 952^{+37}_{-37}\maths{km\,s^{-1}}$ for our GrHa model, which is in agreement. For our best model we find $\sigma_{h2} = 755^{+34}_{-39}\maths{km\,s^{-1}}$ and $\sigma_{h2} = 993^{+37}_{-42}\maths{km\,s^{-1}}$ which is also in agreement with the results from \citet{Richard2014}. Lastly, for the cluster members \citet{Richard2009} find $r_{s,gal} = 9^{+15}_{-8}\maths{kpc}$ and $\sigma_{gal} = 183^{16}_{-16}\maths{km\,s^{-1}}$. For our GrHa model, we find $r_{s,gal} = 86.3^{+51.8}_{-98.7}\maths{kpc}$ which we cannot compare since the truncation radius is greatly underconstrained. We also find $\sigma_{gal} = 237.5^{+15.1}_{-18.4}\maths{km\,s^{-1}}$ which is not in agreement. In general, we find several differences between our results and those from \citet{Richard2014}.

When we look at the results from \citet{Jauzac2014}, who have used the same general approach (two major cluster halos and several cluster-member halos), but with significantly different approach regarding multiple images and cluster members selection, we find that our results are in overall agreement. All values compared here are within $1\sigma$ CL since these are errorbars reported by \citet{Jauzac2014, Jauzac2015}. We see that the position of our cluster halo 1 is further away from the corresponding BCG $(\alpha=0.0\maths{arcsec},\delta=0.0\maths{arcsec})$ with $\Delta_{\alpha} \sim 4.5\maths{arcsec}, \Delta_{\delta} \sim 1.5\maths{arcsec}$ for \citet{Jauzac2015} where we find $\Delta_{\alpha} \sim 8.0\maths{arcsec},\Delta_{\delta} \sim 5.2\maths{arcsec}$ for our GrHa model and $\Delta_{\alpha} \sim 10.8\maths{arcsec},\Delta_{\delta} \sim 7.3\maths{arcsec}$ for our best model. The discrepancies are also evident within $1\sigma$ CL. We also find that our core radius for our best model is somewhat larger than \citet{Jauzac2015}, although our GrHa model is in agreement. We find a major disagreement between their angular position and ours, but since these values are practically perpendicular to each other, and other results suggest that the two cluster halos are similar in shape and angular position, we assume these to be in agreement. Since we know that DM is collisionless but gas is not, the results here suggest that the model from \citet{Jauzac2015} expects less gas and more DM in cluster halo 1, since the cluster member halo is closer to its corresponding BCG.

\begin{table}[!ht]
 \centering
 \caption{Parameter values for our GrHa model in \lenstool coordinates.}
 \label{table:grha-parm-values-lenstool}
 \begin{tabular}{cccccc}
    \toprule
        & Best & Median & $1\sigma$ CL & $2\sigma$ CL & $3\sigma$ CL \\ 
   \midrule
   $x_{\mathrm{h1}}\,(\arcsec)$ & $-8.0$ & $-8.1$ & $^{+1.5}_{-1.3}$ & $^{+3.2}_{-2.7}$ & $^{+5.2}_{-3.8}$ \\ 
   $y_{\mathrm{h1}}\,(\arcsec)$ & $5.2$ & $5.6$ & $^{+1.4}_{-1.5}$ & $^{+2.7}_{-3.0}$ & $^{+3.8}_{-4.4}$ \\ 
   $\varepsilon_{\mathrm{h1}}$ & $0.81$ & $0.79$ & $^{+0.04}_{-0.04}$ & $^{+0.09}_{-0.07}$ & $^{+0.14}_{-0.10}$ \\ 
   $\theta_{\mathrm{h1}}\,(\mathrm{deg})$ & $148.59$ & $148.79$ & $^{+1.67}_{-1.84}$ & $^{+3.22}_{-3.82}$ & $^{+4.85}_{-5.68}$ \\ 
   $r_{\mathrm{c,h1}}\,(\mathrm{kpc})$ & $78.4$ & $82.3$ & $^{+10.0}_{-11.3}$ & $^{+18.6}_{-24.0}$ & $^{+28.1}_{-39.6}$ \\ 
   $\sigma_{\mathrm{h1}}\,(\mathrm{km/s})$ & $765$ & $745$ & $^{+44}_{-44}$ & $^{+86}_{-87}$ & $^{+113}_{-129}$ \\ 
   $x_{\mathrm{h2}}\,(\arcsec)$ & $22.9$ & $23.9$ & $^{+0.8}_{-0.8}$ & $^{+1.6}_{-1.6}$ & $^{+2.6}_{-2.3}$ \\ 
   $y_{\mathrm{h2}}\,(\arcsec)$ & $-42.0$ & $-43.0$ & $^{+1.0}_{-1.0}$ & $^{+2.0}_{-2.1}$ & $^{+3.0}_{-3.5}$ \\ 
   $\varepsilon_{\mathrm{h2}}$ & $0.63$ & $0.61$ & $^{+0.03}_{-0.03}$ & $^{+0.07}_{-0.05}$ & $^{+0.11}_{-0.08}$ \\ 
   $\theta_{\mathrm{h2}}\,(\mathrm{deg})$ & $127.52$ & $127.43$ & $^{+0.72}_{-0.73}$ & $^{+1.48}_{-1.47}$ & $^{+2.27}_{-2.26}$ \\ 
    $r_{\mathrm{c,h2}}\,(\mathrm{kpc})$ & $89.4$ & $89.2$ & $^{+7.0}_{-7.1}$ & $^{+14.3}_{-15.1}$ & $^{+20.5}_{-22.0}$ \\ 
   $\sigma_{\mathrm{h2}}\,(\mathrm{km/s})$ & $952$ & $932$ & $^{+37}_{-37}$ & $^{+76}_{-77}$ & $^{+116}_{-105}$ \\ 
   $r_{\mathrm{s,gal}}\,(\mathrm{kpc})$ & $86.3$ & $143.1$ & $^{+51.8}_{-98.7}$ & $^{+81.6}_{-211.1}$ & $^{+91.5}_{-254.2}$ \\ 
   $\sigma_{\mathrm{gal}}\,(\mathrm{km/s})$ & $237$ & $232$ & $^{+15}_{-18}$ & $^{+27}_{-42}$ & $^{+39}_{-70}$ \\ 
   $\zeta_{r_{\mathrm{s,gal}}}\,$ & $[4.00]$ &  &  &  &  \\ 
   $\zeta_{\sigma_{\mathrm{gal}}}\,$ & $[2.86]$ &  &  &  &  \\ 
   \bottomrule
 \end{tabular}
\end{table}

Interestingly, we see the opposite tendency regarding cluster halo 2. We here see that the position of our cluster halo 2 is closer to the corresponding BCG $(\alpha=20.3\maths{arcsec},\delta=-35.8\maths{arcsec})$ with $\Delta_{\alpha} \sim 2.6\maths{arcsec},\Delta_{\delta} \sim 6.2\maths{arcsec}$ for our GrHa model and $\Delta_{\alpha} \sim 3.1\maths{arcsec},\Delta_{\delta} \sim 6.6\maths{arcsec}$ compared with \citet{Jauzac2015} $(\Delta_{\alpha} \sim 4.2\maths{arcsec},\Delta_{\delta} \sim 8.7\maths{arcsec})$. However, the values are are relatively close considering $1\sigma$ CL which means that although we might expect the model from \citet{Jauzac2015} to put more gas and less DM into the second cluster halo, relative to our models, we cannot say it with certainty from these results. When we compare the radii of the second cluster halo we find that our best model is in agreement with the results from \citet{Jauzac2015}, but our GrHa model shows a significantly smaller core radius for the second cluster halo. For both cluster halos we find that the velocity dispersion from the GrHa and our best model are in agreement with \citet{Jauzac2015}. This suggests that our GrHa model has higher mass concentration in both halos, compared with our best model. Also, it suggests that our GrHa model has a higher concentration in the second cluster halo, compared with \citet{Jauzac2015}. Conversely, our best model shows a lesser concentrated first cluster halo compared with the model from \citet{Jauzac2015}. 

\begin{table}[!ht]
 \centering
 \caption{Parameter values for our best model (2PIEMD + 175(+1)dPIE$_c$ $(M_TL^{-1} = v)$) in \lenstool coordinates.}
 \label{table:best-parm-values-lenstool}
 \begin{tabular}{cccccc}
    \toprule
        & Best & Median & $1\sigma$ CL & $2\sigma$ CL & $3\sigma$ CL \\ 
   \midrule
   $x_{\mathrm{h1}}\,(\arcsec)$ & $-10.8$ & $-11.3$ & $^{+2.2}_{-2.1}$ & $^{+3.5}_{-3.8}$ & $^{+3.7}_{-5.1}$ \\ 
   $y_{\mathrm{h1}}\,(\arcsec)$ & $7.3$ & $7.4$ & $^{+1.7}_{-1.5}$ & $^{+3.1}_{-2.7}$ & $^{+4.1}_{-3.3}$ \\ 
   $\varepsilon_{\mathrm{h1}}$ & $0.88$ & $0.87$ & $^{+0.03}_{-0.03}$ & $^{+0.07}_{-0.05}$ & $^{+0.10}_{-0.07}$ \\ 
   $\theta_{\mathrm{h1}}\,(\mathrm{deg})$ & $149.33$ & $148.51$ & $^{+1.30}_{-1.47}$ & $^{+2.44}_{-2.93}$ & $^{+3.58}_{-4.26}$ \\ 
   $r_{\mathrm{c,h1}}\,(\mathrm{kpc})$ & $91.7$ & $96.9$ & $^{+11.1}_{-12.8}$ & $^{+19.9}_{-25.6}$ & $^{+29.4}_{-40.8}$ \\ 
   $\sigma_{\mathrm{h1}}\,(\mathrm{km/s})$ & $755$ & $770$ & $^{+34}_{-39}$ & $^{+64}_{-80}$ & $^{+87}_{-129}$ \\ 
   $x_{\mathrm{h2}}\,(\arcsec)$ & $23.4$ & $24.2$ & $^{+0.9}_{-0.9}$ & $^{+1.7}_{-1.7}$ & $^{+2.5}_{-2.5}$ \\ 
   $y_{\mathrm{h2}}\,(\arcsec)$ & $-42.4$ & $-43.3$ & $^{+1.3}_{-1.2}$ & $^{+2.7}_{-2.4}$ & $^{+3.9}_{-3.4}$ \\ 
   $\varepsilon_{\mathrm{h2}}$ & $0.68$ & $0.66$ & $^{+0.03}_{-0.02}$ & $^{+0.06}_{-0.05}$ & $^{+0.08}_{-0.08}$ \\ 
   $\theta_{\mathrm{h2}}\,(\mathrm{deg})$ & $126.46$ & $126.07$ & $^{+0.80}_{-0.73}$ & $^{+1.67}_{-1.43}$ & $^{+2.65}_{-2.04}$ \\ 
    $r_{\mathrm{c,h2}}\,(\mathrm{kpc})$ & $107.4$ & $101.4$ & $^{+7.7}_{-8.1}$ & $^{+15.9}_{-15.9}$ & $^{+24.1}_{-22.1}$ \\ 
   $\sigma_{\mathrm{h2}}\,(\mathrm{km/s})$ & $993$ & $951$ & $^{+37}_{-42}$ & $^{+77}_{-82}$ & $^{+116}_{-115}$ \\ 
   $r_{\mathrm{s,gal}}\,(\mathrm{kpc})$ & $105.3$ & $148.0$ & $^{+47.2}_{-74.9}$ & $^{+81.9}_{-184.6}$ & $^{+97.0}_{-244.7}$ \\ 
   $\sigma_{\mathrm{gal}}\,(\mathrm{km/s})$ & $321$ & $321$ & $^{+23}_{-25}$ & $^{+51}_{-52}$ & $^{+79}_{-83}$ \\ 
   $\zeta_{r_{\mathrm{s,gal}}}\,$ & $5.11$ & $3.96$ & $^{+1.45}_{-1.27}$ & $^{+2.33}_{-1.92}$ & $^{+2.66}_{-2.03}$ \\ 
   $\zeta_{\sigma_{\mathrm{gal}}}\,$ & $1.69$ & $1.76$ & $^{+0.17}_{-0.21}$ & $^{+0.32}_{-0.62}$ & $^{+0.48}_{-0.91}$ \\ 
   \bottomrule
 \end{tabular}
\end{table}

Looking at the results from the cluster members we find a significant discrepancy between our models and \citet{Jauzac2015} which we would also expect, given the different approach in the selection of cluster members. \citet{Jauzac2015} find $r_{s,gal} = 29.5^{+7.4}_{-4.3}\maths{kpc}$ and $\sigma_{gal} = 147.9\pm6.2\maths{km/s}$ compared with $r_{s,gal} = 86.4^{+51.8}_{-98.7}\maths{kpc}$ and $\sigma_{gal} = 952^{+15}_{-18}\maths{km/s}$ for our GrHa model and $r_{s,gal} = 105.3^{+}_{-}\maths{kpc}$ and $\sigma_{gal} = 321^{+23}_{-25}\maths{km/s}$ for our best model. This shows that the discrepancy is independent of slope optimization.

\citet{Caminha2016} have produced a more recent analysis of \macs using the same approach to the selection of cluster members and multiple images, but instead using data from the new MUSE instrument. This has provided over 100 confirmed multiple images and almost 200 confirmed cluster members. \citet{Caminha2016} have also found evidence for the necessity of additional halos. They include three smooth DM halos and a galaxy-scale lensing system. The third smooth DM halo is positioned at $\alpha = -34.4\maths{arcsec},\delta = 7.9\maths{arcsec}$ and has been speculated to be spherical in nature. The galaxy-scale lensing system is found to produce several multiple images and hence needs to be added to further constrain the model. The galaxy-scale system consists of two galaxies G1 ($\alpha=64.034084,\delta=-24.066738$) and G2 ($\alpha=64.034191,\delta=-24.067072$).

In general we find that our results and the results from \citet{Caminha2016} are in agreement. The position of the cluster member halos are closer to the corresponding BCG than ours. For halo 1 they find an offset of $\Delta_{\alpha} \sim 2.4\maths{arcsec},\Delta_{\delta} \sim 1.8\maths{arcsec}$, which is much closer to the corresponding BCG than ours. Likewise, for halo 2, we see the same pattern with an offset of $\Delta_{\alpha} \sim 0.7\maths{arcsec},\Delta_{\delta} \sim 0.6\maths{arcsec}$. It would seem that they have found a way to better resolve the gas components from the DM components, even though they have not specifically added them as such. They speculate that the third smooth DM halo explain this difference. They find a smaller cluster halo 1 $r_{c,h1} = 33.6^{+10.2}_{-10.2}\maths{kpc}$ where ours is a little more than twice as large for the GrHa model and almost three times as large for our best model. We also find that our velocity dispersion for halo 1 is a little higher $(\sigma_{h1,\mathrm{GrHa}} = 747^{+113}_{-129}\maths{km/s} \, ; \, \sigma_{h1,\mathrm{best}} = 755^{+87}_{-129}\maths{km/s})$ than \citet{Caminha2016} $(\sigma_{h1} = 707^{+79}_{-83}\maths{km/s})$ but our results are comparable at $3\sigma$ CL. For halo 2 we find a larger halo than \citet{Caminha2016}. They find $r_{c,h2} = 66.8^{+8.5}_{-8.5}\maths{kpc}$ which is comparable to the radius for our GrHa model $(r_{c,h2} = 89.4^{+20.5}_{-22.0}\maths{kpc})$. They also find a relatively higher velocity dispersion $(\sigma_{h2} = 1102^{+47}_{-48}\maths{km/s})$ compared to our GrHa model $(952^{+116}_{-105}\maths{km/s})$ and our best model $(\sigma_{h2} = 993^{+116}_{-115}\maths{km/s})$.

We also find small disagreements regarding the cluster member parameters, although they are not very significant. \citet{Caminha2016} find $r_{t,gal} = 56.1^{+55.0}_{-33.1}\maths{kpc}$ and $\sigma_{gal} = 251^{+48}_{-40}\maths{km/s}$, where we find $r_{s,gal} = 86.3^{+91.5}_{-254.2}\maths{kpc}$ for GrHa and $\sigma_{gal} = 238^{+39}_{-70}\maths{km/s}$ and $r_{s,gal} = 105.3^{+97.0}_{-244.7}\maths{kpc}$ and $\sigma_{gal} = 321^{+79}_{-83}\maths{km/s}$ for our best model. These differences are however not significant at $3\sigma$ CL. What is interesting though is that both our models and the model from \citet{Caminha2016} seems to have difficulties getting the cut radius for the cluster members properly constrained. Especially ours are violently underconstrained, but all the models experience significant differences within $3\sigma$ errorbars. We conclude that the difference is not due to a problem with the model, but from a lack of data. We simply need multiple images around the individual cluster members in order to accurately constrain their parameter values.

To conclude this section we find that we do have a reasonable well constrained mass model of \macs that can serve for further analysis. 

\subsection{MCMC Analysis}
Like \citet{Grillo2015}, \citet{Jauzac2014,Jauzac2015} and \citet{Caminha2016} we also perform a MCMC analysis of our results. One of the advantages of the MCMC method in selecting the best model, is that we get information about the statistical uncertainties in our model The method, however, cannot not take any systematic uncertainties into account. We will here use the term correlated and degenerate interchangeably. The major difference is that degenerate accounts for both correlated and anti-correlated values. Here correlated means that changing a parameter $a$ in a positive direction will also increase the parameter $b$ in a positive direction, whereas anti-correlated means the opposite. Changing $a$ in a positive direction will change $b$ in a negative direction.

One of the problems in exploring the parameter space in strong lensing models is when parameter values are correlated or degenerate. This is of vital importance in order to evaluate whether model parameter values can be viewed as trustworthy.

The great advantage of using the MCMC method for exploring the parameter-space is that it gives direct insight into, not only the statistical probability of a given parameter value, but whether that parameter value is degenerate. These results can be extracted directly from the $10010$ MCMC samples. 

\begin{figure}[h!tb]
 \centering
 \includegraphics[width=0.99\textwidth,keepaspectratio=true]{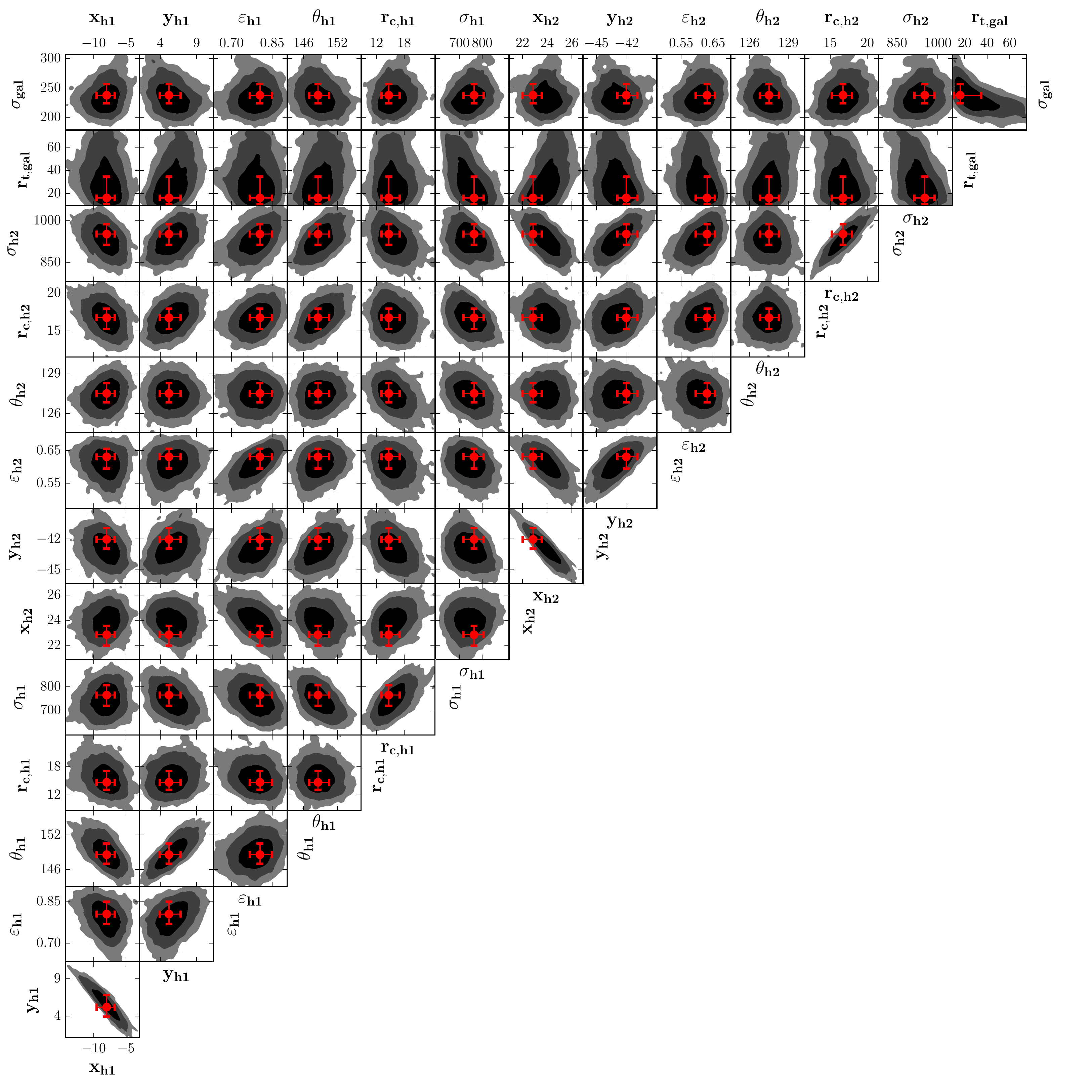}
 \caption{Uncertainty and correlation estimates of our GrHa model. The contours represent from inner to outer, the $1\sigma$, $2\sigma$ and $3\sigma$ confidence regions from the entire MCMC chain (10010 samples). The red dot represents the best value with $1\sigma$ error estimates.}
 \label{fig:degens_grha}
\end{figure}

\begin{figure}[h!tb]
 \centering
 \includegraphics[width=0.99\textwidth,keepaspectratio=true]{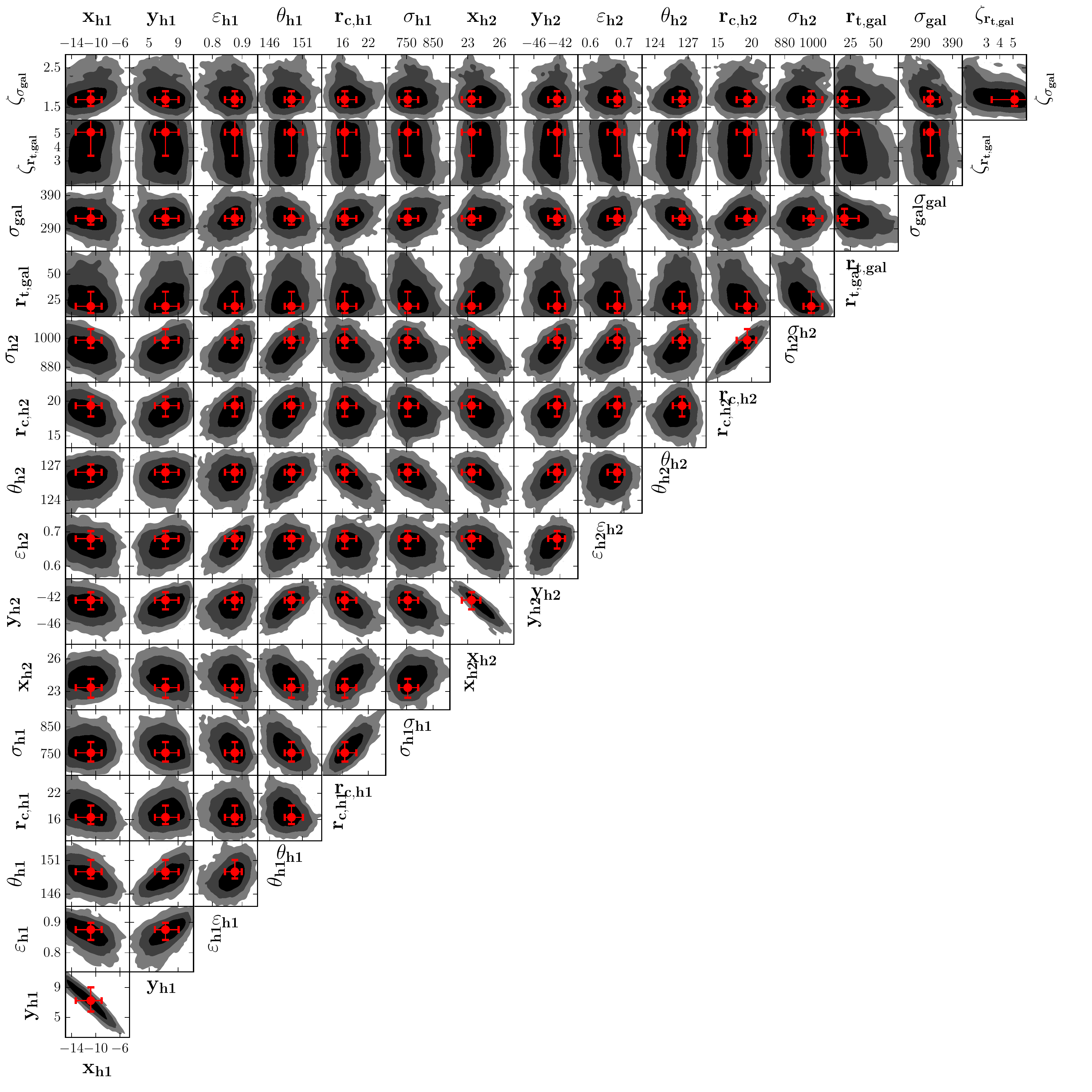}
 \caption{Same as Figure \ref{fig:degens_grha} but for our best model. Here we add uncertainty and correlation estimates of the slope of the velocity dispersion $\zeta_{\sigma_{gal}}$ and the slope of the truncation radius $\zeta_{r_{s,gal}}$.}
 \label{fig:degens_best}
\end{figure}

In order to investigate possible degeneracies, we have plotted the estimated uncertainties from our GrHa model in Figure \ref{fig:degens_grha} and from our best model in Figure \ref{fig:degens_best}.

We can confirm clear signs of anti-correlation between the $x_h$ and $y_h$ positions for both cluster halos \cite{Grillo2015}. This means that the cluster halos are only allowed to move in the North-East or South-West direction. Since we have a cluster with a large flat elliptical core, changing the position of the center along the length of the halo will not change the overall properties of the halo.  We also see clear signs of correlation between the velocity dispersion $\sigma_{h}$ and core radius $r_{c,h}$ for both cluster halos, confirming the findings of \citet{Grillo2015}. This can be explained by appreciating that in order to obtain a given projected mass within a given radius $(R < r_{c,h})$, the value of $\sigma_{h}$ have to increase in order to counterbalance any increase in $r_{c,h}$. Given that these results are evident independently of software used and also evident in both our GrHa (Figure \ref{fig:degens_grha}) and best model (Figure \ref{fig:degens_best}), indicate that these findings are robust.

Besides confirming the findings from \citet{Grillo2015} we also find clear signs of anti-correlation between the positional angle $\theta_{h1}$ and position $x_{h1}$ and and correlation between $\theta_{h1}$ and $y_{h1}$ for one cluster halo, but not for the other ($\theta_{h2}$, $x_{h2}$ and $y_{h2}$). More interestingly, these results are only consistent for the GrHa model, whereas our best model presents degeneracies for all these values. This means that in order to preserve the projected mass of the halo, when moving the halo around, we have to adjust the position-angle in order for the halo to cover the same area. 

We also find evidence of correlation between the mass-ellipticity between both cluster halos ($\varepsilon_{h1}$ and $\varepsilon_{h2}$), correlation between the position and mass-ellipticity ($x_{h1}$ and $\varepsilon_{h1}$) and anti-correlation between $y_{h1}$ and $\varepsilon_{h1}$. The same pattern are noticeable for the cluster halo $h2$. This tells us that as we move the cluster halos further away from the barycenter, we need to compensate by elongating the cluster halo in order to preserve the projected mass. These patterns are repeated in our best model.

Lastly, in Figure \ref{fig:degens_best}, we find no sign of degeneracy between the two slopes, $\zeta_{r_{s,gal}}$ and $\zeta_{\sigma_{gal}}$, while we find a slight sign of degeneracy between the slope of the velocity dispersion for the cluster members $\zeta_{\sigma_{gal}}$ and the velocity dispersion $\sigma_{gal}$. Interestingly, there is no clear sign of degeneracy between the slope of the truncation radius $\zeta_{r_{s,gal}}$ and the truncation radius $r_{s,gal}$. Likewise, we find no sign of degeneracy between the slope of the truncation radius and the velocity dispersion or the slope of the velocity dispersion and the truncation radius.

\section{Multiple Images}
Since the optimization in image plane mode compares the predicted image positions with the observed image positions, we can further investigate whether we have a good model, besides $\chi^2$ and $\log{(E)}$, by both looking quantitatively and qualitatively at the predicted and observed image positions. 

In Table \ref{table:image_error_best} and Table \ref{table:image_error_grha} we show the positional error (rms) and error in image positions for our best and GrHA model, respectively.

\begin{table}[h!p]
 \centering
 \caption{Image error and positions for our best model.}
 \label{table:image_error_best}
 \begin{tabular}{ccccc}
   \toprule
   ID  &  z    &  rms $(\arcsec)$  &  $\delta x \, (\arcsec)$ & $\delta y \, (\arcsec)$ \\
   \midrule
   1.1 & 1.892 &  0.10  &  -0.04 &  0.09 \\
   1.2 & 1.892 &  0.35  &  -0.21 &  0.28 \\
   1.3 & 1.892 &  0.11  &  -0.11 &  0.03 \\
   \textbf{1}  &        &  \textbf{0.22} \\
   2.1 & 1.892 &  0.13  &   0.12 &  0.03 \\
   2.2 & 1.892 &  0.14  &   0.08 & -0.12 \\
   2.3 & 1.892 &  0.12  &   0.01 & -0.12 \\
   \textbf{2}  &        &  \textbf{0.13} \\
   3.1 & 2.087 &  0.18  &   0.17 & -0.06 \\
   3.2 & 2.087 &  0.37  &   0.32 & -0.19 \\
   3.3 & 2.087 &  0.10  &   0.00 & -0.10 \\
   \textbf{3}  &        &  \textbf{0.24} \\
   4.1 & 1.990 &  0.19  &  -0.10 & -0.16 \\
   4.2 & 1.990 &  0.21  &   0.18 & -0.11 \\
   4.3 & 1.990 &  0.21  &   0.21 & -0.03 \\
   \textbf{4}  &        & \textbf{0.20} \\
   5.1 & 1.990 &  0.18  &  -0.12 & -0.13 \\
   5.2 & 1.990 &  0.13  &   0.12 & -0.04 \\
   5.3 & 1.990 &  0.16  &   0.16 & -0.02 \\
   \textbf{5} &         & \textbf{0.16} \\
   6.1 & 3.223 &  0.30  &  -0.15 &  0.25 \\
   6.2 & 3.223 &  0.33  &   0.04 &  0.33 \\
   6.3 & 3.223 &  0.38  &  -0.16 & -0.35 \\
   \textbf{6} &         &  \textbf{0.34} \\
   7.1 & 1.637 &  0.54  &   0.48 & -0.25 \\
   7.2 & 1.637 &  0.41  &   0.25 & -0.32 \\
   7.3 & 1.637 &  0.13  &  -0.10 &  0.08 \\
   \textbf{7} &         & \textbf{0.40} \\
   8.1 & 2.302 &  0.44  &   0.27 & -0.34 \\
   8.2 & 2.302 &  0.31  &   0.02 & -0.31 \\
   8.3 & 2.302 &  0.20  &   0.11 &  0.17 \\
   \textbf{8} &         & \textbf{0.33} \\
   9.1 & 1.964 &  0.09  &   0.04 & -0.08 \\
   9.2 & 1.964 &  0.27  &  -0.26 &  0.08 \\
   9.3 & 1.964 &  0.16  &  -0.10 &  0.13 \\
   \textbf{9}  &        &  \textbf{0.19} \\
   10.1 & 2.218 & 0.25  &  -0.18 &  -0.18  \\
   10.2 & 2.218 & 0.59  &  -0.56 &   0.18  \\
   10.3 & 2.218 & 0.16  &   0.07 &   0.15  \\
   \textbf{10}    &           & \textbf{0.38} \\
   \textbf{Model} &                & \textbf{0.27}   \\
   \bottomrule               
   \end{tabular}
\end{table}                 

\begin{table}[h!p]
 \centering
 \caption{Image error and positions for our GrHa model.}
 \label{table:image_error_grha}
 \begin{tabular}{ccccc}
    \toprule
    ID & z & rms $(\arcsec)$ & $\delta x \, (\arcsec)$ & $\delta y \, (\arcsec)$ \\
    \midrule
    1.1 & 1.892 & 0.17 & -0.15 & 0.07 \\
    1.2 & 1.892 & 0.46 & -0.07 & 0.45 \\
    1.3 & 1.892 & 0.20 & -0.17 & -0.09 \\
    \textbf{1}   &     & \textbf{0.30} \\
    2.1 & 1.892 & 0.13 & -0.09 & 0.09 \\
    2.2 & 1.892 & 0.19 & 0.10 & 0.16 \\
    2.3 & 1.892 & 0.25 & -0.09 & -0.23 \\
    \textbf{2}   &     & \textbf{0.19} \\
    3.1 & 2.087 & 0.59 & 0.53 & -0.26 \\
    3.2 & 2.087 & 0.86 & 0.68 & -0.53 \\
    3.3 & 2.087 & 0.06 & -0.00 & -0.06 \\
    \textbf{3}   &     & \textbf{0.60} \\
    4.1 & 1.990 & 0.18 & -0.04 & -0.17 \\
    4.2 & 1.990 & 0.31 & 0.27 & 0.14 \\
    4.3 & 1.990 & 0.19 & -0.03 & -0.18 \\
    \textbf{4}   &     & \textbf{0.23} \\
    5.1 & 1.990 & 0.16 & -0.04 & -0.15 \\
    5.2 & 1.990 & 0.29 & 0.20 & 0.21 \\
    5.3 & 1.990 & 0.18 & -0.09 & -0.16 \\
    \textbf{5}   &     & \textbf{0.22} \\
    6.1 & 3.223 & 0.17 & 0.00 & 0.17 \\
    6.2 & 3.223 & 0.21 & 0.02 & 0.21 \\
    6.3 & 3.223 & 0.30 & -0.18 & -0.24 \\
    \textbf{6}   &     & \textbf{0.23} \\
    7.1 & 1.637 & 0.58 & 0.40 & -0.42 \\
    7.2 & 1.637 & 0.52 & 0.52 & 0.00 \\
    7.3 & 1.637 & 0.21 & -0.17 & -0.12 \\
    \textbf{7}   &     & \textbf{0.47} \\
    8.1 & 2.302 & 0.42 & 0.34 & -0.25 \\
    8.2 & 2.302 & 0.42 & -0.04 & -0.42 \\
    8.3 & 2.302 & 0.20 & 0.12 & 0.16 \\
    \textbf{8}   &     & \textbf{0.36} \\
    9.1 & 1.964 & 0.27 & 0.08 & -0.25 \\
    9.2 & 1.964 & 0.05 & -0.05 & 0.01 \\
    9.3 & 1.964 & 0.11 & -0.01 & 0.10 \\
    \textbf{9}   &     & \textbf{0.17} \\
    10.1 & 2.218 & 0.27 & -0.25 & 0.10 \\
    10.2 & 2.218 & 0.38 & -0.38 & -0.05 \\
    10.3 & 2.218 & 0.43 & 0.41 & 0.11 \\
    \textbf{10}    &                & \textbf{0.37} \\
    \textbf{Model} &                & \textbf{0.34} \\
    \bottomrule               
 \end{tabular} 
\end{table} 

Here we clearly see that our best model does indeed have a better rms error $(\Delta_{rms} \sim 0.07)$ compared with the GrHa model. We also see the same trend in positional error, where system $6$ and $7$ has the largest rms error and specifically, image $7.1$ has the largest error rms and positional error for both models. Since we do not have the same positional error from \citet{Grillo2015}, we cannot compare the individual values.

\begin{figure}[!ht]
 \centering
 \includegraphics[width=0.16\textwidth,keepaspectratio=true]{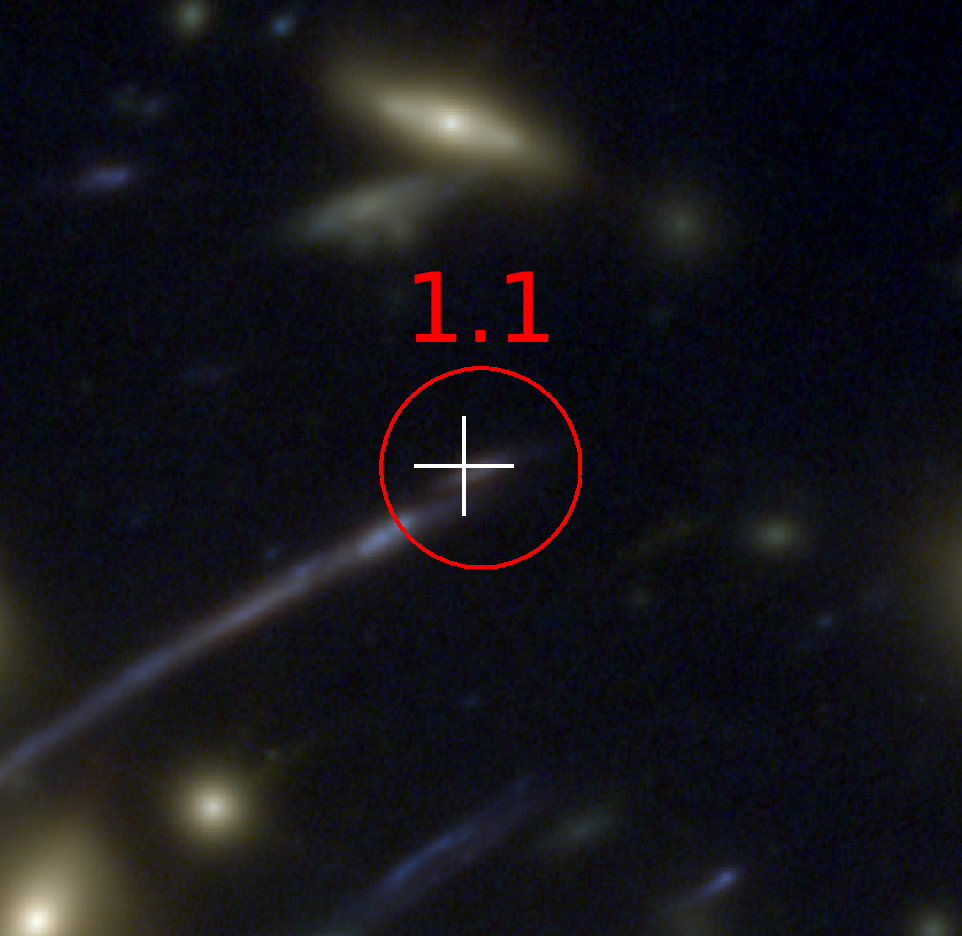}
 \includegraphics[width=0.16\textwidth,keepaspectratio=true]{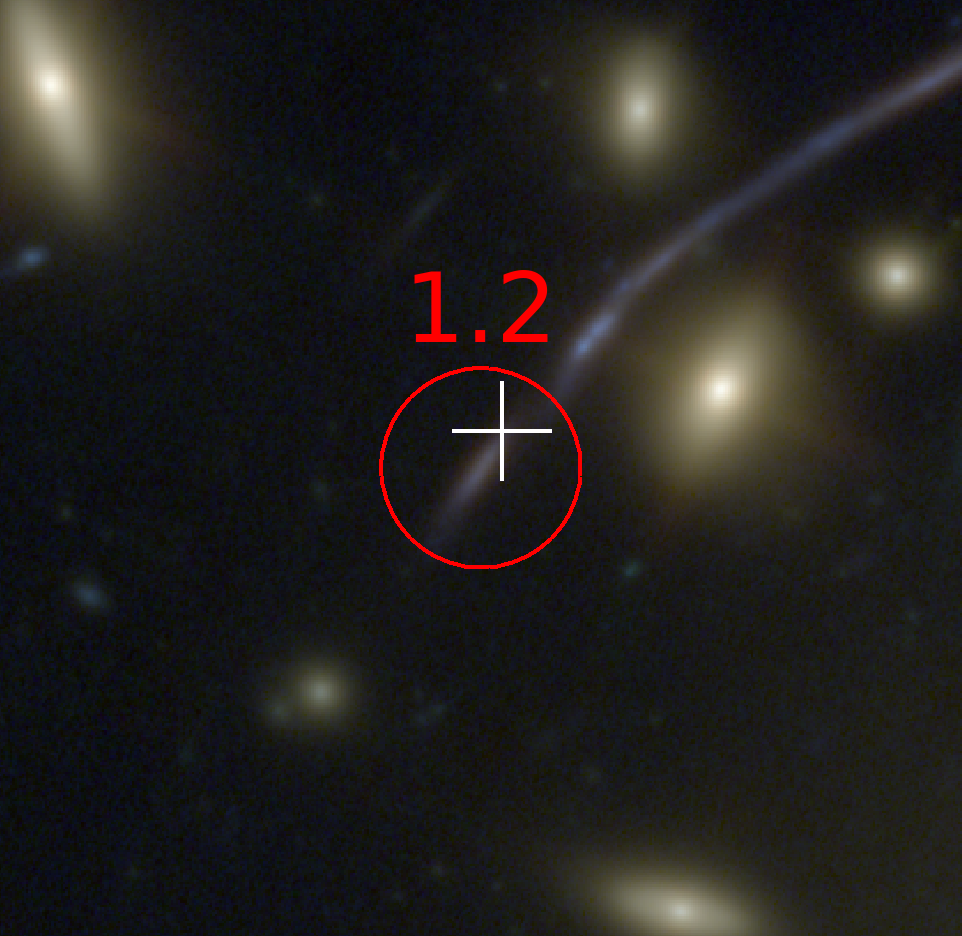}
 \includegraphics[width=0.16\textwidth,keepaspectratio=true]{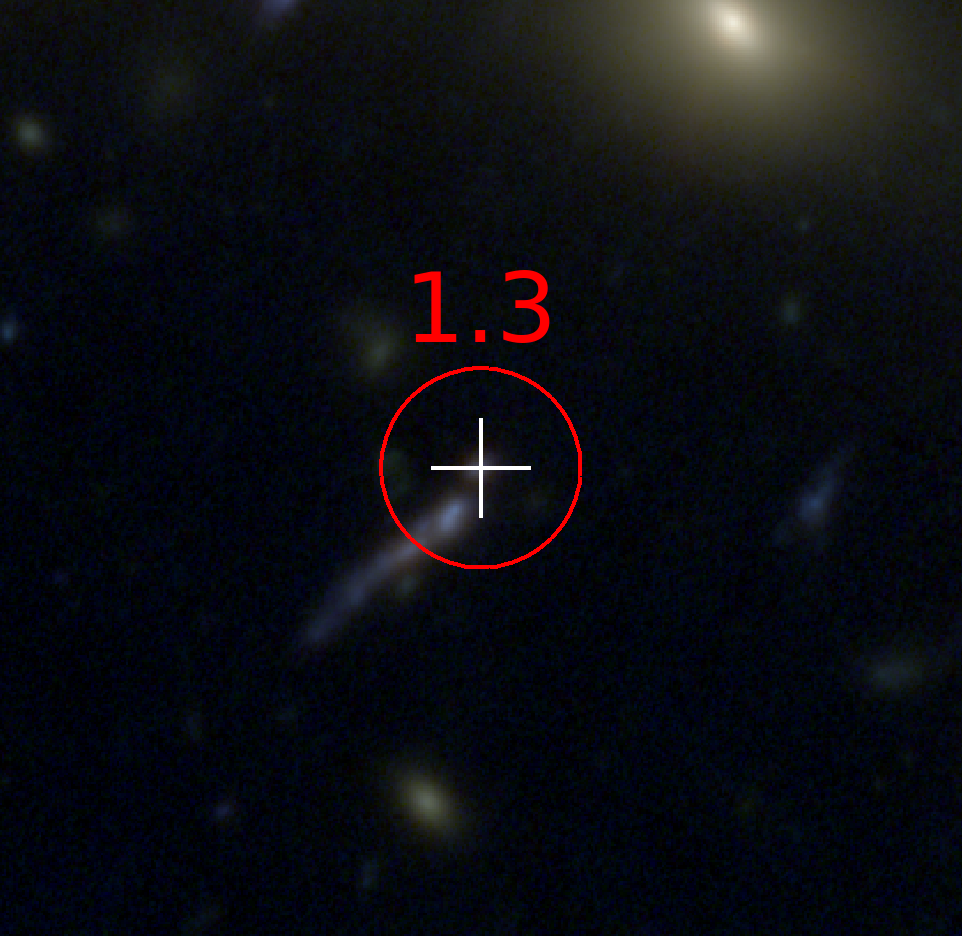} 
 \includegraphics[width=0.16\textwidth,keepaspectratio=true]{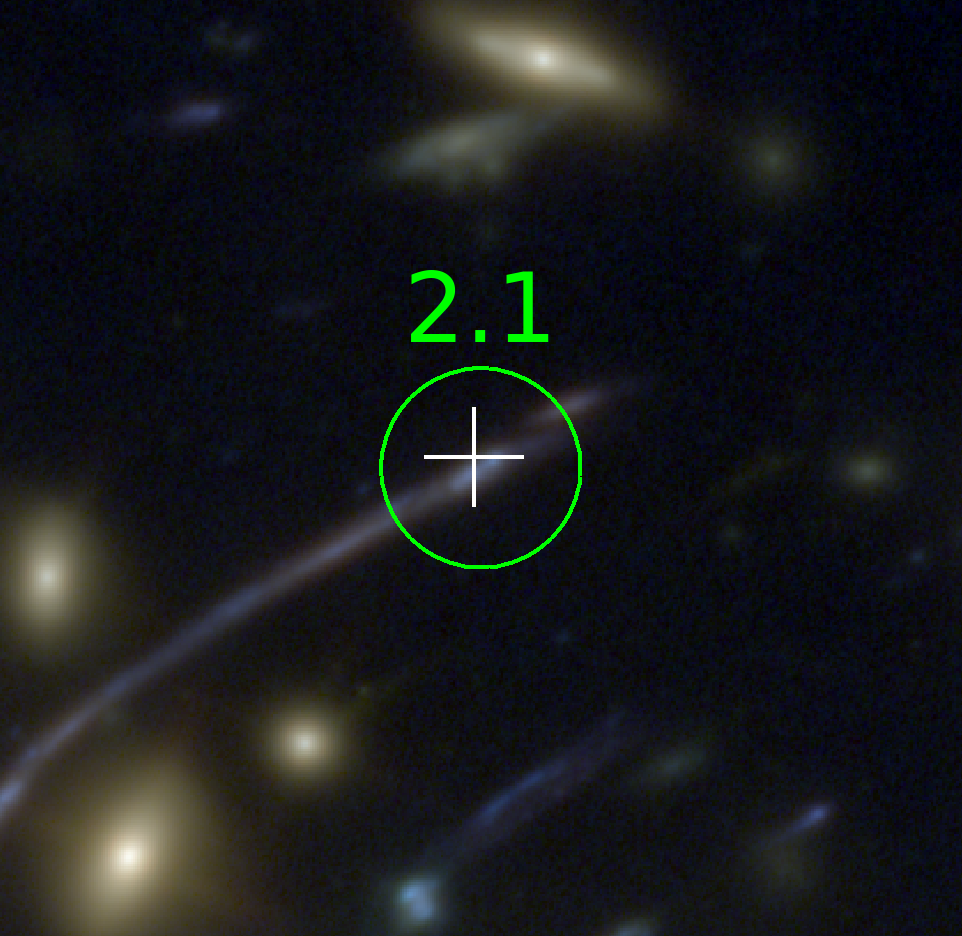}
 \includegraphics[width=0.16\textwidth,keepaspectratio=true]{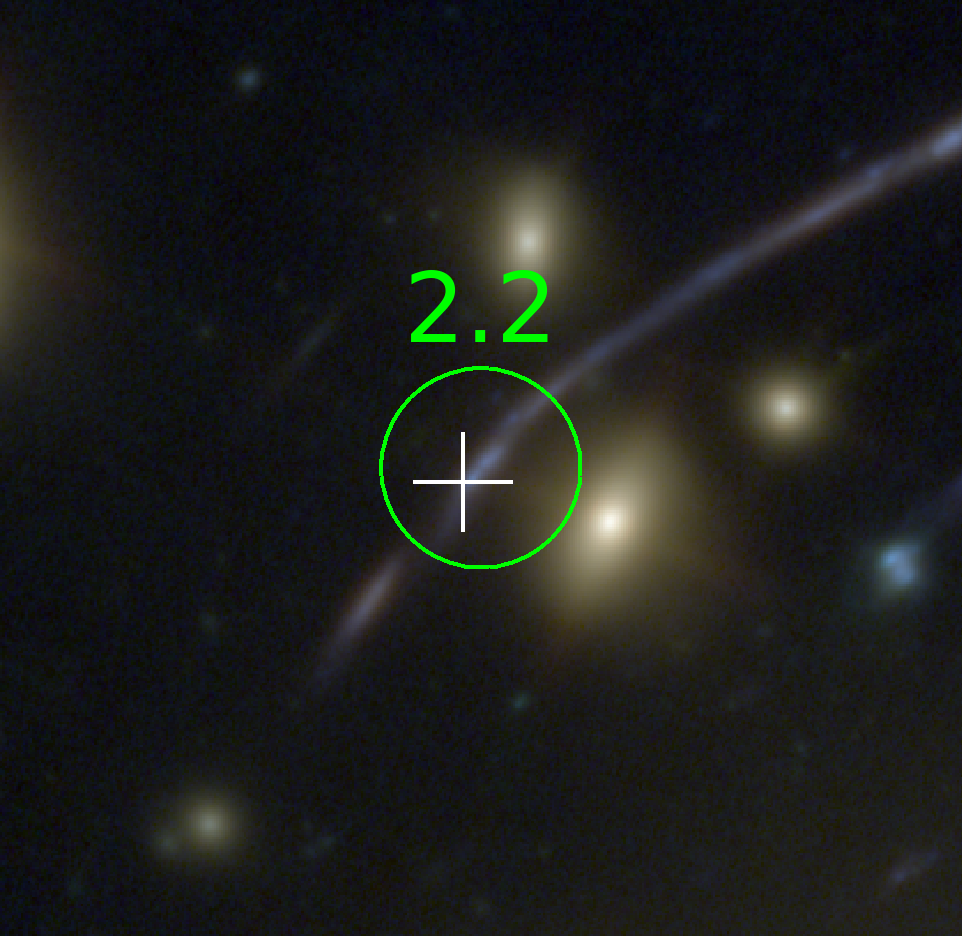}
 \includegraphics[width=0.16\textwidth,keepaspectratio=true]{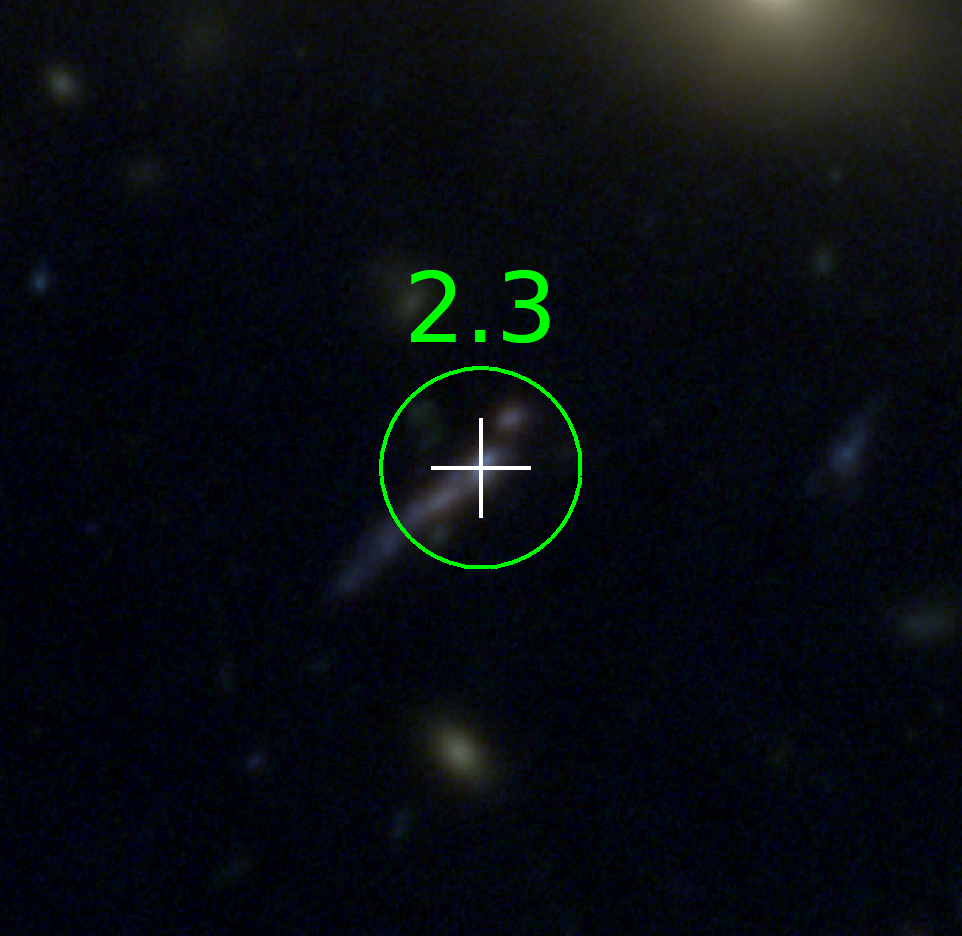} \\
 \includegraphics[width=0.16\textwidth,keepaspectratio=true]{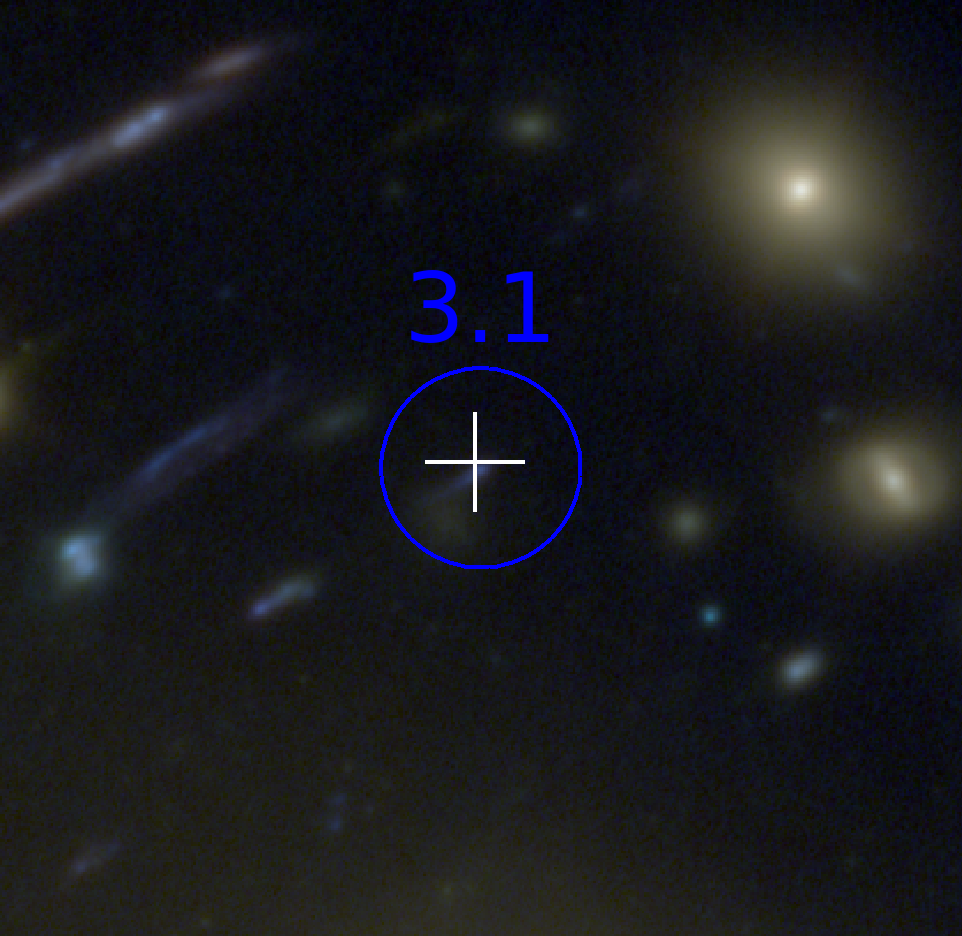}
 \includegraphics[width=0.16\textwidth,keepaspectratio=true]{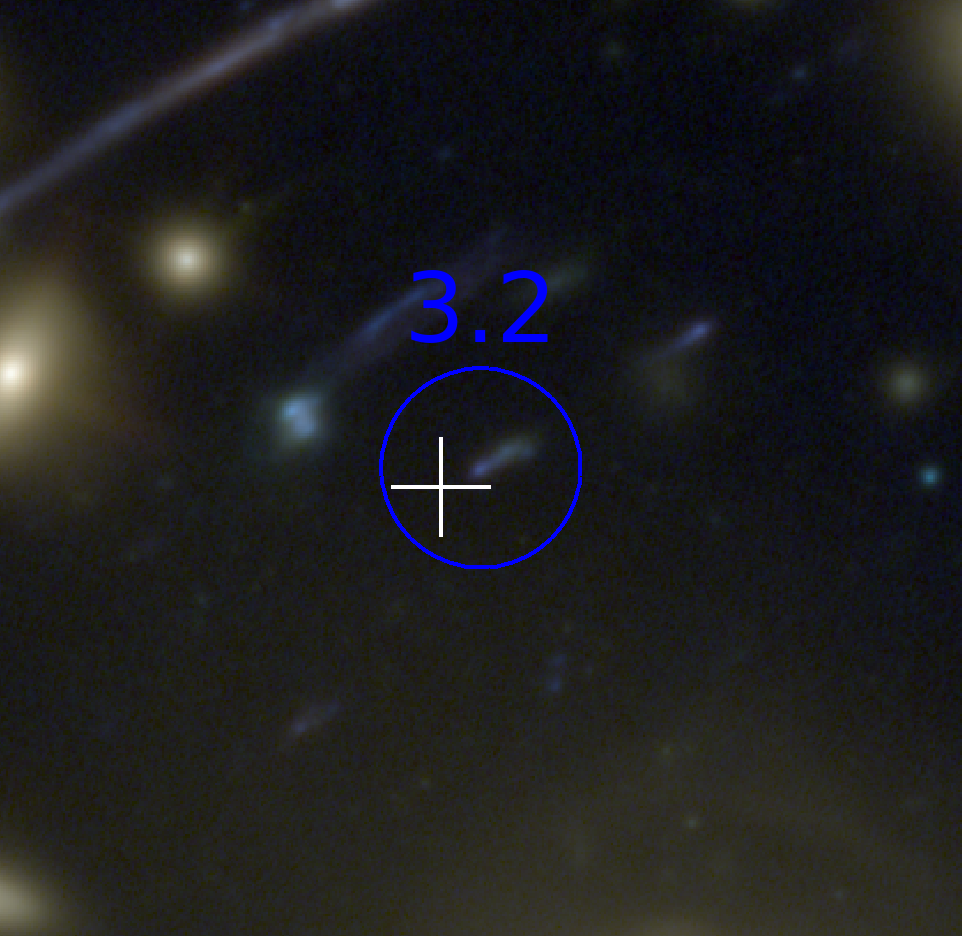}
 \includegraphics[width=0.16\textwidth,keepaspectratio=true]{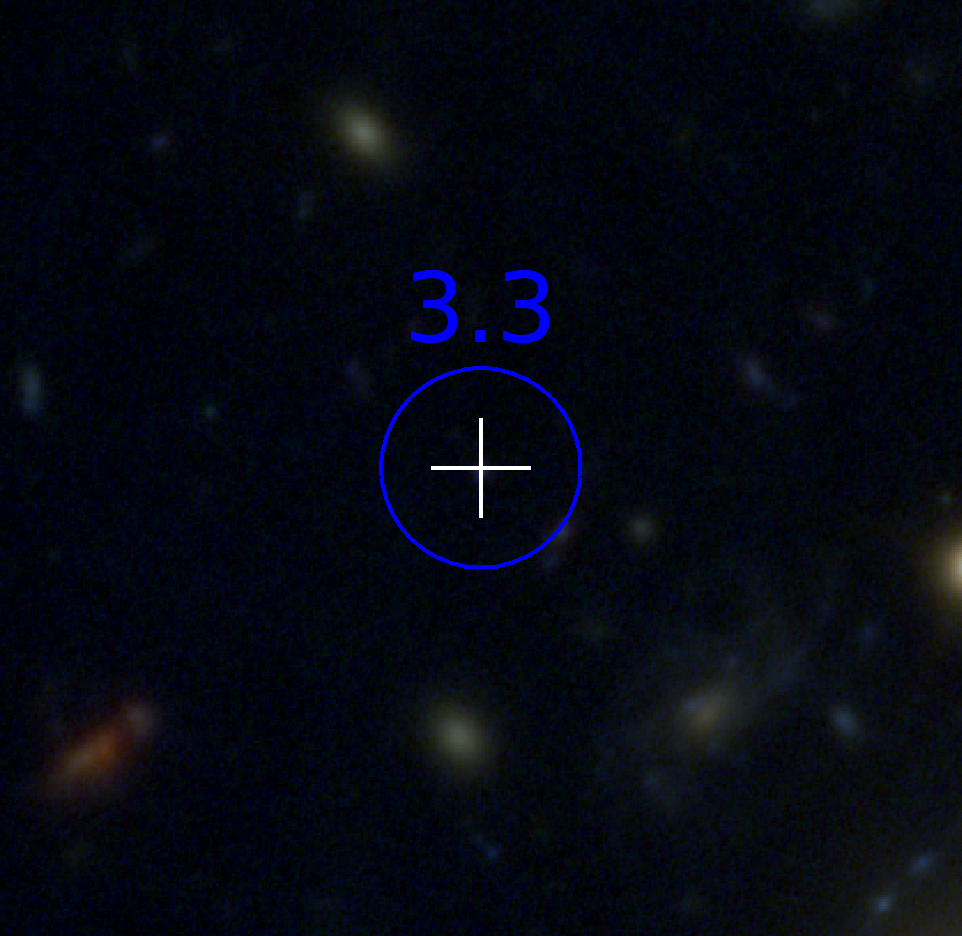} 
 \includegraphics[width=0.16\textwidth,keepaspectratio=true]{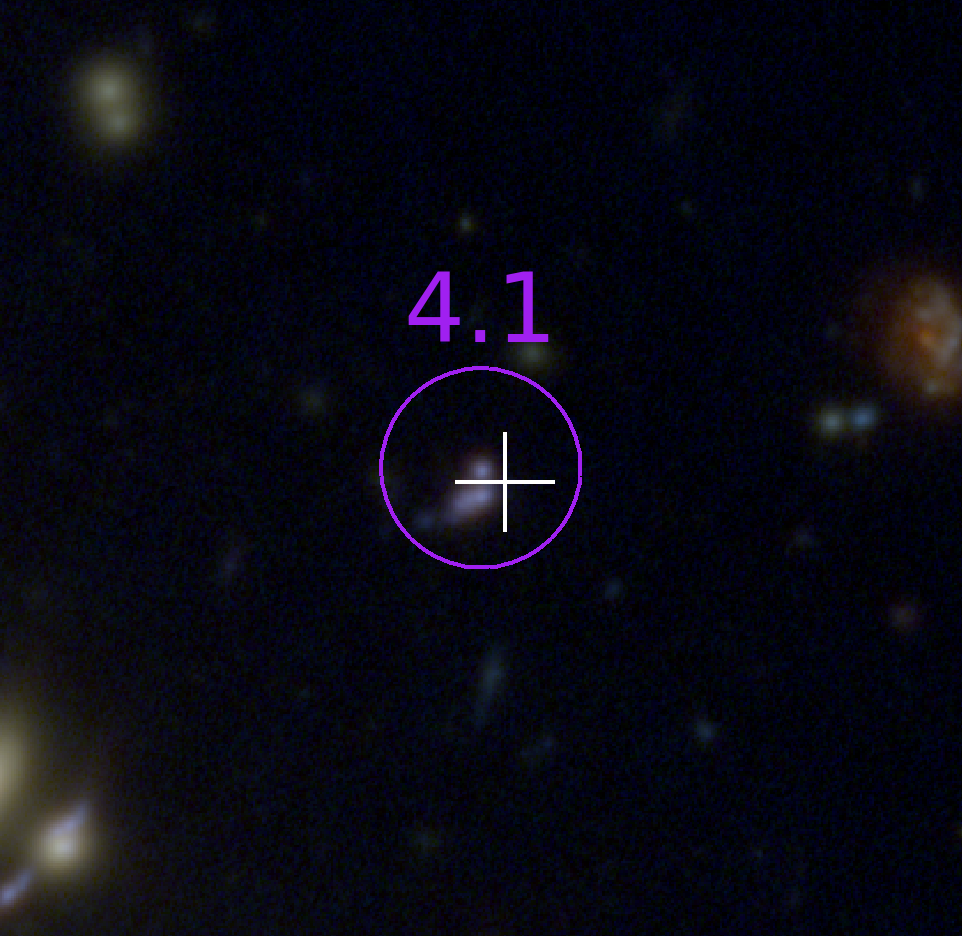}
 \includegraphics[width=0.16\textwidth,keepaspectratio=true]{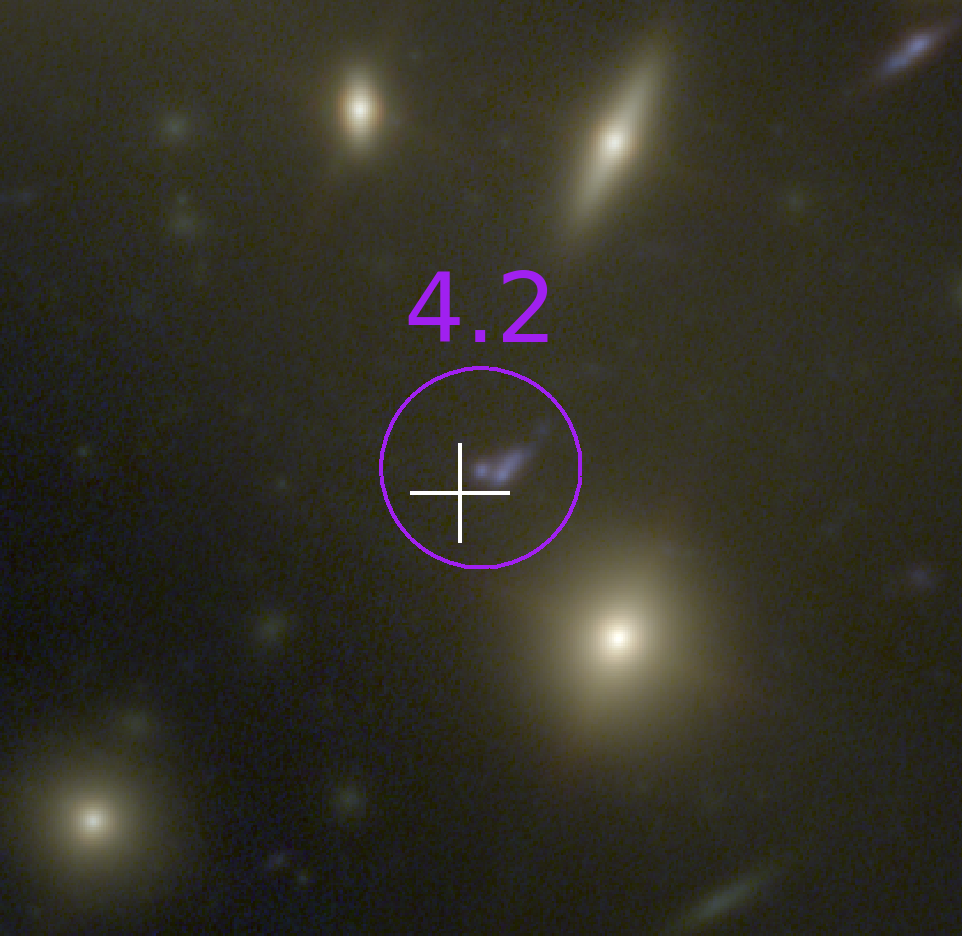}
 \includegraphics[width=0.16\textwidth,keepaspectratio=true]{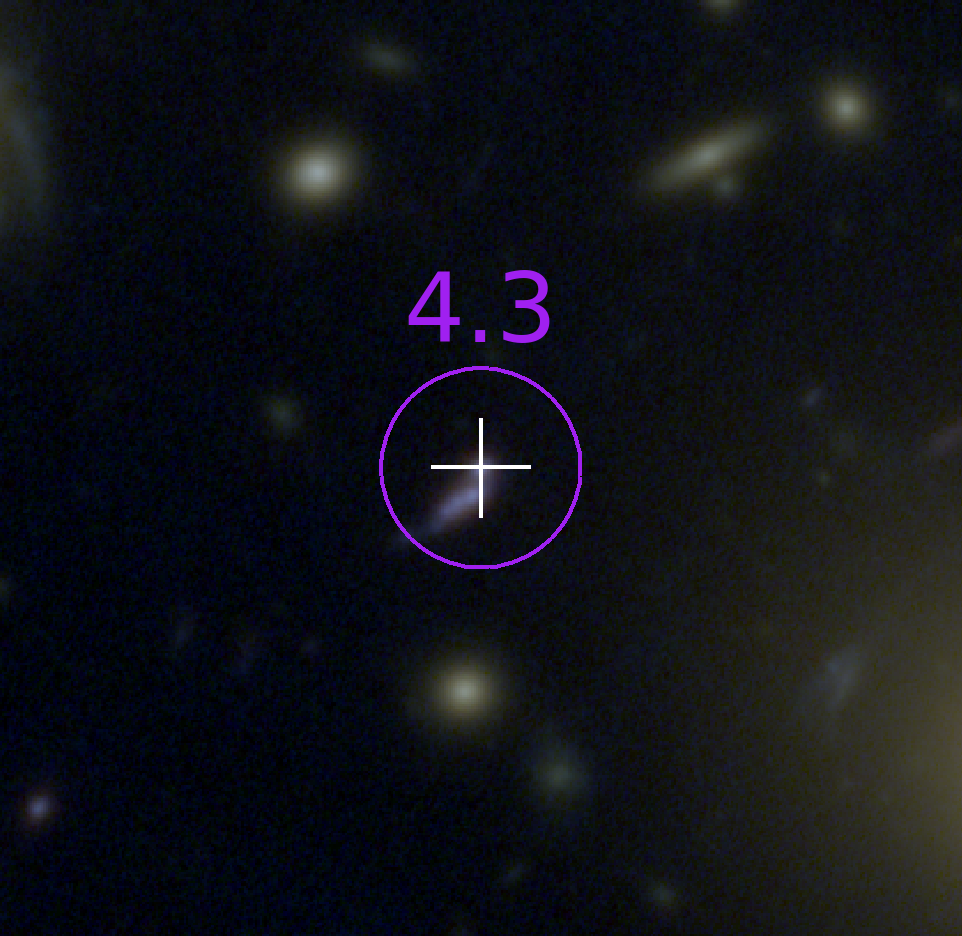} \\
 \includegraphics[width=0.16\textwidth,keepaspectratio=true]{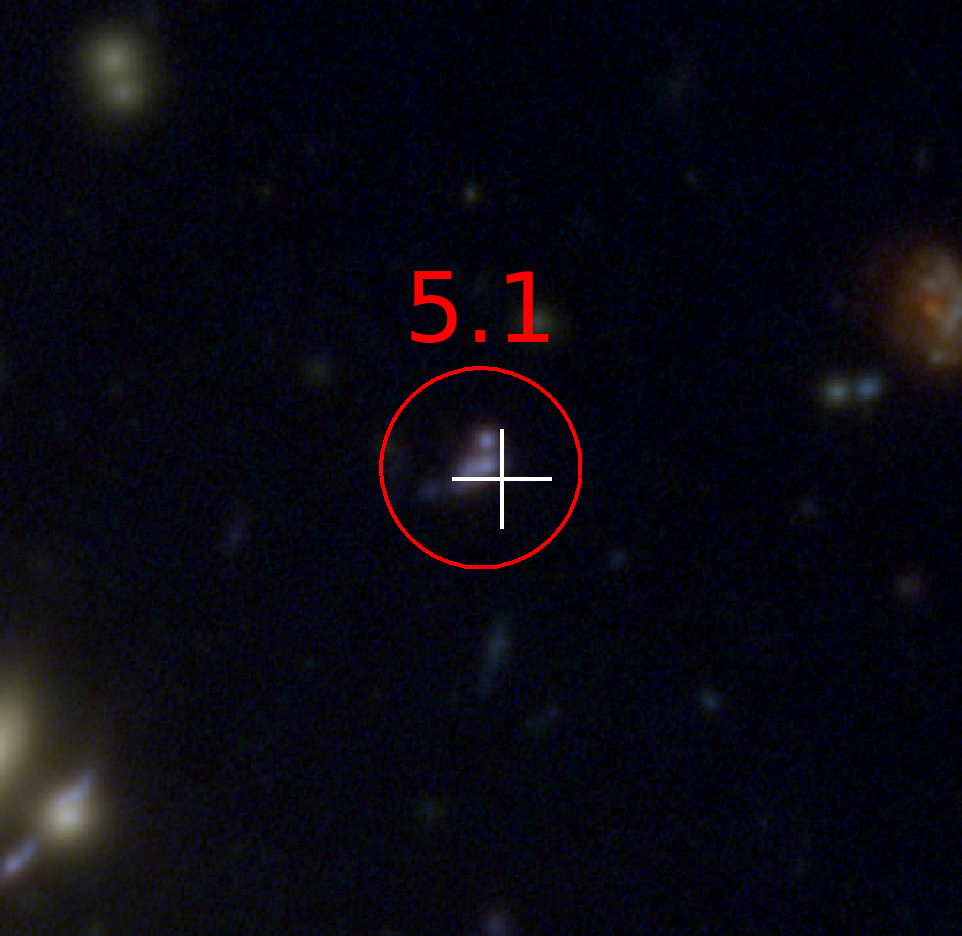}
 \includegraphics[width=0.16\textwidth,keepaspectratio=true]{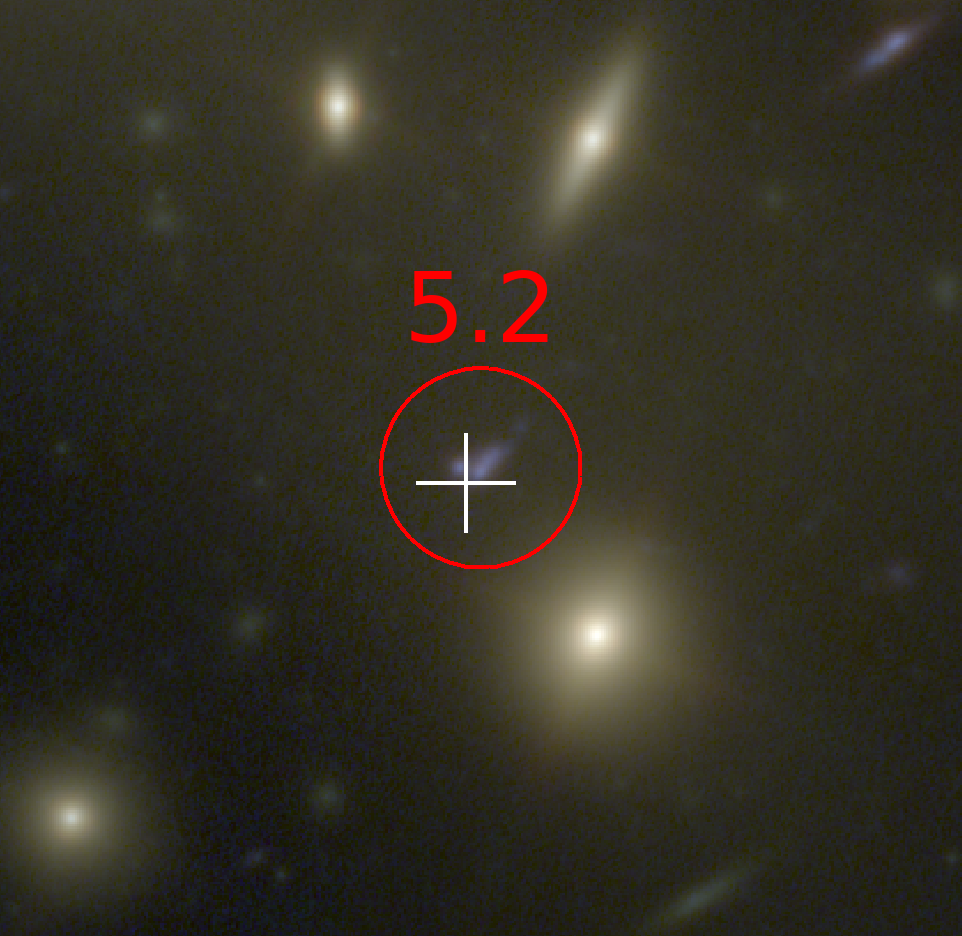}
 \includegraphics[width=0.16\textwidth,keepaspectratio=true]{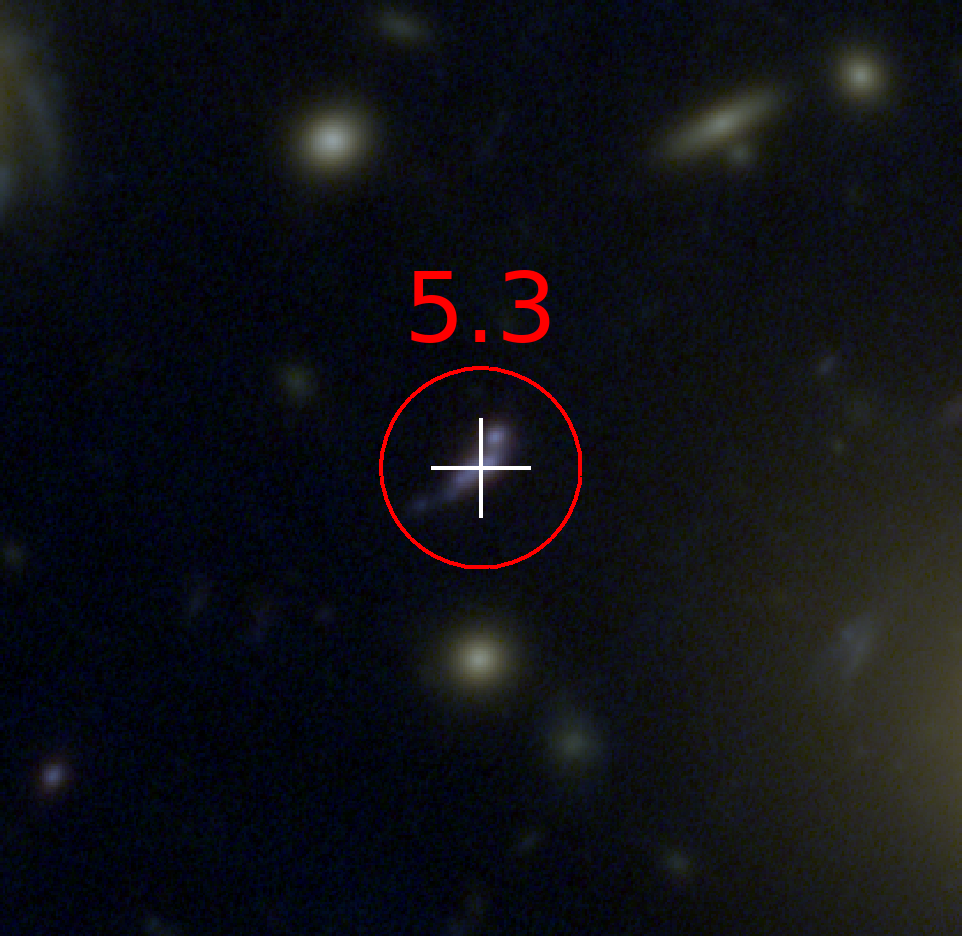} 
 \includegraphics[width=0.16\textwidth,keepaspectratio=true]{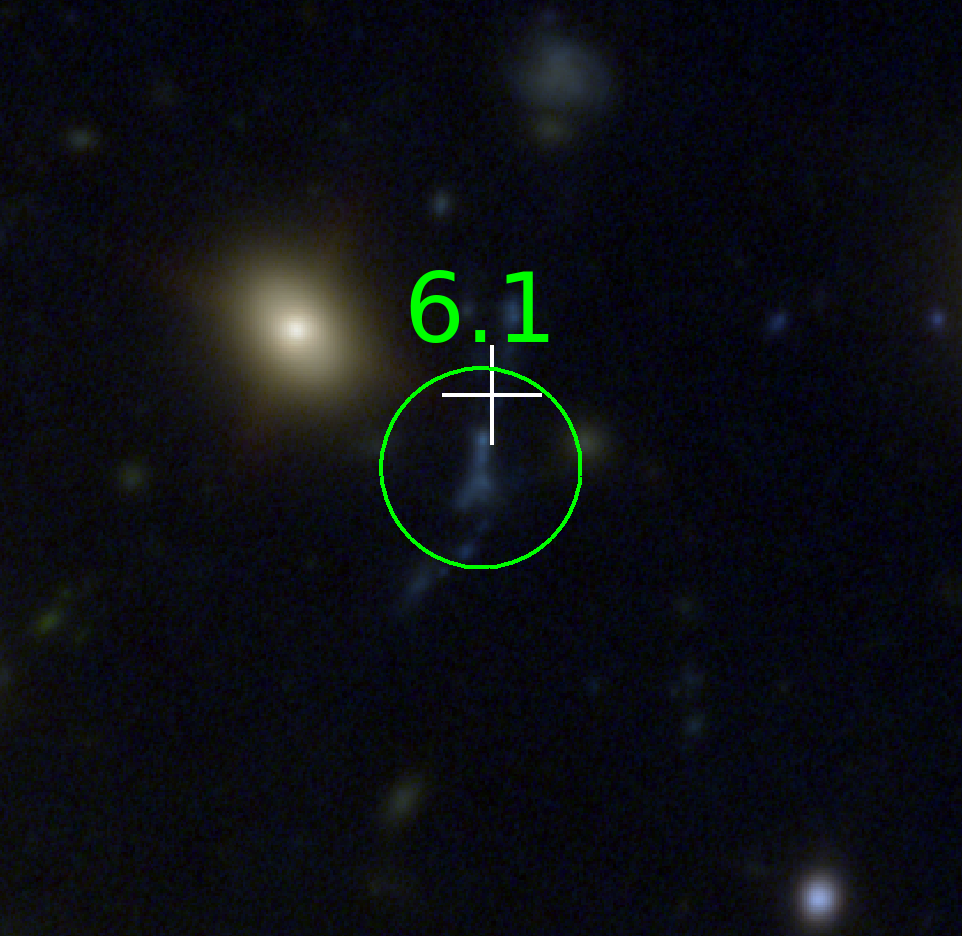}
 \includegraphics[width=0.16\textwidth,keepaspectratio=true]{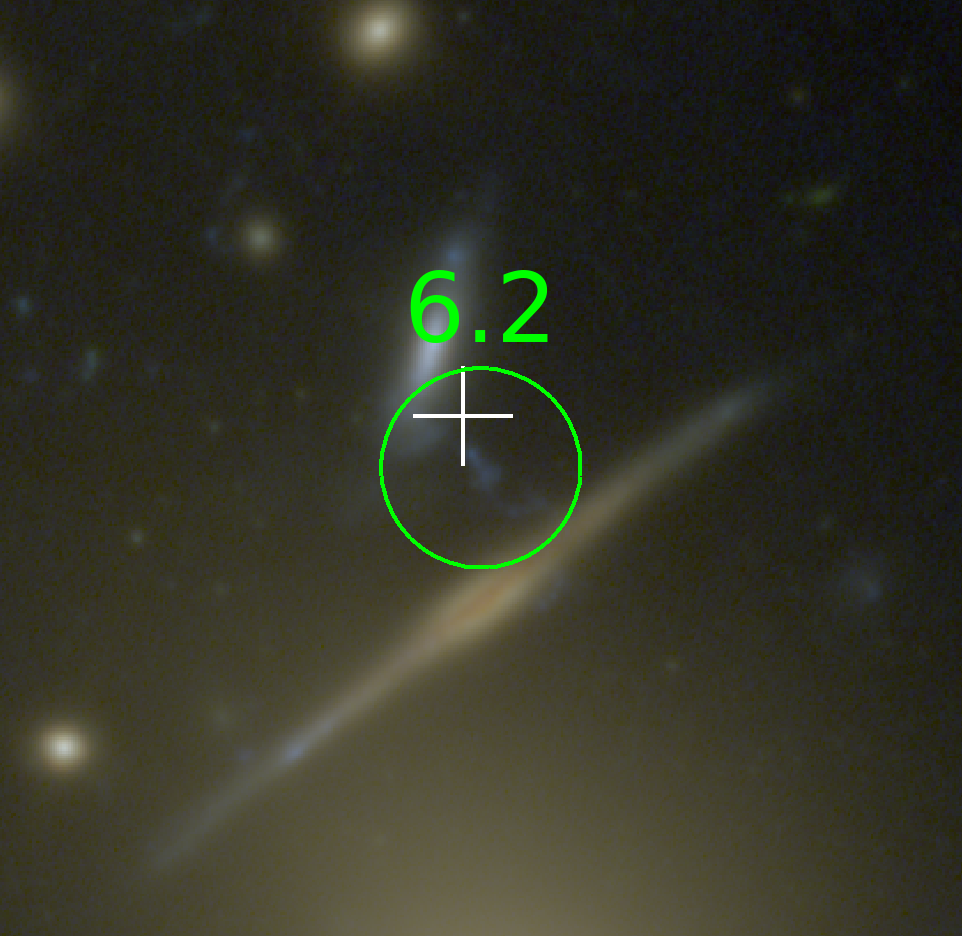}
 \includegraphics[width=0.16\textwidth,keepaspectratio=true]{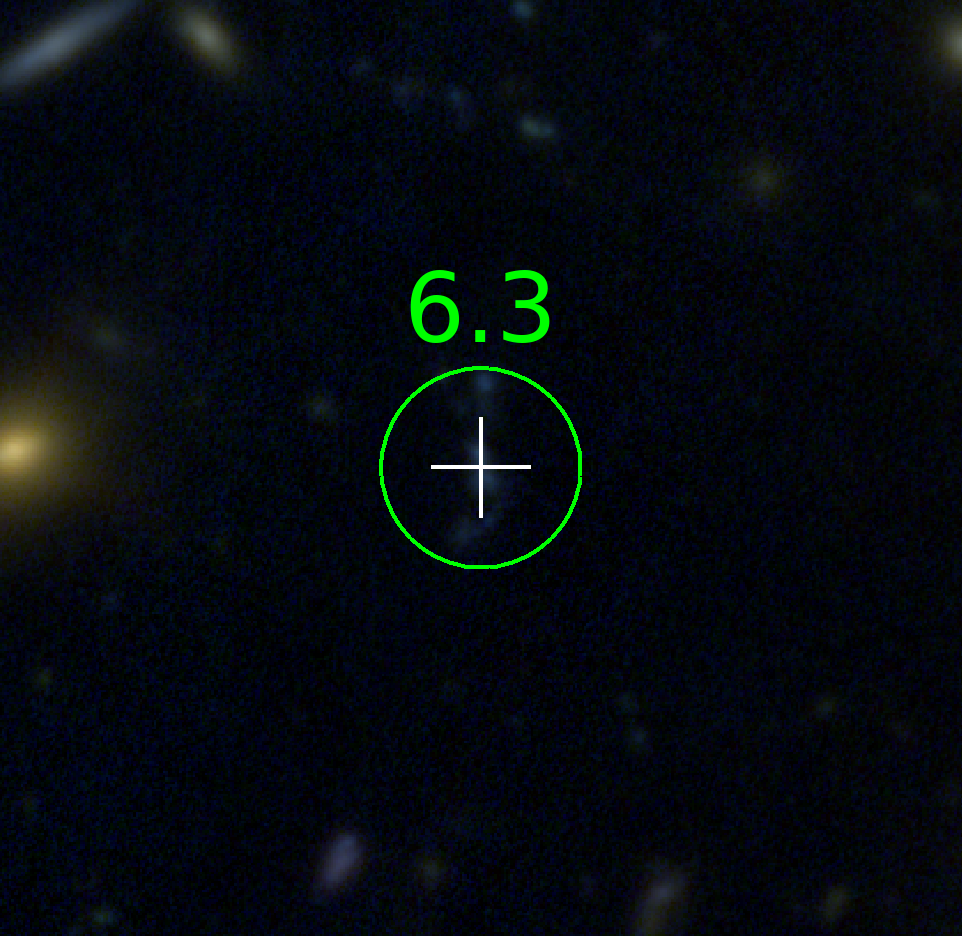} \\
 \includegraphics[width=0.16\textwidth,keepaspectratio=true]{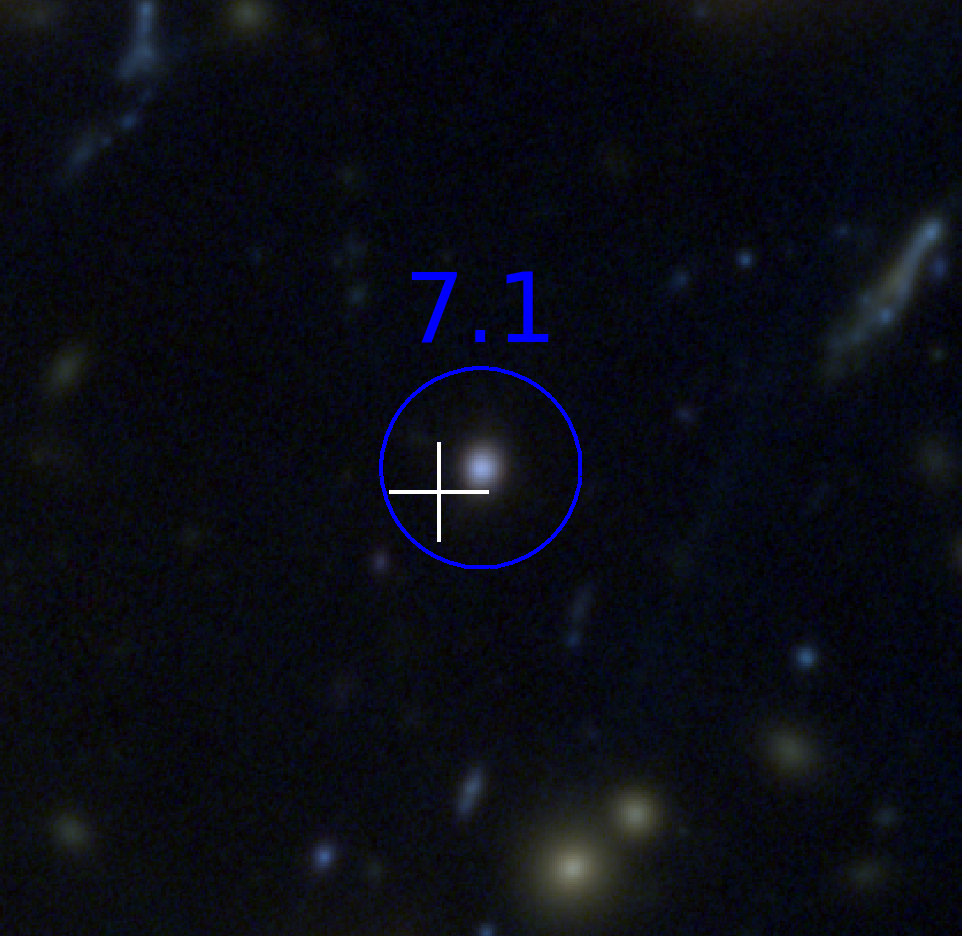}
 \includegraphics[width=0.16\textwidth,keepaspectratio=true]{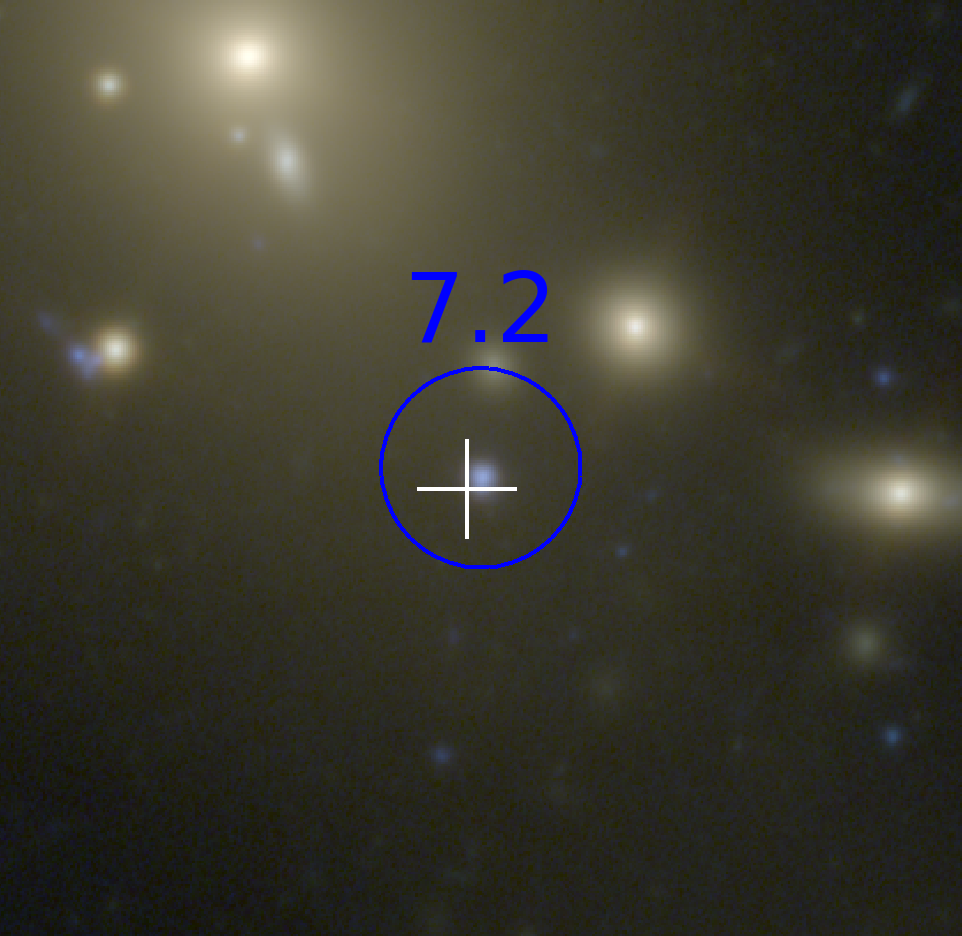}
 \includegraphics[width=0.16\textwidth,keepaspectratio=true]{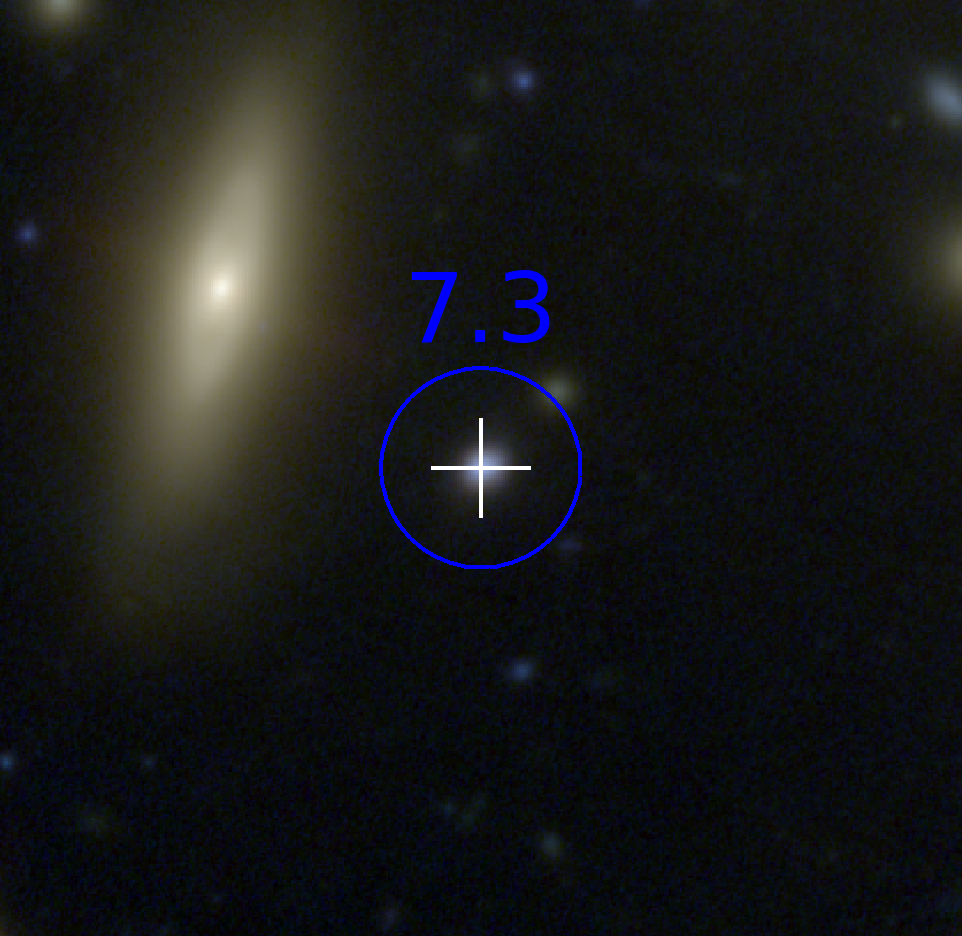} 
 \includegraphics[width=0.16\textwidth,keepaspectratio=true]{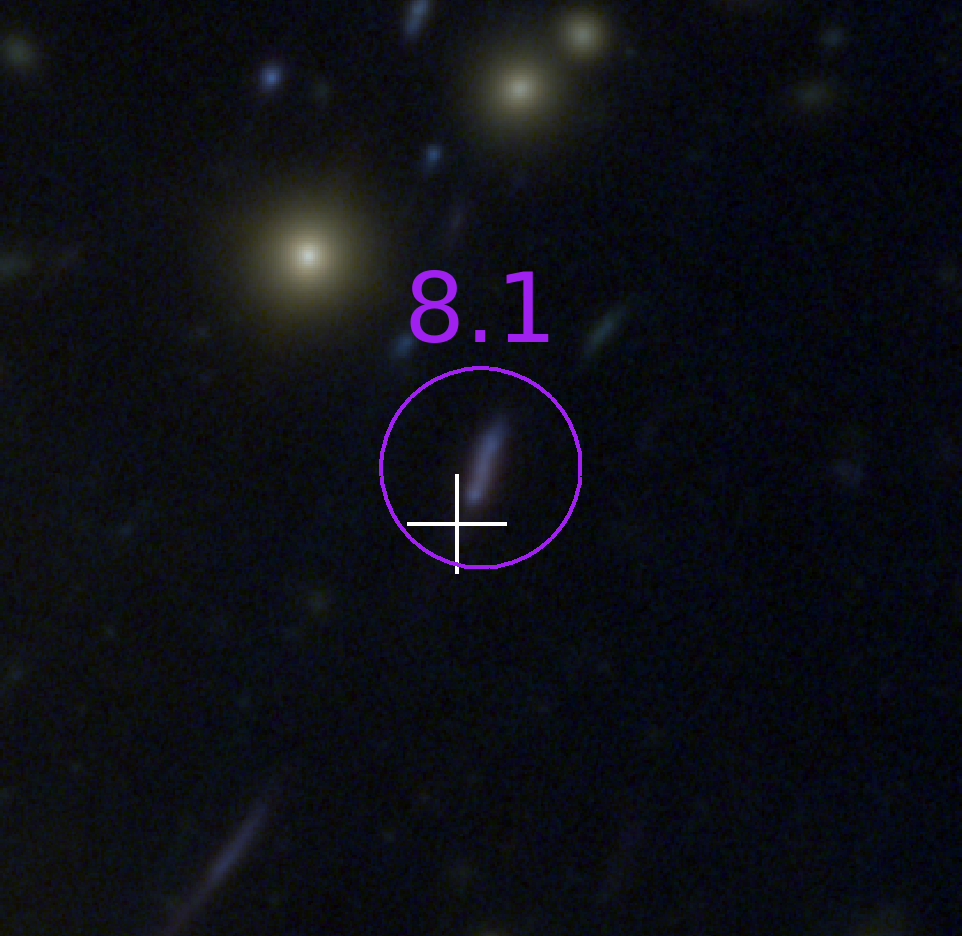}
 \includegraphics[width=0.16\textwidth,keepaspectratio=true]{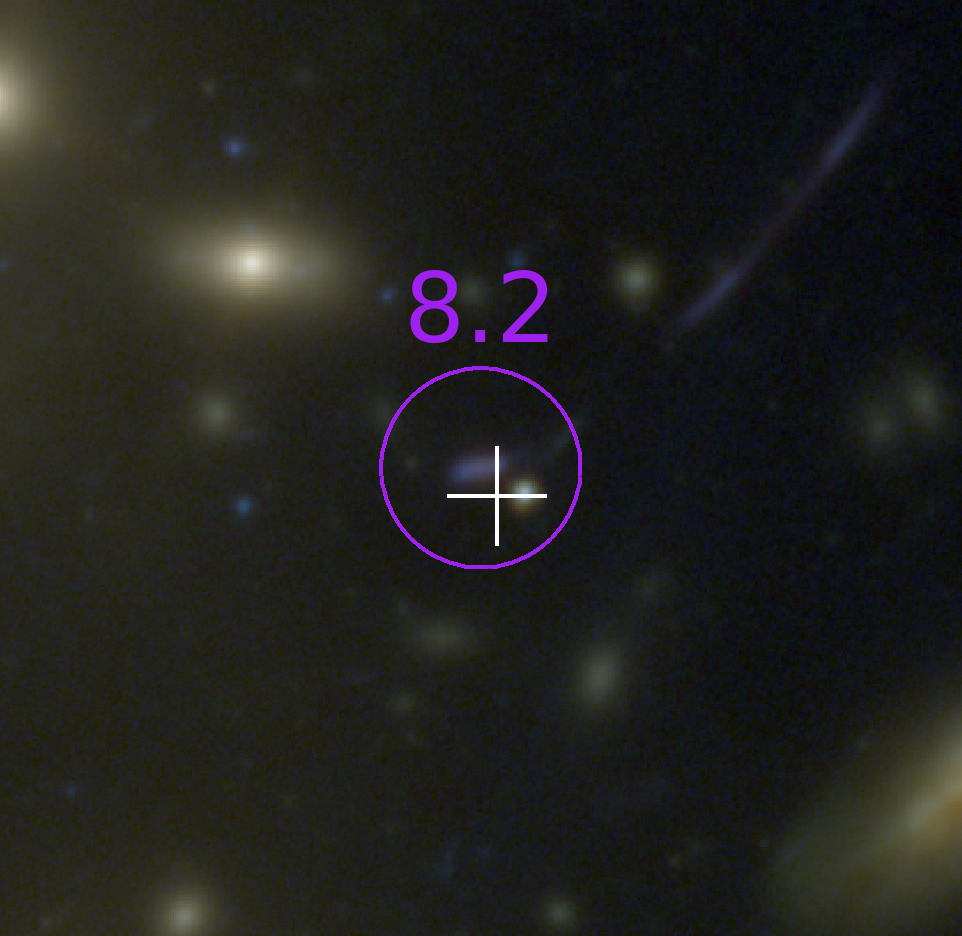}
 \includegraphics[width=0.16\textwidth,keepaspectratio=true]{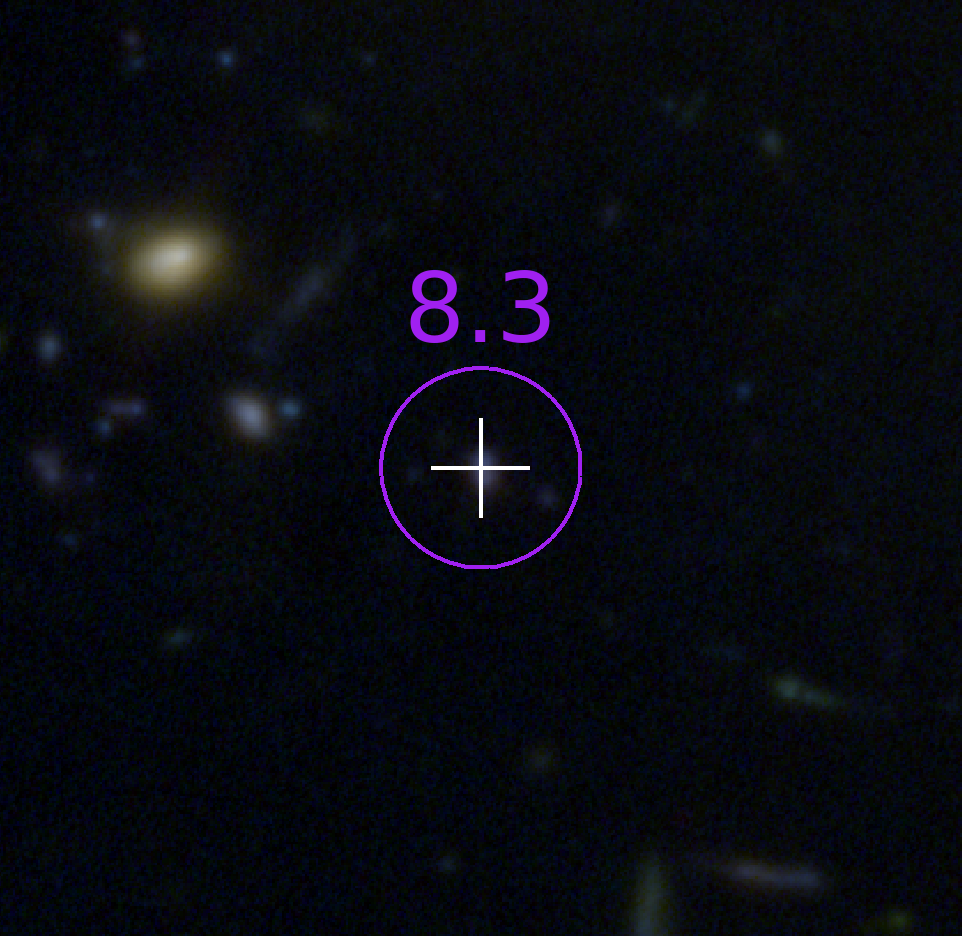} \\
 \includegraphics[width=0.16\textwidth,keepaspectratio=true]{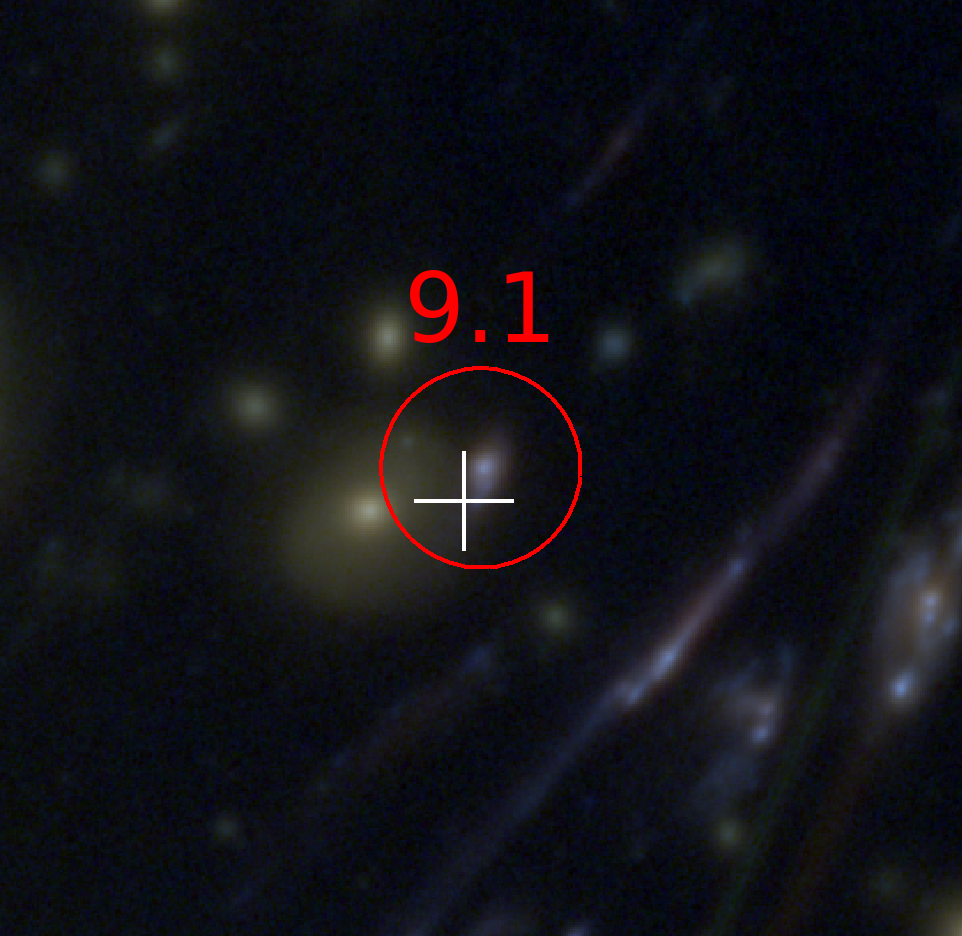}
 \includegraphics[width=0.16\textwidth,keepaspectratio=true]{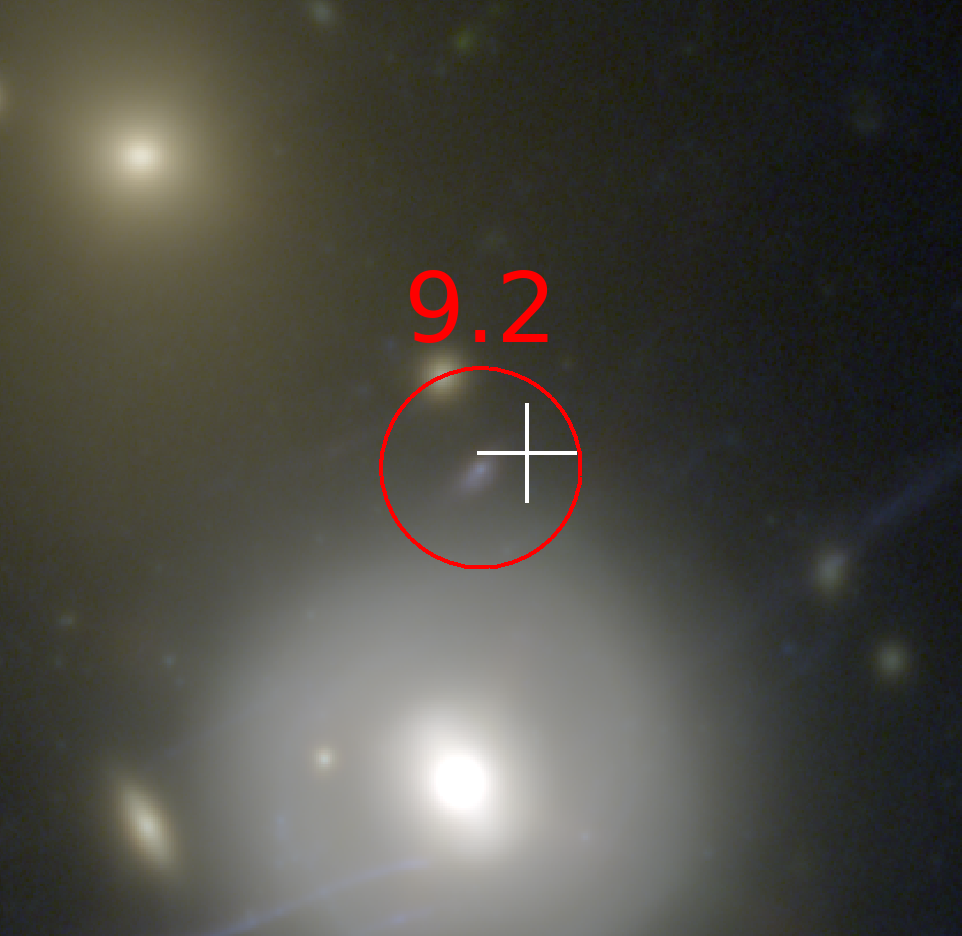}
 \includegraphics[width=0.16\textwidth,keepaspectratio=true]{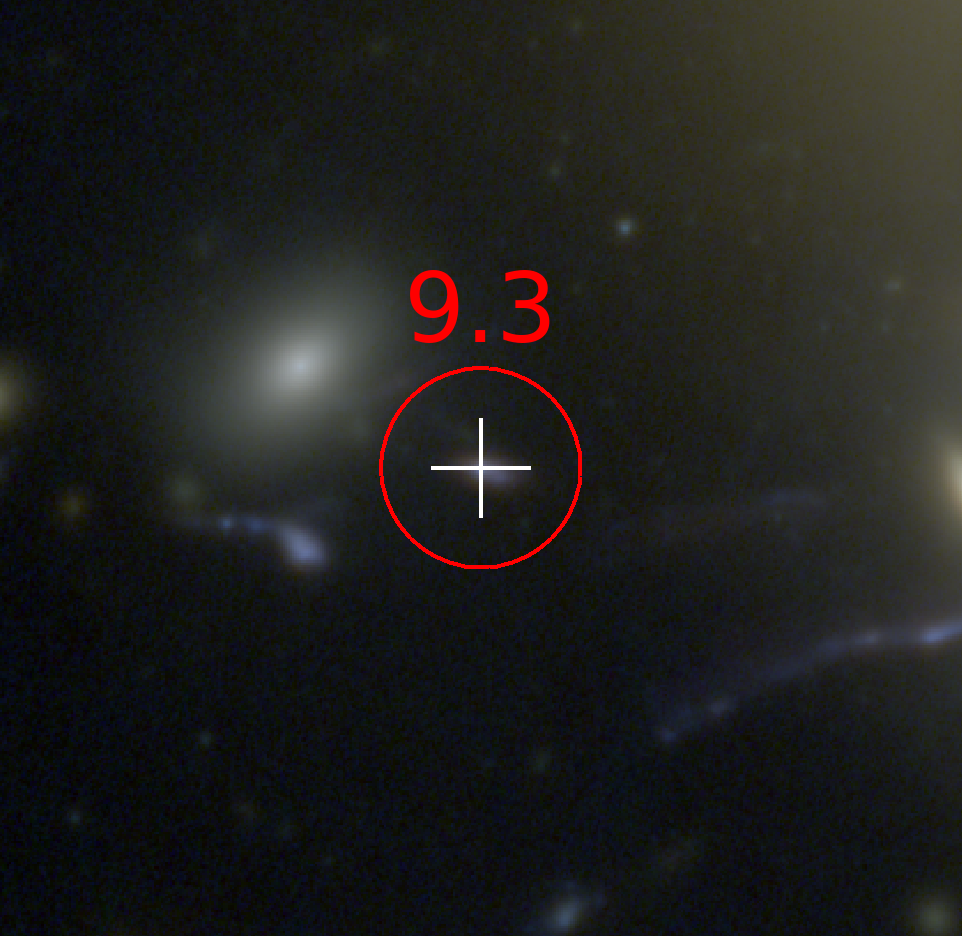} 
 \includegraphics[width=0.16\textwidth,keepaspectratio=true]{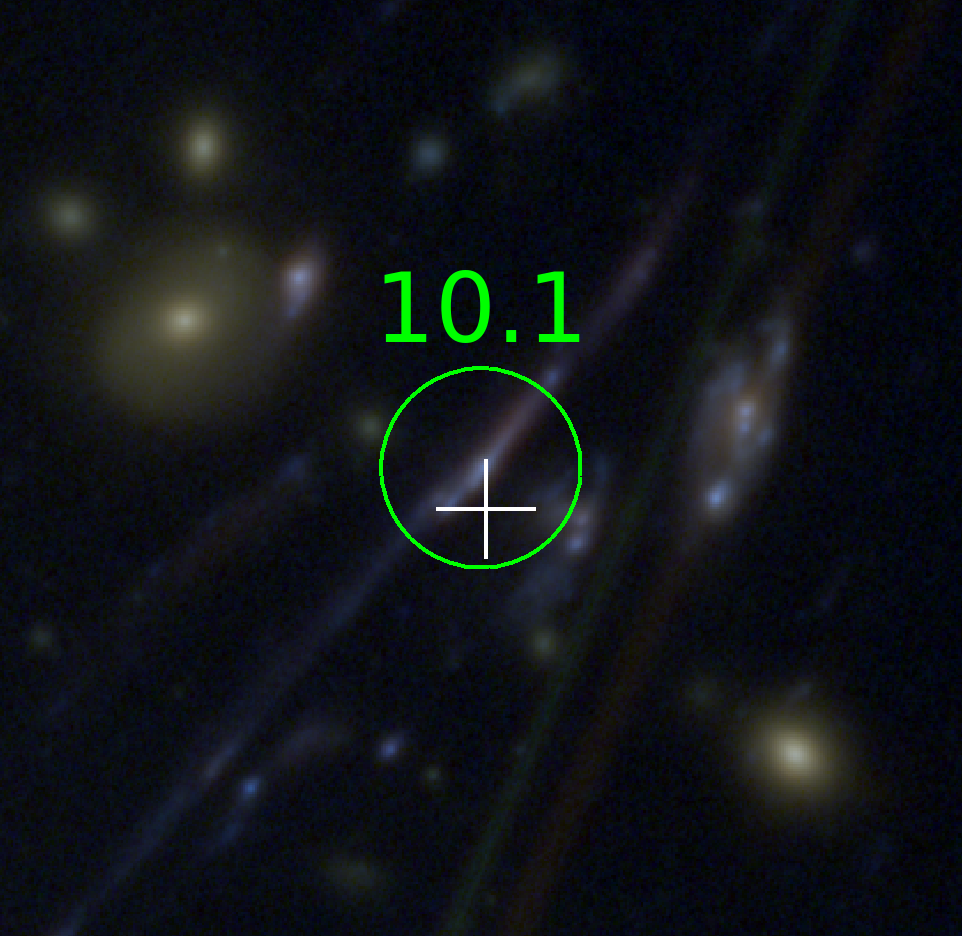}
 \includegraphics[width=0.16\textwidth,keepaspectratio=true]{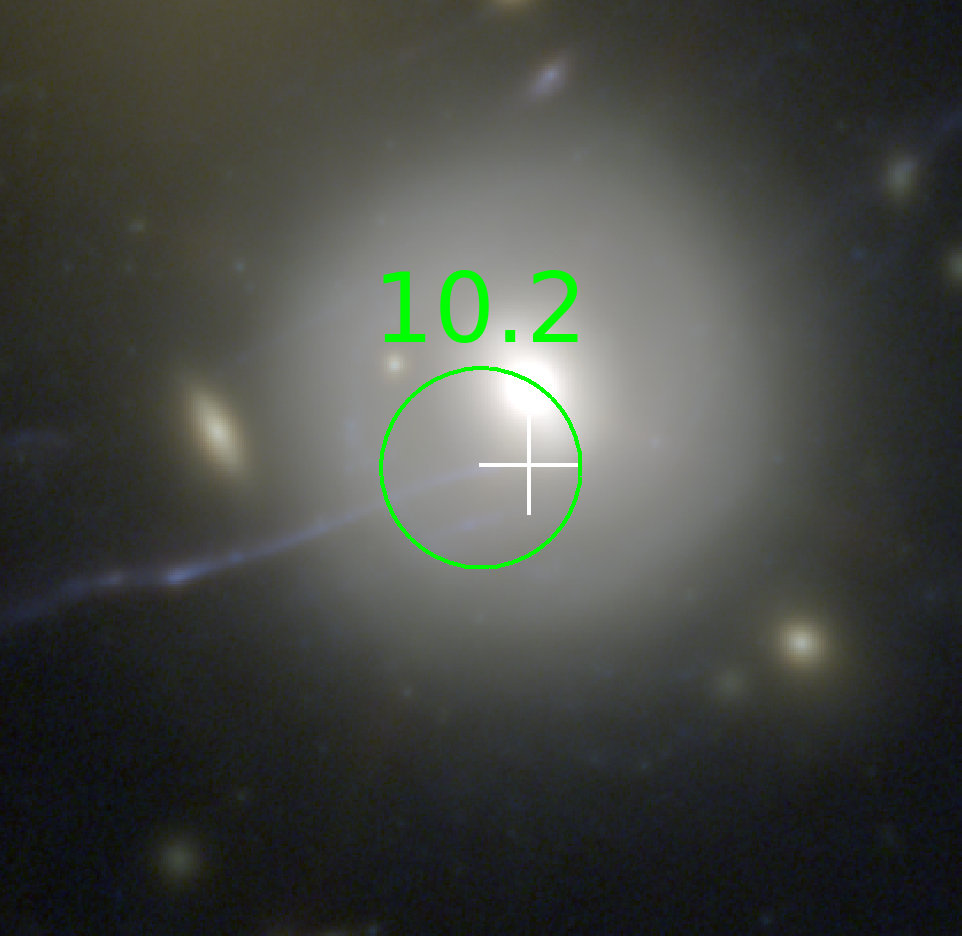}
 \includegraphics[width=0.16\textwidth,keepaspectratio=true]{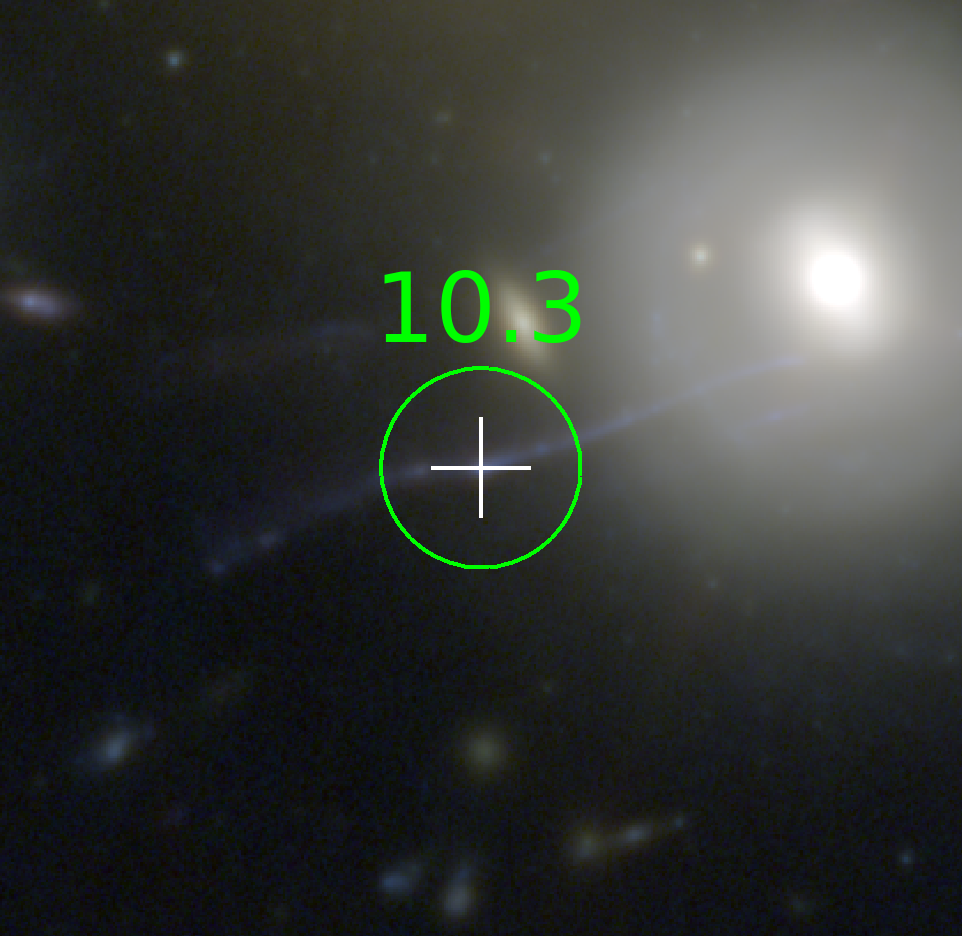}
 \caption{Snapshots (6 arcsecs across) of the multiple images from our best model. Observed image positions are marked with colored circles and predicted model positions with white crosses.}
 \label{fig:image_predictions}
\end{figure}

This is confirmed by a more qualitative approach where we manually compare the predicted image positions with the observed ones. In Figure \ref{fig:image_predictions}, we can see that all our predicted image positions are relatively close to the observed positions, as we would expect. We can also confirm here that systems $6$ and $7$ have larger errors in the predicted values compared with the other systems. 

\begin{figure}[!ht]
 \centering
 \includegraphics[width=0.5\textwidth,keepaspectratio=true]{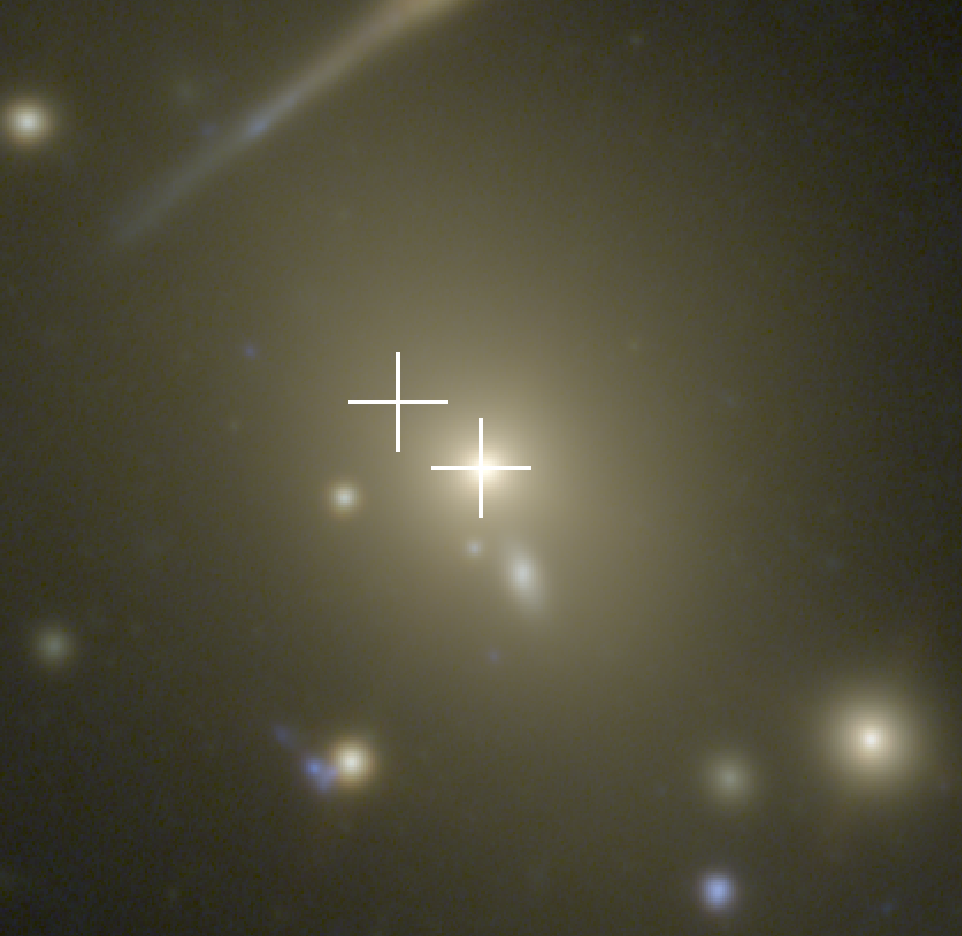}
 \caption{Two new predicted multiple images of system $7$ from our best model.}
 \label{fig:new_images}
\end{figure}

One of the benefits from a robust mass model, is the ability to predict new images. Our best model predicts two new images for system $7$ as can be seen in Figure \ref{fig:new_images} and are listed in Table \ref{fig:new_images}. 

\begin{table}[!ht]
 \centering
 \caption{Predicted new multiple images of system $7$ from our best model.}
 \label{table:new_images}
 \begin{tabular}{cccc}
  \toprule
  ID  & $\alpha$ & $\delta$ \\
  \midrule
  7.a & 64.032306 & -24.077186 \\
  7.b & 64.031978 & -24.077423 \\
  \bottomrule
  \end{tabular}
\end{table}

\section{Magnification, Critical lines and Caustics}
\begin{figure}[!ht]
 \centering
 \includegraphics[width=1.00\textwidth,keepaspectratio=true]{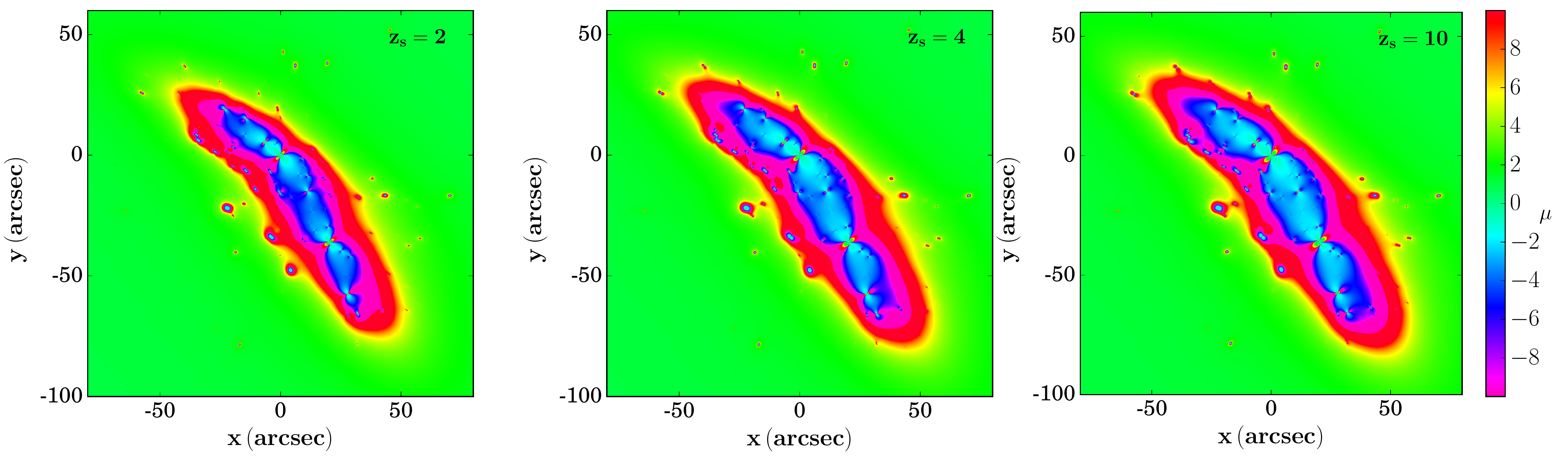}
 \caption{Amplification maps for \macs for the inner region reconstructed from our best model for sources at redshift $z_s = 2$ (left), $z_s = 3$ (middle) and $z_s = 10$ (right). The colors indicate the amplification magnitude on a linear scale extending $-10 \leq \mu \leq 10$. On all panels we have that North is up and East is right.}
 \label{fig:ampl_maps}
\end{figure}

From the three plots of the magnification at $z=2$, $z=4$ and $z=10$ (Figure \ref{fig:ampl_maps}), we see that the magnification at relatively far distances is $\mu \sim 1$ (green area), which is what we would expect. We see that there exists a region with high magnification $(\mu = 9$ - red) followed by a region with high negative magnification $(\mu = -9$ - purple) which then gradually increases to $\mu \sim -2$ inwards.

We also see, as we progress from low redshift $(z = 2)$ to high redshift $(z = 10)$, that the region with high magnification moves outwards, while the inner region with negative magnification increase in size. From a purely qualitative evaluation our results are consistent with previous results \cite{Grillo2015,Jauzac2014,Johnson2014}. We do note that both \citet{Jauzac2014} and \citet{Johnson2014} used $z=9$ for their magnification maps and \citet{Johnson2014} have scaled the magnification from $\mu = 0$ to $\mu = 20$. From these magnification maps we see where we expect images to be magnified.

\begin{figure}[!ht]
 \centering
 \includegraphics[width=0.5\textwidth,keepaspectratio=true]{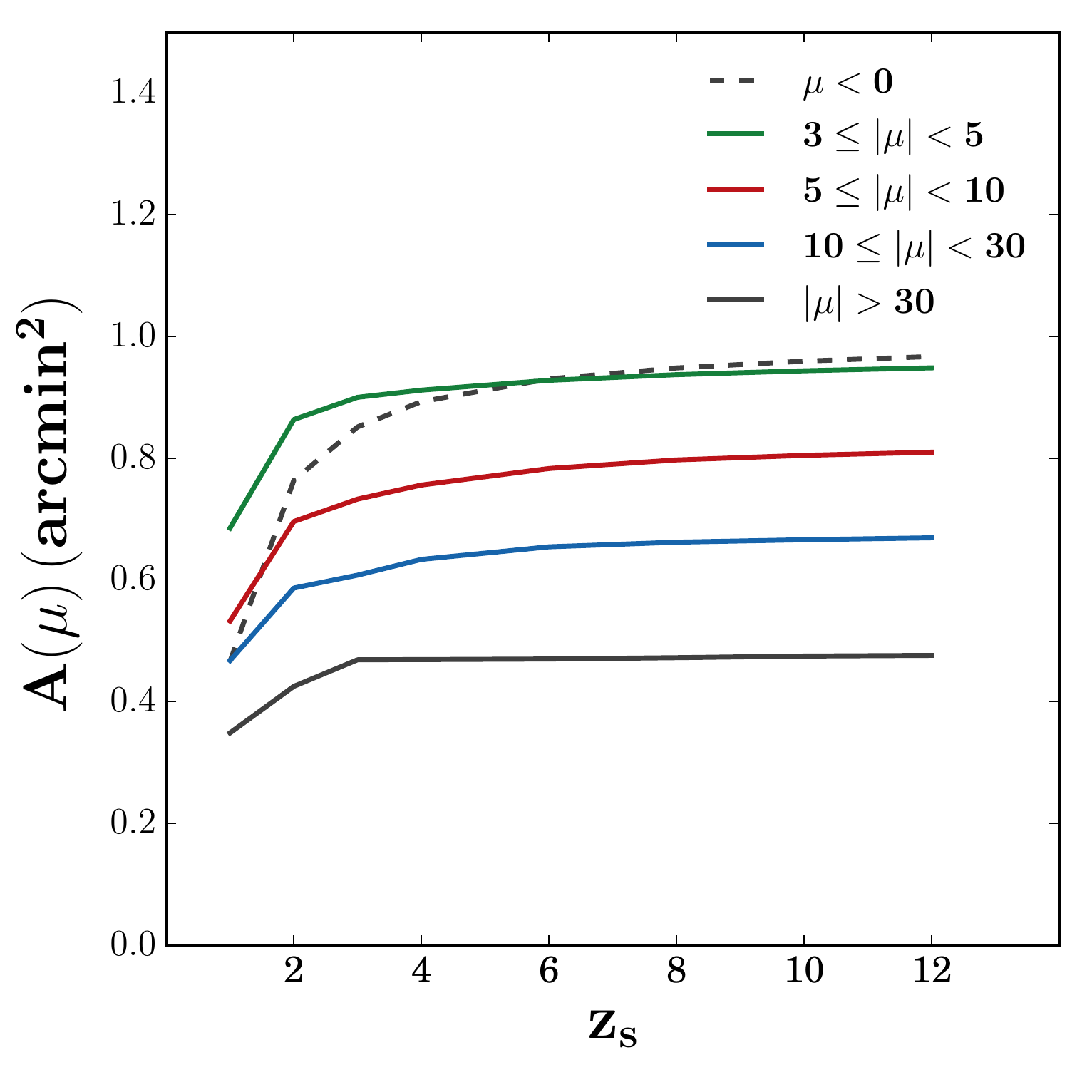}
 \caption{Values of the Area $A$ on the lens plane where the magnification factor $\mu$ is related to a specific source redshift $z_s$.}
 \label{fig:magn_area}
\end{figure}

In order to find the image plane area $A$ from which we expect a given amplification $\mu$ at a given source redshift $z_s$ we created amplification maps from $z=1$ to $z=12$ and counted the number of pixels where the amplification is within a given range and then converted this number to an area in $arcmin^2$. Since we are measuring the value of the individual pixels, we can simply increment our value by one, whenever we find $\mu$ within the specified range. The results from these calculations are listed in Table \ref{table:magn_area} and plotted in Figure \ref{fig:magn_area}. \citet{Grillo2015} find that the area increase with a factor $1.5$ for $A(\mu < 0)$ where we find a factor $\sim 1.3$ for a source redshift increase from $z=2$ to $z=10$. Since \citet{Grillo2015} has chosen the range $z=2$ to $z=10$, we can not compare our values for $z=1$ and $z=12$. For those we find an increase of $\sim 2.1$. Likewise, \citet{Grillo2015} find an increase of $1.3$ for $A(\mu \geq 30)$ where we find $\sim 1.1$, again for $z=2$ to $z=10$. For our full range we find $\sim 1.4$ which is similar to \citet{Grillo2015}.

From the plot in Figure \ref{fig:magn_area} we also see a clear difference. Overall the trend of our plot and the plot from \citet{Grillo2015} is similar. We also find that the amplification area increases with redshift. We do notice that our $A(\mu < 0)$ differs significantly from \citet{Grillo2015}, for which we can offer no explanation. In summary, our values are a bit lower than those from \citet{Grillo2015}.

\begin{table}[!ht]
 \centering
 \caption{Table over the Area $A$ on the lens plane where the magnification factor $\mu$ is related to a specific source redshift $z_s$.}
 \label{table:magn_area}
\begin{tabular}{cccccc}
   \toprule
   $z_s$ & $A(\mu < 0)$ & $A(3 \leq \abs{\mu} < 10)$ & $A(5 \leq \abs{\mu} < 10)$ & $A(10 \leq \abs{\mu} < 30)$ & $A(\abs{\mu} < 30)$ \\ 
   \midrule
   $1$ & $0.47$ & $0.69$ & $0.53$ & $0.47$ & $0.35$ \\ 
   $2$ & $0.76$ & $0.86$ & $0.70$ & $0.59$ & $0.43$ \\ 
   $3$ & $0.85$ & $0.90$ & $0.73$ & $0.61$ & $0.47$ \\ 
   $4$ & $0.89$ & $0.91$ & $0.76$ & $0.63$ & $0.47$ \\ 
   $6$ & $0.93$ & $0.93$ & $0.78$ & $0.65$ & $0.47$ \\ 
   $8$ & $0.95$ & $0.94$ & $0.80$ & $0.66$ & $0.47$ \\ 
   $10$ & $0.96$ & $0.94$ & $0.80$ & $0.67$ & $0.47$ \\ 
   $12$ & $0.97$ & $0.95$ & $0.81$ & $0.67$ & $0.48$ \\ 
   \bottomrule
\end{tabular}
\end{table}

\begin{figure}[!ht]
 \centering
 \includegraphics[width=0.48\textwidth,keepaspectratio=true]{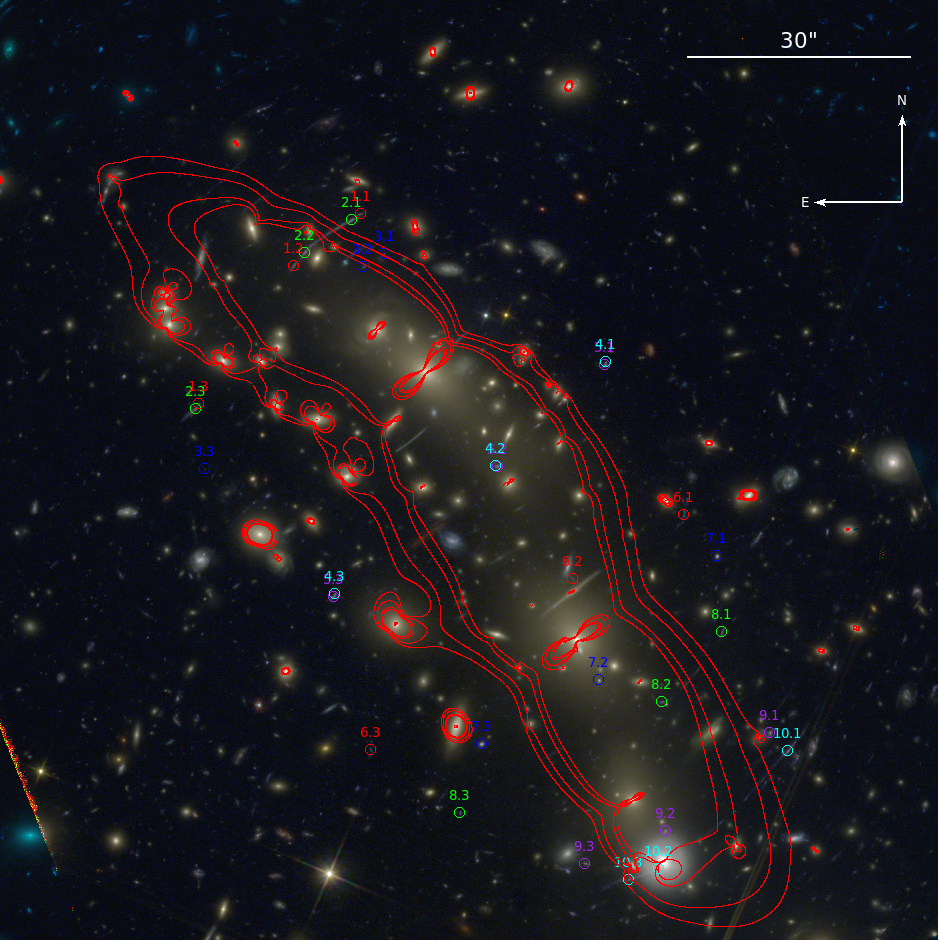}
 \includegraphics[width=0.48\textwidth,keepaspectratio=true]{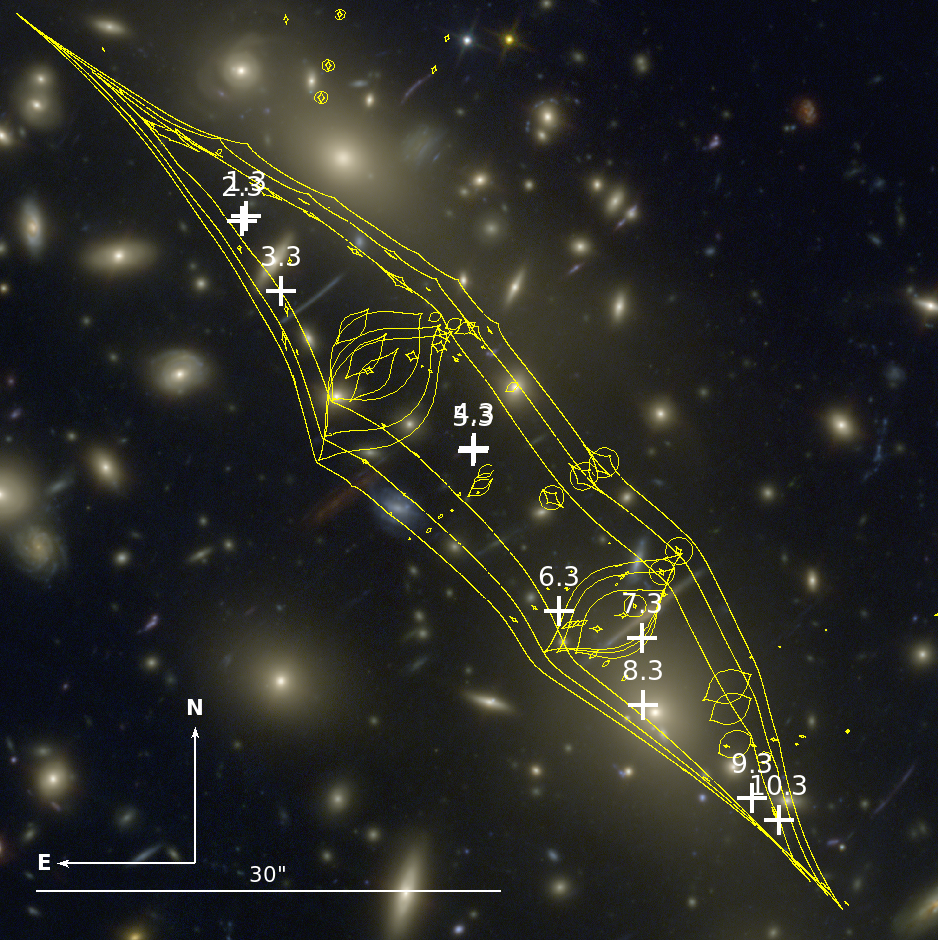}
 \caption{Left: Critical lines and multiple image positions. Right: Caustics and predicted source positions. We see here why the model predicts two additional images for system $7$. }
 \label{fig:crit_caust_results}
\end{figure}

On the map over the critical lines and the caustics in Figure \ref{fig:crit_caust_results} we see why our best model predicts two additional images for system $7$. In the right panel we see that the source position predicted for system $7$ (named $7.3$) is positioned within the inner caustic close to a fold. From this we know that a source positioned within the inner caustic we expect to see five images. Therefore the model predicts two new images. 

By looking at the position of the predicted images in Figure \ref{fig:new_images}, it appears that there are no clearly visible images to be related to the predicted images. One possible explanation could be that the predicted images are highly demagnified. When looking at the predicted images we see that one image has a predicted amplification of $1.040$ and the other $0.004$, compared with the three other images with predicted amplifications of $5.135$, $4.259$ and $3.116$. This means that we would need very deep imaging of \macs in order to discover the true position of these images. Future work might reveal the images by removing the light from the nearby cluster member. Another possible explanation is that the prediction is due to a bad model. This could be suggested by the fact that the GrHa model does not predict new images. 

\section{Mass Distribution}
\begin{figure}[!ht]
 \centering
 \includegraphics[width=0.49\textwidth,keepaspectratio=true]{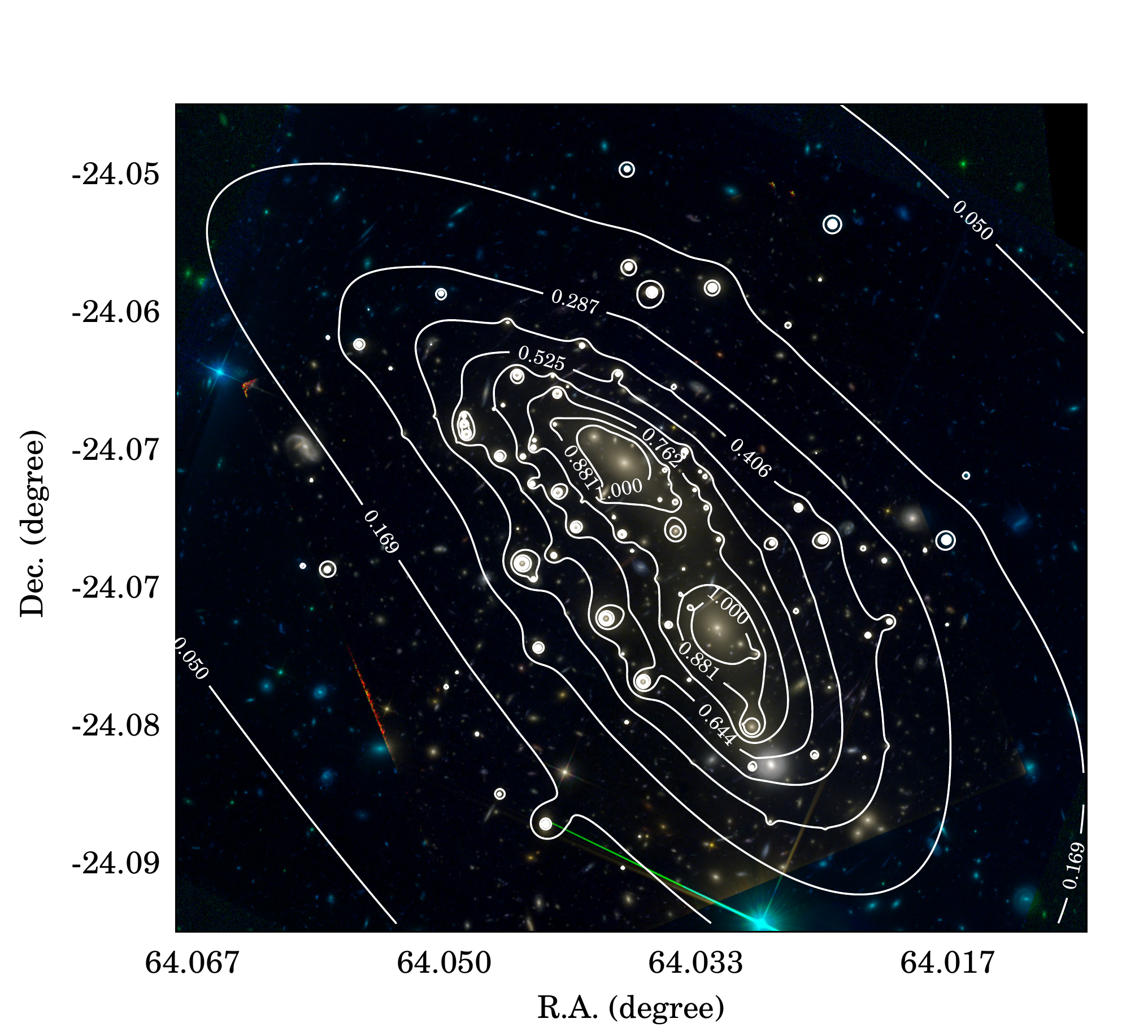}
 \includegraphics[width=0.49\textwidth,keepaspectratio=true]{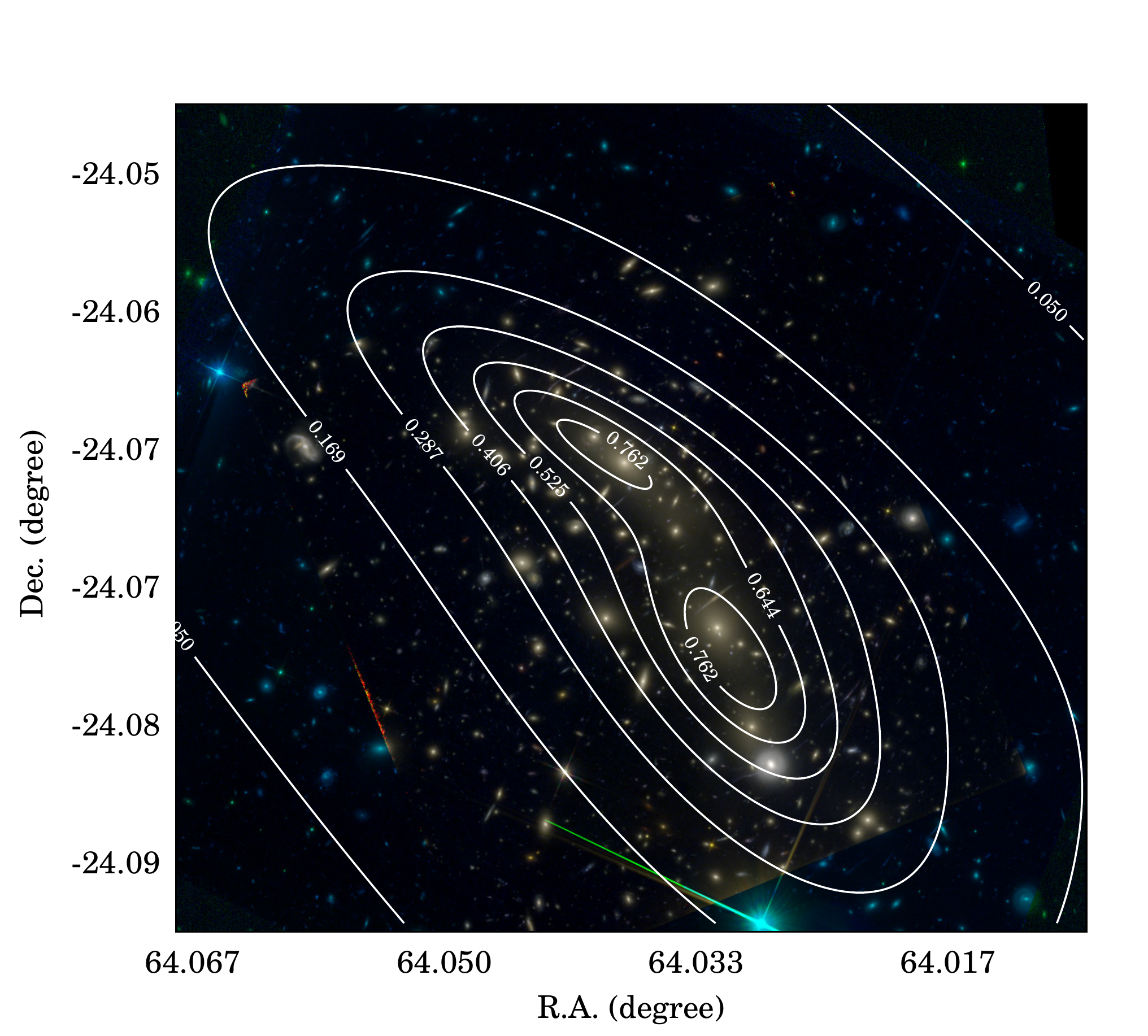}
 \includegraphics[width=0.49\textwidth,keepaspectratio=true]{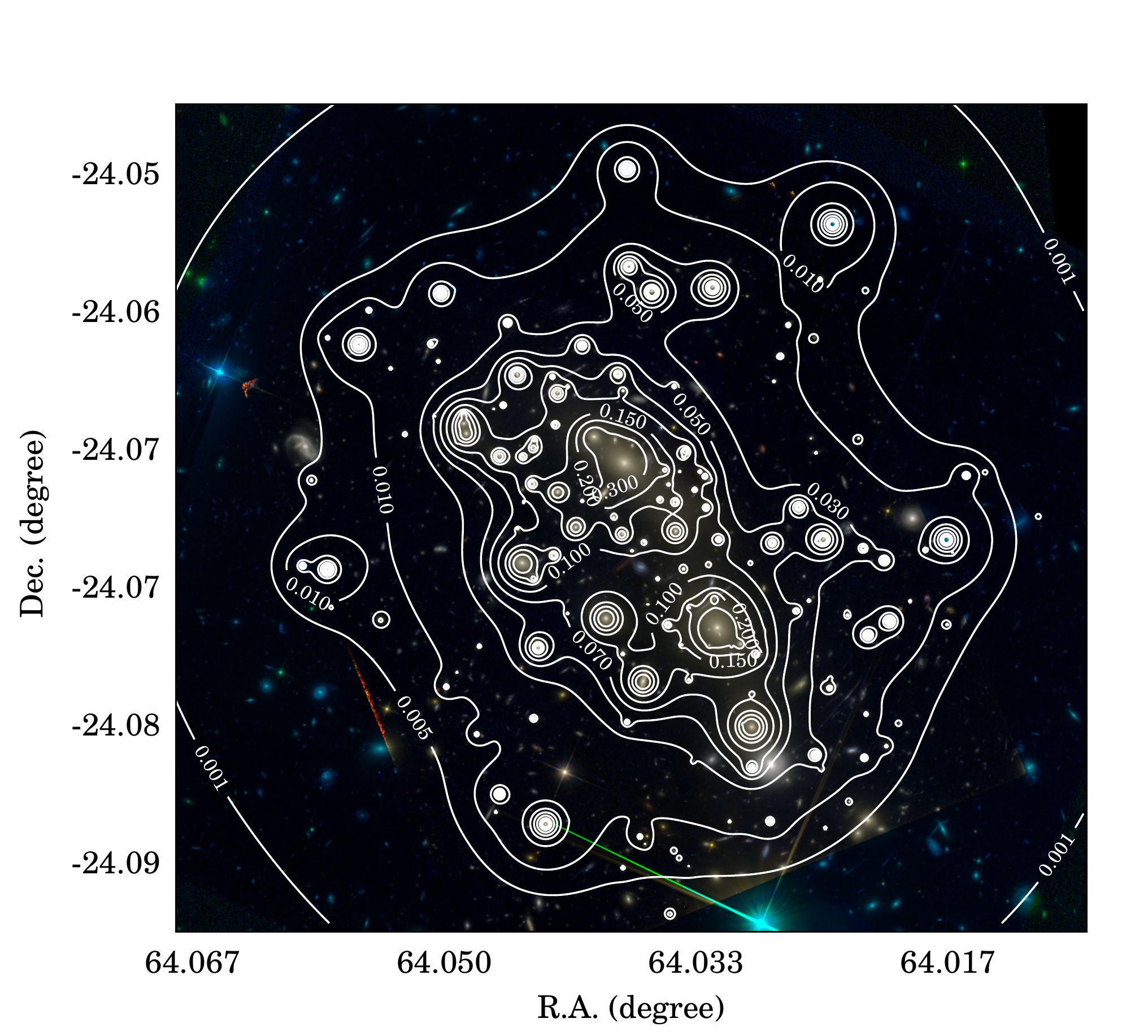}
 \caption{Decomposition of the total surface mass density $\Sigma_T$ (top right) into the surface mass densities of the cluster halos $\Sigma_H$ (top left) and cluster members $\Sigma_G$ (bottom). The contour levels are in $10^{10}\maths{M_{\Sun}}$.} 
 \label{fig:mass_maps}
\end{figure}

We have plotted surface-mass density maps for the all halos $(\Sigma_T)$, cluster halos alone $(\Sigma_H)$ and the cluster-member halos alone $(\Sigma_G)$ in Figure \ref{fig:mass_maps}. As \citet{Johnson2014} we also find that the mass of the foreground galaxy (FG) is unreasonable high when placed at the same redshift as the cluster. By comparing the truncation radius and velocity dispersion of the foreground galaxy $(r_{t,FG} = 504.8^{+847.0}_{-198.0}\maths{kpc} \, ; \, \sigma_{FG} = 110^{+41}_{-71}\maths{km \, s^{-1}})$ with the NE BCG $(r_{t,gal} = 105.3^{+242.4}_{-76.0}\maths{kpc} \, ; \, \sigma_{gal} = 321^{+75}_{-65}\maths{km \, s^{-1}})$, we see that the truncation radius for the foreground galaxy is unreasonably large, even considering the lower velocity dispersion. However, since we are working in a single lensing plane, we cannot model the foreground galaxy at the correct redshift. For that reason we exclude the foreground galaxy from further analysis.

In general when we compare our surface mass-density maps with those from previous studies \cite{Johnson2014,Jauzac2015,Grillo2015}, we find that our results are consistent, despite different modelling methods and software packages. We note that our results are highly consistent with those from \citet{Johnson2014} and \citet{Grillo2015}, where our results differ slightly from \citet{Jauzac2015}. We assume that the differences are due to the fact that \citet{Jauzac2015} used additional halos in their analysis. \citet{Caminha2016} have only added a single iso-contour line for the surface mass-density which makes it difficult to compare. We do notice that our results are consistent, despite minor differences. We see that \citet{Caminha2016} have a slightly less elliptical halo for the northern cluster halo, which we assume also is due to the additional halos. Lastly, we see that our results are directly consistent with \citet{Grillo2015}, independent of the lensing software.

\begin{figure}[!ht]
 \centering
 \includegraphics[width=0.8\textwidth,keepaspectratio=true]{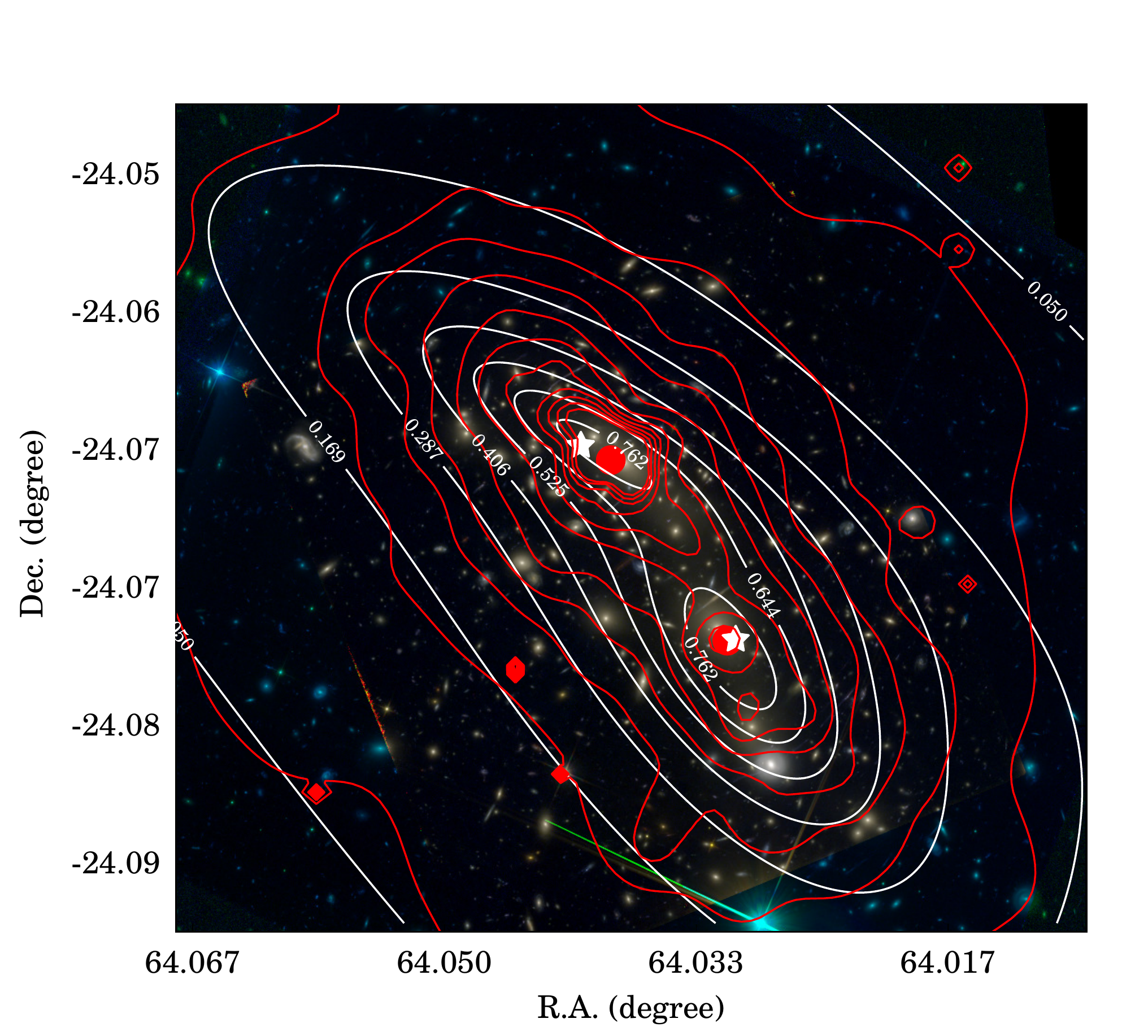}
 \caption{Comparison between \lenstool cluster halos integrated projected mass-density (white contour lines) and gas emission from X-Ray observations (red contours lines). The red filled circles indicate center of the X-Ray emission the and the white stars indicate the center of the gravitational lensing mass-density.}
 \label{fig:mass-density-compare}
\end{figure}

We also make a qualitative comparison between the contours from the X-Ray emission and the cluster halos surface-mass density in Figure \ref{fig:mass-density-compare}. We see clear similarities between the X-Ray contours (red contour lines) and the model contours (white contour lines) which indicates that the total mass trace the gas. We also see that the halo centers from the model (white squares) are consistent with the centers from the gas and offset from the centers of the BCGs. We explain the offset from the fact that dark matter particles are expected to be collisionless whereas gas particles are collisional. Since the model traces the total mass, the model also traces the gas particles and hence should be offset wrt. the BCGs. We speculate that a separation of the DM and gas in the model would result in the DM halos be centered on the BCGs and the gas halos be centered on the X-ray emission center.

\section{Mass Profiles}
\label{sect:mass_profiles}
In order to get a quantitative idea about the projected mass and surface-mass density for \macs, we select 100 models from the MCMC chain to be used to create new parameter files. Since \lenstool creates ten different chains related to the same model, we select random chains for each model. We select the models from a random starting point, but without selecting more than one chain from each model.

This is done by first creating a restricted random starting point, so that we will always have room for selecting a specific number of models\footnote{100 models in our case, but it could be any number up to 1001 models}. From that starting point we select the closest whole number in tens and then select a random number within the starting number and ten more. So if the whole ten number is $10$, we select a random number in the range $10-20$ and so forth. 

Using these random numbers, we select positions within the MCMC chain (\emph{bayes.dat} file) and use these to create 100 individual model parameter files. Since these parameter files resemble the output best parameter file from \lenstool they they are suitable for creating mass-density maps. From these parameter files, we extract 100 maps of the integrated projected mass-density with $500$ px resolution and ranging $200\times200\maths{arcsec}$, where each pixel represents $10^{12}\maths{M_{\Sun}}$. This gives us a spatial resolution of $0.4\maths{arcsec/pixel}$ or $2.1\maths{kpc/pixel}$ (when $1\maths{arcsec} = 5.340\maths{kpc})$ 

These maps are used to calculate the cumulative projected mass and the surface mass density. In order to do that we apply the same method as in \citet{Grillo2015} and find the barycenter. For the GrHa model we apply the same barycenter as \citet{Grillo2015} $(\alpha = 64.035666,\delta = -24.073644)$ but for our best model we find our own barycenter. The barycenter can be found using

\[
 \vec{R}_b \equiv \frac{\int \sum_T (\tilde{\vec{R}})\tilde{\vec{R}}d\tilde{\vec{R}}}{\int \sum_T (\tilde{\vec{R}}) d\tilde{\vec{R}}}
\]

\begin{figure}[!ht]
 \centering
 \includegraphics[width=0.6\textwidth,keepaspectratio=true]{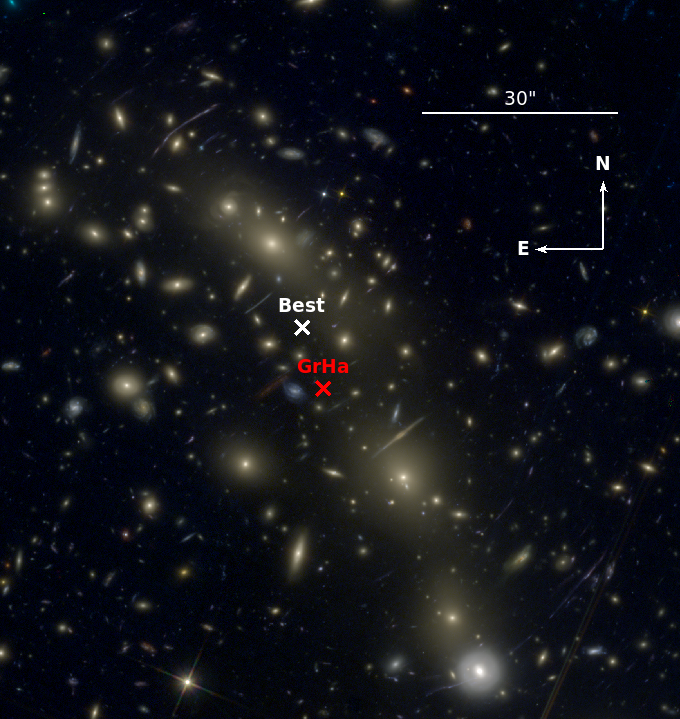}
 \caption{Barycenter for \macs. Red cross indicate barycenter from \citet{Grillo2015} used in our GrHa model and white cross indicate the barycenter from our best model.}
 \label{fig:barycenter}
\end{figure}

We use the following approach to discretize this formula. We will add code snippets from our Python script to better illustrate and explain our approach. We first calculate the total mass of the entire fits file
\begin{lstlisting}
   mTot = simps(simps(m_tot))
\end{lstlisting}
We then create two new arrays as placeholders for the individual positions
\begin{lstlisting}
   massX = np.zeros((fitsDim))
   massY = np.zeros((fitsDim))
\end{lstlisting}
In order to find the barycenter we first calculate the integrated mass along each row and column (x and y rows)
\begin{lstlisting}
   for i in range(0,fitsDim):
      massY[i] = simps(m_tot[i,:])
      massX[i] = simps(m_tot[:,i])
\end{lstlisting}
and we can now find the barycenter by assigning an integrated mass to each pixel and then divide that value, by the total mass in order to find the position with the highest integrated mass pr. pixel.
\begin{lstlisting}
   xb = simps(massX * np.arange(fitsDim)) / mTot
   yb = simps(massY * np.arange(fitsDim)) / mTot
\end{lstlisting}
Using this method we find that the barycenter is located at $(\alpha = 64.036657,\delta = -24.07106)$. Both barycenters are illustrated in Figure \ref{fig:barycenter}. We speculate that \citet{Grillo2015} may have included the foreground galaxy in barycenter calculation, since it is clearly offset towards South-West. On the other hand, inclusion of a single galaxy should not be able to skew the barycenter that much.

By now having the barycenter, we calculate the cumulative projected mass
\[
 M(<R) \equiv \int_0^R \sum (\tilde{R}) 2\pi \tilde{R}\,d\tilde{R}
\]
where $\tilde{R}$ is any given radius. We also calculate the surface-mass density
\[
 \Sigma (<R) \equiv \frac{\int_0^R \sum (\tilde{R}) 2\pi \tilde{R}\,d\tilde{R}}{\pi \tilde{R}^2}
\]
where we see that the surface mass-density basically is
\[
 \Sigma (<R) \equiv \frac{M(<R)}{\pi \tilde{R}^2}
\]
Here we will also add code snippets to explain the approach more fully.

\begin{figure}[!ht]
 \centering
 \includegraphics[width=0.48\textwidth,keepaspectratio=true]{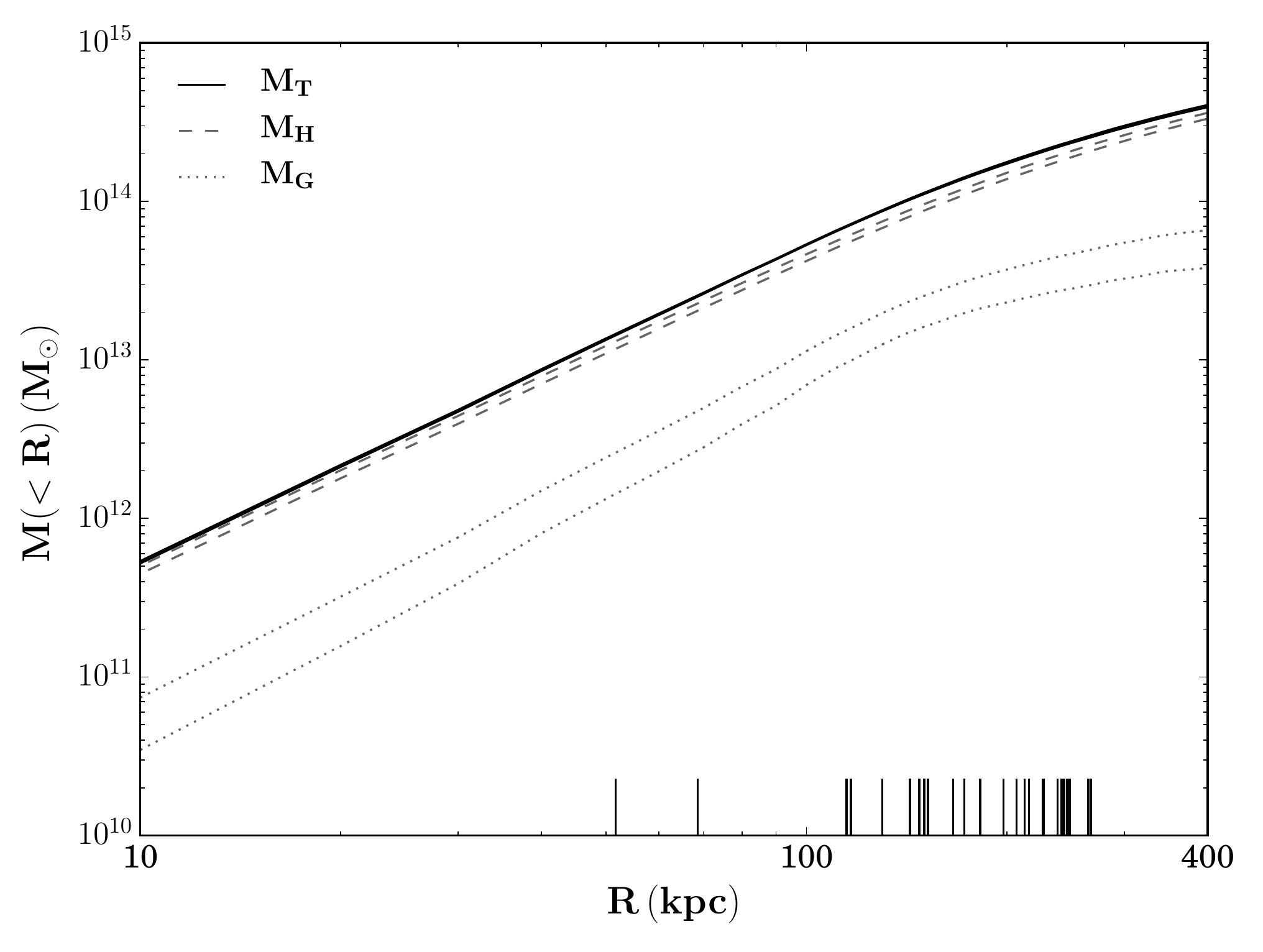}
 \includegraphics[width=0.48\textwidth,keepaspectratio=true]{2dpie_Grillo_averageMassDensity_1sigma.pdf}
 \includegraphics[width=0.48\textwidth,keepaspectratio=true]{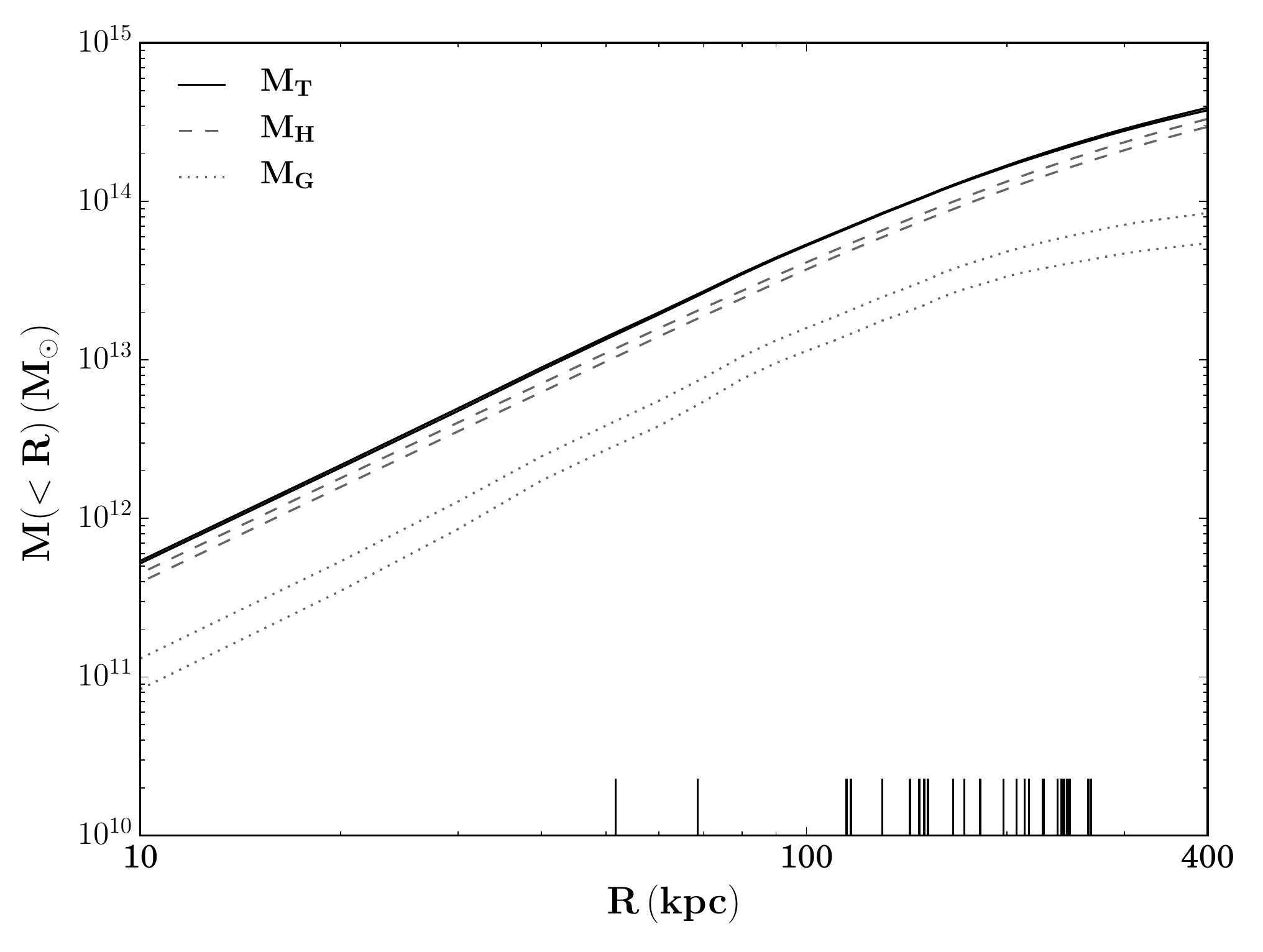}
 \includegraphics[width=0.48\textwidth,keepaspectratio=true]{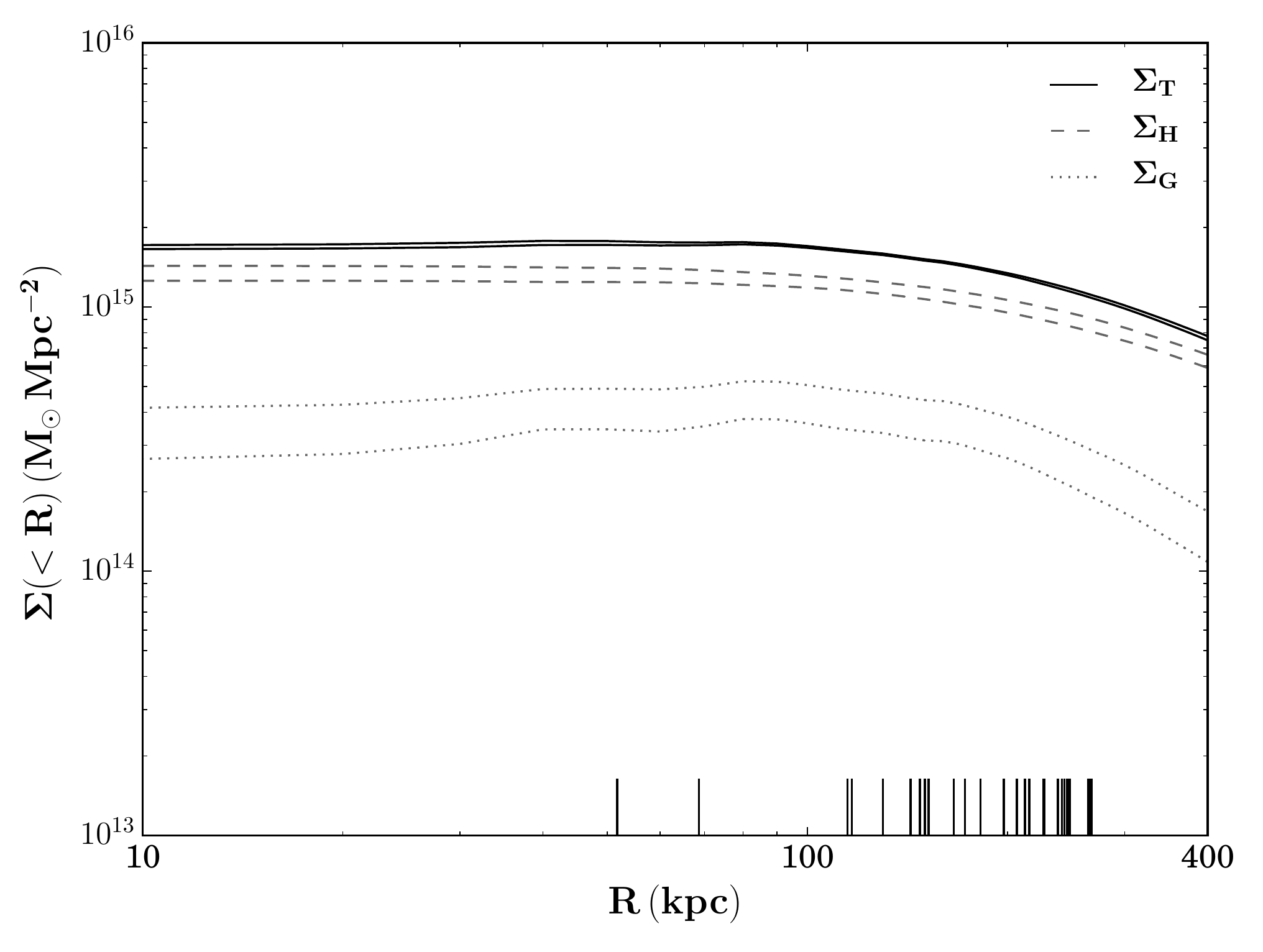}
  \caption{Top row: GrHa model. Bottom row: Best model. Left: The cumulative projected mass-density. Right: The average surface mass-density. The solid lines represent the total cluster, the long dashed lines represents the cluster halos alone and the dotted lines represents the cluster member profiles, all at $1\sigma$ CL. The vertical lines in the bottom represent the relative position of the multiple images.}
 \label{fig:mass-plots}
\end{figure}

First we load our mass-density fits files into a 2D array and convert all the values so they correspond to real mass-density values\footnote{As mentioned previously all pixelvalues are in $10^{12}\maths{M_{\Sun}}$}. Since we have multiple images out to about $270\maths{kpc}$ we set our outer limit roughly the double of that $(R_{\mathrm{max}} = 400\maths{kpc})$, which is also the value chosen by \citet{Grillo2015}. In order to calculate the cumulative projected mass, we want to assign each pixel in our fits file with a specific radius. We do that by first creating a new array to hold the radii
\begin{lstlisting}
    xx, yy = np.meshgrid(np.arange(fitsDim)
                       , np.arange(fitsDim))
\end{lstlisting}
and populate that grid with a radius going from $0-400\maths{kpc}$ from the barycenter. This means that we assign each of our pixels a discrete radius
\begin{lstlisting}
   r = (np.sqrt((yy - yb)**2 + (xx - xb)**2)) * pix2kpc
\end{lstlisting}
where the \emph{pix2kpc} parameter is a conversion value so that each pixel actually represents a distance in kpc and \emph{xb} and \emph{yb} are the barycenter pixel coordinates. We then set up bins for integration and because we want to avoid division by zero we approximate the first value close to zero
\begin{lstlisting}
   bins = np.linspace(0, R, points)
   bins[0] = 0.01
\end{lstlisting}
and we have chosen to divide our radii grid into $81$ individual values. We divide each of our pixels in the xx and yy array for it to be part of a specific radius according to the bins we just created. We do this in order to sum the mass within a given radius
\begin{lstlisting}
   inds = np.digitize(r, bins)
\end{lstlisting}
It is now simply a matter of summing the mass up to $400\maths{kpc}$
\begin{lstlisting}
   for i in range(len(bins)):
      sM_halo[i] = np.sum(m_halo[inds == i])
      sM_gal[i] = np.sum(m_gal[inds == i])
      sM_tot[i] = np.sum(m_tot[inds == i])
      
   massGal = np.cumsum(sM_gal)
   massHalo = np.cumsum(sM_halo)
   massTot = np.cumsum(sM_tot)
\end{lstlisting}
We do that for the cluster halos alone, the cluster members alone and the total mass of all the components. In order to find the average surface mass-density, we simply divide our result with $\pi \tilde{R}^2$ where $\tilde{R}$ is a given radius in our integration bins, plus a dimensional conversion factor applied to $\pi$
\begin{lstlisting}
   avgMassDensGal = massGal / (3.142e-6*(bins)**2)
   avgMassDensHalo = massHalo / (3.142e-6*(bins)**2)
   avgMassDensTot = massTot / (3.142e-6*(bins)**2)
\end{lstlisting}
and the exponent part defines the value as an area. We calculate the cumulative projected mass and average surface mass-density for all 100 models, then sort these 100 different masses from lowest to highest in order to find the median and the $1\sigma$ CL. The results are presented in Figure \ref{fig:mass-plots}.

We clearly see that the average surface-mass density for both models is constant out to about $R = 100\maths{kpc}$ where it then gradually decreases. This is similar for the cluster members and the cluster halos and means that we have a very flat inner core. The small bump at $\approx 130\maths{kpc}$ which is most prominent in the plot from the GrHa model is due to the position of the two BCGs relative to the used barycenter. The shift in position of this bump from the GrHa model to our best model is most likely due to the different barycenter. Similar to  \citet{Grillo2015} we also find evidence that the cluster member and cluster halo components are anti-correlated, although this is less evident in our best model. In order to show this more clearly we compare our models with the one from \citet{Grillo2015}. 

\begin{figure}[!ht]
 \centering
 \includegraphics[width=0.49\textwidth,keepaspectratio=true]{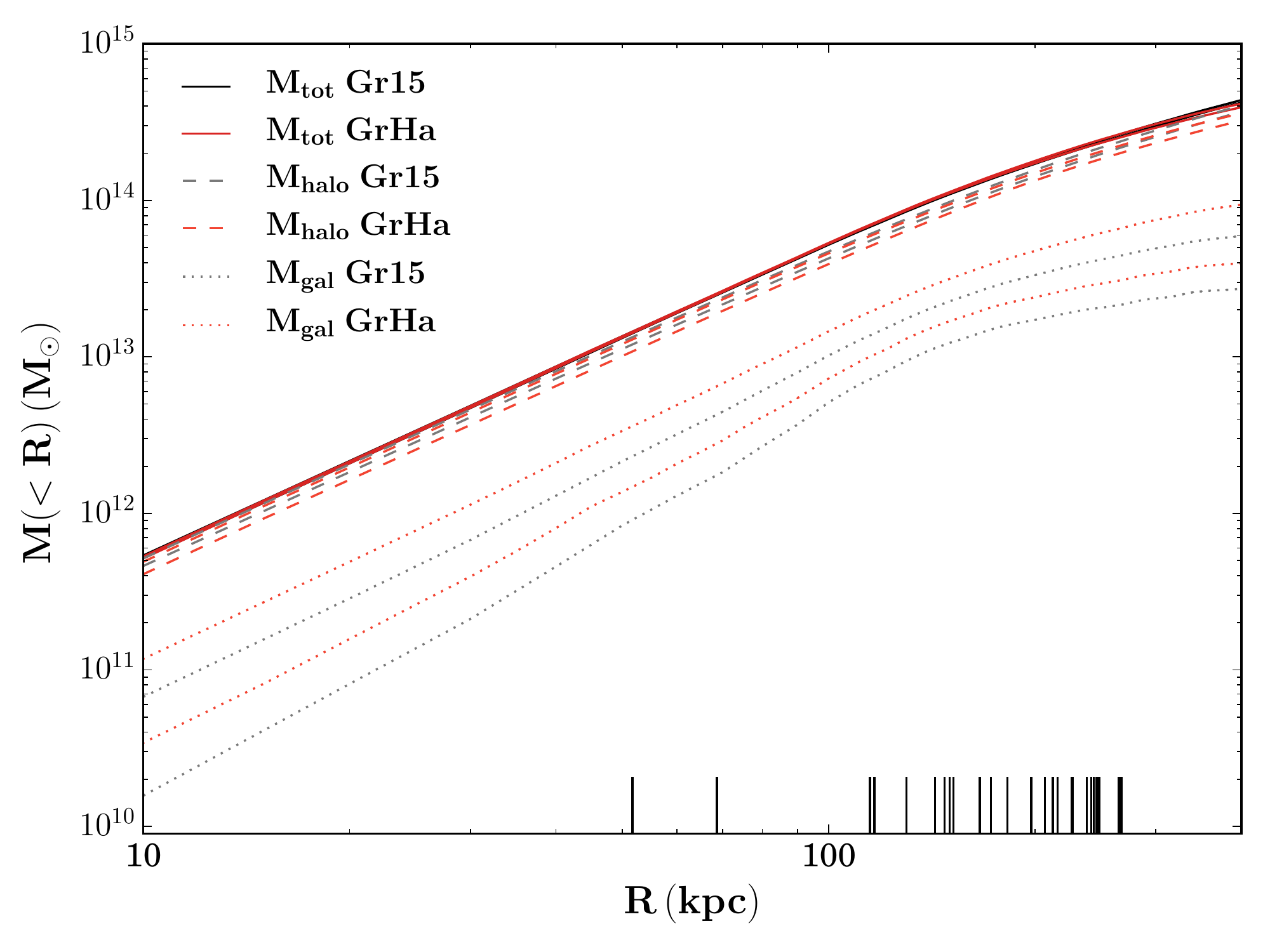}
 \includegraphics[width=0.49\textwidth,keepaspectratio=true]{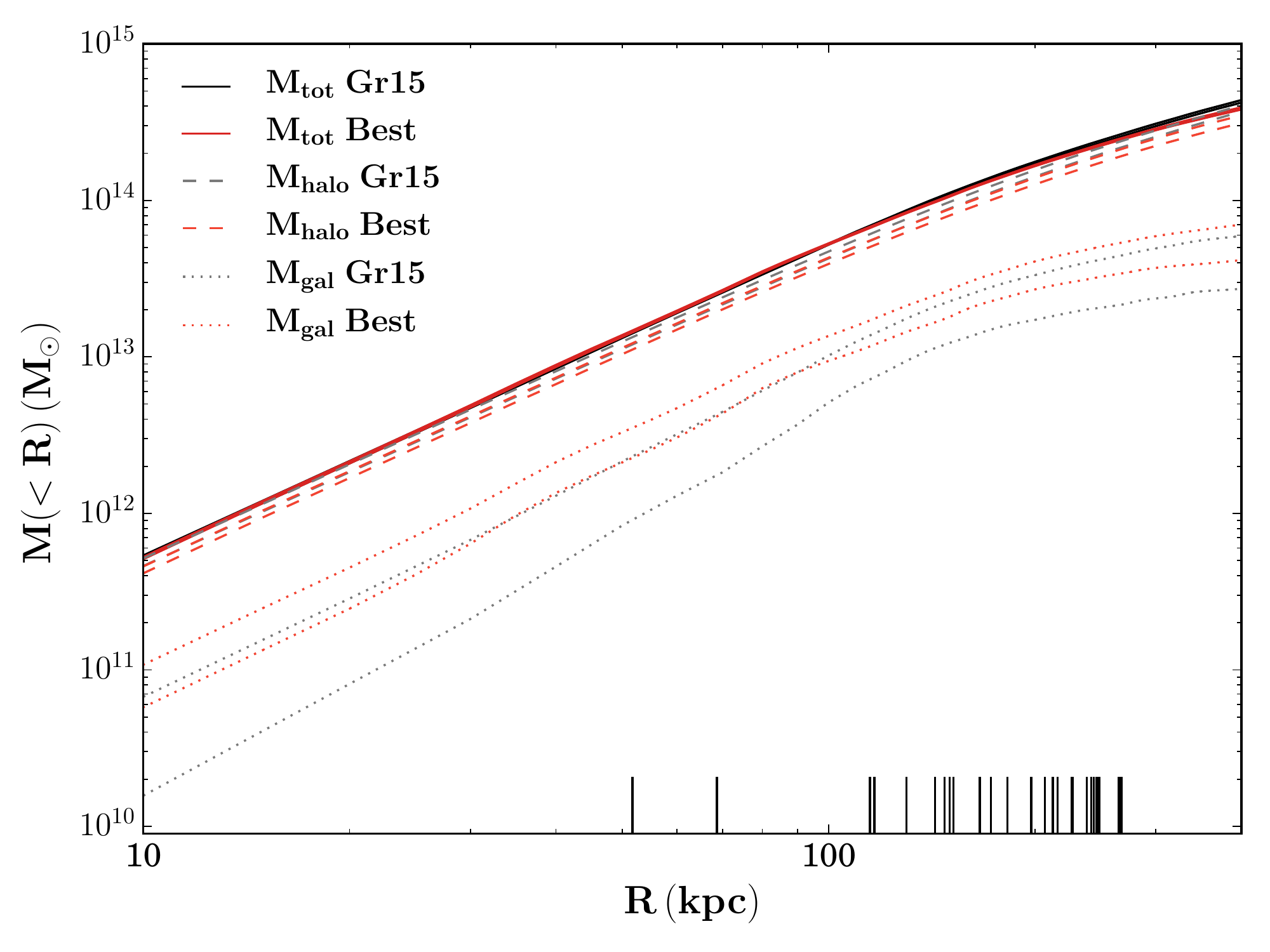}
 \caption{Cumulative projected mass comparison between the best model from \citet{Grillo2015}(Gr15) and our GrHa model (Left) and the Gr15 model and our best model (Right). The solid lines represent the total cluster, the long dashed lines represent the cluster halos alone and the dotted lines represent the cluster member profiles, all at $1\sigma$ CL. The vertical lines in the bottom represent the relative position of the multiple images.}
 \label{fig:mass-comp-Gr15}
\end{figure}

On the left panel of Figure \ref{fig:mass-comp-Gr15} we see a comparison between the best model from \citet{Grillo2015} (Gr15) and our own GrHa model. From these plots we see that our GrHa model puts significantly more mass into the cluster members than the Gr15 model which must mean that our model puts less mass into the cluster halos, since the total masses are comparable. It is only at $R \approx 400\maths{kpc}$ we observe a small deviation where our model has a little less total mass, which is most likely due to the smaller cluster halo mass. On the right panel of Figure \ref{fig:mass-comp-Gr15} we see the same comparison but here between our best model and the Gr15 model. Here we observe a significant difference. The $1\sigma$ CL for our best model cluster members are much smaller than the errors from Gr15. Here we also observe that our model puts more mass into the cluster members, although the difference is most prominent at $R \leq 100\maths{kpc}$. Above $100\maths{kpc}$ we see that the cluster-member mass from our best model approach the cluster-member mass from Gr15. Again we observe a small deviation at $R \approx 400\maths{kpc}$ which indicates that the deviation is not model specific, but may be due to differences in the mass maps themselves.

\begin{figure}[!ht]
 \centering
 \includegraphics[width=0.8\textwidth,keepaspectratio=true]{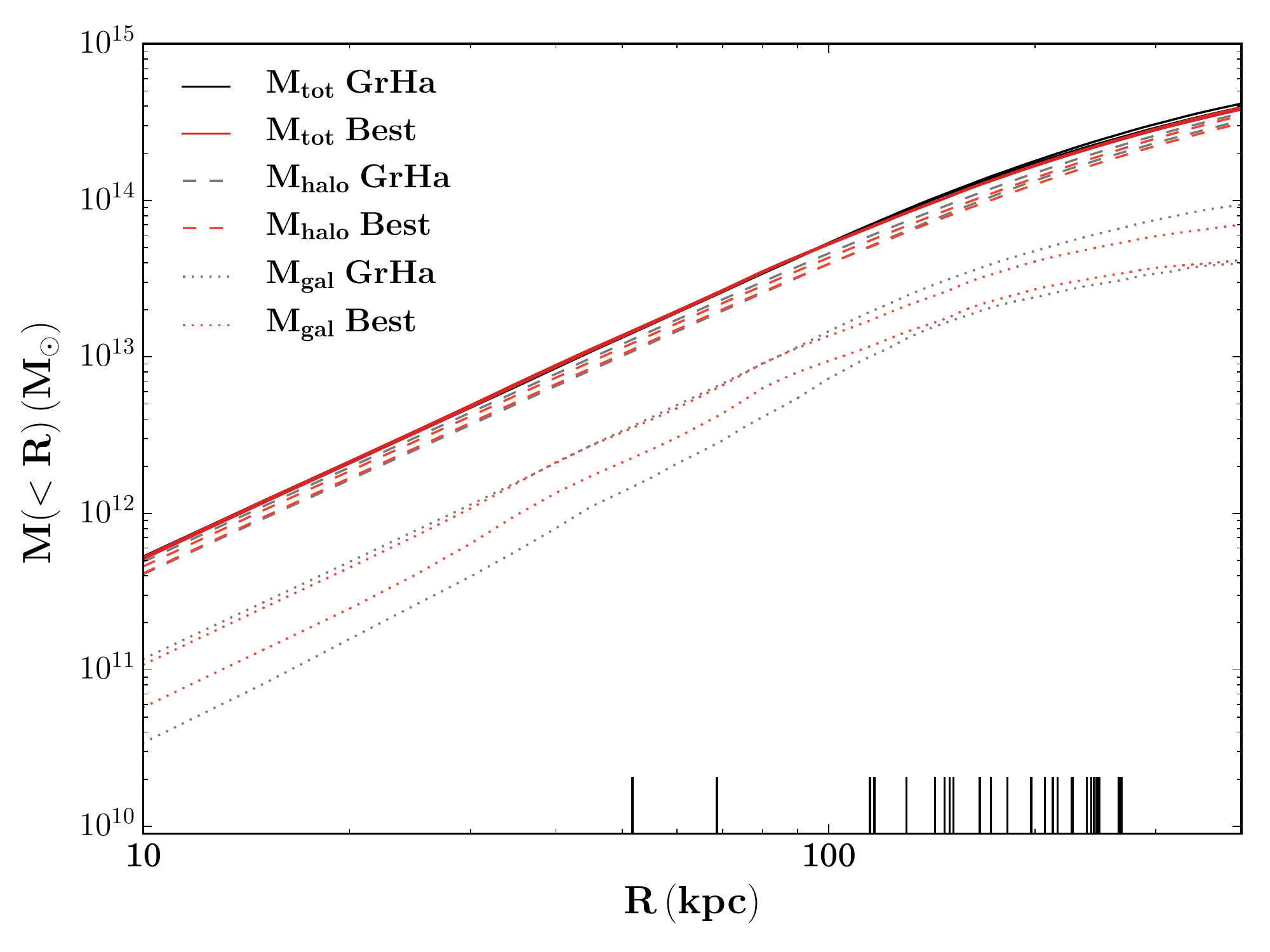}
 \caption{Cumulative projected mass comparison between our best model and our GrHa model. The solid lines represent the total cluster, the long dashed lines represents the cluster halos alone and the dotted lines represents the cluster member profiles, all at $1\sigma$ CL. The vertical lines in the bottom represent the relative position of the multiple images.}
 \label{fig:mass-comp-best-GrHa}
\end{figure}

In order to illustrate the difference between our two models, we compare them in Figure \ref{fig:mass-comp-best-GrHa} where we can see the differences between the two models more clearly. First of all, we see that the cluster member mass is similar out to $R \leq 100\maths{kpc}$ where our best model begins to put less mass into the cluster members. The cluster halo mass for both models is similar and follow each other out to $R = 400\maths{kpc}$. We can therefore conclude the following: The differences in the total mass between the two models must be due to the differences in cluster-member mass. We also see a significant difference in the range of the $1\sigma$ CL, which must be due to the additional optimization of the slopes in our best model, whereas the $1\sigma$ CL are similar in the Gr15 and our GrHa model. This could indicate that the additional optimization of the slopes constrains the mass of the cluster-members better, but is contra-indicated by the fact that both models have significant problems constraining the cluster-member cut radius.

In summary, we can confirm that the total mass distribution in the central regions of \macs is dominated by two highly elliptical and close in projection components, representing two extended massive DM halos. These halos are responsible for the large area on the lens plane with high amplification, as found by \cite{Grillo2015} and earlier studies \cite{Zitrin2013,Johnson2014,Richard2014,Jauzac2015,Caminha2016}. We also find that the inner total mass density profile of the cluster is flat with a core radius of  $R_{core} \sim 100\maths{kpc}$. 

\begin{figure}[!ht]
 \centering
 \includegraphics[width=0.8\textwidth,keepaspectratio=true]{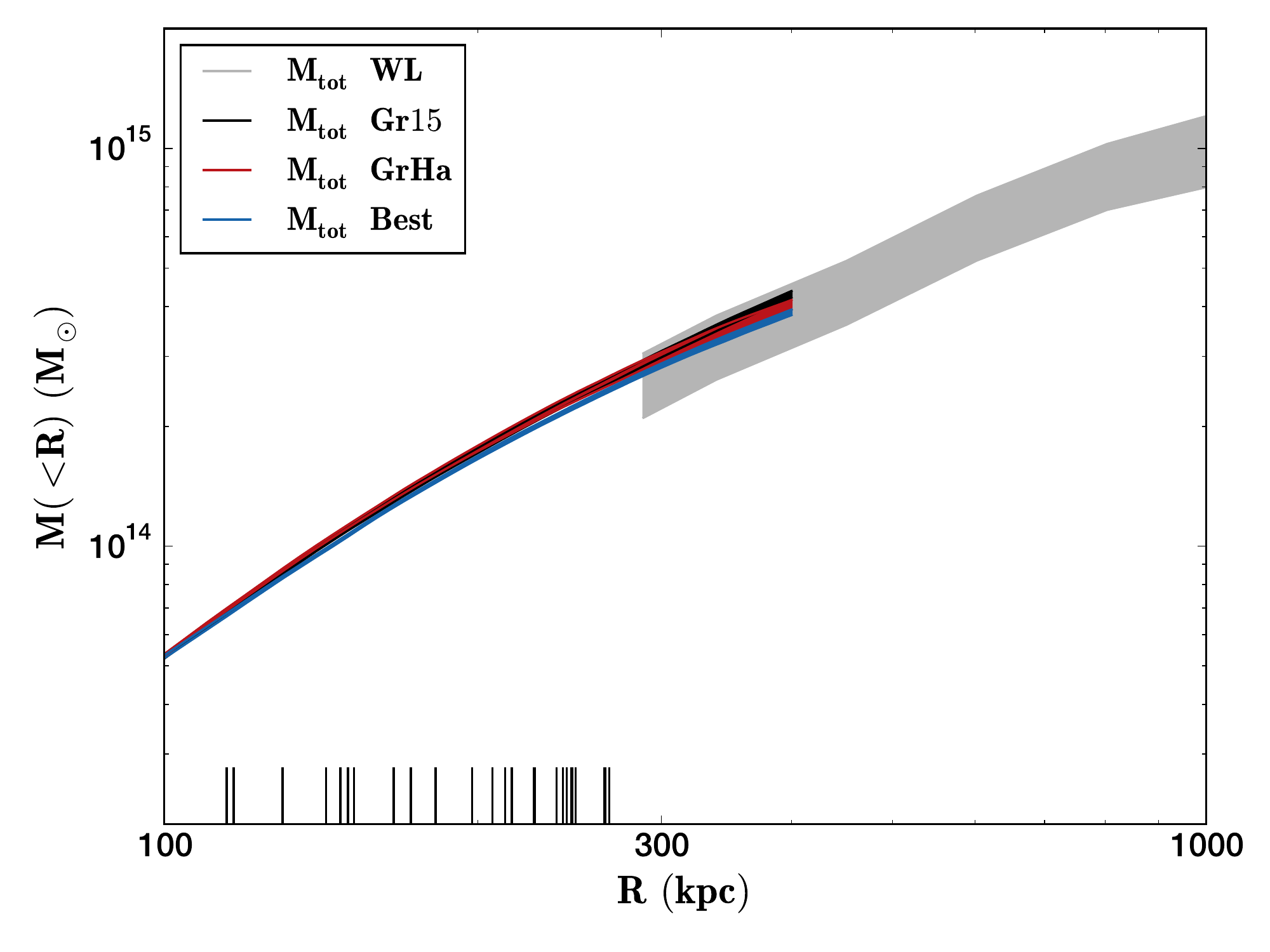}
 \caption{Cumulative projected mass comparison between results from weak lensing \cite{Umetsu2014} and results from strong lensing from the Gr15 model, our GrHa model and our best model. All values are within $1\sigma$ CL. The vertical lines at the bottom represent the relative position of multiple images for strong lensing.}
 \label{fig:mass-comp-wl-sl}
\end{figure}

We perform a comparison between the weak lensing analysis of \macs presented in \citet{Umetsu2014}, the Gr15 model from \citet{Grillo2015} and our GrHa and best model. The results from this comparison is shown in Figure \ref{fig:mass-comp-wl-sl}. Like \citet{Grillo2015} we also find a good correlation between the results from weak lensing and our strong lensing models. 

We do find that our models display a more steeply falling cumulative projected mass around $\sim 400\maths{kpc}$ which we speculate is due to subtle differences between \lenstool and \glee, where \lenstool might put a little less mass in the outer parts of the profiles due to less constraints from multiple images. We arrive at this conclusion since the drop is evident in both our GrHa and our best model. We cannot infer a problem with \lenstool from this data, however, since there might be less mass than predicted in \citet{Grillo2015}. Inclusion of additional multiple images at higher redshift \cite{Caminha2016} might give a more reasonable mass estimate. It should be mentioned however that it is only a few select cluster models that have such good correlation between weak lensing and strong lensing mass estimates \cite{Grillo2015}.

We have performed a direct comparison with specific values from the literature. For the total mass measurements we find that our GrHa model gives $1.73\times 10^{14}\maths{M_{\Sun}} \leq M_T(<200\maths{kpc}) \leq 1.79\times 10^{14}\maths{M_{\Sun}}$, $2.34\times 10^{14}\maths{M_{\Sun}} \leq M_T(<250\maths{kpc}) \leq 2.45\times 10^{14}\maths{M_{\Sun}}$, $3.13\times 10^{14}\maths{M_{\Sun}} \leq M_T(<320\maths{kpc}) \leq 3.31\times 10^{14}\maths{M_{\Sun}}$ and $3.94\times 10^{14}\maths{\MSun} \leq M_T(<400\maths{kpc}) \leq 4.06\times 10^{14}\maths{\MSun}$. 

For our best model we find $1.66\times 10^{14}\maths{M_{\Sun}} \leq M_T(<200\maths{kpc}) \leq 1.69\times 10^{14}\maths{M_{\Sun}}$, $2.24\times 10^{14}\maths{M_{\Sun}} \leq M_T(<250\maths{kpc}) \leq 2.29\times 10^{14}\maths{M_{\Sun}}$, $3.02\times 10^{14}\maths{M_{\Sun}} \leq M_T(<320\maths{kpc}) \leq 3.10\times 10^{14}\maths{M_{\Sun}}$ and $3.76\times 10^{14}\maths{\MSun} \leq M_T(<400\maths{kpc}) \leq 3.90\times 10^{14}\maths{\MSun}$. So we have slightly less total mass in our best model, than in our GrHa model, within $68\%$ CL. 

\citet{Grillo2015} find mass estimates between $1.72\times 10^{14}\maths{M_{\Sun}}$ and $1.77\times 10^{14}\maths{M_{\Sun}}$ for $M_T(<200\maths{kpc})$. We find that our GrHa is comparable to \citet{Grillo2015} whereas our best model presents a lower total mass. \citet{Grillo2015} find between $2.35$ and $2.43\times 10^{14}\maths{M_{\Sun}}$ at $M_T(<250\maths{kpc})$ which is consistent with out GrHa model, but inconsistent with our best model. \citet{Grillo2015}  find $3.23$ to $3.35\times 10^{14}\maths{M_{\Sun}}$ at $M_T(<320\maths{kpc})$ which is also in agreement with our GrHa model, but not our best model. Via private communication we know that \citet{Grillo2015} find $4.21\times 10^{14}\maths{\MSun} \leq M_T(<400\maths{kpc}) \leq 4.38\times 10^{14}\maths{\MSun}$. Here we see that both our models deviate from \citet{Grillo2015}. In general, our best model predicts slightly less total mass than \citet{Grillo2015}.

\citet{Johnson2014} find $2.46^{+0.04}_{-0.08}\times 10^{14}\maths{M_{\Sun}}$ at $M_T(<250\maths{kpc})$, which is consistent with our GrHa model, but inconsistent with our best model. 

\citet{Richard2014} find $1.63\pm 0.03\times 10^{14}\maths{M_{\Sun}}$ at $M_T(<200\maths{kpc})$, which is considerably less than both our models. Our results are therefore inconsistent with \citet{Richard2014}.

\citet{Jauzac2014} find $1.60\pm 0.01\times 10^{14}\maths{M_{\Sun}}$ at $M_T(<200\maths{kpc})$, which is inconsistent with both our models and find $(3.26 \pm 0.03 \times 10^{14}\maths{M_{\Sun}})$ at $M_T(<320\maths{kpc})$ which is agreement with our GrHa model, but not our best model. 

\citet{Jauzac2015} find $3.15 \pm 0.13\maths{M_{\Sun}}$ at $M_T(<320\maths{kpc})$, which is in agreement with both our models. We do note the relatively large error.

\begin{figure}[!ht]
 \centering
 \includegraphics[width=0.6\textwidth,keepaspectratio=true]{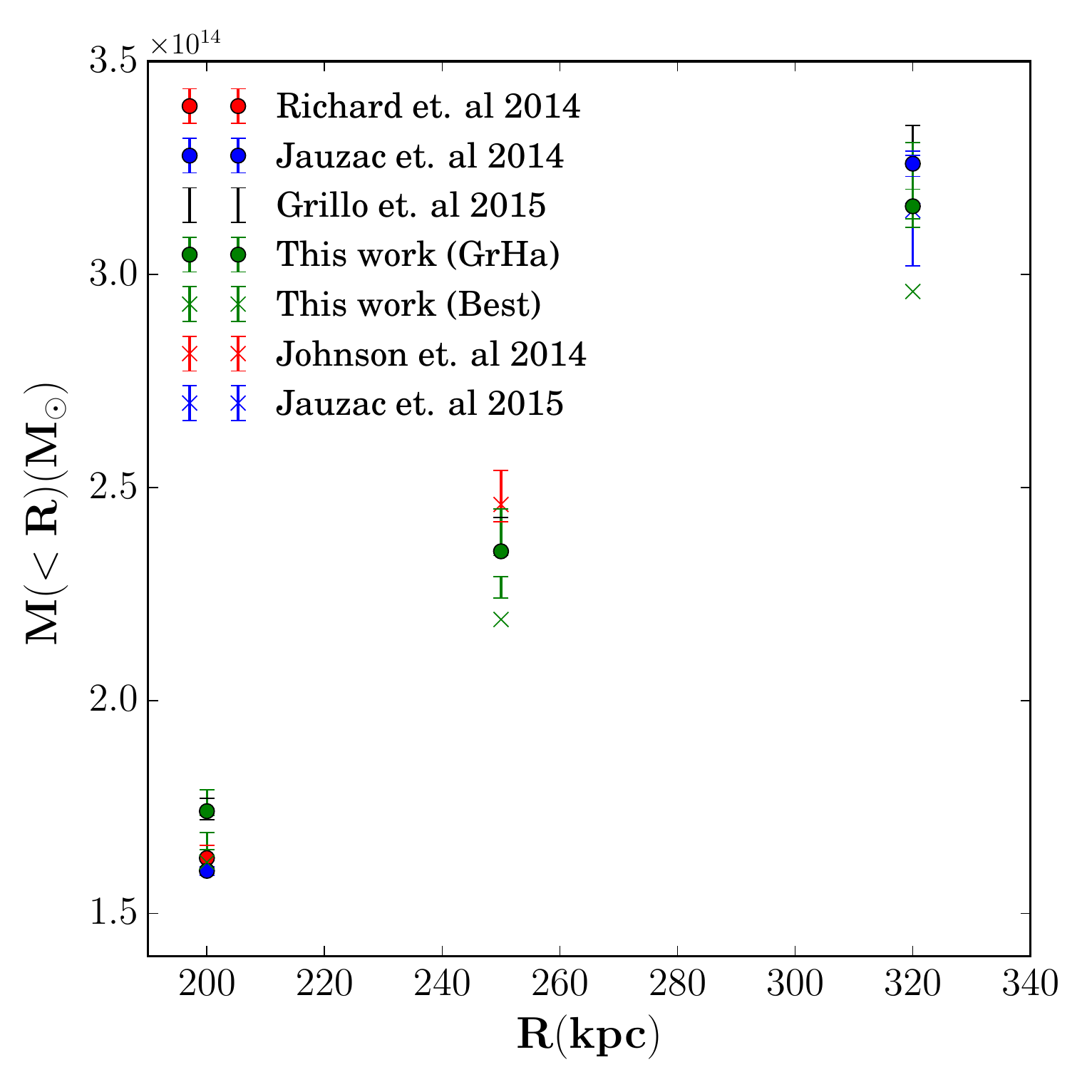}
 \caption{Graphical comparison between the mass calculation from the litterature and this work. Values without dots represent values without any reported best mass value. Dots represent best mass values.}
 \label{fig:litterature_compare}
\end{figure}

We show a graphical representation of the mass estimates from the literature, in Figure \ref{fig:litterature_compare}. We clearly see here that the the total mass calculations agree, but that our best model mass calculations are below the literature calculations. We notice particularly that our cumulative projected total mass $<400\maths{kpc}$ from our best mass maps, are outside the $68\%$ CL. We have no explanation for this.

\begin{figure}[!htb]
 \centering
 \includegraphics[width=0.49\textwidth,keepaspectratio=true]{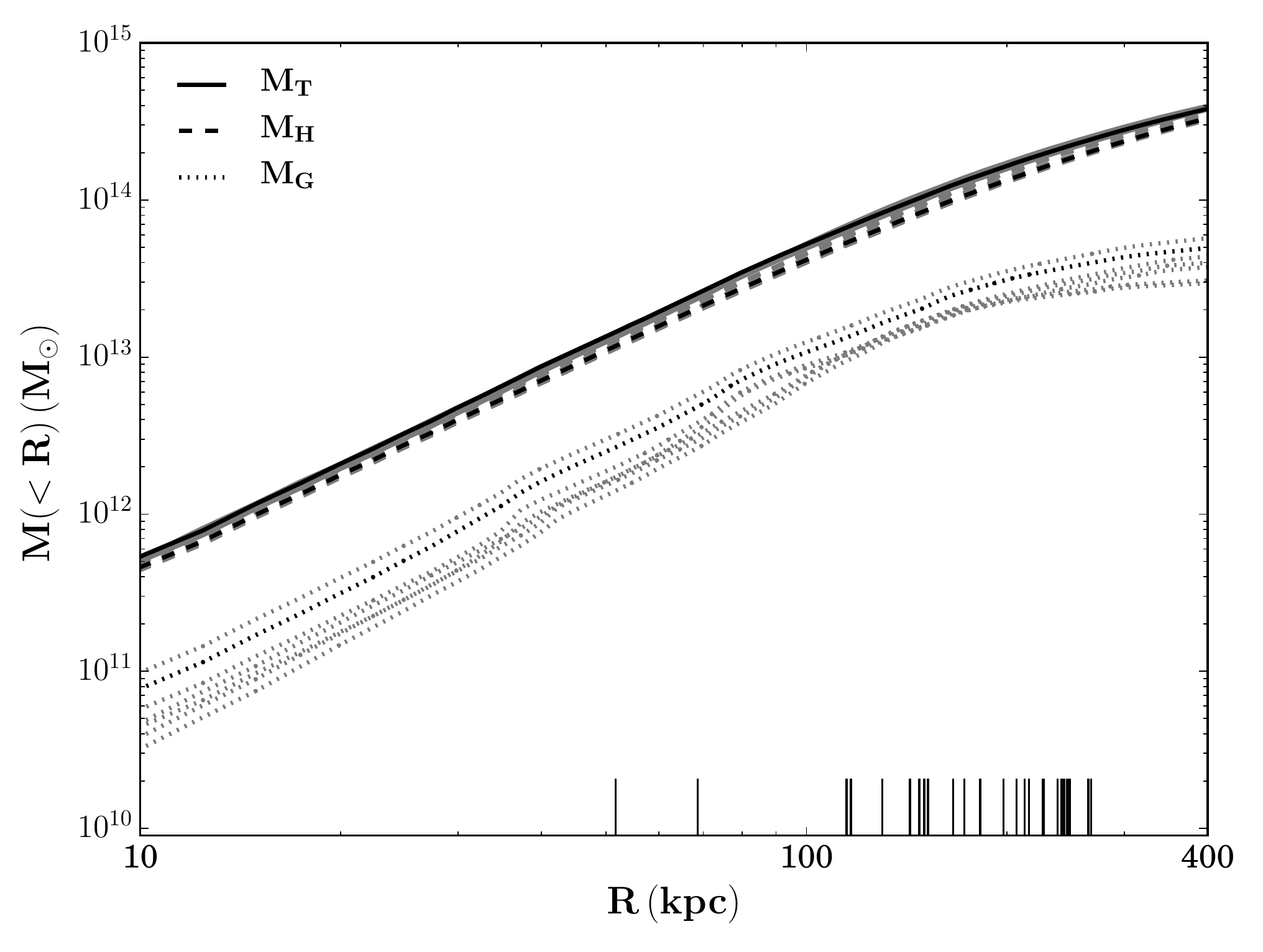}
 \includegraphics[width=0.49\textwidth,keepaspectratio=true]{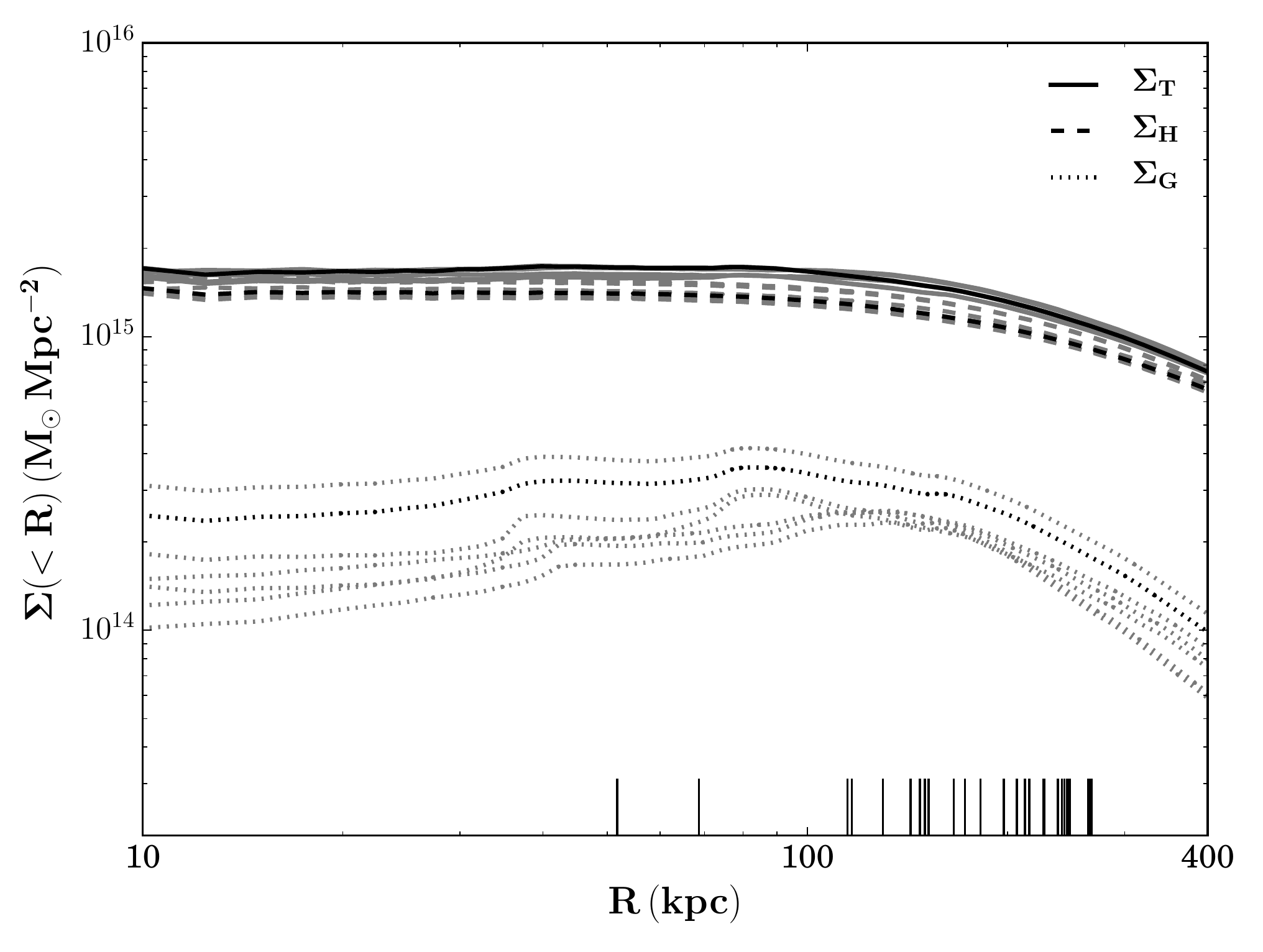}
 \caption{Comparison of the cumulative mass and surface mass-density of all models in Table \ref{table:Results:Image-plane-opt}. The bold lines represents the values from our best model (2PIEMD + 175(+1)dPIE$_c$ $(M_TL^{-1} = v)$).}
 \label{fig:systematic_compare}
\end{figure}

In order to ensure that we have no systematic errors in our method we compare the cumulative mass and surface mass-density for the models we have optimized in the image plane (see Table \ref{table:Results:Image-plane-opt}) in Figure \ref{fig:systematic_compare}. Like \citet{Grillo2015} we find that there are no systematic errors in our model. The total mass estimates and the cluster halo mass estimates are virtually indistinguishable. The major differences are with the cluster member halos, but even these do not shown clear sign of deviation. This further strengthen our reliability in the robustness of our best model.

\section{Cosmological Parameters}
As mentioned in Section \ref{sec:cosmo_models} we initially divided the cosmological parameters into $121$ different values, ranging $0 \leq \Omega_M \leq 1$ and $0 \leq \Omega_{\Lambda} \leq 1$. From these we selected $49$ different models. We will use these models to investigate whether the conditions for comparing mass estimates from strong lensing with mass estimates from X-ray and dynamics, over a range of different cosmologies, is viable.

We have conducted the same statistical calculation as mentioned previously (see Section \ref{sect:mass_profiles}) in order to get the surface mass-density and cumulative total mass for all $49$ cosmological models. We use these results to compare the total mass and surface mass-density to look for any trends. For all calculations regarding probabilities and parameter estimates, we have used the best total mass at $R = 400\maths{kpc}$ unless otherwise stated. Because the size of the cluster change with cosmology we have increased the area of the mass maps to $300\times300\maths{arcsec}$ and increased the resolution of the maps to $600\maths{pixels}$ in order to get a spatial resolution of $0.5\maths{arcsec/pixel}$.

\begin{figure}[h!tb]
 \centering
 \includegraphics[width=0.49\textwidth,keepaspectratio=true]{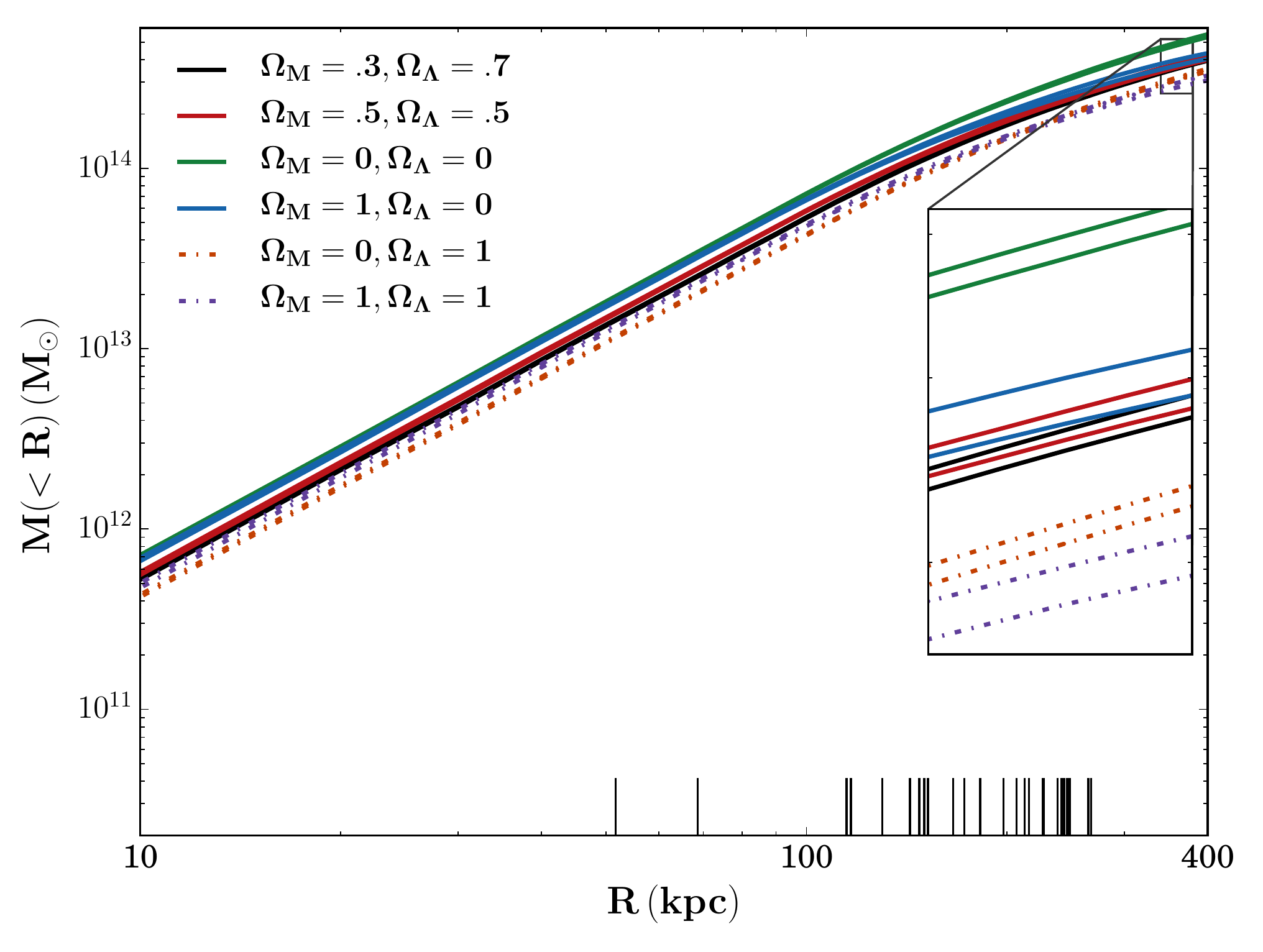} 
 \includegraphics[width=0.49\textwidth,keepaspectratio=true]{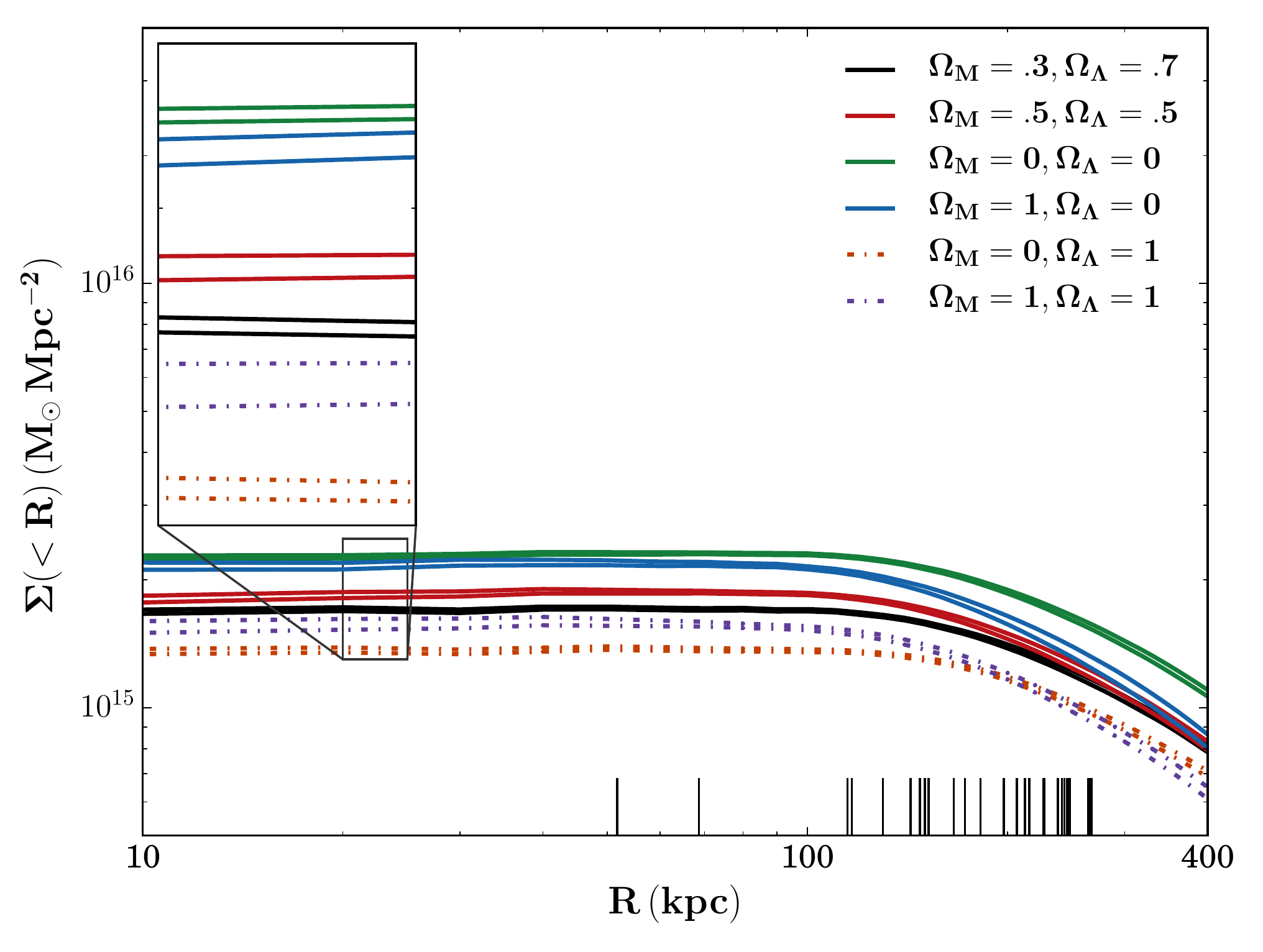} 
 \caption{Selected cumulative mass and surface mass-density plots from our $49$ cosmological models. The figure show what we would expect in relation to the total mass for different cosmological model.}
 \label{fig:cosmo_surfmass}
\end{figure}

In Figure \ref{fig:cosmo_surfmass} we have chosen to plot the most extreme results from our model ensemble. From theoretical predictions we expect to find the largest mass with a $\Omega_M = 0 ; \Omega_{\Lambda} = 0$ cosmology and the lowest mass with a $\Omega_M = 1 ; \Omega_{\Lambda} = 1$ cosmology. From the plots in Figure \ref{fig:cosmo_surfmass} we see exactly that trend. 

We compare the masses directly at $R > 400\maths{kpc}$ to get a more quantitative look. For the $\Omega_M = 0 ; \Omega_{\Lambda} = 0$ model we find $7.29\times10^{14}\maths{\MSun} \leq M_{T} \leq 7.67\times10^{14}\maths{\MSun}$ and for the $\Omega_M = 1 ; \Omega_{\Lambda} = 1$ model we find $3.07\times10^{14}\maths{\MSun} \leq M_{T} \leq 3.27\times10^{14}\maths{\MSun}$ which is clearly less. For the reference model ($\Omega_M = 0.3 ; \Omega_{\Lambda} = 0.7$) we find $3.95\times10^{14}\maths{\MSun} \leq M_{T} \leq 4.09\times10^{14}\maths{\MSun}$ and for the $\Omega_M = 0.5 ; \Omega_{\Lambda} = 0.5$ model we find $3.99\times10^{14}\maths{\MSun} \leq M_{T} \leq 4.19\times10^{14}\maths{\MSun}$ which both are between the two previously mentioned models and the $\Omega_M = 0.5 ; \Omega_{\Lambda} = 0.5$ has slightly more mass than the reference model, which is what we expect.

\begin{figure}[!htb]
 \centering
 \includegraphics[width=0.49\textwidth,keepaspectratio=true]{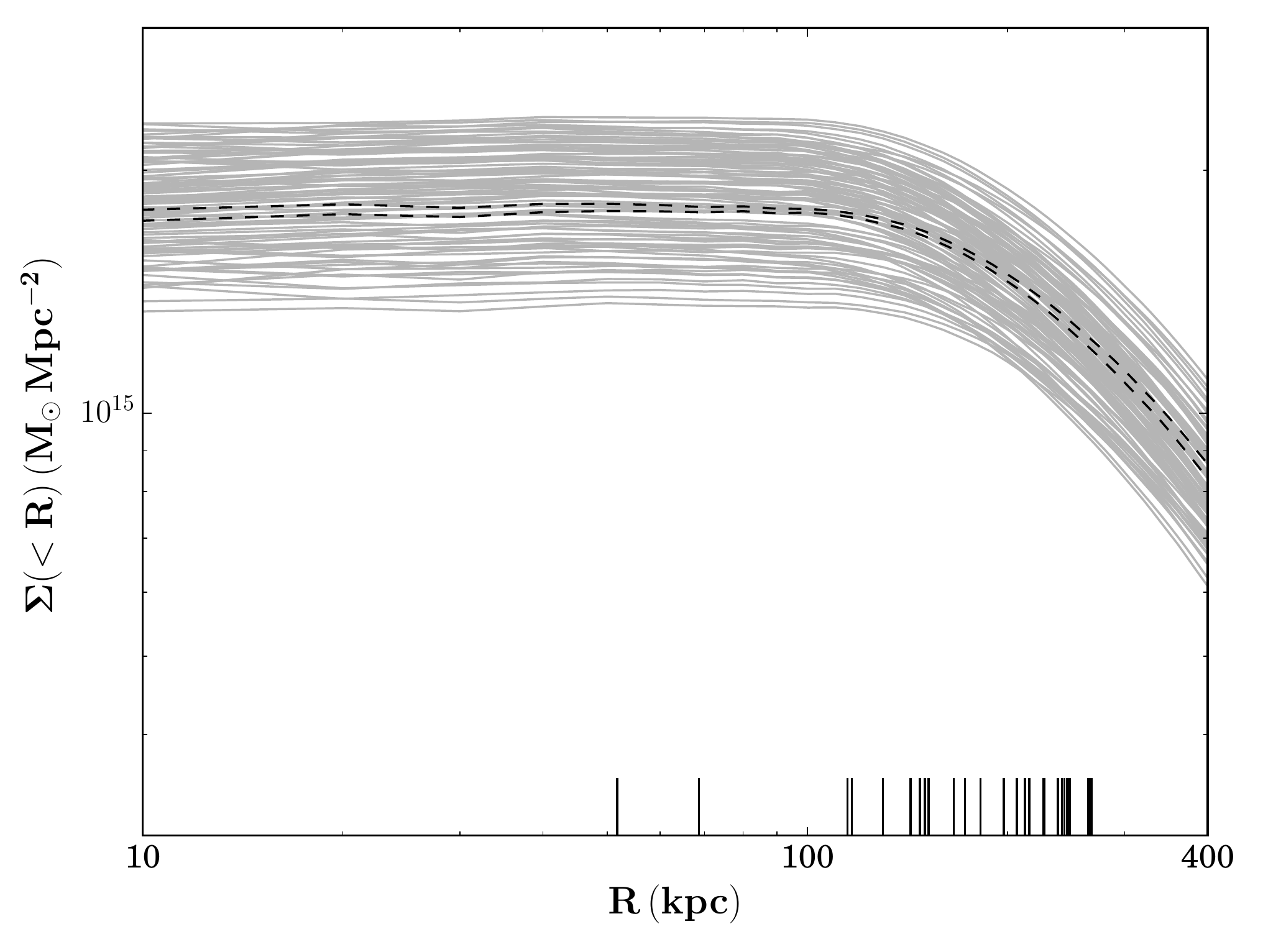} 
 \includegraphics[width=0.49\textwidth,keepaspectratio=true]{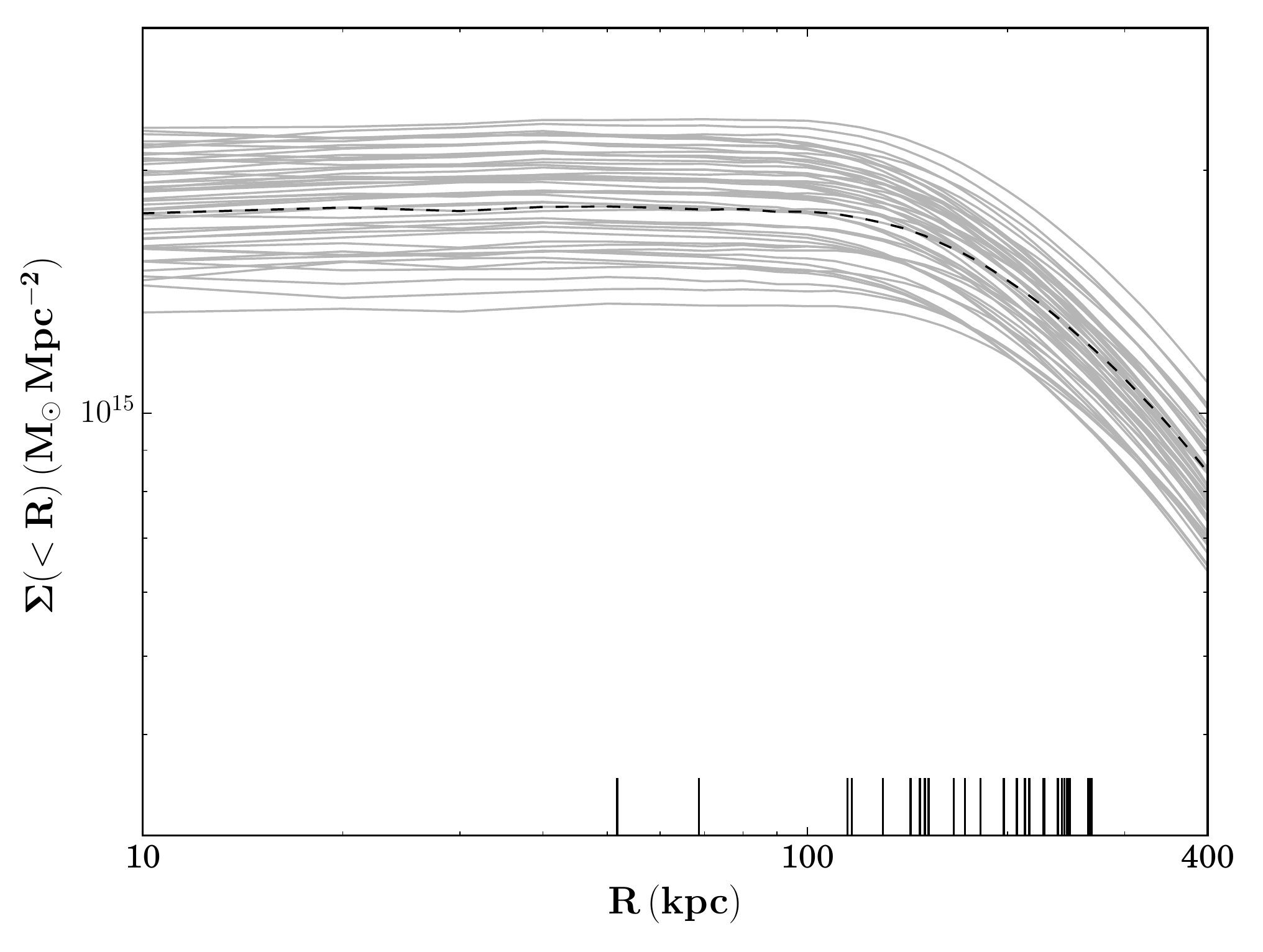} 
 \caption{Surface mass-density plots for all cosmological models. Left: 1$\sigma$ CL surface mass-density. Right: Best fit mass models. GrHa model are marked in thick dashed line.}
 \label{fig:all_cosmo_surfmass}
\end{figure}

In Figure \ref{fig:all_cosmo_surfmass} we show the surface mass-density from all the cosmological models. From these plots we find no obvious deviations, since all plots follow the same trend. This means that we can safely assume that all our models are representative of the true mass at a given set of cosmological parameters.  We also note, which is vital to our investigation, that we have a significant difference in both cumulative mass and surface mass-density which we need in order to compare with estimates from dynamics and X-Ray. Had the cumulative mass and surface mass-density been almost the same we would not be able to distinguish one model from another.

\begin{figure}[!htb]
 \centering
 \includegraphics[width=0.49\textwidth,keepaspectratio=true]{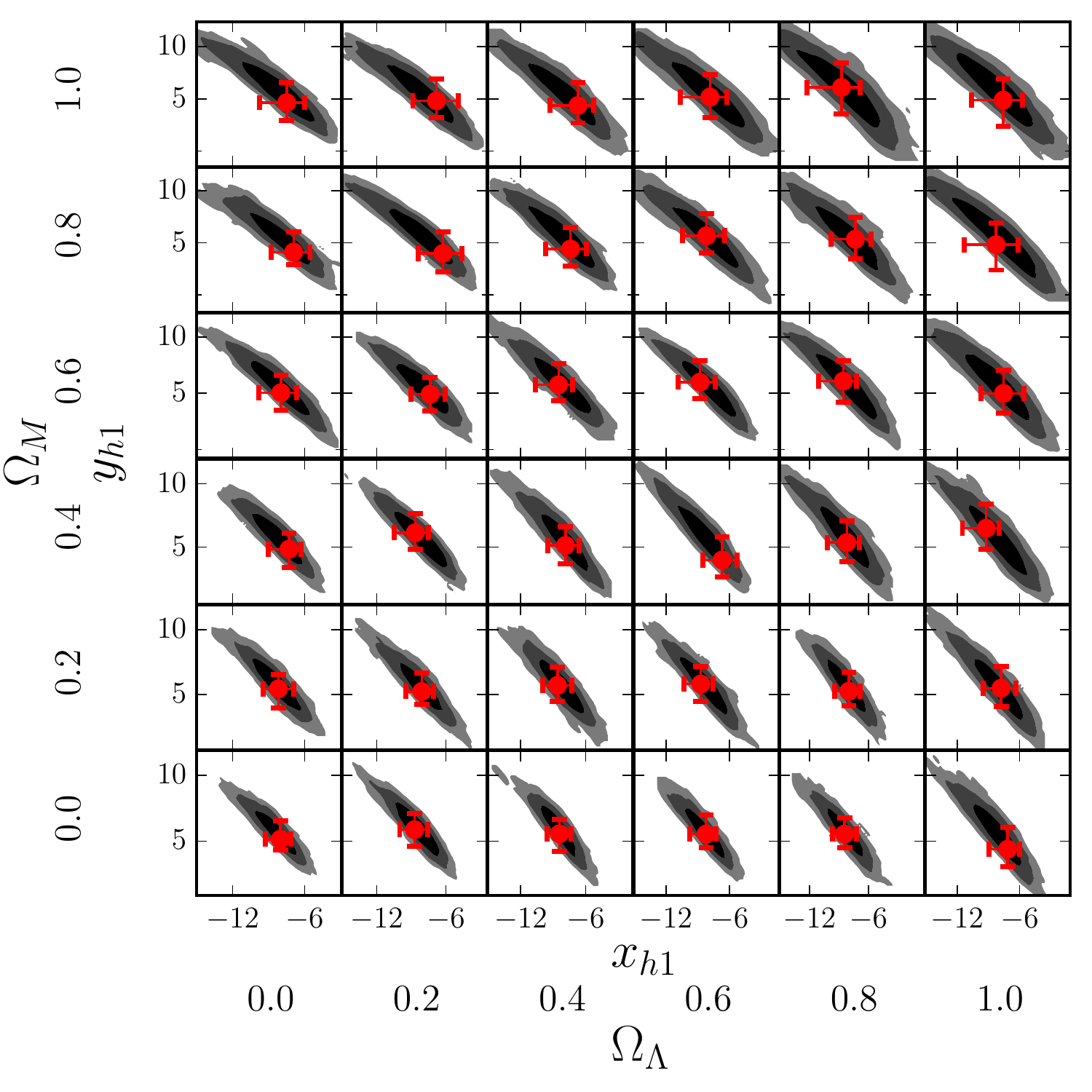}
 \includegraphics[width=0.49\textwidth,keepaspectratio=true]{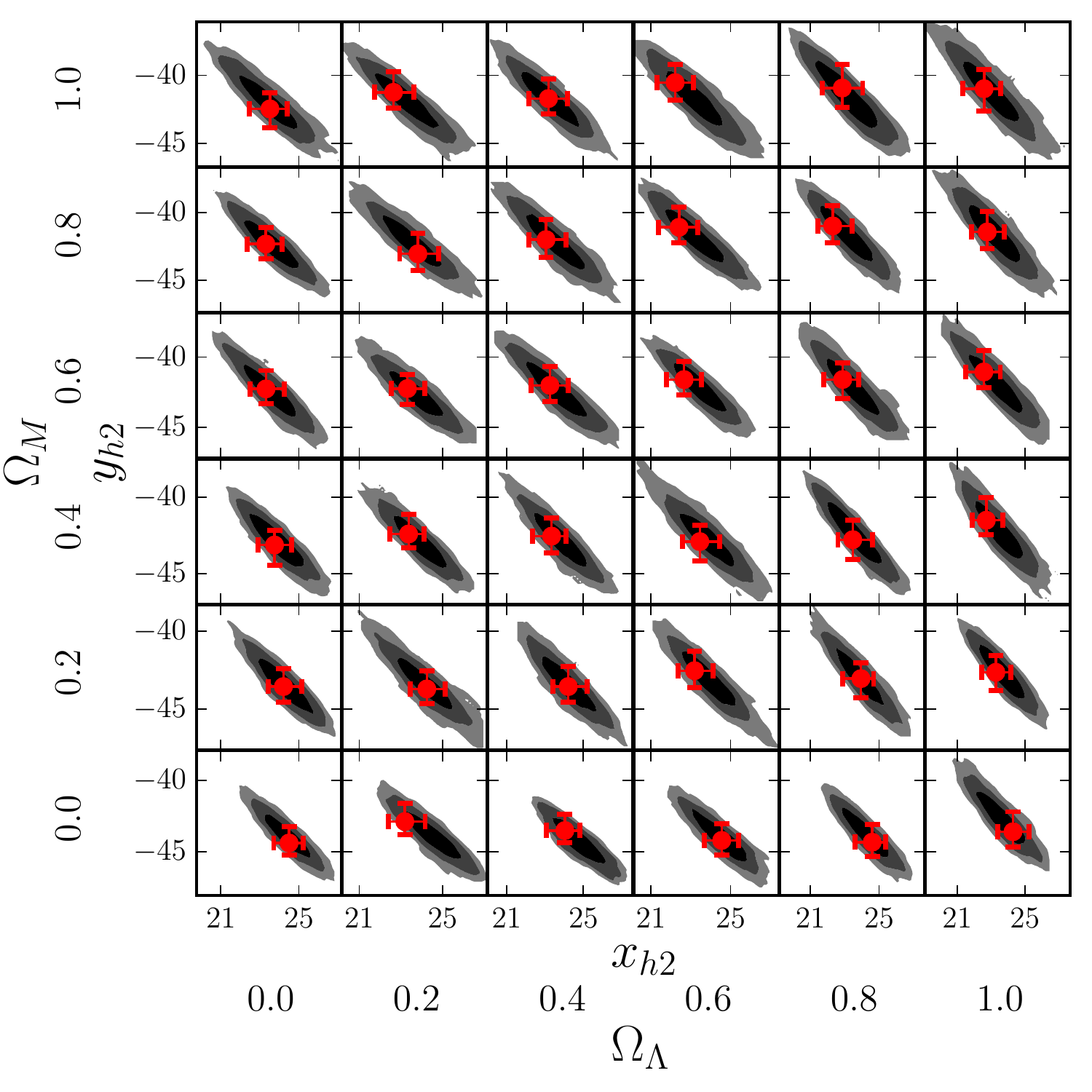}
 \caption{Comparison of uncertainties and correlation of the position parameters for $36$ of our $49$ cosmological models. The uncertainty estimations are given in 1, 2 and 3$\sigma$ regions from inner to outer, respectively. The red mark denotes the best parameter values with $1\sigma$ uncertainty. }
 \label{fig:cosmo_degen_pos}
\end{figure}

We also want to check the degeneracy of our cosmological models with respect to certain crucial parameters. From the 49 models we select 36 models to plot, in order to get a spacing of $0.2$ in the cosmological parameters. We want to see whether the degeneracy in position of the cluster halos ($x_{h1}$ vs. $y_{h1}$ and $x_{h2}$ vs. $y_{h2}$) differ from the degeneracy found in the reference model. As seen in Figure \ref{fig:cosmo_degen_pos}, we find a clear anti-correlated degeneracy between the positions of the two halos, which is also what we expect. 

\begin{figure}[!htb]
 \centering
 \includegraphics[width=0.49\textwidth,keepaspectratio=true]{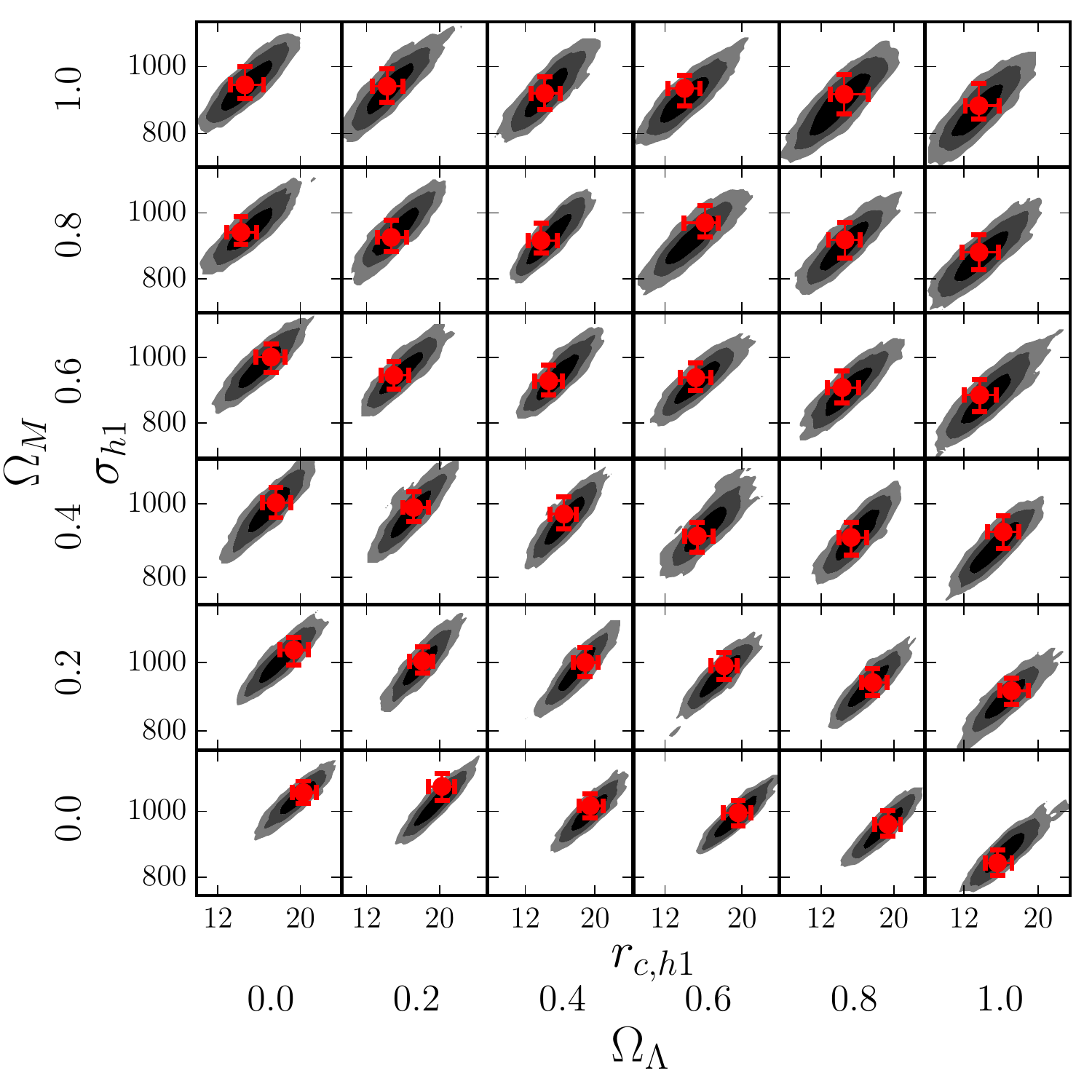}
 \includegraphics[width=0.49\textwidth,keepaspectratio=true]{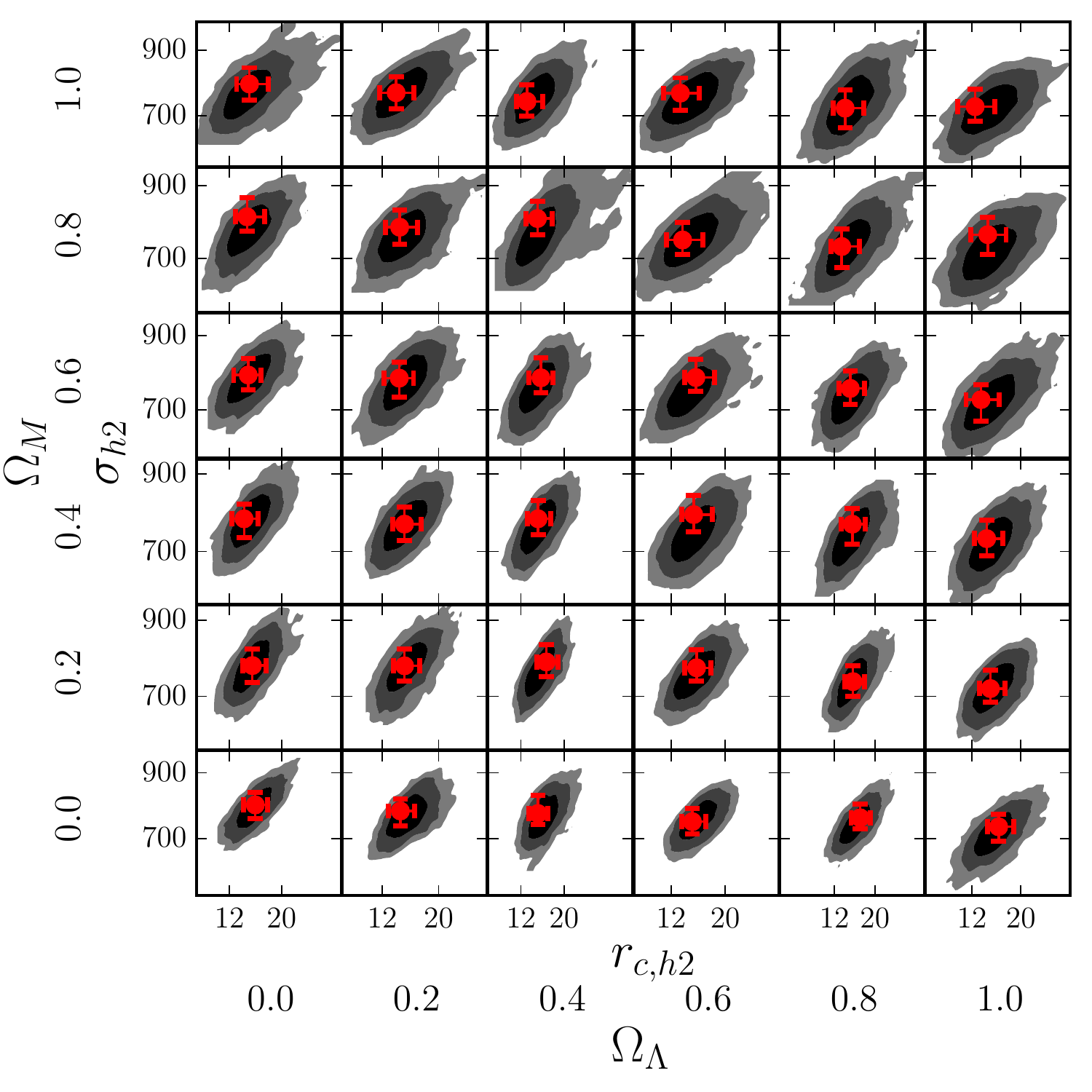}
 \caption{Same as for \ref{fig:cosmo_degen_pos}, only for the velocity dispersion and core radii for the two cluster halos.}
 \label{fig:cosmo_degen_coresigma}
\end{figure}

We expect to see a degeneracy when we compare the velocity dispersion with the the core radius of the two cluster halos. In Figure \ref{fig:cosmo_degen_coresigma} we plot the $r_{c,h1}$ vs. $\sigma_{h1}$ and $r_{c,h2}$ vs. $\sigma_{h2}$. Here we see a clear parameter degeneracy in the correlation comparison between the two cluster halos, although the degeneracy is much clearer for some models, than others. We see a much tighter correlation for halo 1 than for halo 2. We also see that the parameters are correlated, not anti-correlated. This means that in order to preserve the same mass in the cluster, when we increase the core radius, we also need to increase the velocity dispersion. We also expect to see the best value, for the two parameters, differ significantly (indicated by a red dot). This is more evident if we compare the values directly, but since the values are correlated it is enough to compare just one value. For the $\Omega_M = 0 ; \Omega_{\Lambda} = 0$ model we have $\sigma_{h1} =  802^{+109}_{-102}\maths{km\,s^{-1}}$ and for the $\Omega_M = 1 ; \Omega_{\Lambda} = 1$ model we have $\sigma_{h1} = 711^{+132}_{-126}\maths{km\,s^{-1}}$ which clearly differs.

\begin{figure}[!htb]
 \centering
 \includegraphics[width=0.49\textwidth,keepaspectratio=true]{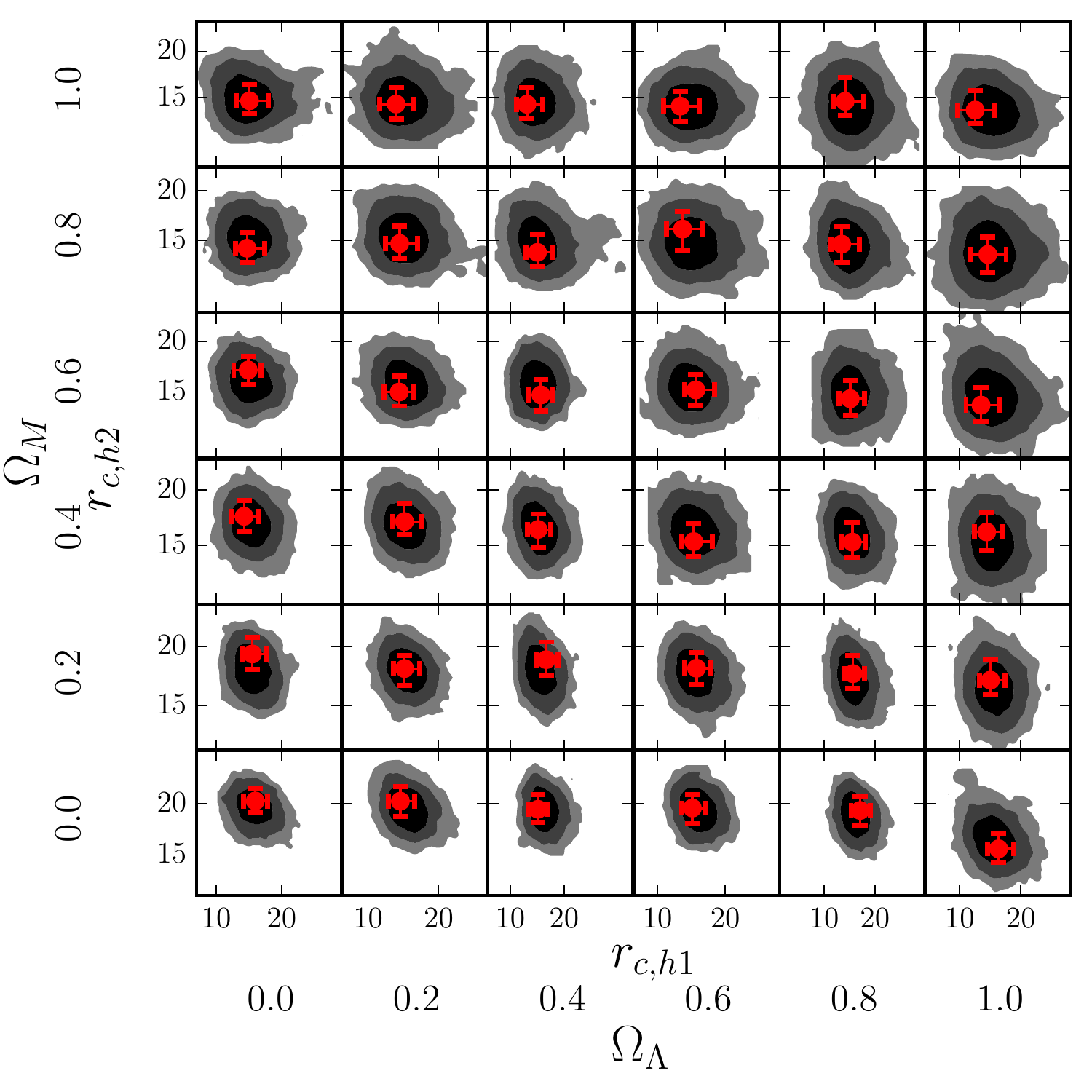}
 \includegraphics[width=0.49\textwidth,keepaspectratio=true]{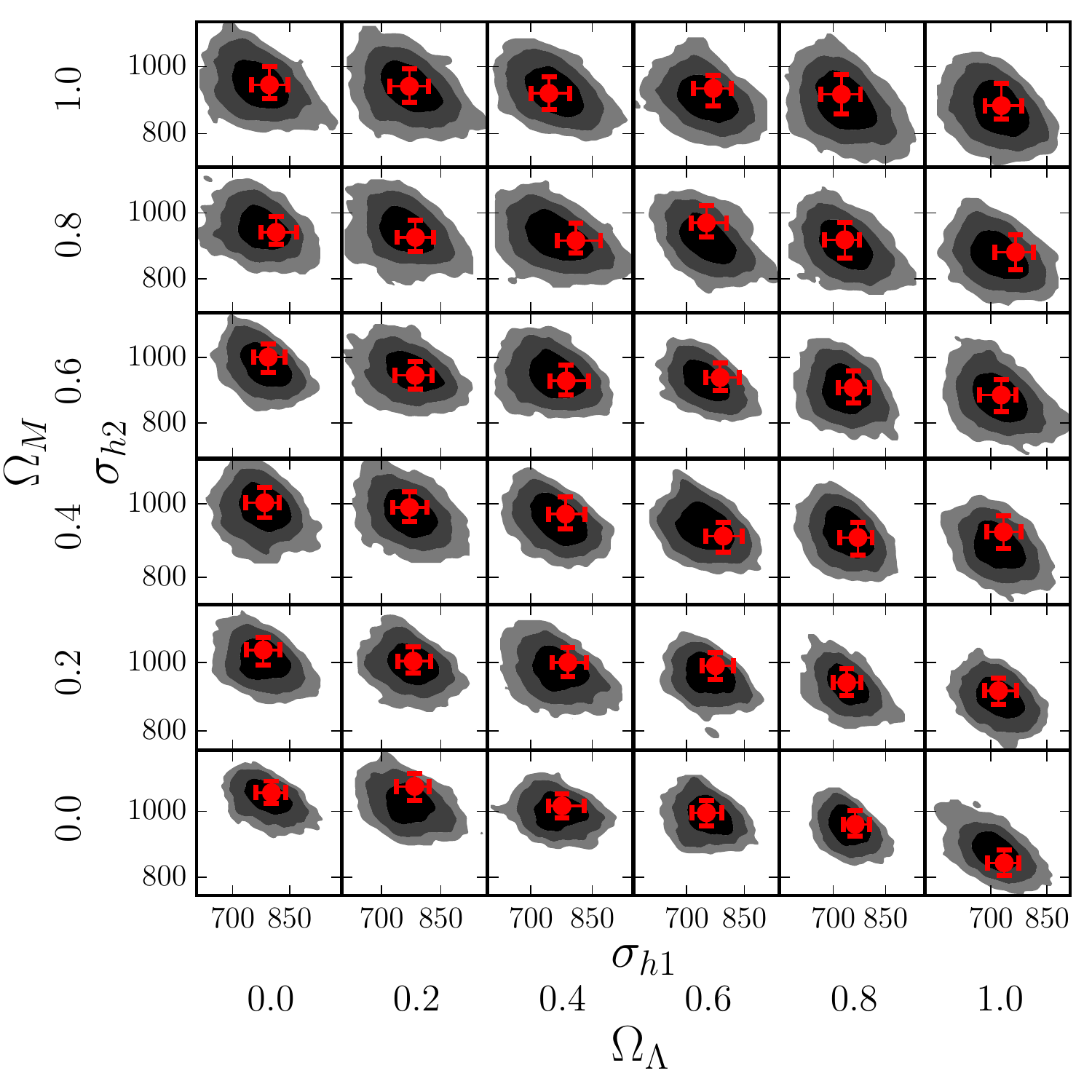}
 \caption{Same as for \ref{fig:cosmo_degen_pos}, only for the velocity dispersion and core radii for the two cluster halos individually.}
 \label{fig:cosmo_degen_corecore}
\end{figure}

We expect to see the same tendency when we cross-compare the core radii and velocity dispersions for the two cluster halos ($r_{c,h1}$ vs. $r_{c,h2}$ and $\sigma_{h1}$ vs. $\sigma_{h2}$), which we show in Figure \ref{fig:cosmo_degen_corecore}. Interestingly we see no clear correlation between the core radii for the two halos, but rather relatively clear anti-correlation between the velocity dispersions. This tells us that the model seem to favour balancing the velocity dispersions rather than the core radii. The velocity dispersion is one of the parameters that we can compare with observations, where it has been used in the determination of cluster members \cite{Rosati2014, Grillo2015} and the calculation of dynamics \cite{Mamon2013, Balestra2016} for \macs. By comparing the core radius of the $\Omega_M = 0 ; \Omega_{\Lambda} = 0$ model $(r_{c,h1} = 16.0^{+5.2}_{-4.6}\maths{arcsec} \, ; \, r_{c,h2} = 20.3^{+3.0}_{-3.1}\maths{arcsec})$ and $\Omega_M = 1 ; \Omega_{\Lambda} = 1$ model $(r_{c,h1} = 12.5^{+8.5}_{-6.5}\maths{arcsec} \, ; \, r_{c,h2} = 13.6^{+5.4}_{-3.7}\maths{arcsec})$ directly, we see the exact same tendency. The former produce markedly larger parameter values than the latter. The same applies to the velocity dispersion. Here we have $\sigma_{h1} =  802^{+109}_{-102}\maths{km\,s^{-1}}\, ; \, \sigma_{h2} = 1057^{+85}_{-94}\maths{km\,s^{-1}}$ for the $\Omega_M = 0 ; \Omega_{\Lambda} = 0$ model and $\sigma_{h1} = 728^{+132}_{-126}\maths{km\,s^{-1}}\, ; \, \sigma_{h2} = 883^{+156}_{-101}\maths{km\,s^{-1}}$ for the $\Omega_M = 1 ; \Omega_{\Lambda} = 1$, which again show a significant different in values. 

\begin{figure}[!htb]
 \centering
 \includegraphics[width=0.49\textwidth,keepaspectratio=true]{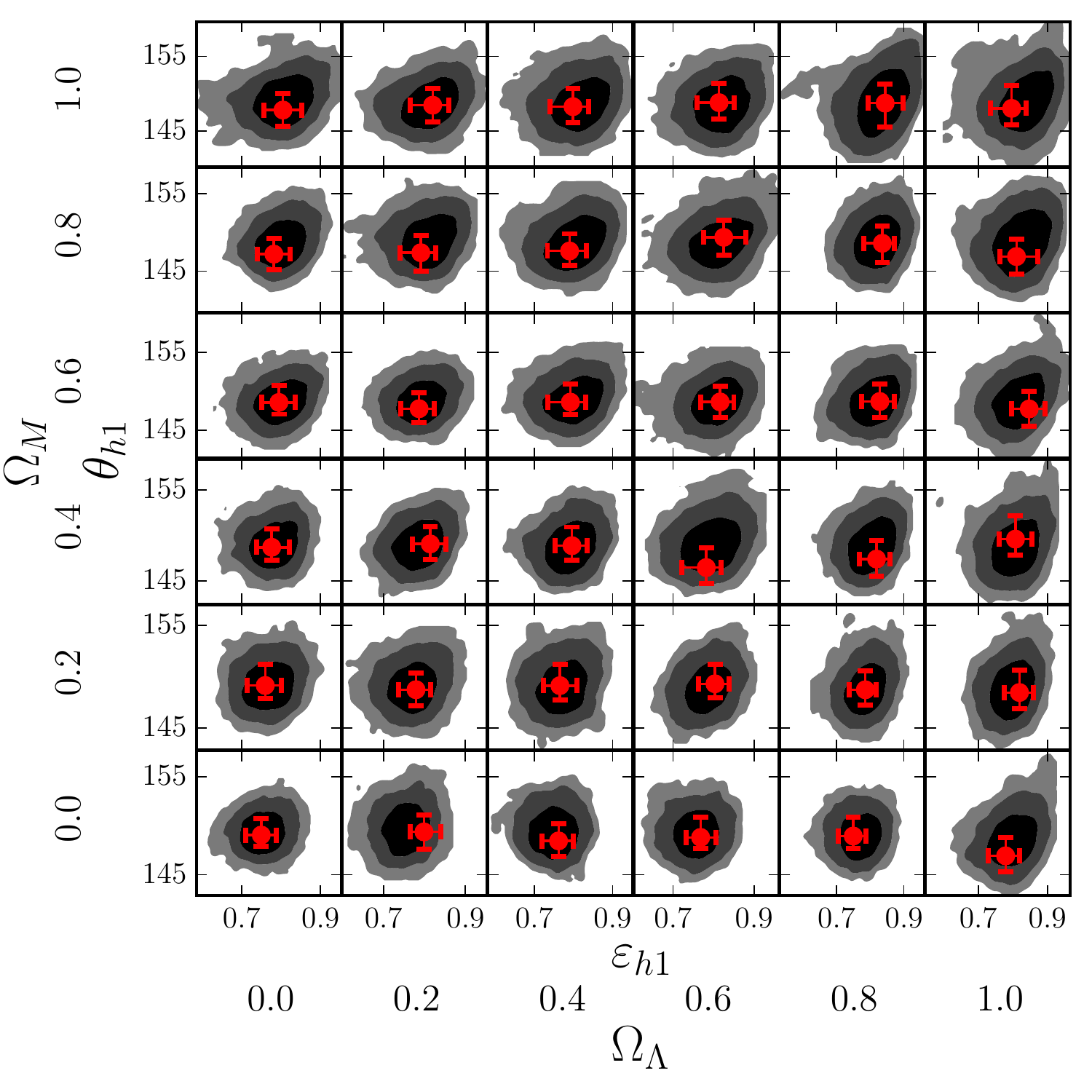}
 \includegraphics[width=0.49\textwidth,keepaspectratio=true]{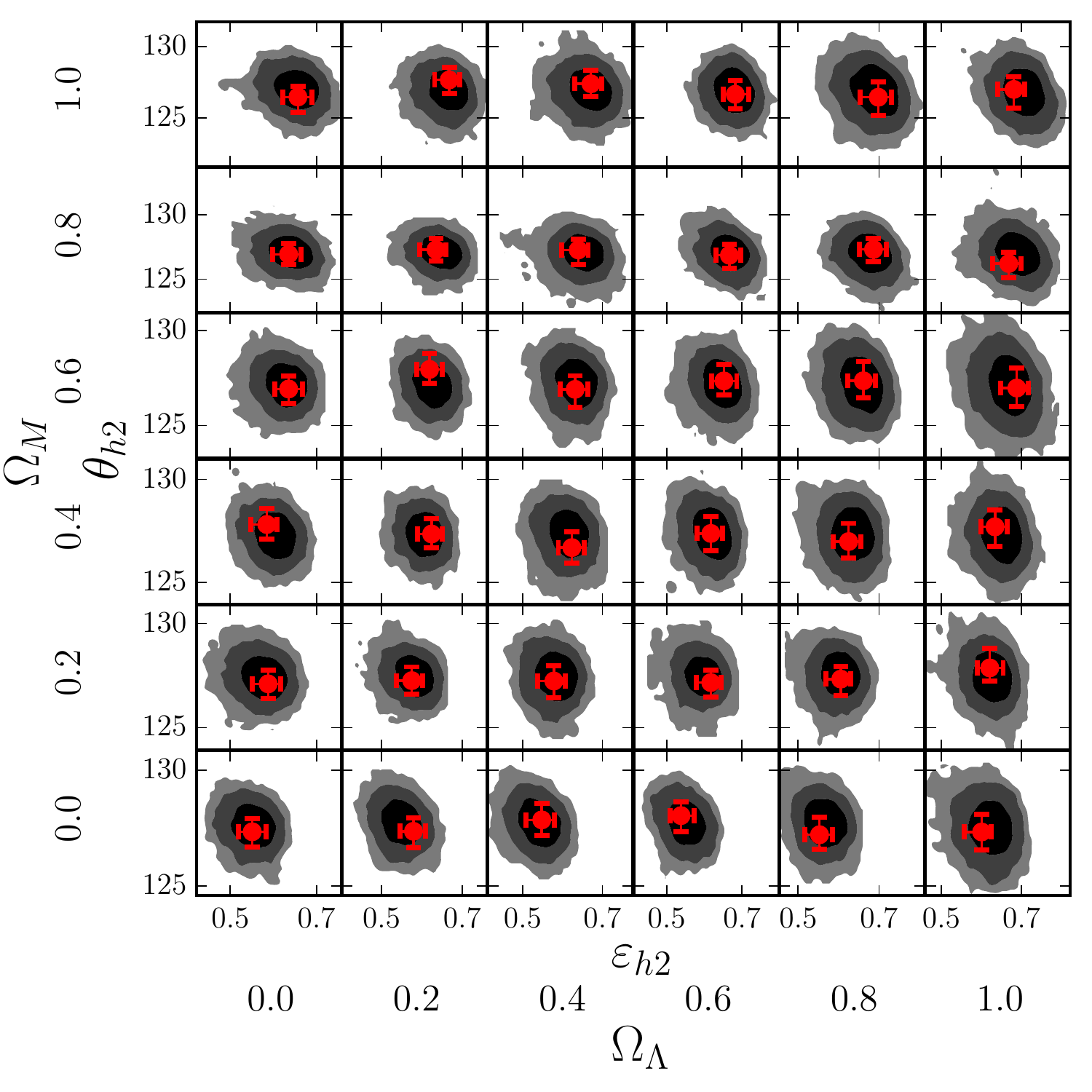}
 \caption{Same as for \ref{fig:cosmo_degen_pos}, only for the ellipticity and position angle for the two cluster halos.}
 \label{fig:cosmo_degen_thetaemass}
\end{figure}

By looking at the ellipticity and position angle of the two clusters individually (see Figure \ref{fig:cosmo_degen_thetaemass}), we see that all models seems to agree fairly on the values of parameters for the two halos within $68\%$ CL. Likewise, there are no clear correlation between the parameters. This is also confirmed by our MCMC analysis of the GrHa model (see Figure \ref{fig:degens_grha}) where we see a slight correlation between $\theta_{h1}$ and $\varepsilon_{h1}$ and a slight anti-correlation between $\theta_{h2}$ and $\varepsilon_{h2}$, but no clear signs of degeneracy. We do note, however, that some models show some degree of degeneracy.

\begin{figure}[!htb]
 \centering
 \includegraphics[width=0.49\textwidth,keepaspectratio=true]{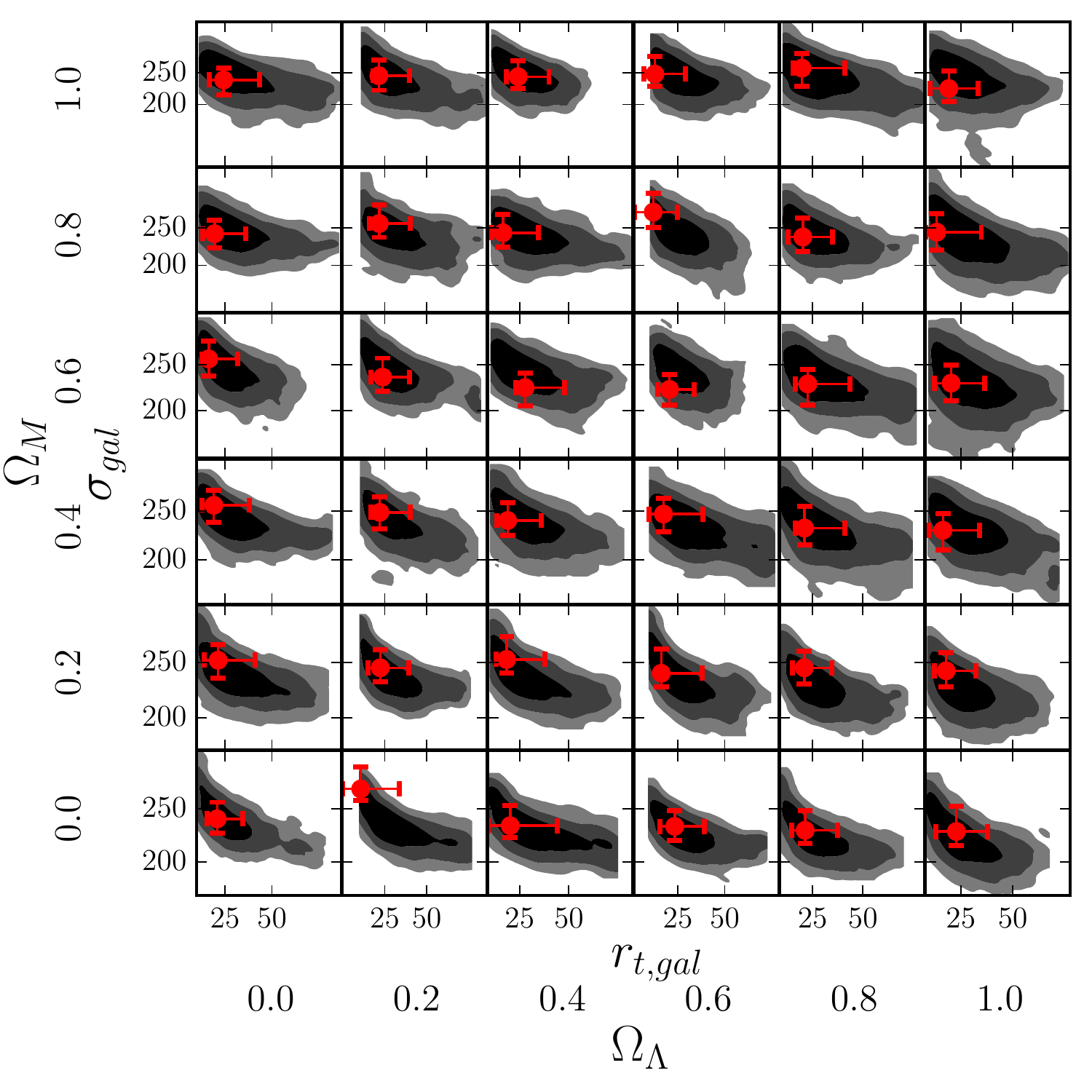}
 \caption{The same as Figure \ref{fig:cosmo_degen_pos}, only for the velocity dispersion $\sigma_{gal}$ and truncation radius $r_{t,gal}$ of the cluster members.}
 \label{fig:cosmo_degen_rtgalsigmagal}
\end{figure}

The velocity dispersion of the individual cluster members can be measured directly via spectroscopy and hence do not depend on cosmology. Therefore we would also like to see how the velocity dispersions of the cluster members $\sigma_{gal}$ depend on the truncation radius $r_{t,gal}$. As can be seen in Figure \ref{fig:cosmo_degen_rtgalsigmagal} while there are no clear signs of degeneracy, there are some differences in the best results from the different models. Some are even at the limit of the confidence region. In general we can say that all the $36$ models we have tested here agree within the limits, on the truncation radius and, although there are stronger fluctuations, also on the velocity dispersion. We do note some indications of anti-correlation between the two parameters, but due to the uneven shape of the confidence regions, we deem it to be insignificant. This means that we, in general, can trust the best parameter values of the velocity dispersion, whereas had there been a clear anti-correlation, we could just as well have had a much lower velocity dispersion followed by a larger truncation radius and still get the same results. Since we cannot measure the truncation radius directly, we have to rely solely on the velocity dispersion. We do note that some of the models have had difficulty constraining the parameter value of the truncation radius, which indicates that it should have been smaller. This is clearly evident in three of the models ($\Omega_{M} = 0.0;\Omega_{\Lambda}=0.2$, $\Omega_{M} = 0.8;\Omega_{\Lambda} =0.6$ and $\Omega_{M} = 1.0;\Omega_{\Lambda} = 0.6$).

In all our calculations so far we have found the cumulative mass within $400\maths{kpc}$ but this method poses a trivial problem in wanting to compare with other results. Since the radius of the selection circle depends on cosmology when measured in physical units, there are one more parameter that might differ, unintentionally. We have chosen to find the cumulative mass within $75\maths{arcsec}$ which represents an fixed aperture in arcseconds equivalent of $400\maths{kpc}$ in the reference cosmology $(\Omega_M = 0.3;\Omega_{\Lambda}=0.7)$. This will make any future comparison more straightforward.

There are some considerations that need to be addressed with this method. In the previous calculations we would always calculate the cumulative mass and surface mass-density of the same parts of the cluster, since the radius scales with cosmology. Using a fixed aperture in arcsec we can no longer expect to have the same parts of the cluster included. This means that we do not actually measure the content of the cluster, but the content within an aperture. 

This is an important consideration to bear in mind since we need a substantial difference in mass-estimates from the different cosmological models. If the masses are almost equivalent, we have no basis to distinguish one cosmological model from the other. 
\begin{figure}[h!p]
 \centering
 \includegraphics[width=0.7\textwidth,keepaspectratio=true]{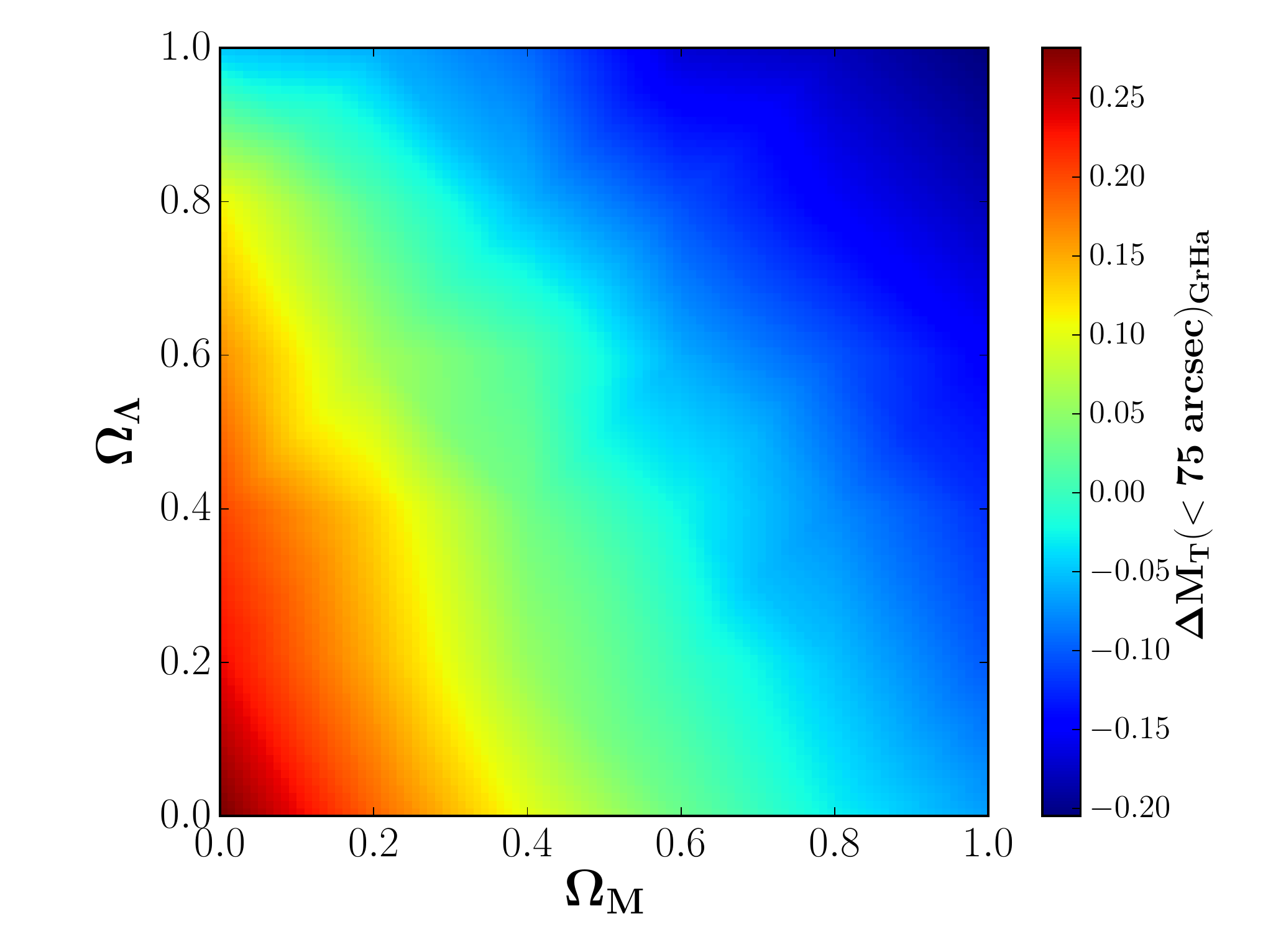} 
 \includegraphics[width=0.7\textwidth,keepaspectratio=true]{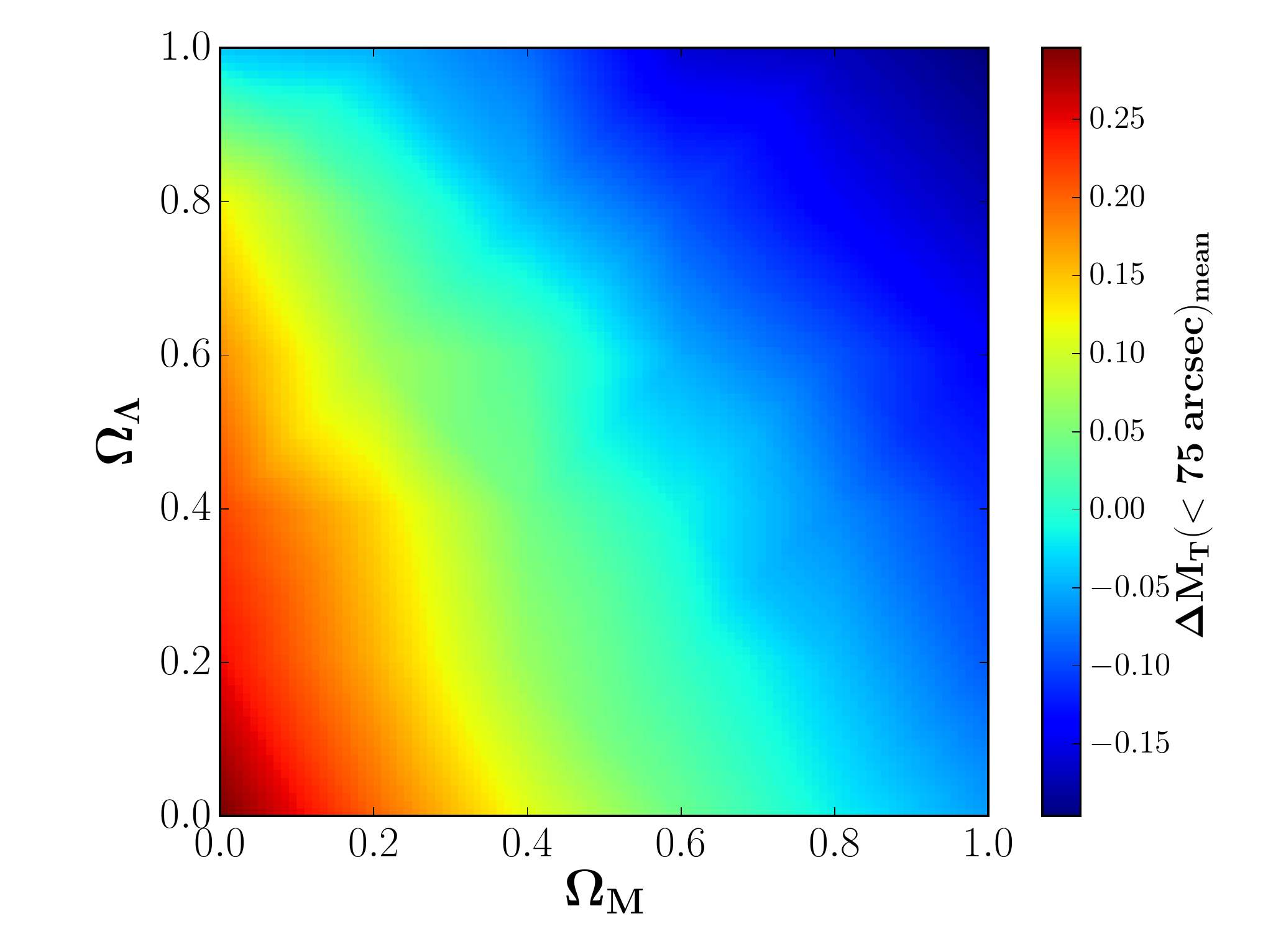} 
 \includegraphics[width=0.7\textwidth,keepaspectratio=true]{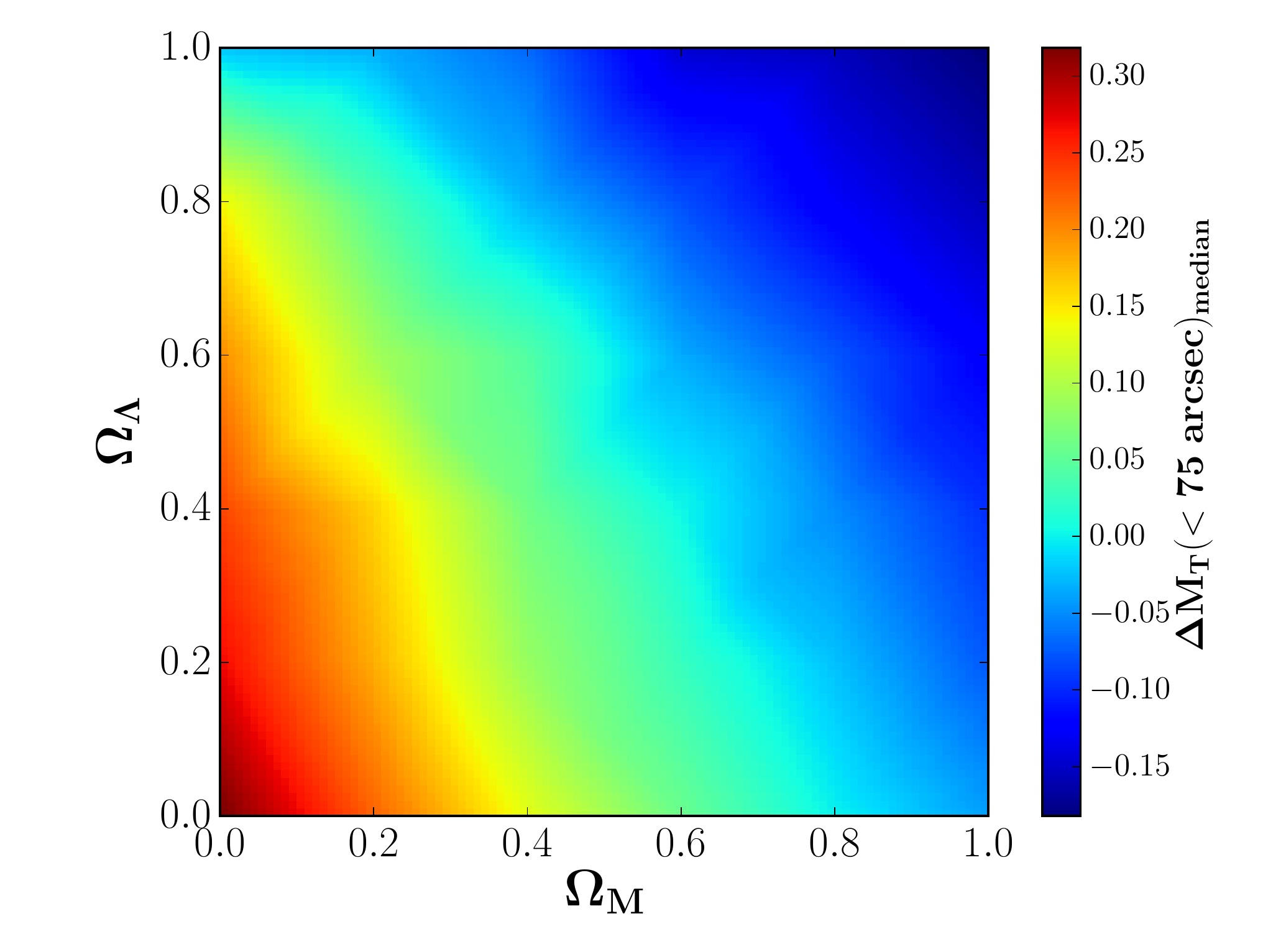} 
 \caption{Surface plots of $\Delta M(<75\maths{arcsec})$ interpolated to a resolution of $0.01$ between the cosmological models and the reference model (top), the mean of the cosmological models (middle) and the median of the cosmological models (bottom).}
 \label{fig:cosmo_surf_plot}
\end{figure}
We have plotted the mass-difference between the cosmological models and the reference model (GrHa), the mean of the cosmological models and the median of the cosmological models, respectively, in Figure \ref{fig:cosmo_surf_plot}. Here we see a difference of $\sim 25\%$ between the reference model and the $\Omega_M=0.0;\Omega_{\Lambda}=0.0$ model and $\sim 20\%$ between the reference model and the $\Omega_M=1.0;\Omega_{\Lambda}=1.0$ model which is also present in the mean and median calculations. We find a total difference of $\sim 49\%$ when comparing with the reference model, $\sim 49\%$ when comparing with the mean and $\sim 50\%$, when comparing with the median. This indicates that going from an aperture in real units $(\maths{kpc})$ to a fixed aperture size do not decrease the cumulative projected total mass significantly. 

In summary we conclude that we have found the prerequisites in this work, to combine the mass estimates from strong lensing with estimates from X-ray and dynamics, in order to estimate the cosmological parameters. But due to lack of time, this comparison has not been developed further.

%% file: discussion.tex
\chapter{Summary}\label{chap:discussion}
In this thesis we have presented our results from an ensemble of different strong lensing models. In the standard cosmological model, we have found that our results, in general, are in agreement with the current results from the literature.

We have improved on previous models in two different ways. The best results in our strong lensing analysis have been found by using elliptical cluster member mass profiles, compared with spherical ones. We do, however, find that the differences are not large enough to justify the significantly increased computational time needed. We have also found that the data are best constrained optimizing the slopes of the cluster member scaling relations, resulting in $\zeta_{r_{s,gal}} = 0.39\, ; \, \zeta_{\vartheta_{E,gal}} = 0.85$ and $\zeta_{r_{s,gal}} = 5.11\, ; \, \zeta_{\sigma_{gal}} = 1.69$ in \glee and \lenstool coordinates, respectively. This indicates that we can exclude the fundamental plane and possibly, also a constant mass-to-light ratio. We do note that adding more multiple images as constraints may very well change these values, even towards fixed parameters. Nevertheless, we conclude that the differences are significant. 

From our results, we can confirm that \macs is a bimodal cluster with a flat inner core. We also confirm that the best strong lensing model includes a significant number of the cluster members and (at least) two cluster-scale halos.

Our primary findings show that we have the necessary prerequisites to combine strong lensing mass estimates with those from X-ray and dynamics, to estimate the values of the cosmological parameters. We have shown that there is a significantly large difference in the estimated projected total mass of the cluster, in the different cosmological models, to continue with this work. We do note one caveat that needs to be addressed. \citet{Caminha2016} recently presented a model, with a significantly increase in the number of multiple images and cluster members, which gave a more refined result. In future work we will have to show that this better constrained model does not reduce the difference in projected total cluster mass significantly between the different cosmological models. Future work will provide us with the answers to this question.

From our results we propose three future studies to be conducted on \macs:
\begin{enumerate}
 \item The separation between dark matter and gas will give a significantly better constrained model, due to the collisional nature of hot gas. For this, we need either a completely new mass density profile, specifically made for the hot gas component, or we need to test an already existing profile to make sure that it can properly constrain the physical properties of hot gas. This accounts, at least, the physical extent of the hot gas halo (scale radius alone, truncation radius alone or both core and truncation radii) and the velocity dispersion of the gas particles. 
 \item A detailed comparison between the results from this work and future results from the \citet{Caminha2016} model, specifically to determine whether our proposed method is  of estimating the values of the cosmological parameters is viable or not. Also, optimizing the \citet{Caminha2016} model with the scaling relation slopes as free parameters.
 \item The implementation of multi-plane optimization in lensing codes, in order to account for significant foreground galaxies, influencing the positions of the multiply lensed images. As of now, the only viable method has been a \emph{hack} method where one models the foreground galaxy as it was a part of the cluster itself. This has proven to be necessary to get reasonable mass estimates, but is in no sense physically accurate. We agree with \citet{Caminha2016} that we have reached the limit on how well we can constrain clusters like \macs using single-plane lensing codes.
\end{enumerate}

%% file: Appendix.tex

\chapter{Appendix}\label{app:Appendix}
\section{List of model priors}

\begin{table}[!htb]
 \centering
 \caption{2PIEMD model priors. Values in square brackets are not optimized.}
 \label{table:only-PIEMD}
 \begin{tabular}{cccc}
  \toprule
  Parameter          					& Value  	& low. limit 	& high. limit 	\\
  \midrule
  $x_{h1} (\arcsec)$ 					& $0.0$  	& $-15.0$    	& $+15.0$    	\\
  $y_{h1} (\arcsec)$ 					& $0.0$  	& $-15.0$    	& $+15.0$     	\\
  $\varepsilon_{h1}$ 					& $0.71$ 	& $0.01$     	& $0.90$      	\\
  $\theta_{h1} (\mathrm{deg})$			& $146.5$  	& $0.0$			& $180.0$	  	\\
  $r_{c,h1} (\mathrm{kpc})$			& $68.8$	& $1.0$			& $500.0$		\\
  $r_{s,h1} (\mathrm{kpc})$			& $[1000.0]$  \\
  $\sigma_{h1} (\mathrm{km\,s^{-1}})$	& $809.0$	& $100.0$		& $2000.0$		\\
  $x_{h2} (\arcsec)$ 					& $20.3$  	& $+5.0$    	& $+35.0$    	\\
  $y_{h2} (\arcsec)$ 					& $-35.9$  	& $-50.0$    	& $-20.0$     	\\
  $\varepsilon_{h2}$ 					& $0.50$ 	& $0.01$     	& $0.90$      	\\
  $\theta_{h2} (\mathrm{deg})$			& $128.5$  	& $0.0$			& $180.0$	  	\\
  $r_{c,h2} (\mathrm{kpc})$			& $74.8$	& $1.0$			& $500.0$		\\
  $r_{s,h2} (\mathrm{kpc})$			& $[1000.0]$  \\
  $\sigma_{h2} (\mathrm{km\,s^{-1}})$	& $1019.0$	& $100.0$		& $2000.0$		\\
  \bottomrule
  \end{tabular}
\end{table}

\begin{table}[!htb]
 \centering
 \caption{2NFW model priors. Values in square brackets are not optimized.}
 \label{table:only-NFW}
 \begin{tabular}{cccc}
  \toprule
  Parameter          					& Value  	& low. limit 	& high. limit 	\\
  \midrule
  $x_{h1} (\arcsec)$ 					& $0.0$  	& $-15.0$    	& $+15.0$    	\\
  $y_{h1} (\arcsec)$ 					& $0.0$  	& $-15.0$    	& $+15.0$     	\\
  $\varepsilon_{h1}$ 					& $0.30$ 	& $0.01$     	& $0.40$      	\\
  $\theta_{h1} (\mathrm{deg})$			& $146.5$  	& $0.0$			& $180.0$	  	\\
  $r_{s,h1} (\mathrm{kpc})$			& $100.0$	& $1.0$			& $1000.0$		\\
  $\alpha_{h1}$							& $1.0$		& $0.1$			& $2.0$			\\
  $c_{h1}$								& $5.0$		& $1.0$			& $10.0$		\\
  $x_{h2} (\arcsec)$ 					& $20.3$  	& $+5.0$    	& $+35.0$    	\\
  $y_{h2} (\arcsec)$ 					& $-35.9$  	& $-50.0$    	& $-20.0$     	\\
  $\varepsilon_{h2}$ 					& $0.30$ 	& $0.01$     	& $0.40$      	\\
  $\theta_{h2} (\mathrm{deg})$			& $128.5$  	& $0.0$			& $180.0$	  	\\
  $r_{s,h2} (\mathrm{kpc})$			& $120.0$	& $1.0$			& $1000.0$  	\\
  $\alpha_{h1}$							& $1.0$		& $0.1$			& $2.0$			\\
  $c_{h1}$								& $5.0$		& $1.0$			& $10.0$		\\
  \bottomrule
  \end{tabular}
\end{table}

\begin{table}[!htb]
 \centering
 \caption{1PIEMD + 175(+1)dPIE$_c$ $(M_TL^{-1} \sim 0.2)$ model priors. Values in square brackets are not optimized.}
 \label{table:1PIEMD+cm}
 \begin{tabular}{cccc}
  \toprule
  Parameter          					& Value  	& low. limit 	& high. limit 	\\
  \midrule
  $x_{h1} (\arcsec)$ 					& $0.0$  	& $-50.0$    	& $+50.0$    	\\
  $y_{h1} (\arcsec)$ 					& $0.0$  	& $-50.0$    	& $+50.0$     	\\
  $\varepsilon_{h1}$ 					& $0.71$ 	& $0.01$     	& $0.90$      	\\
  $\theta_{h1} (\mathrm{deg})$			& $146.5$  	& $0.0$			& $180.0$	  	\\
  $r_{c,h1} (\mathrm{kpc})$			& $68.8$	& $1.0$			& $500.0$		\\
  $r_{s,h1} (\mathrm{kpc})$			& $[1000.0]$  \\
  $\sigma_{h1} (\mathrm{km\,s^{-1}})$	& $809.0$	& $100.0$		& $2000.0$		\\
  \midrule
  $r_{c,gal}(\mathrm{kpc})$			& $[0.15]$ \\
  $r_{s,gal}(\mathrm{kpc})$			&			& $50.0$		& $400.0$		\\
  $\sigma_{gal}(\mathrm{km\,s^{-1}})$	& 			& $100.0$		& $500.0$		\\
  $\zeta_{r_{s,gal}}$					& $[4.0]$	\\
  $\zeta_{\sigma_{gal}}$				& $[2.9]$	\\
  \bottomrule
  \end{tabular}
\end{table}

\begin{table}[!htb]
 \centering
 \caption{2PIEMD + 175(+1)dPIE$_c$ $(M_TL^{-1} \sim 0.2)$ and 2PIEMD + 175(+1)dPIE$_e$ $(M_TL^{-1} \sim 0.2)$ model priors. Values in square brackets are not optimized.}
 \label{table:2PIEMD+cm}
 \begin{tabular}{cccc}
  \toprule
  Parameter          					& Value  	& low. limit 	& high. limit 	\\
  \midrule
  $x_{h1} (\arcsec)$ 					& $0.0$  	& $-15.0$    	& $+15.0$    	\\
  $y_{h1} (\arcsec)$ 					& $0.0$  	& $-15.0$    	& $+15.0$     	\\
  $\varepsilon_{h1}$ 					& $0.71$ 	& $0.01$     	& $0.90$      	\\
  $\theta_{h1} (\mathrm{deg})$			& $146.5$  	& $0.0$			& $180.0$	  	\\
  $r_{c,h1} (\mathrm{kpc})$			& $68.8$	& $1.0$			& $500.0$		\\
  $r_{s,h1} (\mathrm{kpc})$			& $[1000.0]$  \\
  $\sigma_{h1} (\mathrm{km\,s^{-1}})$	& $809.0$	& $100.0$		& $2000.0$		\\
  $x_{h2} (\arcsec)$ 					& $20.3$  	& $+5.0$    	& $+35.0$    	\\
  $y_{h2} (\arcsec)$ 					& $-35.9$  	& $-50.0$    	& $-20.0$     	\\
  $\varepsilon_{h2}$ 					& $0.50$ 	& $0.01$     	& $0.90$      	\\
  $\theta_{h2} (\mathrm{deg})$			& $128.5$  	& $0.0$			& $180.0$	  	\\
  $r_{c,h2} (\mathrm{kpc})$			& $74.8$	& $1.0$			& $500.0$		\\
  $r_{s,h2} (\mathrm{kpc})$			& $[1000.0]$  \\
  $\sigma_{h2} (\mathrm{km\,s^{-1}})$	& $1019.0$	& $100.0$		& $2000.0$		\\
  \midrule
  $r_{c,gal}(\mathrm{kpc})$			& $[0.15]$ \\
  $r_{s,gal}(\mathrm{kpc})$			&			& $50.0$		& $400.0$		\\
  $\sigma_{gal}(\mathrm{km\,s^{-1}})$	& 			& $100.0$		& $500.0$		\\
  $\zeta_{r_{s,gal}}$					& $[4.0]$	\\
  $\zeta_{\sigma_{gal}}$				& $[2.9]$	\\
  \bottomrule
  \end{tabular}
\end{table}

\begin{table}[!htb]
 \centering
 \caption{2PIEMD + 175(+1)dPIE$_c$ $(M_TL^{-1} = v)$ and 2PIEMD + 175(+1)dPIE$_e$ $(M_TL^{-1} = v)$ model priors. Values in square brackets are not optimized.}
 \label{table:2PIEMD+cm+opt}
 \begin{tabular}{cccc}
  \toprule
  Parameter          					& Value  	& low. limit 	& high. limit 	\\
  \midrule
  $x_{h1} (\arcsec)$ 					& $0.0$  	& $-15.0$    	& $+15.0$    	\\
  $y_{h1} (\arcsec)$ 					& $0.0$  	& $-15.0$    	& $+15.0$     	\\
  $\varepsilon_{h1}$ 					& $0.71$ 	& $0.01$     	& $0.90$      	\\
  $\theta_{h1} (\mathrm{deg})$			& $146.5$  	& $0.0$			& $180.0$	  	\\
  $r_{c,h1} (\mathrm{kpc})$			& $68.8$	& $1.0$			& $500.0$		\\
  $r_{s,h1} (\mathrm{kpc})$			& $[1000.0]$  \\
  $\sigma_{h1} (\mathrm{km\,s^{-1}})$	& $809.0$	& $100.0$		& $2000.0$		\\
  $x_{h2} (\arcsec)$ 					& $20.3$  	& $+5.0$    	& $+35.0$    	\\
  $y_{h2} (\arcsec)$ 					& $-35.9$  	& $-50.0$    	& $-20.0$     	\\
  $\varepsilon_{h2}$ 					& $0.50$ 	& $0.01$     	& $0.90$      	\\
  $\theta_{h2} (\mathrm{deg})$ 		& $128.5$  	& $0.0$			& $180.0$	  	\\
  $r_{c,h2} (\mathrm{kpc})$			& $74.8$	& $1.0$			& $500.0$		\\
  $r_{s,h2} (\mathrm{kpc})$			& $[1000.0]$  \\
  $\sigma_{h2} (\mathrm{km\,s^{-1}})$	& $1019.0$	& $100.0$		& $2000.0$		\\
  \midrule
  $r_{c,gal}(\mathrm{kpc})$			& $[0.15]$ \\
  $r_{s,gal}(\mathrm{kpc})$			&			& $50.0$		& $400.0$		\\
  $\sigma_{gal}(\mathrm{km\,s^{-1}})$	& 			& $100.0$		& $500.0$		\\
  $\zeta_{r_{s,gal}}$					& 			& $0.0$			& $6.0$			\\
  $\zeta_{\sigma_{gal}}$				& 			& $0.0$			& $6.0$			\\
  \bottomrule
  \end{tabular}
\end{table}

\begin{table}[!htb]
 \centering
 \caption{2NFW + 175(+1)dPIE$_c$ $(M_TL^{-1}\sim L^{0.2})$ model priors. Values in square brackets are not optimized.}
 \label{table:NFW-Grillo}
 \begin{tabular}{cccc}
  \toprule
  Parameter          					& Value  	& low. limit 	& high. limit 	\\
  \midrule
  $x_{h1} (\arcsec)$ 					& $0.0$  	& $-15.0$    	& $+15.0$    	\\
  $y_{h1} (\arcsec)$ 					& $0.0$  	& $-15.0$    	& $+15.0$     	\\
  $\varepsilon_{h1}$ 					& $0.30$ 	& $0.01$     	& $0.40$      	\\
  $\theta_{h1} (\mathrm{deg})$			& $146.5$  	& $0.0$			& $180.0$	  	\\
  $r_{s,h1} (\mathrm{kpc})$			& $100.0$	& $1.0$			& $1000.0$		\\
  $\alpha_{h1}$							& $1.0$		& $0.1$			& $2.0$			\\
  $c_{200,h1}$							& $5.0$		& $1.0$			& $10.0$		\\
  $x_{h2} (\arcsec)$ 					& $20.3$  	& $+5.0$    	& $+35.0$    	\\
  $y_{h2} (\arcsec)$ 					& $-35.9$  	& $-50.0$    	& $-20.0$     	\\
  $\varepsilon_{h2}$ 					& $0.30$ 	& $0.01$     	& $0.40$      	\\
  $\theta_{h2} (\mathrm{deg})$			& $128.5$  	& $0.0$			& $180.0$	  	\\
  $r_{s,h2} (\mathrm{kpc})$			& $120.0$	& $1.0$			& $1000.0$  	\\
  $\alpha_{h1}$							& $1.0$		& $0.1$			& $2.0$			\\
  $c_{200,h1}$							& $5.0$		& $1.0$			& $10.0$		\\
  \midrule
  $r_{c,gal}(\mathrm{kpc})$			& $[0.15]$ \\
  $r_{s,gal}(\mathrm{kpc})$			&			& $50.0$		& $400.0$		\\
  $\sigma_{gal}(\mathrm{km\,s^{-1}})$	& 			& $100.0$		& $500.0$		\\
  $\zeta_{r_{s,gal}}$					& $[4.0]$	\\
  $\zeta_{\sigma_{gal}}$				& $[2.9]$	\\
  \bottomrule
  \end{tabular}
\end{table}

\begin{table}[!htb]
 \centering
 \caption{2NFW + 175(+1)dPIE$_c$ $(M_TL^{-1} = v)$ and 2NFW + 175(+1)dPIE$_e$ $(M_TL^{-1} = v)$ model priors. Values in square brackets are not optimized.}
 \label{table:NFW+cm}
 \begin{tabular}{cccc}
  \toprule
  Parameter          					& Value  	& low. limit 	& high. limit 	\\
  \midrule
  $x_{h1} (\arcsec)$ 					& $0.0$  	& $-15.0$    	& $+15.0$    	\\
  $y_{h1} (\arcsec)$ 					& $0.0$  	& $-15.0$    	& $+15.0$     	\\
  $\varepsilon_{h1}$ 					& $0.30$ 	& $0.01$     	& $0.40$      	\\
  $\theta_{h1} (\mathrm{deg})$			& $146.5$  	& $0.0$			& $180.0$	  	\\
  $r_{s,h1} (\mathrm{kpc})$			& $100.0$	& $1.0$			& $1000.0$		\\
  $\alpha_{h1}$							& $1.0$		& $0.1$			& $2.0$			\\
  $c_{200,h1}$							& $5.0$		& $1.0$			& $10.0$		\\
  $x_{h2} (\arcsec)$ 					& $20.3$  	& $+5.0$    	& $+35.0$    	\\
  $y_{h2} (\arcsec)$ 					& $-35.9$  	& $-50.0$    	& $-20.0$     	\\
  $\varepsilon_{h2}$ 					& $0.30$ 	& $0.01$     	& $0.40$      	\\
  $\theta_{h2} (\mathrm{deg})$			& $128.5$  	& $0.0$			& $180.0$	  	\\
  $r_{s,h2} (\mathrm{kpc})$			& $120.0$	& $1.0$			& $1000.0$  	\\
  $\alpha_{h2}$							& $1.0$		& $0.1$			& $2.0$			\\
  $c_{200,h2}$							& $5.0$		& $1.0$			& $10.0$		\\
  \midrule
  $r_{c,gal}(\mathrm{kpc})$			& $[0.15]$ \\
  $r_{s,gal}(\mathrm{kpc})$			&			& $50.0$		& $400.0$		\\
  $\sigma_{gal}(\mathrm{km\,s^{-1}})$	& 			& $100.0$		& $500.0$		\\
  $\zeta_{r_{s,gal}}$					& 			& $0.0$			& $6.0$			\\
  $\zeta_{\sigma_{gal}}$				& 			& $0.0$			& $6.0$			\\
  \bottomrule
  \end{tabular}
\end{table}